\documentclass[a4paper,10pt, oneside, twocolumn, balance]{smartthesis}
\usepackage{geometry}
\usepackage{times}
\usepackage{amsmath}
\usepackage{aas_macros}
\usepackage{amssymb}
\usepackage{multicol}
\usepackage{longtable,lscape}
\usepackage{longtable}
\usepackage{multirow}
\usepackage{rotating}
\usepackage{balance}
\usepackage{epsf}
\usepackage{color,graphicx}
\usepackage[small, labelfont=bf,up]{caption}
\usepackage[]{natbib}
\bibpunct{(}{)}{,}{a}{}{,}
\begin{document}

\frontmatter
\begin{titlepage}
{\centerline {\bf {\Huge Spectral and Timing Analysis of the}}
\vskip 0.2cm
{\centerline {\bf {\Huge Prompt Emission of Gamma Ray Bursts}}}}

\vskip 2cm

{\centerline {\Large A Thesis}}

\vskip 2cm

{\centerline {\Large Submitted to the}}
\vskip 0.2cm
{\centerline {\Large Tata Institute of Fundamental Research, Mumbai}}
\vskip 0.2cm
{\centerline {\Large for the degree of Doctor of Philosophy}}
\vskip 0.2cm
{\centerline {\Large in Physics}}
\vskip 1.5cm

{\centerline {\Large by }}
\vskip .7cm
{\centerline {\Large Rupal Basak}}
\vskip 2.5cm
{\centerline {\Large School of Natural Sciences}}
\vskip 0.2cm
{\centerline {\Large Tata Institute of Fundamental Research}}
\vskip 0.2cm
{\centerline {\Large Mumbai}}

\vskip 0.2cm

{\centerline {\Large Final Submission: Aug, 2014}}

\end{titlepage}


\begin{titlepage}
\begin{center}
\vspace*{6cm}
{\LARGE \bf \it{To my Parents}}
\end{center}
\end{titlepage}

%


\setcounter{page}{2}
\tableofcontents
\listoftables
\listoffigures
\chapter{List of Publication}
\section*{\LARGE Publications in Refereed Journals}
\vspace{0.2in}

\begin{large}

\begin{enumerate}

\item ``Time-resolved Spectral Study of Fermi GRBs Having Single
Pulses'', \textbf{Basak, R.} \& Rao, A. R. (2014), MNRAS, 442, 419

\item ``Time Resolved Spectral Analysis of the Prompt Emission of Long Gamma ray Bursts with GeV
Emission'', Rao, A. R., \textbf{Basak, R.}, Bhattacharya, J., Chanda, S., Maheshwari, N., Choudhury, M., Misra,
R. (2014), RAA, 14, 35

\item ``Pulse-wise Amati Correlation in Fermi Gamma-ray Bursts'', \textbf{Basak, R.} \& Rao, A. R. (2013a), MNRAS 436, 3082

\item ``A Lingering Non-thermal Component in the Gamma-ray Burst
Prompt Emission: Predicting GeV Emission from the MeV Emission'', \textbf{Basak, R.} \& Rao, A. R. (2013b), ApJ 775, 31
\item ``A New Method of Pulse-wise Spectral Analysis of Gamma Ray Bursts'', \textbf{Basak, R.} \& Rao, A. R. (2013c), ApJ 768, 187
\item ``Correlation Between the Isotropic Energy and the Peak Energy at
Zero Fluence for the Individual Pulses of Gamma-Ray Bursts: Toward a Universal Physical
Correlation for the Prompt Emission'', \textbf{Basak, R.} \& Rao, A. R. (2012a), ApJ 749, 132
\item ``Measuring the Pulse of GRB 090618: A Simultaneous Spectral and
Timing Analysis of the Prompt Emission'', \textbf{Basak, R.} \& Rao, A. R. (2012b), ApJ 745, 76
\end{enumerate}
\vspace{0.3in}
 
\end{large}

\section*{\LARGE Publications in Conference Proceedings}
\vspace{0.2in}

\begin{large}
\begin{enumerate}
\item ``GRB as luminosity indicator'', \textbf{Basak, R.} \& Rao, A. R. (2014), Proceedings of IAU Vol 9, symposium No. 296, pp 356-357
\item ``A new pulse-wise correlation of GRB prompt emission: a possible cosmological probe'', \textbf{Basak, R.} \& Rao, A. R. (2013b),
39th COSPAR Scientific Assembly, Mysore, India, 2012cosp, 39, 106.
\item ``Pulse spectral evolution of GRBs: implication as standard candle'', \textbf{Basak, R.} \& Rao, A. R. (2013b)
Gamma-Ray Bursts 2012, Munich, Germany, PoS (GRB 2012) [081]
\end{enumerate}
\end{large}
\mainmatter
\chapter{GRBs: The Extreme Transients} \label{ch1}

``... probably hotter, more violent, but what are they? .. We are aware of something we call 
a hypernova ... we got supernova. Bigger, better --- hypernova ...
these flashes are the brightest things in the gamma-ray sky ...''
\begin{flushright}
--- Prof. Jocelyn Bell Burnell\\ (``Star glitter - the story of gold'', Public lecture, TIFR, January 16, 2014)
\end{flushright}
\section{Overview}
Gamma-ray Bursts (GRBs) are fascinating astrophysical objects in many aspects. They are believed to 
be catastrophic events marking the formation of compact objects, most probably stellar mass 
blackholes (BHs). A class of GRBs are associated with the explosive death of a very special kind 
of massive star (``collapsar''; \citealt{Woosley_1993}, or ``hypernova''; \citealt{Iwamoto1998Natur}), 
while another class is suggested to occur via merging of two compact objects, such as a binary 
neutron star (NS), or a NS and a BH. Frequently attributed with superlatives, GRBs are truly the 
most extreme transient phenomenon: 
\begin{itemize}
 \item (i) They are the most efficient astrophysical power house
known to mankind (typical luminosity, $L_{\rm GRB} \sim 10^{52}~ \rm erg~s^{-1}$). 
Their luminosity is many times higher than supernovae, which release the same amount of energy over a much
longer period. 
\item (ii) More remarkably, most of this energy is released during a 
very brief episode (a few milliseconds to hundreds of seconds in observer frame), 
termed as the \textit{prompt emission} phase. During this brief period a GRB radiates 
mostly in the form of $\gamma$-rays --- a few keV to tens of MeV, and its intensity
outshines all other $\gamma$-ray objects combined. 
The burst proper is followed by a longer lasting \textit{afterglow} phase
(observed over a few tens of days to months) in longer wavelengths ranging from x-rays to optical and radio. 
Even during the first day of x-ray and optical afterglow a GRB is about ten thousand times brighter than
the brightest quasars, which in turn are hundred to thousand times brighter than their underlying galaxies.
\item (iii) Due to the high 
luminosity a GRB is visible over a very large distance, corresponding to a very early epoch in cosmic history. 
The highest known redshift ($z$) 
is 9.4 (\citealt{Cucchiara9_4}) which corresponds to only $\sim$ 5\% of the present age of the universe. 
For this reason, GRBs are suggested as the best possible high-$z$ luminosity indicators.
\item (iv) GRBs achieve 
their high luminosity by means of relativistic bulk motion with a \textit{Lorentz} factor ($\Gamma$) reaching $\sim 100-1000$. 
The second most relativistic objects are BL Lacertae with $\Gamma \sim 10-20$, maximally $\Gamma \sim 50$ (\citealt{bllac}).

\end{itemize}

There are two broad aspects of GRB research --- a. understanding the event itself, and b. using GRBs as tools
e.g., studying cosmic star formation history, and using GRBs as luminosity indicators at high $z$.
The prompt and afterglow phase are the most important observables for understanding the physics of GRBs, 
while $z$ measurement, chemical study of the burst environment etc. are essential for using GRBs as tools.
During the prompt emission, a GRB generally has a rapid time variability, while the afterglow 
has a smooth time profile (see Section 1.4 for details).
It is generally believed that the prompt emission of a GRB  has an ``internal'' origin, and the time
variability directly reflects the activity of the central object.
The afterglow phase is more or less related to the ``external'' circumburst medium (Section 1.5). The afterglow phase 
is well studied and the data generally shows excellent agreement with a standard model, known as the ``fireball shock'' model 
(e.g., \citealt{meszaros_rees_1997_afterglow, Reichart_1997_afterglow,Waxman_afterglow, 
Vietri1997_afterglow, Tavani1997_afterglow, Wijers1997_afterglow}). 
It is the prompt emission phase which remains a puzzle. There is no scarcity in the 
number of working models (e.g., \citealt{Meszarosetal_1994_prompt, Rees_and_meszaros_1994_prompt,
Thompson_1994_prompt, Daigne_Mochkovitch_1998_prompt, Pilla_Loeb_1998_prompt, 
Medvedev_Loeb_1999_prompt, Piran_1999_review, Lloyd_Petrosian_2000_prompt, 
Ghisellinietal_2000_prompt, Panaitescu_Meszaros_2000ApJ_prompt, Spruitetal_2001, 
Zhang_Meszaros_2002_prompt, Pe'er_Waxman_2004_prompt, Ryde_2004, Ryde_2005, 
rees_and_meszaros_2005_prompt, Pe'eretal_2005, Pe'eretal_2006}),
but the unavoidable poor spectral quality of the $\gamma$-ray detectors challenges
the correct identification of the fundamental model. As GRBs are very brief, single episode events coming from 
unpredictable directions of the sky, the detectors must have large field of view (and in many cases all-sky)
to detect them. This severely affects the spectral quality as well as source localization. With the advent of 
\textit{Swift} (\citealt{Gehrelsetal_2004_swift}) and \textit{Fermi} (\citealt{Meeganetal_2009_fermi_gbm})
satellites, launched in 2004 and 2008 respectively, GRB research has entered a new era. The \textit{Swift} has enabled
many order better and quicker localization leading to $z$ measurement. The \textit{Fermi} acquires data in a wide
band, with good time resolution. With the current good quality data and large set of GRBs with
known $z$, it is high time for extensive study 
of the prompt emission, identification of the underlying physics, and study GRBs as luminosity indicators. The aims of this thesis are 
(i) developing the most judicial way(s) for using the valuable data to describe the prompt emission, (ii) using prompt emission 
properties in favour of GRBs as luminosity indicators, (iii) developing a method to compare the spectral models
of the prompt emission, and (iv) predicting interesting behaviours of the prompt and early afterglow phase. 

\section{Thesis Organization}
The organization of the thesis is as follows. In this chapter, we shall briefly discuss the history and classification scheme, 
various observables e.g., lightcurve and spectrum, and the working model of GRBs. The next chapter (chapter 2) deals with 
the current instruments used for GRB prompt emission analysis. We shall focus on the \textit{Swift} and the \textit{Fermi}, the 
two main workhorses of modern GRB research. We shall point out the essential features of the satellites and detectors
which make them superior compared to the other GRB experiments. The data analysis technique of these detectors will 
be described, and the issues with the joint \textit{Swift-Fermi} fitting will be discussed. The prompt emission data 
provided by these satellites have very good spectral and timing resolution. However, it is important to use these 
information simultaneously in order to fully utilize the data. Hence, we need a new technique that judiciously 
describes the flux of a GRB as a simultaneous function of time and energy. The third chapter is entirely devoted for 
the new technique, which is developed using the existing empirical models in the literature. We shall see its 
versatile applications e.g., in deriving certain properties of GRBs, and studying GRBs as luminosity indicators. 
For these analyses, we shall choose all \textit{Fermi} GRBs with known redshift (a total of 19 after sample selection).
It is worthwhile to mention that though this model is very promising, 
the fact that it is based on empirical functions puts a limitation on its 
applicability. Hence, in the next chapters (fourth, and fifth), we shall discuss various alternative prompt 
emission models. These alternative models are applied on 5 bright GRBs, and 9 GRBs with single pulses.
In order to test the merits of these different models, a new technique is developed. The analyses show that 
one of the alternative models is indeed a better description of the prompt emission. In Chapter 6, we will see 
some important predictive powers of this model. We shall choose 17 GRBs with very high energy photons to predict 
the high energy features. In chapter 7, we shall summarize the results, draw conclusions and describe the future 
extension of the work presented here.

\section{History And Classification}
\subsection{Discovery, Afterglow and Distance Scale}
The history of GRB research is full of observational and intellectual struggle, development of new techniques, exciting 
turnovers, and outstanding discoveries. Possibly the most important among these is the unambiguous discovery 
of the distance scale, which alone took nearly 30 years. In this section, we shall mention some brief 
historical facts, and refer the reader to the book by \cite{Katz_2002_book} for this exciting story. We shall also 
briefly discuss the classification scheme which is important in order to understand the progenitor of GRBs. 

The GRB research began with the serendipitous discovery by \textit{Vela} satellites on July 2, 1967 
(\citealt{Klebesadeletal_1973}; the burst is named as GRB~670702 following YYMMDD format). These satellites were launched 
by the U.S. Department of Defence to monitor nuclear explosions forbidden by the Nuclear Test Ban Treaty. With 
widely separated four independent satellites, \textit{Vela} team discovered 16 bursts during 1969-1972, with duration 
of less than 0.1 s to $\sim 30$s,
and time-integrated flux $\approx 10^{-5}$-$\sim 2 \times 10^{-4}$ erg cm$^{-2}$ in 0.2-1.5 MeV band 
(\citealt{Klebesadeletal_1973}). From the pulse arrival delay in different satellites, the approximate direction could be found, 
which excluded earth and sun as the possible source. As for distance of the sources, only lower limits (several earth-moon
distance) could be put from the delay analysis. 

For the next 25 years, several bursts were detected without any clue of their distance. The main obstacles for 
the distance measurement were poor localization of the $\gamma$-ray detectors (a few degree radius),
and too brief a duration to look for the signature in other wavelengths. For a long time the distance remained highly debated 
even to the extent of whether the sources are Galactic or extra-galactic. By this time, \textit{Inter-Planetary Network} 
(IPN; e.g., \citealt{Cline_Desai_1976_IPN}) provided a few to hundreds of sq. arcmin localization using several widely spaced 
spacecraft, but unfortunately with a considerable delay (days to months). Hence, no counterpart could be found. 

The failure of direct localization triggered the use of statistical methods to infer the distance. If a large set of GRBs can be 
detected with a few degree of position accuracy, it is good enough to put the sources in the galactic coordinate.
If the sources are extra-galactic, they should have an isotropic distribution. Various statistical tests are
available to test the anisotropy e.g., dipole and quadrupole of the distribution (\citealt{Hartmann_Epstein_1989};
\citealt{Briggs_1995}). For perfect isotropic distribution and isotropic sampling, $\langle$cos $\theta \rangle$=0 
and $\langle$sin$^2 \beta \rangle$=1/3,
where $\theta$ is the angle between the direction of the burst and the Galactic centre, and $\beta$
is the Galactic latitude. \cite{Hartmann_Epstein_1989}, using 88 IPN GRBs, showed that the bursts are isotropic
within the statistical limits (also see \citealt{Mazets_Golenetskii_1981}). Another information comes from the tests of the uniformity of 
space distribution of the bursts e.g., log$N$/log$S$ test (\citealt{Usov_Chibisov_1975, Fishman_1979}), $V/V_{\rm max}$ test
(\citealt{Schmidtetal_1988}). Here, $N$ is the cumulative number of bursts with flux greater than $S$, and $V$ is the volume contained 
within the burst's radial distance. The first test would give a constant slope of -3/2 if the bursts have uniform 
distribution on the Euclidean space. A different slope is expected for fainter bursts if the sources have cosmological distance.
The $V/V_{\rm max}$ test takes the ratio of the actual (unknown) volume and the maximum allowed volume in which the 
burst could have been detected. In doing this it cancels the actual distance and depends only on the ratio of 
the peak ($C_{\rm p}$) and limiting ($C_{\rm lim}$) flux: $V/V_{\rm max}=(C_{\rm p}/C_{\rm lim})^{-3/2}$.
For a uniform space distribution $\langle V/V_{\rm max}\rangle=1/2$. 
The $V/V_{\rm max}$ test is preferred over the log$N$/log$S$ test as it is independent of instrumental sensitivity.
For a limited number of bursts from various experiments, deviations were reported from $\langle V/V_{\rm max}\rangle=1/2$
(\citealt{Ogasakaetal1991}: $0.35\pm0.035$; \citealt{Higdonetal1992}: $0.400\pm0.025$). 
A Galactic model could not reconcile the apparent isotropy and the inhomogeneity of the source distribution. 
However, a large group of researchers generally disbelieved the inferences drawn from a limited number of sources.

The sample size was not the only reason to generally disbelieve the extra-galactic origin, and stick to the Galactic model.
A few attempts to calculate the prompt emission characteristics assuming a cosmological distance could not 
match observations (see Section 1.5). Moreover,
a few earlier discoveries during late 1970's and early 1980's were already pointing towards a nearby origin.
By this time, several models of GRBs were proposed (more than 100; \citealt{Ruderman_1975}; see
\citealt{Nemiroff_1994} for a later review), some of which were related to NSs. 
The Galactic NS model had strong observational evidences --- 
(i) The burst of March 5, 1979 (\citealt{Mazetsetal_1979}) could be associated with a supernova remnant (SNR) 
of the nearby Large Magellanic Cloud (LMC). This burst had a steep spectrum, and a lightcurve with
a strong initial $\sim120$ millisecond pulse followed by very soft pulses. Though the spectrum and 
lightcurve of this burst was quite unusual for a GRB, it was generally attributed 
towards the diversity of GRBs. However, due to the detection of 16 more burst from 
the same source, it was later classified as a soft gamma repeater (SGR) coming from a 
highly magnetized neutron star (magnetar; e.g., \citealt{Kouveliotouetal_1987, Larosetal_1987}). 
Several other such sources were found later (\citealt{Kouveliotouetal_1992, Paczynski_1992}; see \citealt{Harding_2001} for a review). 
SNRs are known to harbour NSs, and the comparatively 
nearby distance of the burst made the Galactic NS origin plausible. 
(ii) In a few cases, cyclotron lines were reported (e.g., \citealt{Murakamietal_1988, Fenimoreetal_1988}; 
also see \citealt{Mazetsetal_1980}), 
which corresponded to $\sim $ a few $\times 10^{12}$ G, typical for a NS. 
(iii) Finally, in order to account for the isotropy, the ``nearby origin'' could be pushed to the extended Galactic halo.
In fact, extended halo origin was strongly supported by the discovery of high transverse velocity of neutron stars 
(NSs) which could populate the extended halo (\citealt{Frailetal_1994}; cf. \citealt{Bloometal_1999}). If GRBs are indeed related 
to NSs, their isotropic distribution supports both the extended halo origin and the NS progenitor.
The cosmological origin was supported only by \cite{Usov_Chibisov_1975}, 
and later by \cite{Goodman_1986}, and \cite{Paczynski_1986}, based on the isotropic source distribution.

In the year 1991, NASA launched \textit{Compton Gamma Ray Observatory} (CGRO), which along with three other instruments, 
carried specialized GRB instrument --- \textit{Burst And Transient Source Experiment} (BATSE; \citealt{Fishman_Meegan_1995}).
The BATSE was designed to detect as many burst as possible, and thereby to rule out statistical bias from the inferences.
It contained 8 NaI (Tl) scintillation detector modules in different directions. Each module consisted of one spectroscopic detector 
(SD), and one large area detector (LAD). The SD was sensitive in 20 keV-10 MeV band (with maximum effective area 126 cm$^2$), 
while the LAD gave a very high effective area (maximum 2025 cm$^2$) in a narrower band (20 keV-2 MeV). All the
detectors were surrounded by plastic scintillators in active anti-coincidence to reduce the cosmic ray background. 
The BATSE had essentially an all-sky viewing so that the GRBs occurring at unpredictable directions could be 
located with some coarse position accuracy. The burst localization
was done by comparing relative flux in the modules, and lay in the range 4$^{\circ}$-10$^{\circ}$. Though this is not an impressive
accuracy, the success of the BATSE lies in the huge number of bursts detected over its lifetime (1991-2000). The final catalogue
contained over 2500 bursts. For the first 1005 BATSE GRBs $\langle$cos $\theta \rangle =0.017 \pm 0.018$, 
and $\langle$sin$^2 \beta -1/3 \rangle=-0.003 \pm 0.009$, which are respectively only $0.9\sigma$ and
$0.3\sigma$ away from perfect isotropy (\citealt{Briggs_1995}). For 520 bursts $\langle V/V_{\rm max}\rangle=0.32 \pm 0.01$
(\citealt{Meeganetal_1994}) which confirmed the inhomogeneity. With the BATSE results the evidence of cosmological distance 
became stronger. However, the extended Galactic halo origin remained an option (\citealt{Li_Dermer_1992}). 
The lower limit on the size of the halo ($> 20$ kpc) can be obtained by the requirement of the observed isotropy. 
With strong evidences in both sides, the famous ``great debate'' 
(\citealt{Lamb_1995_great_debate, Paczynski_1995}) on whether the sources have extended Galactic 
halo or a cosmological origin, led to no final consensus.

The direct distance measurement was crucial, but it seemed difficult solely from $\gamma$-ray
observation, unless by some lucky chance like the burst of March 5, 1979 (which was found in a SNR). The breakthrough
came with the launch of Italian-Dutch satellite --- \textit{Beppo-SAX} (\citealt{Boellaetal_1997_BeppoSAX}). It
contained 2 \textit{Wide Field Camera} (WFC), several \textit{Narrow Field Instruments} (NFIs), and
4 GRB monitors (GRBMs). The detector modules of the WFCs were position sensitive proportional counters (bandwidth: 2-30 keV) 
with a similar size of coded aperture mask (CAM) that provided a very good angular resolution (5 arcmin) and source 
localization (1 arcmin). With a field of view of $20^{\circ} \times 20^{\circ}$ (much lower compared to all-sky BATSE), it could detect 
reasonable number of bursts, essentially with a much better accuracy than the BATSE (a few degree radius).
There were several type of NFIs (including focusing x-ray instruments) having narrow field of view, and very good 
angular resolution (less than 100 arcsec). The GRBMs were open detectors operating in 40-700 keV. A burst detected
both in a GRBM and a WFC could be localized accurately enough to be seen by the NFIs after some delay. The \textit{Beppo-SAX}
succeeded because of the better localization capability of the WFCs, and quicker implementation of 
the high resolution x-ray instruments (NFIs) within hours. This relayed improved position could be used by ground based telescopes to 
observe the burst in optical wavelengths with a much shorter observational delay, an opportunity never provided by 
the previous satellites. On Feb 28, 1997, a burst (GRB 970228) localized by the 
WFC (within 3 arcmin), could be observed in the NFIs (after $\sim8$ hours delay) as a fading x-ray source (\citealt{Costaetal_1997}),
within 50 arcsec error circle. This position accuracy was enough to observe a fading optical source with 4.2-m 
\textit{William Herschel Telescope} (WHT), about 21 hours after the burst, at 23.7 $V$-band and 
21.4 $I$-band magnitude (\citealt{Paradijsetal_1997}). Later observation
using \textit{Hubble Space Telescope} (HST) and 10-m \textit{Keck} telescope revealed a galaxy within the error
circle, with a spectroscopic redshift of $z=0.695$. Of course, inferring the $z$ of the GRB from the galaxy
association could be doubtful as the space coincidence might be a projection effect. However, the detection of 
an absorption spectrum for the next burst (GRB 970508) eliminated this doubt. A direct measurement 
required at least a redshift of $z=0.835$ (\citealt{Metzgeretal_1997}) for this GRB. Later observation of GRBs
with good localization always revealed underlying host galaxies, with some exceptions for
another class of GRBs (see below).

\subsection{Classification of GRBs}
With the direct redshift measurement, there remained no doubt about the cosmological origin of GRBs. With the observed high 
flux and cosmological distance, a typical burst releases $\sim 10^{53}-10^{54}$ erg energy (assuming an isotropic explosion).
The ``central engine'', which liberates this prodigious energy, remains hidden from a direct observation.
However, from the requirement of a variable temporal structure during the prompt emission (see Section 1.5), the inner 
engine is suggested to be a compact object, most likely a blackhole (BH).
In order to produce this energy, a BH requires to accrete $\sim 0.01-0.1 M_{\odot}$ (solar mass), and convert it to pure energy.
There are two popular models to form the central engine --- (i) collapse of a rapidly rotating, massive \textit{Wolf-Rayet} star,
named collasper model (e.g., \citealt{Woosley_1993, MacFadyen_Woosley_1999}), and (ii) coalescence of two NSs, or a NS and a BH
(\citealt{Eichleretal_1989, Narayanetal_1992}). Hence, there are possibly two classes of GRBs. The first phenomenological 
indication of these two classes came from duration-hardness distribution of the bursts. The duration of a GRB is defined
as the time span to accumulate 5\% to 95\% of the total $\gamma$-ray fluence --- $T_{90}$ (cf. \citealt{Koshutetal_1996}). 
\cite{Kouveliotouetal_1993}
have found that the $T_{90}$ distribution has a bimodal structure. GRBs with $T_{90}>2$ s are called long GRBs (LGRBs) and those 
with $T_{90}<2$ s are called short GRBs (SGRBs). Note that the demarcation of 2 s is chosen as phenomenological tool.
In addition to the difference in the duration, the LGRBs are found to have softer spectrum than the SGRBs 
(\citealt{Golenetskiietal_1983, Fishmanetal_1994, Mallozzietal_1995, Dezalayetal_1997, Belli_1999, Fishman_1999, Qinetal_2000, 
Ghirlandaetal_2004, Cuietal_2005, Qin_Dong_2005, Shahmoradi_2013}). 
However, the difference of the prompt emission properties of the two classes are sometimes debated. For example,
the temporal and spectral shapes of a SGRB are broadly similar to those of the first 2 s 
of a LGRB (\citealt{Nakar_Piran_2002, Ghirlandaetal_2004}). 
The spectrum of a LGRB in the first 2 s is as hard as a SGRB. \cite{Liangetal_2002}, however, 
have found some differences, e.g., much shorter variability in SGRBs. Also, LGRBs have higher spectral 
lag (arrival delay between the hard and soft band) than SGRBs (\citealt{Yietal_2006, Norris_Bonnell_2006, Gehrelsetal_2006}).
In Table~\ref{t1}, we have summarized the main differences between these two classes. 

Apart from the differences in the prompt emission properties, various other evidences 
also point towards different population 
of these two classes. These are: (i) supernova connection with only the LGRB class,
and (ii) difference between the host galaxy, and location of the burst in the host.
A classification based on these properties provides important clues for the different progenitors.
The current belief is that LGRBs occur due to the collapse of massive stars, and SGRBs are probably the outcome of mergers.
It is worthwhile to mention that the detection of afterglow and host galaxy 
of SGRBs proved even more difficult than LGRBs. If SGRBs are indeed produced by 
merging NSs, which preferentially reside in the outskirts of the host galaxies 
(low density medium), the afterglow is expected to be dimmer than LGRBs, at least 
by an order of magnitude (\citealt{Panaitescuetal_2001, Perna_Belczynski_2002}). The afterglow and 
host identification of SGRBs is made possible by the \textit{Swift} (\citealt{Gehrelsetal_2004_swift}) 
and the HETE-2 (\citealt{Rickeretal_2003}). The general information from these extensive studies 
are: SGRBs are also cosmological events, and they produce dimmer afterglow than LGRBs 
(see \citealt{Berger_2013} for a review).

\begin{table*}\centering
\caption{Classification of GRBs}

\begin{tabular}{c|c|c}
\hline
\hline
& Long GRBs (Type II) & Short GRBs (Type I)\\
\hline
\hline
& Long duration($T_{90}>2$ s) & Short duration ($T_{90}<2$ s)\\
\cline{2-3}
Prompt & Soft spectrum & Hard spectrum \\
\cline{2-3}
properties & High spectral lag & Low spectral lag\\
\cline{2-3}
& Lower variability & Higher variability\\
\hline
Afterglow & Brighter afterglow & Fainter afterglow\\
\hline
& Associated with   & No supernova association \\
&   Type Ic-BL supernova  & One association of `kilonova'\\
\cline{2-3}
Other & Star forming, low metallicity & All types of host\\
Clues & irregular host galaxy        & including ellipticals\\
\cline{2-3}
& Burst location: very near & Burst location: away from star\\
& to the star forming region &  forming region (sometimes in halo)\\
\hline
Progenitor & Massive \textit{Wolf-Rayet} star collapse & Compact object merger\\
\hline
\end{tabular}
\label{t1}
\end{table*}

\subsubsection{A. The Supernova Connection}
The most direct evidence that at least some GRBs are associated with the collapse of massive stars was 
provided by the watershed discovery of SN 1998bw in space and time coincidence with LGRB 980425 at $z=0.0085$
(\citealt{Galamaetal_1998, Kulkarnietal_1998}). SN 1998bw was a broad-lined Type Ic SN (Ic-BL), with a very 
fast photospheric expansion ($\sim 30000$ km s$^{-1}$) and unusually high isotropic energy ($5\pm0.5 \times 10^{52}$ 
erg, about 10 times higher than a typical core collapse SN). This highly energetic SN was termed as 
a ``hypernova'' (\citealt{Iwamoto1998Natur, Paczynski_1998}). However, the associated GRB had a much 
lower energy (isotropic energy, $E_{\gamma, \rm iso} \sim 10^{48}$ erg) than a typical GRB, and
the proximity of the event raised a doubt --- was it a cosmological GRB at all? 
Moreover, the identification of the x-ray counterpart of the GRB was controversial (\citealt{Pianetal_2000}).
After about five years, the discovery of another LGRB (031329) in association with SN 2003dh at $z=0.1685$ cleared any doubt 
about the association (\citealt{Staneketal_2003, Hjorthetal_2003}). This event had similar SN properties as the SN 1998bw, and
had a typical bright GRB ($E_{\gamma, \rm iso} \sim 1.3 \times 10^{52}$ erg). The spectrum of both
the SNe had remarkable similarity, proving that the 1998 event was in fact a real SN-GRB (\citealt{Hjorthetal_2003}).
Subsequently, a few more such events were found with strong spectroscopic evidences. These are: GRB 031203/SN 2003lw at $z=0.105$ 
(\citealt{Malesanietal_2004}), GRB 060218/SN 2006aj at $z=0.0331$ (\citealt{Modjazetal_2006, Pianetal_2006,
Sollermanetal_2006, Mirabaletal_2006, Cobbetal_2006}), GRB 100316D/SN 2010bh at $z=0.0.0591$
(\citealt{Chornocketal_2010}), GRB 120422A/SN 2012bz at $z=0.28$ (\citealt{Wiersemaetal_2012, Malesanietal_2012}).
It is worth mentioning that the SN observation in association with LGRBs have some observational challenges e.g., 
unfavourable observation condition of sky region, the amount of dust along the line of sight, redshift of the event, and
the luminosity of the underlying host galaxy (\citealt{Woosley_Bloom_2006}). Hence, the number of events with secured 
spectroscopy is only handful, and all lie below $z=0.28$. Other than the secured spectroscopy, 
some SNe are detected (up to $z= 1.058$, till date) as optical bumps in the afterglow lightcurve,
and a few with some spectroscopic evidences (see \citealt{Hjorth_Bloom_2012} for a list of all events).
In all cases found so far, the associated GRB is either a LGRB, or a x-ray flash (XRF), a softer 
version of a GRB. With these definite associations, the collapsar model of LGRBs seems reasonable.
As the spectrum of the associated SN have no H and He (a feature of Ib/c), the progenitor is most likely 
a massive \textit{Wolf-Rayet} star (\citealt{Smith_Owocki_2006}). 
It is interesting to note that not all Type Ib/c SNe produce LGRBs. In fact, the average estimated
rate of cosmological LGRBs ($\sim 1.1~\rm Gpc^{-3}~yr^{-1}$, \citealt{Guetta_Della_Valle_2007}) is much lower than 
the SN Ib/c rate ($2.58_{-0.42}^{+0.44}\times 10^4~\rm Gpc^{-3}~yr^{-1}$; \citealt{Lietala_2011}). 
If GRBs are jetted events then the event rate may increase. However, radio observations of a few 
SN Ic revealed no signature of off-axis jet (\citealt{Soderbergetal_2006}).
In a comparative study, \cite{Modjazetal_2008} have found a clear difference of metallicity 
between the hosts of SN Ic with and without GRB. Hosts of SN Ic with GRB preferentially occur in low metallicity 
environments, which can be one of the reasons for the lower event rate.

On the other hand, deep search of SN in the optical afterglow of SGRBs with low $z$ have found no connection down to 
at least 4 mag lower peak flux (\citealt{Hjorthetal_2005ApJ, Hjorthetal_2005Nat, Foxetal_2005, Bloometal_2006}).
A recent discovery of a ``kilonova'' (\citealt{Li_Paczynski_1998, Metzgeretal_2010MNRAS, Barnes_Kasen_2013})
associated with a SGRB (130603B) is advocated as the ``smoking gun'' signature of the merging of 
compact objects scenario (\citealt{Bergeretal_2013, Tanviretal_2013}). A kilonova is a near-infrared (IR)
transient powered by r-process radioactive elements which are believed to be produced in the  
neutron-rich environment of a merger. The remarkable agreement of the observation with the 
predicted band (near-IR), time scale ($\sim$ 1 week) as well as the flux ($M_J \sim -15$ mag) 
makes this a strong case in favour of merging model of SGRBs.
 
\subsubsection{B. Differences In The Host Galaxy}
Apart from the supernova connection, the extensive studies of the host galaxies of these two classes 
provide further clues for their origin. This has become possible because
of the afterglow observations facilitated by the \textit{Swift} satellite. 
LGRBs are always found in blue, sub-luminous, irregular, low metallicity dwarf galaxies, which are undergoing active star formation
(\citealt{Paczynski_1998ApJ, Hogg_Fruchter_1999, Bergeretal_2003, Leetal_2003, Christensen_2004, Fruchteretal_2006,
Staneketal_2006, Savaglioetal_2006}). On the other hand, SGRBs are found in all types of galaxies including elliptical 
galaxies (\citealt{Bergeretal_2005, Foxetal_2005, Gehrelsetal_2005, Bloometal_2006}), which are much older systems. 
A significant fraction of SGRBs are found in star forming galaxies. But, we know that a large fraction of 
type Ia SNe are also found in star forming spiral galaxies. Hence, one cannot expect
that SGRBs should be exclusively populated in old systems (\citealt{Prochaskaetal_2006}).
In fact, the star-forming hosts of SGRBs have different properties than the LGRB hosts (\citealt{Fongetal_2010}).
Important differences between the two classes can be drawn by comparing the host luminosity, star formation rate, 
metallicity, and age. \cite{Berger_2009} have shown the following distinctions.
In the same redshift range ($z\lesssim 1.1$), (i) the LGRB hosts are sub-luminous 
with a median value of $M_B \sim 1.1$ mag fainter than the SGRB hosts. (ii) The star formation 
rate (SFR) of the LGRB hosts $\approx 0.2 - 50 M_{\odot} \rm yr^{-1}$, with specific SFR (SSFR=SFR/L) 
$\approx 3 - 40 M_{\odot} \rm yr^{-1} L_{*}^{-1}$ and a median SSFR $ 10 M_{\odot} \rm yr^{-1} L_{*}^{-1}$.
The SFR of SGRB hosts $\approx 0.2 - 6 M_{\odot} \rm yr^{-1}$, with a median value of SSFR one order lower 
than the LGRB hosts. (iii) The metallicity of the SGRB hosts (12+log(O/H) $\approx$ 8.5-8.9; $Z \approx 0.6-1.6 Z_{\odot}$)
are higher than LGRB hosts by $\sim 0.6$ dex. (iv) \cite{Leibler_Berger_2010}, using single stellar population models, have shown that the 
median SGRB host mass ($\langle {\rm log~} (M_{*}/M_{\odot})\rangle \approx 10.1$) and age ($\langle \tau_{*}\rangle \approx$ 0.3 Gyr)
are higher than those of the LGRB hosts ($\langle {\rm log~}(M_{*}/M_{\odot})\rangle \approx 9.1$, and
$\langle \tau_{*}\rangle \approx$ 0.06 Gyr). All these evidences point towards massive star 
progenitor for LGRBs, while merging scenario is more reasonable for SGRBs.

The location of a GRB in its host is also important for the classification scheme. LGRBs are always 
found near the star forming location (projected) of the host (\citealt{Bloometal_2002, Fruchteretal_2006}). 
Some of the SGRBs are found in the outskirts of their elliptical hosts (\citealt{Gehrelsetal_2005, Barthelmyetal_2005,
Bergeretal_2005, Bloometal_2006}). Though some SGRBs are found in star forming galaxies, their 
locations have large physical offsets (\citealt{Foxetal_2005, Soderbergetal_2006_sgrb}). NSs generally receive
large ``natal kicks'' (\citealt{Bloometal_1999}), which is consistent with the SGRB locations. Recently, \cite{Fongetal_2010}
have suggested that the host-normalized offsets (which are advocated to be better measurements of the 
host-GRB distances than the physical offsets) of SGRBs are similar to those of LGRBs, owing to the larger size of SGRB hosts.
However, they also suggest that the median of the offsets of SGRBs ($\sim$ 5 kpc) is consistent with the binary NS distribution.
Moreover, analysis of the brightness distribution of the two classes of GRBs show opposite behaviour. While the
concentration of LGRBs is biased towards the brightest location, SGRBs under-represent the host light 
distribution (\citealt{Fongetal_2010}).

\begin{figure*}\centering
\includegraphics[width=5.0in]{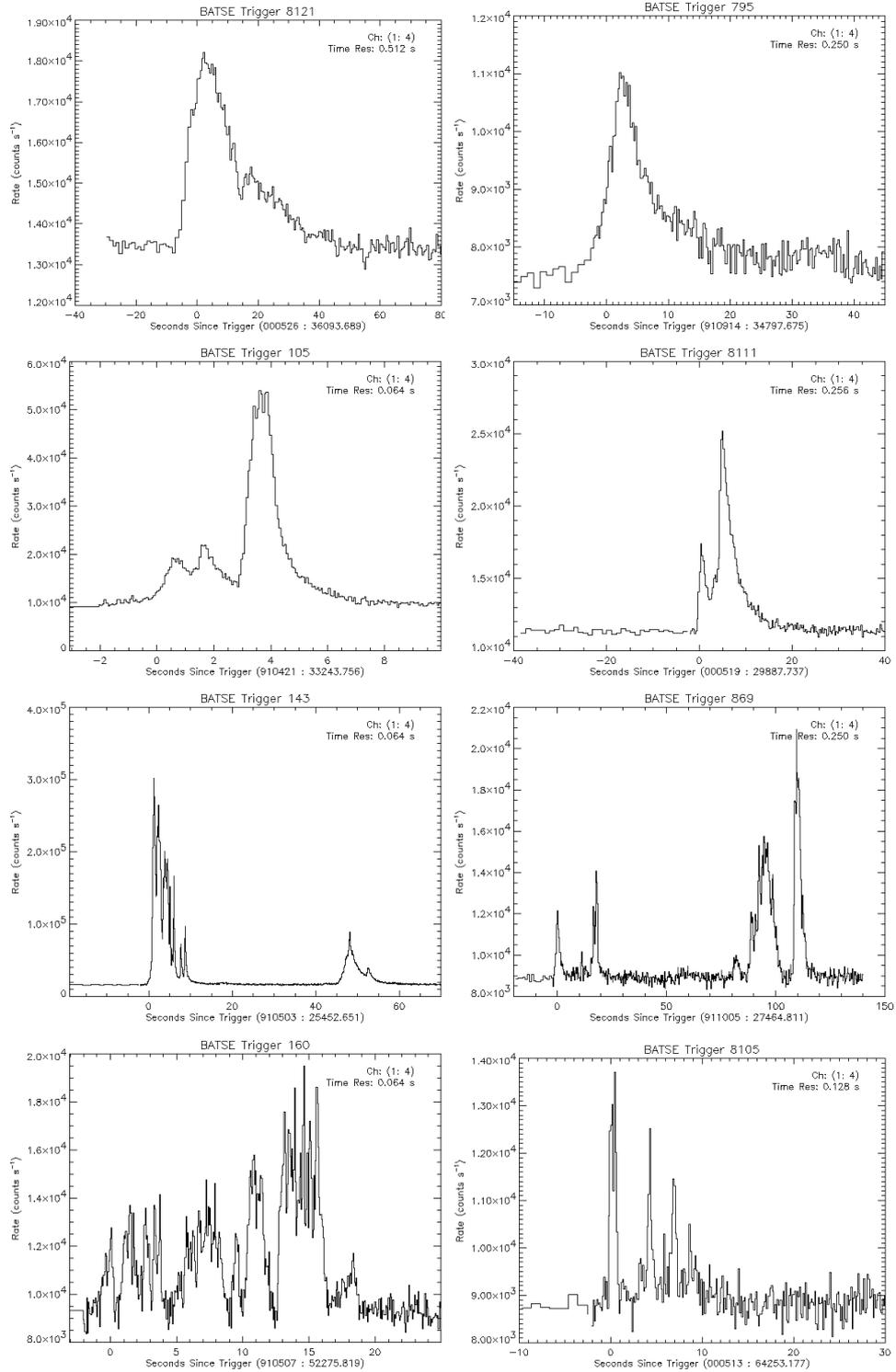}
\caption[Lightcurve of a few BATSE GRBs during prompt emission]{Lightcurve of a few GRBs during prompt emission 
(Compiled from BATSE catalogue: http://www.batse.msfc.nasa.gov/batse/grb/lightcurve/)}
\label{fig1}
\end{figure*}

\subsubsection{C. Controversial Cases: A New Classification Scheme}
The classification scheme as described above got serious challenges from a few observations.

(i) The observation of extended emission in nearly 1/3 rd of the SGRB sample (\citealt{Lazzatietal_2001, 
Connaughton_2002, Norris_Bonnell_2006}) renders the definition of $T_{90}$ uncertain. 
The general feature of extended emission is an initial hard spike, followed by softer emission.
The extended emission can last for tens of sec, and thus blurring the 
demarcation of $T_{90}$ between the LGRB and SGRB class. 

(ii) Discovery of two nearby LGRBs (060505, 060614) and one XRF (040701; \citealt{Soderbergetal_2005}) 
with no SN association made the matter even more complicated. GRB 060505 (\citealt{Fynboetal_2006}) was 
classified as a nearby ($z=0.089$) LGRB (duration: $4-5$ s, spectral lag: 0.36 s,
host: star forming galaxy), but without an observable SN down to deep limit (similar to SGRBs).
GRB 060614 (\citealt{Gehrelsetal_2006}) was also a nearby ($z=0.125$) LGRB (duration $\sim 100$ s), without a SN
(\citealt{Fynboetal_2006, Della_Valleetal_2006, Gal-Yam_2006}). \cite{Hjorth_Bloom_2012} suggests that 
this GRB may be a ``failed SN'' (\citealt{Woosley_1993}). Alternatively, this can be a short GRB with 
extended emission (\citealt{Zhangetal_2007_classification}). This GRB, except for a longer duration, showed all the 
characteristics of a short GRB, e.g., negligible delay (\citealt{Gehrelsetal_2006}), low star forming host
(\citealt{Fynboetal_2006, Della_Valleetal_2006, Gal-Yam_2006}), offset position (\citealt{Gal-Yam_2006}).
In fact, \cite{Zhangetal_2007_classification} have generated a ``pseudo-burst'' with 8 times lower energy and showed that 
the synthetic burst is remarkably similar to a short GRB (050724) with extended emission. 

(iii) Observation of very distant LGRBs are also puzzling. For example, the LGRB 090423 (\citealt{Tanviretal_2009,
Salvaterraetal_2009}) have a spectroscopic $z=8.2$. At this high redshift 
the cosmological $T_{90}=T_{90}/(1+z)$ would make this a SGRB. However, one should be careful that 
the classification is based on the observed $T_{90}$, which is a purely phenomenological scheme. 

In view of these issues \cite{Zhangetal_2007_classification} have suggested a new classification scheme. The classical
short-hard GRBs are termed as Type I, while the long-soft GRBs are termed Type II. 
This classification roughly follows the standard SN classification scheme. However, 
one cannot rule out the possibility of a third category of objects. Incidentally,
based on the properties like non-repetition and harder spectrum, GRBs are possibly 
a different class from SGRs. However, some of the SGRBs may be giant SGR flares from 
relatively nearby galaxies (\citealt{Palmeretal_2005, Tanviretal_2005, Abbottetal_2008, Ofeketal_2008}).

\section{Observables}
Though there are two classes of GRBs, the radiation properties of them are remarkably similar.
Of course, the SGRBs are shorter, harder, and they show dimmer afterglow emission than the LGRBs, which 
are in fact the distinguishing feature of SGRBs. But, the emission mechanisms are probably similar,
and the differences in the observed emission arise due to different environments. Both of them have an initial prompt 
emission (keV-MeV) phase, followed by softer afterglow in x-rays to optical and radio wavelengths.
The prompt and afterglow data of LGRBs have certain advantages over SGRBs: (i) the LGRB sample is
much larger than SGRB sample (about 3:1), (ii) LGRBs are brighter and provides better statistics
for data analysis, and (iii) due to the longer duration and higher flux, LGRBs are better suited
for time-resolved spectroscopy, which is more important than a time-integrated study. Moreover,
the analysis done on LGRBs are also applicable for SGRBs, because the emergent spectral and timing 
properties are likely to be similar. Hence,
we shall discuss about LGRBs when analyzing the data. It is worth mentioning that some of the 
prompt emission properties of these two classes differ, e.g., the spectral lag of SGRBs are much
smaller compared to LGRBs (\citealt{Yietal_2006, Norris_Bonnell_2006, Gehrelsetal_2006}). Also, SGRBs 
do not follow the ``lag-luminosity'' correlation of LGRBs (\citealt{Norrisetal_2000}). These differences may 
provide useful insight, and can be used for the classification scheme (\citealt{Gehrelsetal_2006}).

\subsection{Prompt Emission Characteristics}
\subsubsection{A. Lightcurve}
During the prompt emission a GRB has a variable lightcurve (LC). Figure~\ref{fig1} shows some 
LCs compiled from the BATSE website (full BATSE band). Each burst is different from the other.
This is in direct contrast with SNe, which have broad similarity of LCs in a class. Close inspection 
of the LCs in Figure~\ref{fig1} shows multiple pulses in each of them. Following \cite{Fishman_Meegan_1995}, 
we can broadly classify the bursts based on their underlying pulse structure as follows.
\begin{enumerate}
 \item single pulse (or sometimes a spike) events (see the upper most panels)
 \item bursts with smoothly overlapping pulses (second panels from the top)
 \item bursts with widely separated episodic emission (third panels)
 \item bursts with very rapid variability (bottom panels) 
\end{enumerate}

\begin{figure}\centering
\includegraphics[width=3.0in]{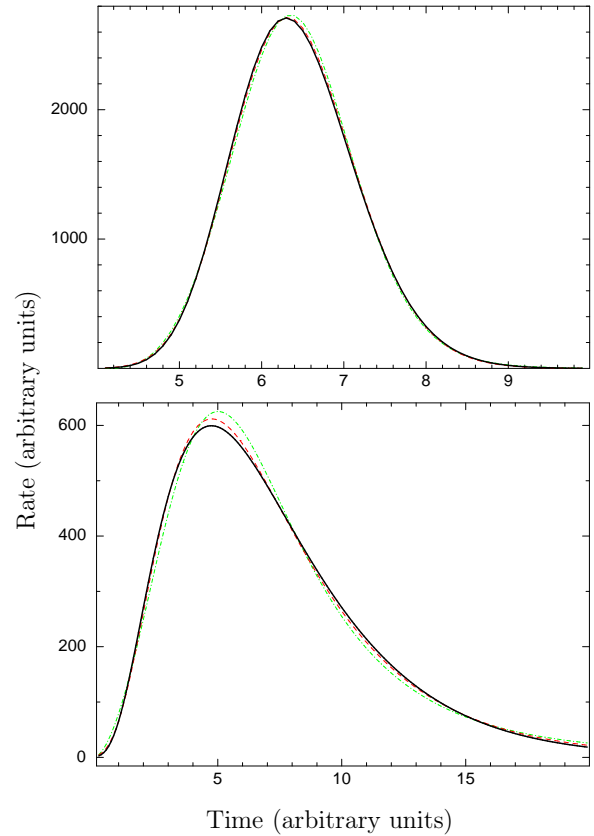}
\caption[Comparison of different pulse models (FRED, Exponential, Lognormal)]
{Comparison of FRED (green dot-dashed line), Exponential (black solid line) and lognormal (red dashed line)
functions. \textit{Upper Panel:} Pulse with higher symmetry, \textit{Lower Panel:} Pulse with higher asymmetry.}
\label{fig2a}
\end{figure}

It is clear from the LCs that there are broadly two variability scales: slowly varying component, 
and fast varying component on top of the individual broad variability. Except for the single pulse GRBs, 
the broad variability time scale is smaller than the total duration of a burst ($T_{\rm GRB}$). 
The definitions of the variability timescales are rather empirical. A rough estimate 
of the fast variability can be found by dividing $T_{\rm GRB}$ by the number of peaks in a burst 
(\citealt{Piran_1999_review}, cf. \citealt{Li_Fenimore_1996}). This variability timescale has important 
implication for the working model of a GRB (\citealt{Kobayashietal_1997}). In our description of GRBs, 
we generally assume the rapid variability as ``weather'' on top of the broad pulse. 
In a recent study, \cite{Xu_Li_2014} have simulated LC for both single variability of 
Lorentz factor ($\Gamma$) and two-component variability (see Chapter 3 for detail).
They have concluded that the latter is preferred to explain the observation of GRB 080319B in both
$\gamma$-ray and optical band. The LC in both these bands can be consider as a superposition
of a slow varying and a fast varying component. These components are possibly 
related to the refreshed activity, and the dynamical time scale of the central engine, respectively.

The LC of GRBs are so diverse that no general description is possible.
This situation can be simplified by considering the broad constituent pulses.
It is suggested that pulses are the basic building blocks of a GRB (\citealt{Norrisetal_2005, Hakkilaetal_2008}). 
These pulses are (possibly) independently generated in a broad energy band (\citealt{Norrisetal_2005}),
and have self-similar shape (\citealt{Nemiroff_2000}). More importantly, GRB pulses have some 
universal features (\citealt{Golenetskiietal_1983, Norrisetal_1986, Norrisetal_1996, Pendletonetal_1994,
Ramirez-Ruiz_Fenimore_2000, Norris_2002, Kocevski_Liang_2003, Norrisetal_2005, 
Ryde_2005, Hakkilaetal_2008}) e.g., the pulses are generally asymmetric, with a sharp rise 
and slow decay (\citealt{Kocevski_Liang_2003}). Spectrum in a pulse generally exhibit a 
``hard-to-soft'' evolution (e.g., \citealt{Pendletonetal_1994}).
An alternative description of the same behaviour is a negative spectral lag of the hard band
with respect to the soft band (i.e., a soft delay). GRB pulses also follow ``lag-luminosity'' correlation 
(\citealt{Norrisetal_2000}; see below). In view of these properties, the description of 
a GRB is equivalent to the description of the constituent pulses. A few empirical 
functions are available to describe the individual pulses. These are:

\begin{itemize}

\item (i) Fast Rise Exponential Decay model (FRED; \citealt{Kocevskietal_2003}):
This pulse shape signifies the phenomenological pulse asymmetry.

\begin{small}

\begin{equation}
F(t)=F_m \left(\frac{t}{t_m}\right)^{r} \left[ \frac{d}{d+r}+\frac{r}{d+r}\left(\frac{t}{t_m}\right)^{r+1}\right]^{-\frac{(r+d)}{(r+1)}}
\end{equation}

\end{small}

Here $F_m$ is the maximum flux at time, $t_m$; $r$ and $d$ are the characteristic indices of the rising and decaying phase, respectively .

\item (ii) Exponential model (\citealt{Norrisetal_2005}):

\begin{equation}
F(t)=A_n\lambda {\rm exp}\{-\tau_{1}/(t-t_{s})-(t-t_{s})/\tau_{2}\}\label{Norris1}
\end{equation}

for $t>t_{s}$. Here $\mu=\left(\tau_{1}/\tau_{2}\right)^{\frac{1}{2}}$ and
$\lambda=exp\left(2\mu\right)$. $A_n$ is defined as the pulse amplitude, $t_{s}$
is the start time, while $\tau_{1}$, $\tau_{2}$ characteristic times
or the rising and falling part of a pulse. One can derive various 
parameters from these model parameters, e.g., the peak position ($\tau_{\rm peak}$), 
pulse width ($w$), which is measured as the interval between the two times 
where the intensity falls to e$^{-1}$, and asymmetry of the pulse ($\kappa$).

\item (iii) Lognormal distribution (\citealt{Bhatetal_2012}):

\begin{equation}
F(t)=\frac{A_{\rm L}}{\sqrt{2\pi} (t-t_{\rm L}) \sigma} {\rm exp} \left[ - \frac{({\rm log} (t-t_{\rm L})-\mu)^2}{2\sigma^2} \right]
\end{equation}
 
for $t > t_{\rm L}$, where $t_{\rm L}$ is the threshold of the lognormal function. 
Here, $A_{\rm L}$ is the pulse amplitude, $\mu$ and $\sigma$ are respectively the sample mean and standard 
deviation of log$(t-t_{\rm L})$. The pulse rise time and decay time can be derived from these 
quantities. The lognormal distribution is motivated by the fact that a parameter, in general, 
tend to follow a lognormal function if it can be written as a product of $\geq3$ random variables.
It is shown that various GRB parameters follow a lognormal distribution e.g., 
successive pulse separation (\citealt{McBreenetal_1994, Li_Fenimore_1996}), break energies 
of spectra (\citealt{Preeceetal_2000}), and duration of pulses (\citealt{Nakar_Piran_2002}).

\end{itemize}

In Figure~\ref{fig2a}, the functions are plotted with arbitrary time axis for a roughly
symmetric (upper panel) and an asymmetric pulse (lower panel). Note that though for 
asymmetric pulse, the FRED profile (green dot-dashed line) tends to have lower width than the 
other functions, the three models are generally very similar both for a symmetric and asymmetric 
pulse profile. For asymmetric pulse, the FRED pulse is deliberately drawn with a slightly higher 
normalization to show its marginal deviation at different parts. However, the deviation is much 
lower compared to rapid variability of a GRB pulse. Hence, any model can be used as an empirical pulse 
description. We have chosen Norris model (shown by solid black line) for our later analysis.
This function is very similar to the lognormal function. 

\subsubsection{B. Spectrum}

A GRB produces high energy $\gamma$-ray photons in a broad band ($\approx$ 10 keV-10 MeV, with a $\nu F_{\nu}$ peak $\sim 250$ keV).
The prompt emission spectrum has a non-thermal shape, or more precisely, the spectrum is not a blackbody (BB).
This is generally described by a decaying power-law with photon index $\sim -1$ ($N(\nu)\sim \nu^\alpha$, with $\alpha\sim -1$). 
The spectrum at higher energy can be modelled either as an exponential break or a steeper photon index ($\beta \sim -2.5$)
than at the lower energies. Though a GRB has a universal spectral shape, the spectral parameters have a wide
range of values.

\begin{figure}\centering
\includegraphics[width=3.2in]{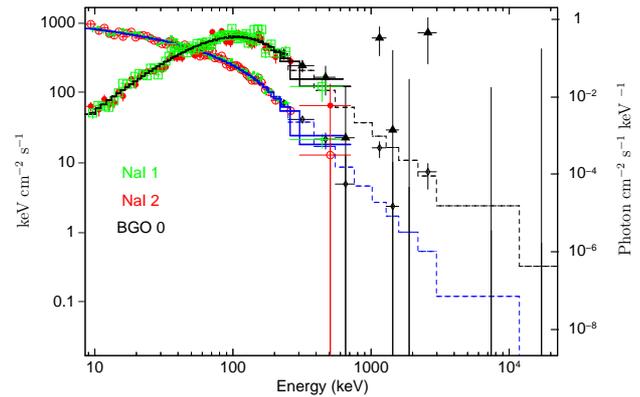}
\caption[Time-resolved spectrum (20-23 s post trigger) of GRB~081221 fitted with Band function]
{Time-resolved spectrum (20-23 s post trigger) of GRB~081221. Both the photon spectrum (shown by blue histogram fitting) 
and the corresponding $\nu F_{\nu}$ representation (shown by black histogram fitting) are shown. 
This spectrum is extracted from \textit{Fermi} data. The different
detectors are shown by different colours. The data is fitted using Band function (Band et al. 1993)
with $\alpha=-0.32^{+0.08}_{-0.08}$, $\beta=-3.65^{+0.40}_{-1.2}$ and $E_{\rm peak}=106^{+4.3}_{-4.0}$ 
keV. The errors quoted are at nominal 90\% confidence.}
\label{fig2}
\end{figure}

\begin{itemize}
\item The simplest function to describe a GRB spectrum is a power-law (PL; e.g., \citealt{Fishman_Meegan_1995}). 
Though this can fit a data with a low value of flux, it is generally inapplicable for a spectrum with a high flux. 
A first order correction to this model is a cut-off power-law (CPL) which has an exponential cut-off at $E_0$. 

\begin{equation}
 N(E)={\rm cons}\times E^{\alpha} {\rm exp}(-E/E_{0})
\end{equation}

\item Band function (\citealt{Bandetal_1993}): It is shown that a large number of BATSE GRBs (with high flux) generally
require another power-law at higher energy. \cite{Bandetal_1993} provides a universal empirical function as:

\begin{small}

\begin{equation}\label{ch1_band1}
N(E) = \left\{ \begin{array}{ll}
 A_{b}\left[\frac{E}{100}\right]^{\alpha}exp\left[\frac{-(2+\alpha)E}{E_{peak}}\right] \\
    $\rm if$ ~~E\le [(\alpha-\beta)/(2+\alpha)]E_{peak}\\
\\
  A_{b}\left[\frac{E}{100}\right]^{\beta}exp\left[\beta-\alpha\right]\left[\frac{\left(\alpha-\beta\right)E_{peak}}
     {100\left(2+\alpha\right)}\right]^{(\alpha-\beta)} \\
   {\rm otherwise}
      
       \end{array} \right.
\end{equation}
 
\end{small}

This function has two PL indices $\alpha$ and $\beta$ for the lower and higher energies, respectively.
The two PL join smoothly (with an exponential roll-over) at a spectral break energy, 
$E_{\rm break}=[(\alpha-\beta)/(2+\alpha)]E_{peak}$. Apart from the normalization $A_{b}$ and the two photon indices, the break energy
or equivalently, the peak energy of $\nu F_{\nu}$ representation
is the fourth parameter of the model. Band function is the most discussed spectral function, and it
is extensively used to fit GRB spectrum, whether it is time-integrated or time-resolved (e.g., \citealt{Kanekoetal_2006,
Navaetal_2011}). In Figure~\ref{fig2}, we have shown a typical time-resolved (20-23 s bin)
spectrum of a GRB (081221), fitted with the Band function (blue line). The $\nu F_{\nu}$ 
representation (black line) shows the peak of the spectrum ($E_{\rm peak}$). 
The Band function has some interesting properties: (i) in the limit $\beta \rightarrow -\infty$, the spectrum 
approaches a CPL function, (ii) as $\alpha \rightarrow \beta$, the spectrum approaches a PL, and (iii) the value
of $\alpha$ can be directly used to interpret the possible radiation mechanism. 

\item Other than these functions, a few more functions are discussed in the literature. For example, a broken 
PL (\citealt{Schaeferetal_1992}), lognormal (\citealt{Pendletonetal_1994}), optically thin bremsstrahlung
spectrum with a PL, and smoothly broken PL (SBPL; \citealt{Preeceetal_1996}). However, the Band function
is regarded as the most appropriate standard function of GRBs.

\end{itemize}

In recent years, a few new functions are suggested for the spectral description, e.g.,
a BB with a PL (BBPL; \citealt{Ryde_2004}), multicolour BB with a PL (mBBPL; \citealt{Pe'er_2008}), 
BB+Band (\citealt{Guiriecetal_2011}), two BB with a PL (2BBPL; \citealt{Basak_Rao_2013_parametrized}).
The general feature of these models is that the spectrum is decomposed into a 
thermal (either one BB, or mBB, or two BBs) and a non-thermal (PL, Band etc.) component.
The aim of these models is obtaining a physical insight of the spectrum, which is 
not provided by an empirical Band function. We shall discuss more about the alternative models in chapter 4.

\subsubsection{C. Spectral evolution}
It is found that a GRB spectrum rapidly evolves during the prompt emission. Hence, time-resolved 
spectral study is more important than the time-integrated study. Within the individual 
pulses of most of the bursts, $E_{\rm peak}$ evolves from high to low values, commonly 
described as a ``hard-to-soft'' (HTS) evolution (e.g., \citealt{Pendletonetal_1994, Bhatetal_1994,
Fordetal_1995, LK_1996, Kocevski_Liang_2003, Hakkilaetal_2008}; see \citealt{Hakkila_Preece_2014} 
for a recent discussion). It is suggested that the HTS spectral evolution 
is a pulse property. However, this feature is questioned in a few studies. For example,
\cite{Luetal_2012} have studied $E_{\rm peak}$ evolution in a set of 51 long and 11 short GRBs.
Though they have found HTS pulses, a substantial number of GRBs also show
a ``intensity tracking'' (IT) behaviour. They have found that the first pulses are generally 
HTS, but later pulses tend to follow an IT behaviour rather than a HTS
evolution. They have suggested that some of these IT pulses might have 
contamination effect from an earlier HTS pulse. However, the fact that some of the single pulse
GRBs also show IT behaviour cannot be explained by an overlapping effect. 
Note that IT evolution can be considered as a ``soft-to-hard-to-soft'' evolution.
These issues will be discussed in Chapter 4, when we will deal with alternative 
spectral models and their evolution.

\subsection{Generic Features Of Afterglows} 
The prompt emission is followed by an afterglow phase, which progressively becomes visible in x-ray, optical 
and some times in radio wavelengths. These emissions last on time scales of days to months, with longer duration
in the longer wavelengths. Unlike the prompt emission, which has rapid variability in the LC, the 
afterglow is a smooth function of time, decreasing as a PL 
($F_{\nu}(t)=t^{\alpha_A}\nu^{\beta_A}$, with $\alpha_A=-1.1$ to $-1.5$ and $\beta_A=-0.7$ to $-1.0$; 
\citealt{Meszaros_2006}). A wealth of x-ray afterglow data is provided by the \textit{Swift}/XRT 
(\citealt{Burrowsetal_2005_XRT}; also see \citealt{Evansetal_2007, Evansetal_2009}). One of the most important discoveries 
of the XRT is finding a canonical behaviour for all bursts, from as early as 100\,s till a few days 
(\citealt{Chincarinietal_2005, Nouseketal_2006, Zhangetal_2006, O'Brienetal_2006}), which consists of three
phases (see \citealt{Zhang_2007, Zhang_2007_review} for detail): a steep decay ($10^2-10^3$ s with an 
index $ \gtrsim -3; ~ F_X(t) \sim t^{\alpha_{XA}}$), 
followed by shallow decay ($10^3-10^4$, with an index -0.5), and then a normal decay phase 
($\alpha_{XA} \sim -1.2$). Other than these phases, an occasional ``post jet break phase'', and 
in nearly 50\% cases, one or more x-ray flares are seen (\citealt{Burrowsetal_2005}). 
\cite{Kannetal_2010} have compiled a list of optical afterglow data and have shown that 
the data is consistent with the standard model. \cite{Chandra_Frail_2012} have 
reported a total of 304 bursts which were observed with very large array (VLA). They found 
$\sim$ 31\% having radio, $\sim$ 65\% having optical and $\sim$ 73\% having detected x-ray afterglow.
The reason that many GRBs do not show radio afterglow is attributed to synchrotron self absorption 
in the initial phase of the radio afterglow. 

An important feature of the afterglow lightcurve is an achromatic break (\citealt{Fruchteretal_1999,
Kulkarnietal_1999, Staneketal_1999, Harrisonetal_1999, Frailetal_2001}).
This break is claimed as the smoking gun signature of a jet. In other words, a GRB is probably a collimated 
event rather than a fully isotropic emission, and they are detected only when the 
jet points towards the observer. Note that, this assumption reduces the energy requirement, and 
increases the population of GRBs, both by a factor of $\approx100$.

\subsection{GeV Emission} 
Apart from the keV-MeV emission and later x-ray, optical and radio emission, GRBs are also accompanied by 
very high energy (GeV) emission (\citealt{Hurleyetal_1994}). GeV emission appears during the prompt
phase, generally starting with a delay with respect to the keV-MeV emission, and lasts longer 
than the prompt emission phase. The later evolution of GeV flux resembles an afterglow behaviour.
In fact, \cite{Kumar_Duran_2010} have shown that the evolution of GeV flux can correctly
predict the later x-ray and optical afterglow flux. However, the origin of this emission is 
debatable (\citealt{meszaros_rees_1994_GeV, Waxman_afterglow, Fan_Piran_2008, Panaitescu_2008,
Zhang_Pe'er_2009, meszaros_rees_2011_GeV}), and remains unknown during the prompt emission
phase. 

The spectrum containing the full MeV-GeV band has either an overall Band functional shape (\citealt{Dingusetal_1998}), 
or sometimes an additional PL is required to fit the spectrum (e.g., \citealt{Gonzalezetal_2003})
For the last 5 years, a great amount of GeV data is being provided by \textit{Fermi}/LAT
(\citealt{Atwoodetal_2009_LAT}). The LAT detects high energy (GeV) emission in a wide band of 30 MeV to 300 GeV 
(see Chapter 2 for details). More than 35 GRBs with GeV photons have been detected by the LAT so far 
(\citealt{Abdoetal_2009_sourcelist, Ackermannetal_2013_LAT}). The broadband data of the 
\textit{Fermi}/GBM and the \textit{Fermi}/LAT has provided various clues both in the GRB science and 
in basic physics, e.g., (i) detection of distinct spectral components and their evolution
(\citealt{Abdoetal_2009_080916C_Sci, Abdoetal_2009_090902B}), (ii) detection of $\gamma$-ray photons 
of energy of up to 30 GeV constraining the bulk \textit{Lorentz} factor ($\Gamma$) to be greater than $\sim1000$ 
(\citealt{Abdoetal_2009_080916C_Sci, Ghirlandaetal_2010_GeV}), and (iii) discovery of GeV photons in the SGRB 090510 which 
helps to put a stringent limit on the possible violation of \textit{Lorentz} invariance (\citealt{Abdoetal_2009_quantumgravity}).

\subsection{GRB Correlations} 
One of the most important, and promising ingredients in understanding the physics of a GRB is 
empirical correlations during the prompt emission. Though the underlying physical reason of the 
correlations are not always clear, any prompt emission model should successfully reproduce the 
data driven empirical correlations. Hence, correlation study can put some constraint on the possible models.
Another more ambitious goal of the correlation study is using GRBs as cosmological luminosity indicators.
One of the greatest discovery in modern cosmology is the finding of the accelerated expansion of the universe,
using high-$z$ type Ia SN (SN Ia) as standard candle (\citealt{Schmidtetal_1998, Riessetal_1998, Perlmutteretal_1999}; 
2011 Nobel Prize in Physics). From the theoretical absolute luminosity and the observed luminosity the luminosity 
distance is derived to measure the acceleration, the amount of dark and baryonic matter ($\Omega_m$), 
and dark energy ($\Omega_\Lambda$) in $\Lambda$ cold dark matter cosmology model ($\Lambda$-CDM). 
However, due to the absorption of optical light, SN cannot be seen
at high $z$ (maximum $z\leqslant 1.755$; \citealt{Riessetal_2007}). On the other hand, 
GRBs being very luminous in $\gamma$-rays, are visible from very large distances.
Hence, they are considered as potential luminosity indicators  beyond  this redshift limit. 
However, unlike SN Ia, the GRB energetics is not standardized. Though \cite{Frailetal_2001}
and \cite{Bloometal_2003}, using the pre-\textit{Swift} data have shown that GRBs are standard energy reservoir,
this is discarded with systematic observations by \textit{Swift} (\citealt{Willingaleetal_2006, 
Zhangetal_2007_standard_reservoir}). In addition, the radiation mechanism of a GRB is also 
uncertain. Hence, with no other options in hand, the empirical correlations are the 
only way to study GRBs as a cosmic ruler.

GRB correlations are studied either in time or in energy domain. For example, the $\nu F_{\nu}$ peak energy
($E_{\rm peak}$) correlates with the $\gamma$-ray isotropic energy ($E_{\gamma, \rm iso}$), known as
Amati correlation (\citealt{Amatietal_2002, Amatietal_2006, Amatietal_2009}). $E_{\rm peak}$ also correlates 
with isotropic peak luminosity ($L_{\rm iso}$; \citealt{Schaefer_2003_HD, Yonetokuetal_2004}), and 
collimation-corrected energy ($E_{\gamma}$; \citealt{Ghirlandaetal_2004_correlation}). In the time domain, the correlations are e.g.,
spectral lag ($\tau_{\rm lag}$) - $L_{\rm iso}$ (\citealt{Norrisetal_2000}), variability (V) - $L_{\rm iso}$ 
(\citealt{Fenimore_Ramirez_2000, Reichartetal_2001}), and rise time ($\tau_{\rm rise}$) - $L_{\rm iso}$ 
(\citealt{Schaefer_2007_HD}). It is worthwhile to mention that several apparent correlations can arise due to 
the instrumental selection effect (e.g., \citealt{Nakar_Piran_2005, Band_Preece_2005}). One argument against the 
selection bias is to show that the correlation exists within the time-resolved data of a given burst (e.g., 
\citealt{Ghirlandaetal_2010}). In chapter 3, based on a new pulse model, we shall introduce a new 
GRB correlation (\citealt{Basak_Rao_2012_correlation}). We shall primarily discuss the correlations 
studied in the energy domain. We shall also discuss about how the new correlation exists against the selection bias.
In Table~\ref{t2}, we have summarized the correlations. Here, N is the number of bursts, $\rho$, $r$ are 
Spearman rank, Pearson linear correlation, respectively. P is the chance probability. The corresponding
relations are shown in the last column.

The quantities are defined as follows.

\begin{table*}\centering
\caption{Correlations in GRBs}

\begin{tabular}{ccccc}
\hline
\hline
Correlation & N & $\rho$/$r$ & P & Relation\\
\hline
\hline

 & & & &\\

$^{(a)}E_{\rm peak}$-$E_{\gamma, \rm iso}$ & 9 & $\rho=0.92$ & $5.0 \times 10^{-4}$ & $\frac{E_{\rm peak}}{1 ~\rm keV}$=$(105\pm11) \left[\frac{E_{\gamma,\rm iso}}{10^{52} \rm erg}\right]^{0.52\pm0.06}$\\
 & & & &\\
\hline
 & & & &\\
$^{(b)}E_{\rm peak}$-$E_{\gamma, \rm iso}$ & 41 & $\rho=0.89$ & $7.0 \times 10^{-15}$ & $\frac{E_{\rm peak}}{1 ~\rm keV}$=$(81\pm2) \left[\frac{E_{\gamma,\rm iso}}{10^{52} \rm erg}\right]^{0.57\pm0.02}$\\
 & & & &\\
\hline
 & & & &\\
$^{(c)}E_{\rm peak}$-$L_{\rm iso}$ & 16 & $r=0.958$ & $5.31 \times 10^{-9}$ & $\frac{10^{5} L_{\rm iso}}{10^{52} \rm erg~s^{-1}}$=$(2.34_{-1.76}^{+2.29})\left[\frac{E_{\rm peak}}{1 ~\rm keV}\right]^{2.0\pm0.2}$\\
 & & & &\\
\hline
 & & & &\\
$^{(d)}E_{\rm peak}$-$E_{\gamma}$ & 27 & $\rho=0.80$ & $7.6 \times 10^{-7}$ & $\frac{E_{\rm peak}}{1 ~\rm keV}$=$(95\pm7) \left[\frac{E_{\gamma}}{10^{52} \rm erg}\right]^{0.40\pm0.05}$\\
 & & & &\\

\hline

\end{tabular}
\label{t2}

\begin{footnotesize}
 $^a$The original Amati correlation (\citealt{Amatietal_2002}).
 $^b$\cite{Amatietal_2006}, $^c$\cite{Yonetokuetal_2004}, $^d$\cite{Ghirlandaetal_2004_correlation}
\end{footnotesize}

\end{table*}

Let us assume that the observed fluence (time integrated flux) is $S_{\rm obs}$ and peak flux is $P_{\rm obs}$.
The k-corrected bolometric fluence and peak flux are:

\begin{equation}
S_{\rm bolo}=S_{\rm obs} \times \frac{\int_{1/1+z}^{10^4/1+z} E \times N(E) dE}{\int_{E_{\rm min}}^{E_{\rm max}} E \times N(E) dE} ~~~\rm erg~cm^{-2}
\end{equation}

and

\begin{equation}
F_{\rm bolo}=P_{\rm obs} \times \frac{\int_{1/1+z}^{10^4/1+z} E \times N(E) dE}{\int_{E_{\rm min}}^{E_{\rm max}} E \times N(E) dE} ~~~\rm erg~cm^{-2}~s^{-1}     
\end{equation}

Here, the integration in the numerator are done in the energy band 1 keV to $10^4$ keV in the source frame. The spectral
function $N(E)$ is generally taken as the Band function. In the denominator, the integration is done over the instrument band width.
The ratio of these fluxes give the bolometric k-correction.
The $\gamma$-ray isotropic energy ($E_{\gamma, \rm iso}$) and isotropic peak luminosity ($L_{\rm iso}$) are defined as:

\begin{equation}
E_{\gamma, \rm iso}=4\pi d_{\rm L}^2 \frac{S_{\rm bolo}}{1+z}~~~\rm erg
\end{equation}

and

\begin{equation}
L_{\rm iso}=4\pi d_{\rm L}^2 F_{\rm bolo}~~~\rm erg~s^{-1}
\end{equation}

Here $d_{\rm L}$ is the luminosity distance of the source, which is dependent on $z$ and the version of cosmology in use.
Generally, a $\Lambda$-CDM cosmology with a zero curvature ($\Omega_K=0$), ($\Omega_m$, $\Omega_{\Lambda}$)=(0.27, 0.73), and
Hubble parameter, $H_0=70$ km s$^{-1}$ Mpc$^{-1}$ is used.

If a GRB is a jetted event, then the energy of the source is corrected for the collimation. If $\theta_j$
is the half opening angle, then the collimation corrected energy is 

\begin{equation}
E_{\gamma}=(1-{\rm cos}\theta_j)  E_{\gamma, \rm iso}
\end{equation}

\section{A Working Model for  GRBs}
In this section, we shall briefly discuss the working model of GRBs. The basic ingredients of this model 
are known, however, quantitative calculations are still lacking. In addition, several modifications of the 
radiation process, emission region, and even completely different models are also proposed. For a detailed 
description of the standard model see reviews by \cite{Piran_1999_review} and \cite{Meszaros_2006}. 
For later purpose, we shall use notation $q_{x}$ to denote a quantity $q$ in the cgs units of $10^x$. 
For example, $E_{52}$ is $E$ in the units of $10^{52}$ erg.

\subsection{Compactness And Relativistic Motion}
From the discussion of the prompt emission features, we know that a GRB has rapid observed variability 
($\delta t_{\rm obs} \approx$ 1 s down to 10 ms), and an enormous observed flux. A cosmological 
distance translates the observed flux to high luminosity ($L_{\rm iso} \sim 10^{53}~ \rm erg s^{-1}$). 
This particular combination has a very important consequence, known as compactness problem.
From the variability argument, the emission radius has an upper limit, $R_{\rm s} < c\delta t_{\rm obs} \sim 10^8$ cm.
Hence, the compactness parameter, $\sigma_{\rm T} L_{\rm iso}/R_{\rm s} m_{\rm e} c^3\gg1$. In other words, a huge number of photons 
are created in a small volume of space. Hence, the photons will pair produce leading to a spectral
cut-off precisely at 512 keV, the rest frame energy of an electron ($m_e c^2$). This is in direct 
contradiction with the observed spectrum, which contains many photons in the range 0.5 MeV-10 MeV 
(sometimes extending to even GeV energies). 

The compactness will lead to rapid pair production ($\gamma \gamma \rightarrow e^-e^+$), and hence a 
high optical depth is attained. If $f_{\rm p}$ is the fraction of photon pairs which satisfy
the pair production criteria, then for a source with an observed flux $F$, distance $d$, and variability $\delta t$
has an average optical depth (\citealt{Piran_1999_review}),

\begin{equation}
 \tau_{\gamma \gamma}=10^{13} f_{\rm p} \left[ \frac {F}{10^{-7}~{\rm erg~ cm^{-2}}}\right] \left[\frac{d}{3 Gpc}\right]^2 \left[\frac{\delta t}{10~\rm ms}\right]^{-2}
\end{equation}

Note that the compactness problem is an unavoidable consequence of the cosmological origin of GRBs --- 
the severity of the situation essentially increases with increasing distance. In fact, this was 
one of the most important arguments against the cosmological origin 
(\citealt{Ruderman_1975, Schmidt_1978, Cavallo_Rees_1978}).
During late 1950's to mid-1960's, astronomers faced a similar inconsistency in quasars. From the observed 
line ratios in the optical spectrum the distance was found to be cosmological. The variability of a typical quasar
is $\sim$ day, which indicates a compact region. With this high compactness the sources should lead to 
``Compton catastrophe'', and no radiation should be seen. 
This even led to the proposal of discarding the extra-galactic origin of quasars (\citealt{Hoyle_1966}).
The cosmological distance scale of quasars was saved by implementing ultra-relativistic expansion of the 
sources (\citealt{Woltjer_1966, Rees_1967}). Later observation with \textit{Very Large Baseline Array} (VLBA)
confirmed the apparent superluminal motion of quasar jets with bulk \textit{Lorentz} factor, $\Gamma \approx 2-20$.
Similar situation with even more severity occurred for GRBs. The solution was also similar (see \citealt{Katz_2002_book} for detail).

The relativistic motion with a bulk \textit{Lorentz} factor has three effects which help GRBs bypassing the compactness problem.

\begin{itemize}
\item (i) Due to the relativistic effect an observed photon will have a blue shifted observed energy ($h\nu_{\rm obs}$).
The energy in the source frame can be obtained as $h\nu_{\rm obs}/\Gamma$. This will enormously reduce 
the fraction of photons eligible for pair production in the source frame. This factor is $\Gamma^{2\beta}$, 
where $\beta$ is the high energy spectral index. Note that for simplicity, we have neglected the cosmological 
redshift correction of the source frame energy, which is much smaller compared to the correction discussed above.

\item (ii) For an observer, the arrival time of two successive pulses will be 
compressed as the source generating the pulses moves with high $\Gamma$. 
Hence, the emission radius is allowed to be larger than 
$R_{\rm s} < c\delta t_{\rm obs}$ by a factor $\Gamma^2$. 

\item (iii) The pair production threshold for a head-on interaction for two photons with energy 
$E_1$ and $E_2$ is $\sqrt{E_1 E_2} > m_e c^2$. For a relativistic source, the radiation will be beamed, and the photons 
will interact only in grazing angles with each other. This increases the pair production threshold to an
arbitrary higher value (\citealt{Katz_2002_book}).

\end{itemize}

The optical depth using these correction factors is $\tau^{\rm (ac)} \sim \frac{\tau_{\gamma \gamma}}{\Gamma^{4+2\beta}}$.
Putting the actual numbers, one gets a lower limit on $\Gamma>100$ (Piran 1999). An alternative 
approach also gives a similar $\Gamma>100[(E_1/10~\rm GeV)(E_2/\rm MeV)]^{1/2}$, where $E_1$ is the 
energy (in 10 GeV) of a high energy photon that escapes annihilation with a lower energy target photon of energy $E_2$ MeV.
Following \cite{Meszaros_2006}, the following notations will be used:

\begin{itemize}
\item $K_* \rightarrow $ Origin of the explosion (lab frame), $K' \rightarrow $ Co-moving frame of the gas (fireball),
$K \rightarrow $ Observer frame

\item $dr_*=dr=\frac{dr'}{\Gamma} \rightarrow$ Usual length contraction

$dt_*=dt'\Gamma \rightarrow$ Usual time dilation

$dt_*=\frac{dr_*}{\beta c} \approx \frac{dr_*}{c}\rightarrow$ Time separation between successive events. $\beta=v/c$

$dt=dt_*(1-\beta) \rightarrow$ Time separation between events (in observer frame)

\item Transformation from $K'$ to $K$ is done by Doppler factor which is defined as
$D\equiv\left[\Gamma(1-\beta \mu)\right]^{-1}$. Here, $\mu=\rm cos\theta$, where
$\theta$ is the angle between the expansion direction of the ejecta and the line of sight towards the observer.
Some examples of transformations are: time transformation: $dt=D^{-1}dt'$ (combining second and fourth relations), 
frequency transformation: $\nu=D\nu'$. For $\Gamma \gg 1$, and $\mu \rightarrow 1$ (approaching), $D\approx2\Gamma$,
and $\mu \rightarrow -1$ (receding), $D\approx1/2\Gamma$

\end{itemize}

\subsection{``Fireball Model'' And Radiation Mechanism}

\subsubsection{A. Photon-Lepton Fireball}
The first major attempt to explain the consequences of a cosmological distance on the 
dynamics and spectral features during the prompt emission of a GRB was proposed in two 
independent papers by \cite{Goodman_1986} and \cite{Paczynski_1986}. 
Note that these papers were published even before the BATSE was launched. Both the authors start with 
the assumption that GRBs have cosmological origin, and hence from the observed flux their luminosity must be high, in fact 
much higher than the Eddington luminosity ($L_{\rm E} = 1.25 \times 10^{38} (M/M_{\odot}) ~{\rm erg~s^{-1}}$).
Hence, the radiation pressure largely exceeds the self-gravity, leading to an expansion of the source. In this 
model, the ejecta is considered to be a purely photon-lepton ($\gamma-e^-/e^+$) opaque plasma, referred to as
a ``fireball''. 

Let us assume that in a region of size $r_{\rm i}$, a huge energy $E_{\rm i}$ is created, and the mass of the system,
$M_{\rm i} \ll E_{\rm i}/c^2$. As the fireball expands, the co-moving temperature ($T'$) decreases due to 
an adiabatic cooling. As the fireball is radiation dominated, the adiabatic index, $\gamma_a=4/3$. Hence, the 
cooling law is, $T' \propto r^{-1}$, where $r$ is the radius of the fireball measured from observer frame.
Note that $r$ is the same as measured from the stationary lab frame, $K_*$. As temperature decreases, 
the random \textit{Lorentz} factor ($\gamma_{\rm r}$) also drops ($\gamma_{\rm r}\propto r^{-1}$). Hence, from
energy conservation, the internal energy per particle is continuously supplied for expansion energy, i.e.,
$\gamma_r \Gamma=\rm constant$. Hence, $\Gamma$ increases linearly with $r$. However, the acceleration 
cannot go on for ever. When the bulk kinetic energy ($\Gamma M_{\rm i} c^2$) becomes 
equal to $E_{\rm i}$, the value of $\Gamma$ ceases to increase. From the relation, $\Gamma_{\rm max} M_{\rm i} c^2=E_{\rm i}$
one can obtain this coasting value as $\Gamma_{\rm max} \sim \eta \equiv E_{\rm i}/ M_{\rm i}c^2$.
Here, $\eta$ is called the dimensionless entropy of the fireball. The value of $\eta$ determines the coasting 
value of $\Gamma$. The time evolution of $\Gamma$ can be written as follows (see \citealt{Meszaros_2006}):

\begin{eqnarray}
\Gamma(r) \sim \left\{ \begin{array}{ll}
 (r/r_{\rm i}) &\mbox{ for $r < r_{\rm s}=\eta r_{\rm i}$} \\
  \eta &\mbox{ otherwise}
       \end{array} \right.
\label{gamma}
\end{eqnarray}

The comoving temperature can be shown to vary as follows:

\begin{eqnarray}
kT'(r) \sim \left\{ \begin{array}{ll}
 (r/r_{\rm i})^{-1} &\mbox{ for $r < r_{\rm s}$} \\
 (r/r_{\rm i})^{-2/3} &\mbox{ otherwise}
       \end{array} \right.
\label{temp1}
\end{eqnarray}

Here $r_{\rm s}$ is called the saturation radius (where $\Gamma$ attains the maximum value and the acceleration stops).
Another important parameter is the photospheric radius ($r_{\rm ph}$), which is defined as the radius where the photons 
decouple from the matter. Within $r_{\rm ph}$, the energy of the photons is continuously converted
into the kinetic energy of the fireball. The fireball remains optically thick (optical depth, $\tau > 1$) below this radius.
The optical depth to \textit{Thomson} scattering in the radial direction from a radius ($r$) to infinity is

\begin{equation}
 \tau=\int_{r}^{\infty} {\rho'(r')\kappa D^{-1}dr'}
\label{tau}
\end{equation}

Here, $\kappa=\sigma_{\rm T}/m_p$ ($\sigma_{\rm T}$ is \textit{Thomson} scattering cross section,
$m_p$ is proton mass) is the total mass opacity, and $\rho'$ is the co-moving density. The value of 
$r_{\rm ph}$ can be found by putting $\tau = 1$ in equation~\ref{tau}. If $\dot{M}$ is the mass injection rate, then 
$\rho'=\frac{1}{\Gamma} \frac{\dot{M}}{4 \pi r^{2} \beta c}=\frac{L}{\Gamma 4 \pi r^2 \beta \eta c^3}$
(as $\eta=E_{\rm i}/ M_{\rm i}c^2=L/\dot{M}c^2$). Using these values in equation~\ref{tau}, and
putting $\tau=1$ for $r=r_{\rm ph}$, we get

\begin{equation}
 r_{\rm ph}=\frac{\sigma_{\rm T} L}{8 \pi \eta^3 m_p c^3}
\label{rph}
\end{equation}

The observer frame temperature ($kT$) can be found by $kT\sim\Gamma kT'$. Using equation~\ref{gamma} and \ref{temp1}, we obtain,

\begin{eqnarray}
kT(r) \sim \left\{ \begin{array}{ll}
 {\rm constant }&\mbox{ for $r < r_{\rm s}$} \\
 (r/r_{\rm i})^{-2/3} &\mbox{ otherwise}
       \end{array} \right.
\label{kT'}
\end{eqnarray}

If the photosphere occurs higher than the saturation radius ($r_{\rm ph}>r_{\rm s}$), an observer 
will see a hard-to-soft (or rather a hot-to-cold) evolution. However, the spectral peak and luminosity will be degraded.
On the other hand, for $r_{\rm ph}>r_{\rm s}$, a break is expected in the $E_{\rm peak}$ (or rather kT) evolution. However, note that 
the spectrum predicted by the model is a blackbody (BB --- Planckian spectrum), rather than a Band function. 
A BB has a photon index $+1$, in the lower energy, while a typical GRB has a $-1$ photon index. 
In other words, a BB spectrum is too hard for GRBs. Also, in the high energy part of the spectrum,
while a BB has an exponential fall off, the Band function falls only as a PL with an index $\sim-2.5$. 
A typical GRB has a wider peak than a BB. \cite{Goodman_1986} proposed a geometric broadening of the BB 
spectrum due to the finite size of the photosphere. However, the proposed modified BB cannot account for the very different 
shape of a GRB spectrum. Though this model cannot explain the spectral features, it gives the essential ingredient to 
achieve relativistic motion in a GRB. In recent years, modified forms of photospheric emission has received considerable 
attention (e.g., \citealt{Ryde_2004, Ryde_2005, Pe'eretal_2005, Pe'eretal_2006}; see also \citealt{Mizutaetal_2011, Lazzatietal_2013}). 
We shall discuss more about these models 
in chapter 4.

\subsubsection{B. Baryon Loading And Internal-External Shock Paradigm}

The failure of the photon-lepton fireball led researchers to try some modifications of the basic assumptions. The 
first logical step was to introduce baryons in the otherwise pure photon-lepton plasma (\citealt{Shemi_Piran_1990}).
A baryon loaded fireball has two major modifications on the original fireball: (i) As the $M_{\rm i}$ is larger, the coasting 
\textit{Lorentz} factor, $\eta=E_{\rm i}/M_{\rm i}c^2$ is lower. Of corse, the baryon load should 
not be so high that the ultra-relativistic motion (which is required by observation) is killed.
Such a situation occurs in supernova explosion, where the baryon load leads to a non-relativistic 
motion of the ejecta ($<0.1c$). (ii) If the baryon 
load $M_{\rm b}$ exceeds the value $M_{\rm critical}=2 \times 10^{-7} M_{\odot} E_{52}^{2/3}r_{\rm i7}^{2/3}$
(where $E_{52}$ is $E$ in units of $10^{52}$ erg, $r_{\rm i7}$ is $r_{\rm i}$ in units of $10^{7}$ cm),
then the fireball becomes matter dominated before it reaches the photosphere (in this case,   
a baryonic photosphere rather than a pair photosphere is formed). The internal 
energy will be mostly converted into kinetic energy of the baryons, and no radiation will be seen
at the photosphere. Hence, the outcome of a baryon loaded fireball is a ``clothed fireball''.

It is worthwhile to mention that if the baryon load is less than $\sim 10^{-12} M_{\odot} E_{52}^{1/2}r_{\rm i7}^{1/2}$,
the baryon will have negligible effect. With $M_{\rm b} < M_{\rm critical}$, still the
fireball will be radiation dominated, with a degraded temperature. Certainly, in order 
to avoid producing a thermal spectrum the feasible choice is $M_{\rm b} > M_{\rm critical}$. 
But, the energy is then drained by the baryons leading to no radiation. However, this 
energy is available in the form of kinetic energy of the baryons. Hence, in order to produce 
a GRB, the kinetic energy must be made available in the form of radiation. \cite{Rees_Meszaros_1992_ES} 
and \cite{Meszaros_Rees_1993_ES} proposed a mechanism to reconvert the kinetic energy into radiation. This model involves
shock generation in the external circumburst medium, and it is known as the ``External shock'' (ES) model. 
The essential idea is that the fireball plasma cannot move with constant velocity for ever, and when it 
eventually plunges into the external medium, it heats up the gas, ``sweeps up'' mass and decelerates. 
The external medium can be either a pre-ejected wind of the progenitor, or the interstellar medium (ISM).
The interaction process is mediated via ``collisionless'' turbulent plasma shock wave rather than 
direct particle collision. This shock is commonly referred to as the ``External shock''. This shock is expected 
because of the discontinuity of density, temperature and pressure between the fireball plasma 
and the ISM plasma. The detailed mechanism of a turbulent plasma shock is incalculable, but it is 
believed that a part of the energy is used to accelerate electrons to very high $\gamma_e$, and 
another part is converted to magnetic field. The electrons can gyrate in the magnetic field 
producing synchrotron radiation. As a synchrotron spectrum is non-Planckian with a wider 
peak in the $\nu F_{\nu}$ representation, it naturally explains the observed spectrum of a GRB. 
This process will produce a single pulse with a fast rise and a slow decay. The complex LC of a GRB 
can be produced by assuming density fluctuations (clumps) in the ISM. The requirement 
of this hypothesis is that the clumps should be small and sparsed, otherwise the temporal
fluctuation will not be observed (\citealt{Piran_1999_review}). However, the clumps are required to be 
so small and sparsed that it will be very unlikely to have multiple collisions in the line of sight. Hence, ES
can make only a faint burst with a single pulse by interacting with a clumpy ISM. Even if continuous collisions happen
(e.g., in a relatively uniform density ISM), due to the deceleration of the fireball (decreasing $\Gamma$): 
(i) the spectrum should have a hard-to-soft evolution, and (ii) the later sub-pulses should be more stretched out 
in the observer frame. Though a hard-to-soft evolution is common in a GRB, the sharpness of the individual 
sub-pulses is independent of their time sequence. Note that this model was proposed 
before the afterglow era. Following the detections of afterglows (from 1997 onwards), 
it became apparent that the predictions of ES model are compatible with afterglow features. Hence,
it was suggested that the ES can give rise to the afterglow rather than the prompt emission phase. 
In fact, the features of the afterglow data matches quite well with the ES predictions --- hard-to-soft 
evolution, and longer duration in lower energies (e.g., \citealt{meszaros_rees_1997_afterglow, Reichart_1997_afterglow,
Waxman_afterglow, Vietri1997_afterglow, Tavani1997_afterglow, Wijers1997_afterglow}). In addition to the 
forward external shock as described above, a ``reverse shock'' may generate which propagates back into the 
material behind the shock front (\citealt{Sari_Piran_1999_prediction}). The signature of such a shock is 
found as an optical/UV flash during the afterglow for a handful of GRBs (e.g., \citealt{Sari_Piran_1999_observation}).

\begin{figure*}\centering
\includegraphics[width=5.9in]{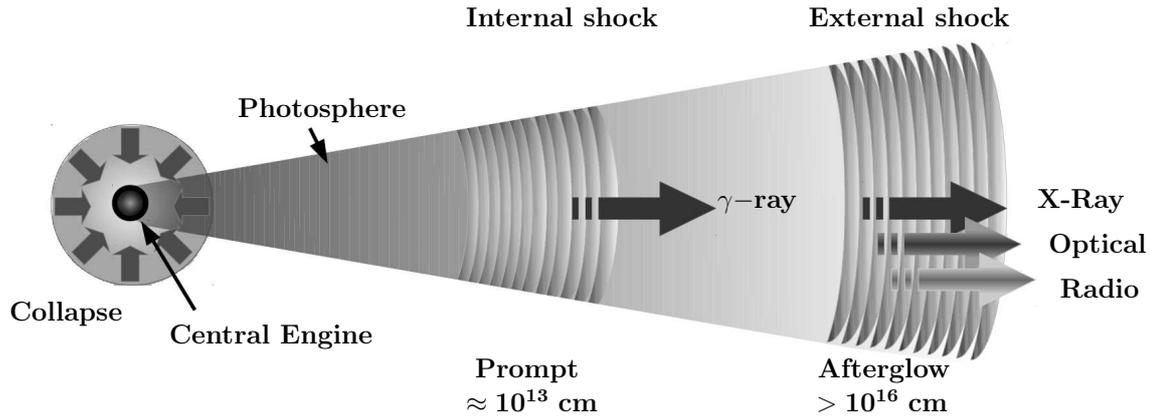}
\caption{Model of Gamma-Ray Bursts, re-drawn following \cite{Meszaros_2001}}
\label{fig3}
\end{figure*}

\cite{Rees_and_meszaros_1994_prompt} proposed another possible site for energy dissipation. They pointed out that 
the central compact object, which has a dynamical time scale $\sim$ milliseconds, releases fireshells 
with varying \textit{Lorentz} factors ($\Gamma$), instead of a steady ejecta.
If a fast moving shell catches up with a slower one, it generates ``internal shock'' (IS). 
If two successive shells of same mass but different \textit{Lorentz} factors ($\Gamma_1$ and $\Gamma_2$) 
are ejected on a timescale $\delta t_{\rm var}$ (in the lab frame), then the radius where ISs build up is 
$\sim c\delta t_{\rm var} \Gamma_1\Gamma_2$. The ISs accelerate electrons which will generate synchrotron 
spectrum. The IS model can explain the complex LC, and the non-thermal spectrum of a GRB. However, 
it does not predict anything about the spectral evolution. Also, due to the fact that both the 
shells are moving out in the same direction, their relative velocity is small, and consequently,
the efficiency of energy conversion is low compared to the ES case.
If $\Gamma_{\rm f}$ and $\Gamma_{\rm s}$ are a pair of fast moving and slow moving shells which collide 
to form a final shell with $\Gamma_{\rm t}=\sqrt{\left(\Gamma_{\rm f} \Gamma_{\rm s} \right)}$, then the efficiency
is $\epsilon=\left(\Gamma_{\rm f}+\Gamma_{\rm s}-2\sqrt{\Gamma_{\rm f}\Gamma_{\rm s}}\right)/(\Gamma_{\rm f}+\Gamma_{\rm s}) \sim 20\%$.

In Figure~\ref{fig3}, we have shown the schematic view of the IS-ES model (\citealt{Meszaros_2001}). This is the most 
widely discussed model of GRB radiation (e.g., \citealt{Rees_Meszaros_1992_ES, Rees_and_meszaros_1994_prompt, 
Katz_1994_grb_model, Sari_Piran_1997}).

Some modifications of the ingredients of the prompt emission mechanism do exist. For example, a magnetically dominated ejecta
is expected to produce prompt emission via magnetic reconnection (\citealt{Usov_1992, Usov_1994, Metzgeretal_2007,
Metzger_2010}; see also \citealt{Zhang_Yan_2011}: Internal-Collision-induced MAgnetic Reconnection and Turbulence, ICMART).  
Such a mechanism will have different emission radius and spectral evolution. The prompt emission is likely to 
be followed by a similar afterglow due to ES of the standard model. The only difference would be that due to 
a very high Alfv\'{e}n speed of the ejecta, the reverse shock will be absent (or weak). The prompt emission is expected 
to be highly polarized. 
A completely different model, called ``cannon-ball model'' (CB), is proposed by \cite{Dadoetal_2002, Dadoetal_2007, 
Dar_Rujula_2004, Dar_2006}. This model advocates particle interaction, rather than shock wave generation.
The central engine shoots out CBs of ordinary matter which produce prompt emission via bremsstrahlung, and afterglow via 
inverse Compton (IC) of the ambient photon field.  

\subsubsection{C. Locations}
In the standard fireball model the radiation can arise from several regions. In addition to the standard IS 
and ES regions the photosphere can also contribute to the radiation (see chapter 4).
By putting the actual numbers, the locations of these emissions can be estimated (\citealt{Meszaros_2006}) as follows.

\begin{itemize}
 \item The photospheric radius,
\begin{equation}
 r_{\rm ph} \sim 6 \times 10^{11} L_{51} \eta_{2}^{-3}~\rm cm
\label{rph2}
\end{equation}

The baryonic photosphere can occur below or above this radius depending on the baryon load.
 \item The radius where ISs develop is,

\begin{equation}
 r_{\rm IS} \sim c \delta t_{\rm var} \eta^2 \sim 3 \times 10^{13} \delta t_{\rm var,-1} \eta _2^2~\rm cm
\label{rIS}
\end{equation}

Here $\delta t_{\rm var,-1}$ is $\delta t_{\rm var}$ in units of 0.1 s.
 \item The radius where ESs develop is,

\begin{equation}
\begin{split}
r_{\rm ES} &\sim (3 E_{i}/4\pi n_{ext}m_p c^2 \eta^2)^{1/3} \\
           &\sim 5.5 \times 10^{16} E_{53}^1/3 n_{ext}\eta_{2.5}^{-2/3}~\rm cm
\end{split}
\label{rES}
\end{equation}

\end{itemize}

\subsubsection{D. Achromatic Breaks And Evidence Of Jet}
One of the most important observations of GRB afterglow is the achromatic break in the LC. This is claimed as the 
evidence of GRB jets (\citealt{Fruchteretal_1999, Kulkarnietal_1999, Staneketal_1999, Harrisonetal_1999, 
Frailetal_2001}; cf. \citealt{Sarietal_1999_Jet}). In fact, a LC break is predicted for a jet due to the following 
relativistic effect (\citealt{Rhoads_1997, Rhoads_1999}). If $\Gamma$ is the bulk \textit{Lorentz} factor of the 
source, then an observer can see only $1/\Gamma$ portion of the ejecta. As the ejecta decelerates in the 
external medium, $\Gamma$ decreases. Thus an observer tends to see more an more portion of the ejecta i.e.,
the light-cone becomes wider. If the ejecta is totally isotropic, then the observed flux decreases steadily as a combination 
of decreasing flux and increasing accessible area that an observer sees. However, for a physical jet with an 
opening angle $\theta_j$ (or a solid angle $\Omega_j$), this situation will be different. If an observer 
is within $\Omega_j$ and as long as $\Gamma \gtrsim \Omega_j^{-1/2}$, the observer is unaware of the physical 
structure. But, as soon as $1/\Gamma$ becomes larger than $\theta_j$ ($\Gamma \lesssim \Omega_j^{-1/2}$), the 
accessible area does not increase any further leading to a change in the observed flux evolution law. 
In addition, the jet expands sideways, which also affects the observed flux. A combination of these two 
effects leads to a steeper flux decay law ($F_{\nu}(t) \sim t^{-2}$) than a normal afterglow decay (index 1.1-1.5). 
As this effect is purely relativistic, the expected break should be achromatic. Hence, the observation supports that 
GRBs are jetted events. The assumption of jet also helps reducing the energy, and makes GRB rate higher.

\subsubsection{E. Radiation Mechanism}
Note that (equation \ref{rIS} and \ref{rES}), the IS and ES are produced at a radius where the source is optically thin.
If IS-ES are indeed the dominant process to make a GRB, then the major radiation mechanism should be an optically thin synchrotron 
emission (\citealt{Piran_1999_review}; also see \citealt{Granotetal_1999, Granot_Sari_2002}). The electrons are Fermi accelerated 
in the shock. Hence, electron energy
has a power-law distribution. The parameters of synchrotron radiation are: magnetic field strength ($B$),
the electron power-law index, p and the minimum \textit{Lorentz} factor ($\gamma_{\rm e, min}$). If the 
bulk \textit{Lorentz} factor in the shocked region is $\Gamma_{\rm sh}$, then the electron energy distribution 
can be written as

\begin{equation}
 N(\gamma_{\rm e})=\gamma_{\rm e}^{-p} , ~~~{\rm for}~ \gamma_{e}>\gamma_{\rm e, min}=\frac{m_{\rm p}}{m_{\rm e}}\frac{p-2}{p-1}\epsilon_{\rm e}\Gamma_{\rm sh}
\label{synchrotron_distribution}
\end{equation}

Here, $\epsilon_{\rm e}$ is the fraction of energy in the random motion of electrons in the shocked region. One can also define 
the fraction of energy in the magnetic field as $\epsilon_{\rm B}$. The value of $p$ can be directly found from the 
high energy index of a typical GRB spectrum, $\beta$, and typical value is $p=2.5$. The synchrotron frequency and the 
power emitted by a single electron in the fluid frame are,

\begin{equation}
 \nu_{\rm syn}(\gamma_{\rm e})= \gamma_{\rm e}^2 \left(\frac{q_{\rm e} B}{2\pi m_{\rm e} c } \right),~~~
 P_{\rm syn}= \frac{4}{3}\sigma_{\rm T} c \gamma_{\rm e}^2 B^2/8\pi
\label{synchrotron}
\end{equation}

As the electrons cool one can define a \textit{Lorentz} factor, $\gamma_{\rm e, c}$, which 
cools on a hydrodynamical timescale, $t_{\rm hyd}$. If the emitting material moves 
with a bulk \textit{Lorentz} factor, $\Gamma_{\rm E}$, then the timescale can be found as the
observer time which an electron with energy, $\gamma_{\rm e, c} m_{\rm e}c^2$ takes to cool down at a 
rate $P_{\rm syn}(\gamma_{\rm e, c})$, i.e., 
$t_{\rm dyn}=\gamma_{\rm e, c} m_{\rm e}c^2/\Gamma_{\rm E} P_{\rm syn}(\gamma_{\rm e, c})$

\begin{equation}
\gamma_{\rm e, c}=\frac{6\pi m_{\rm e}c}{B^2\sigma_{\rm T} \Gamma_{\rm E}t_{\rm hyd}}
\end{equation}

The synchrotron spectrum of a single particle with \textit{Lorentz} factor, $\gamma_{\rm e}$
is $F_{\nu}\propto\nu^{1/3}$ up to $\nu_{\rm syn}$ (equation \ref{synchrotron}), and exponential decay thereafter.
An energetic electron with $\gamma_{\rm e}>\gamma_{\rm e, c}$, rapidly cools to $\gamma_{\rm e, c}$. 
Hence, in the range $\nu_{\rm syn}(\gamma_{\rm e, c})<\nu<\nu_{\rm syn}(\gamma_{\rm e})$, the spectrum is $F_{\nu}\propto\nu^{-1/2}$
For an electron distribution as in equation \ref{synchrotron_distribution}, we have to integrate 
over all $\gamma_{\rm e}$. At low frequency, the spectrum has 1/3 slope till $\nu_{\rm syn}(\gamma_{\rm e, min})$ 
(cf. \citealt{Katz_1994_spectrum}).
At the highest frequency, the electrons will have rapid cooling leading to a synchrotron spectral slope $-p/2$.
Depending on $\gamma_{\rm e, c}$, the spectrum will have different intermediate slope.
Let us define, $\nu_{\rm m}\equiv\nu_{\rm syn}(\gamma_{\rm e, min})$, $\nu_{\rm c}\equiv\nu_{\rm syn}(\gamma_{\rm e, c})$,
and the highest observed peak flux $F_{\rm \nu, max}$.

\begin{itemize}
 \item Case I: $\gamma_{\rm e, c}<\gamma_{\rm e, min}$: Fast cooling

All electrons above $\nu_{\rm c}$ cool rapidly. Thus the spectrum is
\begin{footnotesize}
 \begin{eqnarray}
\hspace{-0.5in}
\frac{F_{\nu}}{F_{\rm \nu, max}} \propto \left\{ \begin{array}{ll}
 {(\nu/\nu_{\rm c})^{1/3} }&\mbox{ $\nu < \nu_{\rm c}$} \\
 (\nu/\nu_{\rm c})^{-1/2} &\mbox{ $\nu_{\rm c}<\nu < \nu_{\rm m}$} \\
 (\nu_{\rm m}/\nu_{\rm c})^{-1/2}(\nu/\nu_{\rm m})^{-p/2} &\mbox{ $\nu > \nu_{\rm m}$}
       \end{array} \right.
\label{fast_cooling}
\end{eqnarray}
\end{footnotesize}

 \item Case II: $\gamma_{\rm e, c}>\gamma_{\rm e, min}$: Slow cooling

Only the highest energy electrons cool rapidly. Above $\gamma_{\rm e, min}$, the
synchrotron spectrum, generated by PL electrons (index p) has the slope $-(p-1)/2$
till $\nu_{\rm c}$

\begin{footnotesize}

\begin{eqnarray}
\hspace{-0.5in}
\frac{F_{\nu}}{F_{\rm \nu, max}} \propto \left\{ \begin{array}{ll}
 {(\nu/\nu_{\rm m})^{1/3} }&\mbox{ $\nu < \nu_{\rm m}$} \\
 (\nu/\nu_{\rm m})^{-(p-1)/2} &\mbox{ $\nu_{\rm m}<\nu < \nu_{\rm c}$} \\
 (\nu_{\rm c}/\nu_{\rm m})^{-(p-1)/2}(\nu/\nu_{\rm c})^{-p/2} &\mbox{ $\nu > \nu_{\rm c}$}
       \end{array} \right.
\label{slow_cooling}
\end{eqnarray}
\end{footnotesize}

\end{itemize}

In order to have high efficiency, and variable LC, GRBs are expected to be in the fast cooling regime (\citealt{Piran_1999_review}).
In addition, synchrotron self-absorption (SSA) can give a steeper slope in low frequency (radio) during the afterglow phase
(e.g., \citealt{Granotetal_1999_SSA}).
Another important contributor of the spectrum is Inverse Compton (IC). As the $\gamma_{\rm e}$ is high, only 
single episode events will occur due to the rapidly declining IC cross section at higher energies.
The Comptonization parameter, $Y\sim \epsilon_{\rm e}/\epsilon_{\rm B}$ for $\epsilon_{\rm e}\ll\epsilon_{\rm B}$,
or $\sim \sqrt{\epsilon_{\rm e}/\epsilon_{\rm B}}$ for $\epsilon_{\rm e}\gg\epsilon_{\rm B}$ for fast cooling
(\citealt{Sarietal_1996}). For $Y<1$, IC can be neglected. For $Y>1$, and typical values of $B$, $\Gamma_{\rm E}$ and
$\gamma_e$, the IC spectrum occurs at $\sim 10$ MeV. If this is outside the detector bandwidth, even then 
the IC can reduce the cooling timescale and affect the energy budget.

\subsection{Central Engine And Progenitor}
From the requirement of energetics, and the observational signature of achromatic LC break, the evidence of 
a GRB jet is strong. It is believed that a nascent central engine accretes the surrounding matter and 
channels it in the confined jet. However, this is just a speculation as the central engine of a GRB remains 
hidden from direct view. For example, we infer about the formation of a SN not only by observing the radiation 
from the shocked gas, but observation of a SN remnant (SNR) sometime reveals the presence of a pulsar at the centre of explosion.
Though in general a SN is not an engine driven explosion, but the presence of a pulsar gives important clues on 
the formation process. On the other hand, the inference for a GRB is not so strong. Observationally, the activity 
of a GRB engine is reflected in the emission processes. Based on the current understanding of the emission mechanisms, 
it is reasonable to assume that the prompt emission and early afterglow are directly related to the central engine, 
while the afterglow reflects the environment of the progenitor.

From the discussion of Section 1.5.1, we know that the time separation between events in the observer frame ($K$)
is related to that in the burst rest frame ($K_*$) as $dt=dt_*(1-\beta)=dt_*/D\Gamma \approx dt_*/2\Gamma^2$ (where Doppler 
factor, $D\equiv\left[\Gamma(1-\beta \mu)\right]^{-1}\approx2\Gamma$, for approaching gas). In other words, the time 
of a distant observer is ``compressed''. As the fireball moves with $\beta=v/c\approx 1$, the distance of IS region
expressed in terms of the $K_*$ time is $\delta r=c\delta t_*$, where we have replaced $dt_*$ with $\delta t_*$ to denote finite 
time interval. Hence for a distant observer, 

\begin{equation}
 \delta t_{\rm obs} \approx \frac{\delta r}{2c \Gamma^2}
\label{delta_t}
\end{equation}
 
But, from the discussion of IS, $\delta r$ expressed in terms of lab frame time is $\delta r\approx c\delta t_{\rm var}\Gamma^2$.
Hence, $\delta t_{obs}\approx\delta t_{\rm var}$. In other words, though there is a time compression 
in the observer frame, the variability and duration of a GRB directly denotes the activity of the central engine.
Hence, the central engine must satisfy certain properties which conform with the observed timescales.
(i) It should release $\sim 10^{50}-10^{52}$ erg energy (collimation-corrected) in 1\,ms to tens of sec. 
(ii) It should have a short dynamical timescale to account for the variability timescale.
Based on these criteria, the best candidates are BHs and NSs. In case of a LGRB, this is formed during collapse 
of a massive progenitor. The low metallicity environment, and the association with SN Ic's have made their 
case stronger. In all SN-GRBs, the calculated main sequence mass of the progenitor is found in the range 20-50 $M_{\odot}$ 
(e.g., \citealt{Larssonetal_2007, Raskinetal_2008}). Hence, the progenitor must be massive star, which forms a central 
engine (most probably a BH) during collapse. For SGRBs, the merging scenario is tentatively supported by 
the host properties, and the ``kilonova'' association. Gravitational wave (GW) signature from a few 
SGRBs in advanced LIGO and VIRGO may shed light on this matter.

An alternative suggestion of the central engine is a ``protomagnetar'' with a high magnetic field 
($\sim10^{14}-10^{15}$ G), and small spin period ($P\sim 1$ ms) at the birth (\citealt{Metzgeretal_2007,
Metzger_2010}; cf. \citealt{Usov_1992, Usov_1994}). A fraction of the rotational energy ($\sim2 \times 10^{52}$ erg) can be
made available through highly magnetized wind via interaction with the high magnetic field. 
The observational signature would be a highly polarized prompt emission.

\section{GRB Research}
This thesis primarily addresses the prompt emission of GRBs from a phenomenological point of view.
This topic is only a subset of the vast area of active research in GRB science. As the \textit{Swift} 
and the \textit{Fermi} satellites are in orbit, and providing a stream of prompt emission data, such a research is 
timely. In addition, we shall use the data of GeV and early afterglow emission to establish a coherent 
connection of the prompt emission with emission processes in other time and energy domain. Such an extensive 
data analysis may shed light on the ongoing research in this field. In the following we have listed some of 
the research topics in GRBs.

\begin{itemize}
 \item (i) \textit{Prompt emission mechanism and its connection with the afterglow:} With the large set of quality data available from 
the \textit{Swift} and the \textit{Fermi} satellites, it is probably the best time to extensively study the prompt emission properties in all 
possible combinations. The \textit{Swift}/XRT also provides useful data as a connection between the prompt and afterglow phase.
As we shall discuss in chapter 3, the data can be used in a clever way to extract as much information as possible.
Also, in order to pin down the physical mechanism, one has to study the time-resolved spectrum to the finest possible bin,
but also respecting the statistics. In chapter 5, we shall develop a new technique of spectral analysis in order 
to compare different prompt emission models.

 \item (ii) \textit{GeV emission and its connection with the prompt emission:} GeV emission in GRBs remains a puzzle.
As discussed earlier, GeV emission generally starts with a delay during the prompt phase, and shows a longer lasting 
behaviour. The flux evolution at the later phase shows afterglow features. Having the characteristics of both prompt and 
afterglow phase, GeV emission can give important clues for GRB physics in general. In some bursts with high GeV emission, 
a separate spectral component is required to describe the spectrum spanning the full keV to GeV band. However, in some 
bursts a Band only function is sufficient in this wide spectrum. Whether this difference is only due to lower statistics 
or for some unknown and more fundamental reason remains an open question. It is also not clear why the GeV emission is 
delayed than keV-MeV emission in some cases, and simultaneous in others. In addition, the poor correlation between 
the two emission components indicates that GRBs with similar keV-MeV brightness may or may not produce high GeV 
emission. The fundamental difference between these two classes of GRBs is not addressed. In chapter 6, we shall 
discuss these issues using a set of GRBs detected by the \textit{Fermi}/LAT.

 \item (iii) \textit{Progenitor and central engine:} As discussed in Section 1.5.3, the progenitor and the central 
engine of GRBs are only speculative, and do not have any direct observational evidence. The current research in this area 
is done in two ways: (a) a BH/magnetar is assumed to be formed, and a jet is launched (e.g., \citealt{Aloyetal_2000, 
Progaetal_2003, McKinney_2005, McKinney_2006, Proga_Zhang_2006, McKinney_Narayan_2007, Komissarov_Barkov_2007, 
Barkov_Komissarov_2008, Nagataki_2009}). The launching mechanism is assumed to be either a Blandford-Znajek process 
(\citealt{Blandford_Znajek_1977}), or via magnetic field interaction. (b) some independent simulations involve 
studying the collapser to form BH/magnetar (e.g., \citealt{MacFadyen_Woosley_1999, 
MacFadyenetal_2001, Fryeretal_2001, Hegeretal_2003, Woosley_Heger_2006, Takiwakietal_2009, Sekiguchi_Shibata_2011}). 

\item (iv) \textit{SN-GRB connection:} The GRB rate is found to be a tiny fraction of SN Ic rate.
This could be due to the fact that GRBs are collimated. However, \cite{Soderbergetal_2006} have found 
no evidence for off-axis event connected to SN Ic-BL, which disfavours the intuition that SN Ic-BL
are relativistic and engine driven. However, objects like mildly relativistic SN 2009bb 
(\citealt{Soderbergetal_2010_2009bb}) may provide an intermediate class. In this regard, it is also interesting to 
have objects like SN-less GRBs, which might be ``failed-SN'' (\citealt{Woosley_1993}). The study of 
GRB-SN opens up an opportunity to understand the mechanism of massive stellar death in general. 
However, as \textit{Swift} ``sees'' at higher $z$, the detection of these connection has become 
harder (\citealt{Woosley_Bloom_2006}).

\item (v) \textit{Correlation in prompt emission:} As discussed earlier, GRB correlations can give further 
constraints on GRB models. Also, these can be used to study GRBs as cosmological luminosity indicators.
In chapter 3, we shall discuss more about GRB correlations, and in particular, we shall propose a new correlation.

\item (vi) \textit{Study of chemical evolution of host galaxies:} As LGRBs are always associated with 
star forming irregular galaxies, they can be used to study the chemical properties of late type galaxies.
In fact, GRBs act like light beacons from distant universe to help in finding very distant faint galaxies, which 
could not have been detected otherwise. 

\item (vii) \textit{GRBs as tracers of cosmic star formation history:} As LGRBs are directly related 
with the death of the massive stars, they are important tracers of cosmic star formation.

\item (viii) \textit{SGRBs as sources of gravity wave:} Finally, SGRBs are the primary sources 
to study gravitational waves (GW) from a pair of inspiraling NSs. GWs are likely to be detected 
in the advanced LIGO and Virgo project in the next decade.

\end{itemize}

\section{Books And Review Articles }
Following is an inexhaustive list of books and review articles. 

\begin{itemize}
\item \textit{Books:} \cite{Katz_2002_book, book2003, book2004, book2005, book2009, Vedrenne_2009, Bloom_2011_book, new_book}
\item \textit{General reviews:} \cite{Higdon_Lingenfelter_1990, Paczynski_1991_review, Piran_1992_review,
Fishman_Meegan_1995, Zhang_Meszaros_2004, Meszaros_2006, Gehrels_Meszaros_2012}
\item \textit{Fireball model:} \cite{Piran_1999_review, Piran_2000_review, Piran_2004_review, Meszaros_2002_review, Meszaros_2006}
\item \textit{Prompt and Afterglow:} \cite{Paradijs_2000_review, Gaoetal_2013_review, Zhang_2014_review}
\item \textit{Swift and Fermi era:} \cite{Zhang_2007_review, Gehrelsetal_2009_review} (\textit{Swift}). 
\cite{Granotetal_2010, Gehrels_Razzaque_2013_review} (\textit{Fermi})
\item \textit{SN connection:} \cite{Woosley_Bloom_2006, Hjorth_Bloom_2012}
\item \textit{Progenitor:} \cite{Woosleyetal_2002_review, Woosley_Heger_2006}
\end{itemize}

\chapter{Instruments And Data Analysis} \label{ch2}

\section{Overview}
This chapter gives a brief overview of the instruments and data analysis procedure of the 
dedicated GRB satellites, namely \textit{Swift} and \textit{Fermi}. We shall briefly discuss the 
detectors, observations, data archive, reduction techniques,  software, and statistical methods 
used in the thesis. As a requirement of the prompt emission analysis, archival data provided by 
\textit{Swift}/Burst Alert Telescope (BAT) and \textit{Fermi}/Gamma-ray Burst Monitor (GBM) are 
extensively used in this study. In addition, soft x-ray data of \textit{Swift}/X-Ray Telescope (XRT), 
and very high energy $\gamma$-ray data of \textit{Fermi}/Large Area Telescope (LAT) are also used 
as required for data interpretation. The next two sections describe the \textit{Swift} and 
the \textit{Fermi} satellites, their instrument designs, and the detectors. In section~\ref{ch2_s4}, we 
discuss the data analysis technique, required softwares and the usage of statistics. Finally, 
section~\ref{ch2_s5} gives an overview of the science perspective of the two satellites.

\section{The \textit{Swift} Satellite}\label{ch2_s2}
The \textit{Swift} (\citealt{Gehrelsetal_2004_swift}) is a medium-sized explorer (MIDEX), launched 
by Delta 7320 launch vehicle of \textit{NASA} in Novemeber, 2004. The satellite is orbiting in a low earth 
orbit (LEO) at an altitude of about 600 km, and an inclination of $<22^{\circ}$. The satellite has well 
passed the nominal targeted mission life (2 year, orbital life $>5$ year), and continues to provide a wealth 
of data right from the prompt emission to the early afterglow phase. The \textit{Swift} carries three instruments:
Burst Alert Telescope (BAT), X-Ray Telescope (XRT), and UV-Optical Telescope (UVOT). The BAT is the primary GRB 
instrument with a large field of view (FOV), while the XRT and the UVOT are high precision focusing instruments. The main 
objective of the \textit{Swift} is to quickly localize the position of a burst, and \textit{swiftly} employ 
the focusing instruments to facilitate quicker and more accurate position measurement to be usable by 
ground based optical telescopes. To achieve this requirement the spacecraft platform (3-axis stabilized)
is built with an autonomous, and enhanced slew rate ($0^{\circ}$ to $50^{\circ}$ in $20-70$ s). 
Based on an automatic trigger and following the constraints to avoid sun, moon and earth limb, 
the Attitude Control System (ACS) automatically initiates the slewing, with a success of $\sim90\%$.
In the following, we shall discuss the BAT and the XRT, which are used to study prompt and early 
afterglow emission. 

\subsection{Burst Alert Telescope (BAT)}
The Burst Alert Telescope (BAT; \citealt{Barthelmyetal_2005_swift}) is a large field of view (FOV) instrument 
with a coded-aperture mask (CAM). The CAM is used to obtain a ``shadow'' of a point source, and thus reconstruct 
the source position with an arcmin accuracy. The main components of the BAT are: (i) detector plane with Detector Modules (DM), 
(ii) Block Controller and Data Handler (BCDH), (iii) Coded Aperture Mask (CAM), (iv) Image Processor (IP),
and (v) supporting components. In Table~\ref{ch2_t1}, the specification of the BAT instrument is shown.

\begin{table*}\centering
\caption{Specification of the \textit{Swift}/BAT instrument}

\begin{tabular}{l|l}
\hline
Parameter & Specification \\
\hline
\hline
Detectors & CdZnTe (CZT)\\
Individual Detector Dimension & $4.00~{\rm mm}\times4.00~{\rm mm}\times2.00~{\rm mm}$ \\
Detector Arrangement & Hierarchical \\
Detector Area & 5240 cm$^2$ \\
Effective Imaging Area & Max. $\sim1400$ cm$^2$ (on-axis) \\
Energy Range & 15-150 keV \\
Energy Resolution & $7$ keV (average FWHM)\\
Timing Resolution & 100 $\mu s$ \\
\hline
Instrument Dimension & $2.4~{\rm m}\times1.2~{\rm m}\times1.2~{\rm m}$\\
Telescope Type & Coded Aperture Mask (50\% open) \\
Coded Mask Cell & $5.00~{\rm mm}\times5.00~{\rm mm}\times1.00~{\rm mm}$ (Pb Tiles) \\
Field of view & 1.4 sr ($50\%$ coded)\\
Telescope PSF & 17 arcmin (FWHM) \\
Position Accuracy & 1-4 arcmin \\
Sensitivity & $\sim2\times10^{-10}$ erg cm$^{-2}$ s$^{-1}$ $\left(\frac{T}{20~{\rm ks}}\right)^{-0.5}$ ($5\sigma$)\\
\hline
Operation & Photon counting \\
\hline
\end{tabular}
\label{ch2_t1}
\end{table*}

\subsubsection{I. Detector Array Plane (DAP), Detector Modules (DM) and Blocks}
The detectors of the BAT are 32,768 pixels of CdZnTe (CZT), each with $4.00~{\rm mm}\times4.00~{\rm mm}$ 
area and $2.00$ mm thickness. The detector area is 5240 cm$^2$. The effective area of the BAT is 
maximally $\sim1400$ cm$^2$ (on-axis) in a range 30-80 keV. The energy range where the effective 
area is at least 50\% of the peak value is 15-150 keV. The lower end of the energy range is 
determined by the level of electronic discriminator, while the upper end is determined 
both by the level of transparency of the CAM and the limit of CZT. For fabrication purpose, electronic 
control, and data handling, the CZT pixels are arranged in a hierarchical manner. A unit of $16\times16$ 
CZT pixels makes a detector module (DM). A unit of $4\times2$ DMs makes a Block. 16 
such Blocks are mounted in a $2\times8$ configuration to make the Detector Array Plane (DAP). 
The CZT pixels are mounted with a pitch of 4.2 mm, and the gaps between the DMs and Blocks are 
adjusted to be integral multiples of the pixel pitch so that these are easily handled by the 
image reconstruction process.

Though the pixels are the basic detectors of the BAT, the commercial unit is a DM.
Each DM has two Application Specification IC or ASIC (called XA1) controlling half of the pixels
($8\times16$) of a DM, called Sandwich. The pixel elements of the BAT have planar electrodes, with 
a typical bias voltage -200 V. The bias voltage is commandable for each DM in (0 to -300 V). The anode
of each pixel of a Sandwich is AC-coupled to XA1 ASIC. Each ASIC has 128 channels (for each of the 
$8\times16$ pixels) of charge-sensitive pre-amplifier (CSPA), shaping amplifiers, and discriminators.
An ASIC is a self-triggering device. Each channel can be individually commanded, and disabled to handle 
noisy pixels. The ASIC recognizes an event (over a supplied threshold) from one of the 
128 input channels, and temporarily blocks the other channels. The pulse height of the corresponding 
event is digitized (0.5 keV quantization) by an ADC unit coupled to the ASIC. The information of 
pulse height and detector number of each event is then transmitted to a Block Controller and Data 
Handler (BCDH). The full process takes 100 $\mu s$. The XA1 ASIC is linear up to $\sim 200$ keV, 
which is well above the energy range of the BAT (up to 150 keV). For calibration (offset, gain and 
linearity), each DM contains a commandable electronic calibration pulser circuit. It generates 
charge pulses of specified number and level into each channel, when commanded. In addition, 
two $\alpha-$ tagged Am241 sources (60 keV photon) are used for absolute calibration of energy 
scale and efficiency of the individual CZT detectors. The calibration events are flagged 
to separate out from the actual events. 

\subsubsection{II. Coded Aperture Mask (CAM)}
Due to the requirement of a good position measurement along with a large FOV, the BAT uses coded 
mask technique, and reconstruct the point source position from the shadow pattern. The Coded 
Aperture Mask (CAM) is a $2.4~{\rm m}\times 1.2~{\rm m}$ Pb tile mounted in a light honeycomb panel at a height 
of 1 m above the detector plane. There are a total of $\sim52,000$ mask elements, each with a size of
$5.00~{\rm mm}\times5.00~{\rm mm}\times1.00~{\rm mm}$. The mask has a completely random pattern of 
50\% open and 50\% closed filling factor. This configuration provides a $100^{\circ}$ 
by $60^{\circ}$ (half-coded) FOV, and a Point Spread Function (PSF) $\sim17$ arcmin (FWHM).

\subsubsection{III. Supporting Units And Data Processors}
The BAT contains the following supporting units. In order to reduce the background 
the space below DAP, and between DAP and CAM is shielded with four layers of Pb, 
Ta, Sn and Cu, called graded-Z shield. It reduces $\sim95\%$ of the background due to
cosmic diffused x-ray, and earth albedo. To regulate the operating temperature 
($20^{\circ}$ desired) and a low spatial and temporal thermal gradients (maximally 
$0.5^{\circ}$) a thermal control unit is used. 

For data handling, the BAT uses two steps. The data of a block is handled by Block Controller and Data 
Handler (BCDH). The mechanical structure of the electronics unit along with the Block sits on the 
DAP with 8 DMs. It multiplexes a single serial data stream containing the information of each photon 
event, flagged calibration data, an identification number of DM, and a time tag (100 $\mu s$ quantization) 
of each event. BCDH acts as a data concentrator to reduce the burden of the highest level data handler, 
Image Processor (IP). The IP of the BAT does the major tasks for event analysis and GRB trigger. The major 
tasks are: (i) analyzing the event data in order to get an indication of a burst, (ii) constructing 
a sky image for such indication, and scan for new source, (iii) determining the Figure Of Merit (FOM)
to decide for a slew, (iv) accumulating Detector Plane Histograms (DPHs) for survey mode, (v) gathering house 
keeping (HK) information, (vi) handling and processing of commands from spacecraft, (vii) sending 
telemetry data from the BAT to the spacecraft. 

\subsubsection{IV. Operation And Burst Detection}
The BAT works in two modes of operation: hard x-ray survey mode and burst mode. In the survey mode spectral 
data from each pixel is collected in every $\sim5$ minutes, and the Blocks are periodically calibrated.
The BAT performs an all-sky hard x-ray survey with a sensitivity of $\sim1$ mCrab (\citealt{Tuelleretal_2010}).
Detection of a GRB in the BAT is determined by certain trigger algorithms. The algorithm looks for excess 
counts in the detector compared to that from constant sources and background. Note that in a low earth 
orbit, the background is variable by a factor of two in an orbit (90 minutes). This is the main obstacle 
for a large FOV instrument like the BAT. Though the graded-Z shield helps reducing the background by a large factor,
still a typical background rate is $\approx10,000-12,000$ counts per sec (1 Crab $\sim10\%$ of the background).
In addition, the durations of GRBs have a very wide range of values. Hence, the trigger algorithm must be able
to correctly extrapolate the background to compare it with the event rate with a variety of timescales and 
in many energy bands.

The trigger algorithm of the BAT involves two level of testing. The first test is based on the excess count rate 
over the background. The algorithm continuously applies a number of criteria that specify the following: 
(i) pre-burst background intervals (0-100 sec), (ii) degree of polynomial for background extrapolation,
(iii) duration of the burst emission test interval, (iv) illuminated portion of the detector plane, and 
(v) the energy range. Apart from the Rate Trigger algorithm, an image construction test (Image Trigger) 
is applied. The image reconstruction involves an FFT-based cross-correlation of the count rate array in the detectors 
and the CAM pattern. This employs ray-tracing of all possible pattern of shadow due to all possible source 
location in the sky. Each shadow pattern is multiplied with the detected counts, and results are summed. The source 
location is found by noting an excess in the correlation value. In every 64 sec, the count rate map 
in the detector array is processed through this algorithm and searched for a new source by comparing 
against an on-board catalogue. 

The algorithm takes 7 s to generate $1024\times512$ pixel image with the location of the transient 
source. the position of the excess in the image is found within a single 17-arcmin sky pixel. Depending 
on the significance, the BAT checks for subsequent increase in rate to make a stronger image. With the 
approximate location, a back-projection algorithm is employed to produce the image with typically 
1 arcmin pixel size. From the centroid of the peak the source location can be determined with 
1-3 arcmin accuracy, depending on the burst intensity. \cite{Tuelleretal_2010} provides an empirical 
source localization error of the BAT as 

\begin{equation}
 {\rm Error~radius} = \sqrt{\left[ \frac{30}{S/N-1}\right]^{2}+[0.25]^2}
\end{equation}
\vspace{0.1in}
 
where 0.25 arcmin is added to account for the systematic error. From the constructed image, 
a Figure of Merit (FOM) algorithm decides whether the source is worth for a slew maneuver 
of the spacecraft.

\begin{table}\centering
\caption{Specification of the \textit{Swift}/XRT instrument}

\begin{tabular}{l|l}
\hline
Parameter & Specification \\
\hline
\hline
Telescope Type & Wolter I, 3.5m focal length, 12 shells  \\
Field of view & $23.6\times23.6$ arcmin\\
Telescope PSF & 18 arcsec HPD at 1.5 keV \\
Position Accuracy & 3 arcsec \\
Sensitivity & $2\times10^{-14}$ erg cm$^{-2}$ s$^{-1}$ (in $10^4$ s) \\
\hline
Detectors & e2v CCD-22\\
Effective Area & $\sim125$ cm$^2$ \\
Detector Format & $600\times600$ pixels \\
Pixel Size & $40\mu{\rm m} \times 40\mu{\rm m}$ \\
Pixel Scale & 2.3 arcsec/pixel \\
Energy Range & 0.2-10 keV \\
Energy Resolution & $140$ eV at 5.9 keV (at launch)\\
Timing Resolution & 100 $\mu s$ \\
\hline
Operation & Autonomous \\
Readout Modes & Image (IM) Mode \\
              & Photodiode (PD) Model \\
              & Windowed Timing (WT) Mode \\
              & Photon-Counting (PC) Mode \\
\hline
\end{tabular}
\label{ch2_t2}
\end{table}

\subsection{X-Ray Telescope (XRT)}
The X-Ray Telescope (XRT; \citealt{Burrowsetal_2005_XRT}) of the \textit{Swift} is a focusing soft x-ray telescope operating in a 
range 0.2-10 keV. There are 12 gold-coated confocal shells, arranged in a grazing incidence Wolter I geometry to focus 
soft x-rays at a focal length of 3.5 m. The FOV is $23.6\times23.6$ arcmin, and the PSF is 18 arcsec Half-Power 
Diameter (HPD) at 1.5 keV. The instrument can localize a point source with 3 arcsec accuracy. 
The detectors are thermoelectrically cooled \textit{XMM-Newton}/EPIC MOS CCD (e2v CCD-22). 
In the following, we shall briefly discuss the essential features of the XRT.

\subsubsection{I. Requirements}
The design of the XRT is driven by the requirements of (i) rapid, and accurate position determination
(At least 5 arcsec accuracy within $\sim100$ s of the BAT trigger), (ii) lightcurve with high time 
resolution, and (iii) spectrum with moderate resolution. The mirror of the XRT focusing system 
has a PSF of 15 arcsec HPD. To get a uniform PSF in the entire FOV, it is slightly defocused. The 
instrument PSF is 18 arcsec HPD at 1.5 keV. The late prompt emission flux of a GRB in 0.2-10 keV 
is likely to be in the range of 0.5-5 Crabs. The XRT can localize such a source in 1-3 arcsec (better 
for a brighter source) within 5 s of target acquisition. To minimize the alignment uncertainty of 
the XRT with star tracker the trackers are mounted on the XRT forward telescope tube, while a Telescope 
Alignment Monitor (TAM) is used to measure the flexing of the tube with sub-arcsec accuracy.

The \emph{Swift} requires to provide photometric data with 10 ms time resolution. Two modes of operation are 
designed for this purpose. (i) In photodiode (PD) mode 0.14 ms accuracy is provided by integrating 
the count rate of the entire CCD. No spatial information is provided in this mode. This mode is 
suitable for uncrowded field photometry, and can measure a source brightness up to 65 Crabs. (ii)
Windowed Timing (WT) mode provides data with 2 ms time resolution and 1-D spatial resolution.
This mode is similar to the corresponding mode in \textit{Chandra}/ACIS and \textit{XMM}/EPIC 
MOS camera. The typical flux error in lightcurve is 10\%. In addition to the position and timing 
data, the XRT provides reasonable spectrum in 0.2-10 keV energy band with $140$ eV at 5.9 keV resolution
(at the time of launch). The readout modes are designed for a maximum flux of $\sim 6\times10^{-8}$ 
erg cm$^{-2}$ s$^{-1}$ (0.2-10 keV). For brighter source, the pile up effect requires great caution 
in the region extraction. Table~\ref{ch2_t2} shows the important specifications of the XRT instrument.

\begin{table*}\centering
\caption{Specification of the \textit{Swift}/UVOT instrument}

\begin{tabular}{l|l}
\hline
Parameter & Specification \\
\hline
\hline
Telescope Type & Ritchie-Cretien  \\
Telescope Diameter & 30 cm \\
Field of view & $17.0\times17.0$ arcmin$^2$\\
Telescope PSF & 0.9 arcsec FWHM at 350nm \\
f-number & 12.7 \\
Filters & 11 \\
Position Accuracy & 0.3 arcseconds ($2\sigma$) \\
Sensitivity & 24th magnitude (in 1000 s) \\
Maximum Source Brightness & 8th magnitude \\
\hline
Detectors & Microchannel-intensified CCD\\
Detector Format & $2048\times2048$ pixels \\
Pixel Scale & 0.5 arcsec/pixel \\
Spectral Range & $170 - 600$ nm \\
Timing Resolution & 11 milliseconds \\
\hline
Operation & Autonomous \\
\hline
\end{tabular}
\label{ch2_t3}
\end{table*}

\subsubsection{II. Structural Specification}
The structure of the XRT can be divided into three parts: (i) an aluminium \textit{Optical Bench Interface Flange}
(OBIF), (ii) \textit{Telescope Tube}, (iii) \textit{Door}. OBIF is the primary structural element.
It supports the telescope tubes, mirror module, electron deflector, TAM optics and camera. Telescope 
tube is a graphite fiber/cyanate ester tube with 508 mm diameter. The graphite fiber is chosen in order 
to minimize temperature gradient and thus to preserve the focus and alignment. The composite material 
is lined with aluminium foil vapour barrier to guard the interior from outgassing and epoxy contamination.
The telescope tube has two parts: the forward telescope encloses the mirrors and supports the star 
trackers, while the aft telescope supports the Focal Plane Camera Assembly (FPCA). A door, attached to the 
forward telescope, protects the x-ray mirrors from any contamination.

\subsubsection{III. Optics}
The XRT mirror assembly consists of x-ray mirror module, a thermal baffle, spacers, and 
an electron deflector. The mirror module contains 12 concentric gold-coated electroformed Ni shells
with 600 mm length and 191-300 mm diameters. As discussed, the arrangement is deliberately 
made slightly defocused to get uniform PSF in the FOV. A thermal baffle, placed in front 
of the mirrors, prevents temperature gradients. A electron deflector, having 12 rare
earth magnets, is placed behind the rear side of mirror assembly to prevent background 
electrons from reaching the CCD. 

\subsubsection{IV. Focal Plane Camera Assembly (FPCA)}
The FPCA provides a vacuum interior for the CCD detector and blocking filter, radiation shield against 
trapped particles, and cool environment. FPCA has a sun shutter to provide safety for the CCD and the 
filter from accidental solar illumination in case of attitude control failure. For calibration, two 
sets of $^{55}$Fe sources (5.9 keV and 6.5 keV) are used. 

The CCD of the XRT is designed by e2v and named CCD-22. The energy band is 0.2-10 keV with resolution  
$140$ eV at 5.9 keV. Each CCD contains $600\times600$ pixels of $40\mu{\rm m} \times 40\mu{\rm m}$
size. Each pixel corresponds to 2.3 arcsec in the focal plane. the quantum efficiency of CCD 
is calculated with Monte Carlo code, and it is used to obtain the spectral response. In a low 
earth orbit, the CCD is most likely to be degraded by high proton flux. The protons generate 
electron traps in the silicon lattice which degrade the energy resolution over time. The degraded 
resolution can reach $\sim300$ eV at 6 keV in 3-year time.

\subsection{UV-Optical Telescope (UVOT)}
The UV-Optical Telescope (UVOT; \citealt{Romingetal_2005}) has a modified (30 cm) Ritchey-Chr\'{e}tien design 
with Microchannel-intensified CCD detectors operating in a range $170-600$ nm with good time resolution (11 milliseconds)
and position accuracy (0.3 arcseconds, $2\sigma$). The specification of the instrument are shown in Table~\ref{ch2_t3}.
The UVOT is designed to study the afterglow data in UV and Optical band from as early as the XRT observation.

\section{The \textit{Fermi} Satellite}\label{ch2_s3}
The \textit{Fermi} satellite was launched on June 11, 2008 in a low earth orbit (565 km, $25.6^{\circ}$ inclination).
The detectors onboard this satellite are successors of the BATSE and the EGRET instruments of the CGRO. The detectors have 
extended the spectroscopic capability of the previous instruments by many orders. The \textit{Fermi} contains two 
instruments: Gamma-ray Burst Monitor (GBM), and Large Area Telescope (LAT). The GBM provides the most useful spectral 
data of the prompt emission in a wide keV-MeV energy band. The high energy detectors of the GBM surpass the energy 
coverage of the spectroscopic detector of the BATSE. On the other hand, the LAT has a wider band, and a factor of 
5 larger effective area than the EGRET of CGRO. In the following, we shall discuss the essential features of 
both these instruments.

\begin{table*}\centering
\caption{Specification of the \textit{Fermi}/GBM instrument}

\begin{tabular}{l|l|l}
\hline
Parameter & Low-Energy Detector & High-Energy Detector \\
\hline
\hline
Material & NaI & BGO  \\
Number & 12 & 2 \\
Area & 126 cm$^2$ & 126 cm$^2$ \\
Thickness & 1.27 cm & 12.7 cm \\
Energy Range & 8 keV to 1 MeV & 150 keV to 30 MeV\\
Energy Resolution & 12\% FWHM at 511 keV & Same \\
Time Resolution & 2$\mu$s & Same \\ 
Field of View & 9.5 sr & Same \\
On-board Localization & $\sim 10^{\circ}$ in 1 s & --- \\
\hline
\end{tabular}
\label{ch2_t4}
\end{table*}

\subsection{Gamma-ray Burst Monitor (GBM)}
The Gamma-ray Burst Monitor (GBM; \citealt{Meeganetal_2009_fermi_gbm}) is the major dedicated instrument for the prompt 
emission spectroscopy of GRBs. It contains two types of scintillation detectors: sodium iodide 
(NaI) and bismuth germanate (BGO). In Table~\ref{ch2_t4}, the parameters of the detectors are given.

\subsubsection{I. Scientific Requirements}
The primary scientific requirement of the GBM is to provide a wide energy band in the keV-MeV range.
The secondary objective is to obtain a rough location of the burst as quickly as possible. This feature 
is used to re-point the very high energy detector (LAT) to observe the delayed emission in GeV energies.
The GBM can obtain a GRB position with $\sim 10^{\circ}$ uncertainty in 1 s. The software on-board the GBM 
performs several trigger algorithms. If a trigger occurs, the information is sent 
to the LAT and to the ground in real time. For particularly strong bursts the spacecraft is oriented to employ 
the LAT for $\sim2.5-5$ hr. The GBM also gets triggered on solar flares, soft gamma repeaters (SGRs), and 
terrestrial gamma flashes (TGFs). While not in trigger mode, the GBM acquires background data in 
hard x-ray and provides the useful information for other guest observations. Using the background data 
variable x-ray sources are monitored by earth occultation technique which was previously done using the BATSE.

\subsubsection{II. Detectors and Data system}
The GBM contains 12 NaI and 2 BGO detectors. The NaI detectors are Tl activated. They cover the lower 
energy part, 8 keV to 1 MeV. Each detector is made in the form of a disk of 12.7 cm diameter and 
1.27 cm thickness. As NaI is hygroscopic, the crystals are packed inside a hermetically sealed 
light-tight aluminum housing with a glass window (thickness 0.6 cm). The glass is attached to the 
aluminum housing by white Araldite. The entrance window is 0.2 mm thick Be sheet. For mechanical 
reason a silicone layer of 0.7 mm thickness is placed at the front side of the crystals. This
determines the lower limit of the energy band. To increase the light output, Tetratec and Teflon 
materials are used to cover the crystals from front window and circumference, respectively.
The NaI detectors are placed in various axes. This configuration is used to calculate an 
approximate direction of a burst from the ratio of observed flux in different detectors. 

The BGO detectors are placed on two opposite sides of the spacecraft. The detectors cover the 
higher energy part of the GBM, $\sim200$ keV to $\sim40$ MeV, which overlaps with the energy bands 
of both the NaI detectors (in the lower part) and the LAT (in the higher part). The crystals have diameter of 12.7 cm, 
and thickness of 12.7 cm. These are polished to mirror quality on the circular glass side
window. The cylindrical surface is roughened. This is done to get diffuse reflection of the 
generated photons. The BGO crystals are packed inside a carbon-fibre reinforced plastic 
housing, held by titanium rings on both sides. The ring material is chosen based on its 
similar thermal expansion coefficient as the BGO crystal. This arrangement provides light 
tightness and mechanical stability. The rings act as holders for the two PMTs used on either 
side of the crystals. PMTs with commandable high voltage (735-1243 V) are used to collect 
scintillation light from both types of crystals. The pulses detected by PMT are fed to 
Front End Electronics (FEE), which amplifies, shapes and sends the pulse to Data Processing 
Unit (DPU). The data are sent to the ground, where they are packaged in FITS format.

\begin{table*}\centering
\caption{Specification of the \textit{Fermi}/LAT instrument}

\begin{tabular}{l|l}
\hline
Parameter & Specification \\
\hline
\hline
Telescope Type & Pair-conversion Telescope \\
Detector Effective Area & 9500 cm$^2$ \\
Energy Range & 20 MeV - 300 GeV \\
Energy Resolution & 9\%-15\% (100 MeV - 1 GeV, on axis) \\
(Gaussian $1\sigma$) & 8\%-9\% (1 GeV - 10 GeV, on axis)\\
 & 8.5\%-18\% (10 GeV - 300 GeV, on axis)\\
 & $\leq 6\% $($>10$ GeV, $>60^{\circ}$ incidence)\\
Angular resolution & $\leq0.15^{\circ}$ ($>10$ GeV) \\
(Single photon) & $0.6^{\circ}$ (1 GeV) \\
                & $3.5^{\circ}$ (100 MeV) \\
Field of View   & 2.4 sr \\
Time Resolution & $<10\mu$s \\
On-board GRB location accuracy & $0.1^{\circ}-0.5^{\circ}$ \\
GRB notification time to spacecraft & $<5$ s \\
Position Accuracy (point source) & $<0.5'$ \\
Sensitivity (point source) & $3\times10^{-9}$ photon cm$^{-2}$ s$^{-1}$ \\
\hline
\end{tabular}
\label{ch2_t5}
\end{table*}

\subsection{Large Area Telescope (LAT)}
The Large Area Telescope (LAT; \citealt{Atwoodetal_2009_LAT}) of the \textit{Fermi} is designed to 
detect $\gamma$-rays at very high energy (GeV) by using pair-production method. 
Compared to the EGRET (20 MeV-30 GeV) of CGRO, the LAT has much wider energy coverage 
(20 MeV-300 GeV). More importantly, the EGRET has an effective area 1500 cm$^2$ till 
1 GeV, and lower at higher energies. In comparison, the LAT has an effective area is 
9500 cm$^2$ throughout. In Table~\ref{ch2_t5}, the features of the LAT are shown. 
 
\subsubsection{I. Scientific Requirements}
The LAT is designed in accordance with the following requirements. (i) It should 
have a large FOV in order to cover a large sky. (ii) It should quickly localize GRBs 
with good accuracy. For other sources, a better localization accuracy is required. 
(iii) The LAT should provide a good effective area in a large energy band. (iv) Specifically
for GRBs, the LAT should be able to measure $\gamma$-rays over a short time interval.
(v) To observe persistent sources over many years, the LAT instrument should not degrade.
(vi) It should correctly reject most of signals generated by cosmic ray particles, 
which can mask the low signal from a $\gamma$-ray source.

\subsubsection{II. Technique For Detection of $\gamma$-rays}
There are four subsystems in the LAT which work together for $\gamma$-ray detection and the 
rejection of cosmic ray particles. The subsystems are: (i) Tracker, (ii) Calorimeter, 
(iii) Anticoincidence Detector (ACD), (iv) Data Acquisition System (DAQ). The procedure is as follows.

\begin{enumerate}
\item When a $\gamma$-ray enters, it does not produce any signal in the ACD.
It directly passes to the pair-converter. 
\item The pair-converter interacts to form pairs. 
\item The tracker measures the path of each pair, and thus helps in determining the photon arrival direction.
\item The energy of the pairs are measured in the calorimeter to get the energy of the incident $\gamma$-ray.
\item Unwanted cosmic ray particles are rejected by DAQ on the basis of their signal in the ACD. This procedure 
rejects $\approx99.97\%$ of the signal. The DAQ also rejects the unwanted photons from non-source on the basis 
of the arrival direction.

\end{enumerate}

\subsubsection{III. Detector and Instruments}

The subsystems of the LAT are described in the following.

\begin{itemize}
\item \textit{(i) Tracker:} It consists of a $4\times4$ array of tower modules. Each of the
tower modules contains 16 planes of silicon-strip detectors (SSDs) with interleaved tungsten 
converter foils. The tungsten foils convert the incident $\gamma$-ray photons into electron-positron pairs.
The SSDs act as particle trackers. The signature of pair conversion also helps in rejecting 
background due to cosmic ray particles. The path of the produced pairs is used to reconstruct 
the source location. the PSF is limited by multiple scattering of pairs and bremsstrahlung 
production. To get an optimal result, the SSDs have high efficiency and placed very close 
to the converter foils. One of the important design aspects of the conversion foils is dictated
by the trade off between thin foil for good PSF at lower energy and thick foil for maximizing 
the effective area at higher energies. To achieve a resolution, the ``front'' 12 planes have
thin tungsten each with a thickness of 0.03 radiation length, while the ``back'' 4 foils are 
$\sim6$ times thicker. The high thickness of foils costs the angular resolution by less than 
a factor of two (at 1 GeV). 

\item \textit{(ii) Calorimeter:} The converted pairs pass through the calorimeter which is a 
CsI scintillation detector. The flash of light produced by the pairs is detected by a PMT which 
generates a characteristic pulse. The pulse height measures the energy of the pairs. The cosmic 
ray particles are vetoed on the basis of their different pulse shape.

\item \textit{(iii) Anticoincidence Detector (ACD):} The ACD of the LAT consists of specially formulated
plastic tiles on top of the tracker. It  helps in reducing the cosmic ray background.
While a $\gamma$-ray does not produce a signal in passing through ACD, the cosmic ray particles due 
to their charge produces a signal in the ACD. One of the major improvement of the LAT ACD over the 
older EGRET is its ability to retain high efficiency against backsplash. High energy $\gamma$-ray
($\sim10$ GeV) produces electromagnetic shower in the calorimeter. These shower particles can hit 
the ACD creating signal of ``false'' cosmic ray. The ``good'' $\gamma$-ray events, which would have 
been accepted otherwise, are rejected by the backsplash effect. EGRET suffered about $50\%$ efficiency 
degradation (at 10 GeV, compared to 1 GeV) due to backsplash effect. The LAT uses position sensitive 
segmented ACD to ignore hits which are far from the reconstructed entry point. The size and thickness 
of the ACD segments are optimized by simulation.

\item \textit{(iv) Data Acquisition System (DAQ):} The information of the detected signals from all the 
regions are processed in the Data Acquisition System (DAQ) of the LAT. The DAQ makes the distinction between 
the cosmic ray events and the real events, it finds the source location and relays the information 
of ``real'' $\gamma$-ray events to the ground. The system also performs an on-board search for GRBs.

\end{itemize}

\section{Data Analaysis}\label{ch2_s4}
In the following, we shall describe the essential steps for the data reduction.
A step by step analysis procedure for all the instruments are provided in \textit{http://grbworkshop.wikidot.com/}.
\subsection{\textit{Swift}/BAT}
The BAT data is accessible from \textit{http://heasarc.gsfc.nasa.gov/cgi-bin/W3Browse/swift.pl}, or from
\textit{http://swift.gsfc.nasa.gov/cgi-bin/sdc/ql?}, for a GRB which is less than 7 day old.
The BAT data contains information of the event, house keeping data, and some non-GRB products e.g., 
survey data, rate data. The event folder contains the most relevant data for GRB analysis. Note that 
the BAT instrument sends raw data to the ground which contains all the events from source, particle 
background, bad/noisy detectors. The user requires to perform ``mask weighting'' to extract the true 
events. The relevant procedure for time-resolved spectral extraction are as follows.

\begin{enumerate}
\item \textit{Energy calibration:} It is a good idea to calibrate the provided event file with the latest 
calibration database (CALDB). The task is {\tt bateconvert}. 
\item \textit{Creating DPI:} The calibrated event file is used to form detector plane image (DPI). The 
task is {\tt batbinevt}.
\item \textit{Known Problematic Detectors:} The known problematic detectors are retrieved from the 
information provided in house keeping data. The task is {\tt batdetmask}.
\item \textit{Noisy Pixels:} Using the task {\tt bathotpix}, the noisy pixels are found.
\item \textit{Mask Weighting:} The fundamental operation for the timing and spectral analysis is 
mask weighting of the detector count. It is done by the task {\tt batmaskwtevt}. It also generates an 
auxiliary file for ray-tracing.
\item \textit{Spectrum:} Spectrum, in a specified time interval, can be extracted using the task {\tt batbinevt}. 
This task can be used for timing analysis as well.
\item \textit{Correction:} The spectrum is corrected for the ray-tracing using {\tt batupdatephakw}. The 
known systematic errors are added using {\tt batphasyserr}.
\item \textit{Response file:} The response matrix is created from the spectral file to be usable by {\tt XSPEC}. 
The task is {\tt batdrmgen}. Response is created for each time interval during a time-resolved spectroscopy.

\end{enumerate}
 
\subsection{\textit{Swift}/XRT}
The XRT data is accessed from \textit{http://heasarc.gsfc.nasa.gov/cgi-bin/W3Browse/swift.pl}.
For new bursts the quicklook site is: \textit{http://swift.gsfc.nasa.gov/cgi-bin/sdc/ql}.
The XRT has four modes of operation. These are shown in Table~\ref{ch2_t2}. The specification of the 
modes are as follows:
\begin{itemize}
\item \textit{Imaging (IM) Mode:} Imaging data. Exposure 0.1-2.5 s, No spectroscopic data.
\item \textit{Photodiode (PD) Mode:} No spatial information, high time resolution (0.14 ms).
\item \textit{Windowed Timing (WT) Mode:} 1-D imaging, 1.7 ms time resolution, spectral data.
\item \textit{Photon Counting (PC) Mode:} 2.5 s time resolution, spectral data.
\end{itemize}

As we are interested in the XRT spectrum, the WT and PC modes are the relevant data types for our 
purpose. We shall discuss mainly WT data analysis as this mode is used for early afterglow phase which we 
are interested in. The procedure is as follows.

\begin{enumerate}
\item \textit{Pipeline:} The XRT data is available in three levels. The Level 1 data products are 
directly obtained by converting the telemetry data into FITS file. Hence, no data is lost. 
For change of mode, data is lost in that time span in the telemetry itself. For PD and WT mode, 
there is an intermediate Level 1a, which converts the frame time to proper arrival time, and assigns 
grade and PHA values to the events. The Level 1 data are calibrated and screened through a standard 
screening to get Level 2 data. User can choose either Level 1, 1a, or 2 according as the need. There 
is a pipeline script, named {\tt xrtpipeline}, in which several parameters can be set by the user for 
the processing. 

\item \textit{Xselect:} The next step is to use {\tt xselect} to extract image, lightcurve and spectra
with desired specifications. 

\item \textit{Pile-up:} For high count rate events more than one photon can hit a single or adjacent 
pixel before the charge is read out. This is called pile-up effect, and it affects the spectrum. The 
pile-up is handled by using annular extraction region, the radius of which is determined by fitting 
PSF profile (count/sq arcmin/s) with radius (arcsec) with ``King function''. The pile-up correction 
is important mainly for PC mode, however, WT data can have pile-up effect (\citealt{Evansetal_2009}).

\item \textit{Exposure Map:} Position of a source over bad column leads to loss of flux. This is 
corrected by using {\tt xrtexpomap} task. With this map, {\tt xrtmkarf} task can produce the Ancillary 
response file. The RMF is located in the CALDB.

\end{enumerate}
 
\subsection{\textit{Fermi}/GBM}

\begin{table*}\centering
\caption{Specification of the \textit{Fermi}/GBM data type}

\begin{tabular}{cccc}
\hline
Data Type & Purpose & Energy Resolution & Time Resolution \\
\hline
\hline
CSPEC & Continuous high & 128 channels      & Nominal: 4.096 s \\
      &  spectral resolution                &              & Burst: 1.024 s \\
      &                                     &              & Adjustable: 1.024-32.768 s \\

\hline
CTIME & Continuous high &  8 channels   & Nominal: 0.256 s \\
      &  time resolution                &              & Burst: 0.064 s \\
      &                                 &              & Adjustable: 0.064-1.024 s \\
\hline
TTE   & Time-tagged events &  128 channels   & $2\mu$s from -30 to 300 s \\
\hline
\end{tabular}
\label{ch2_t6}
\end{table*}

\begin{figure}\centering
\includegraphics[width=3.0in]{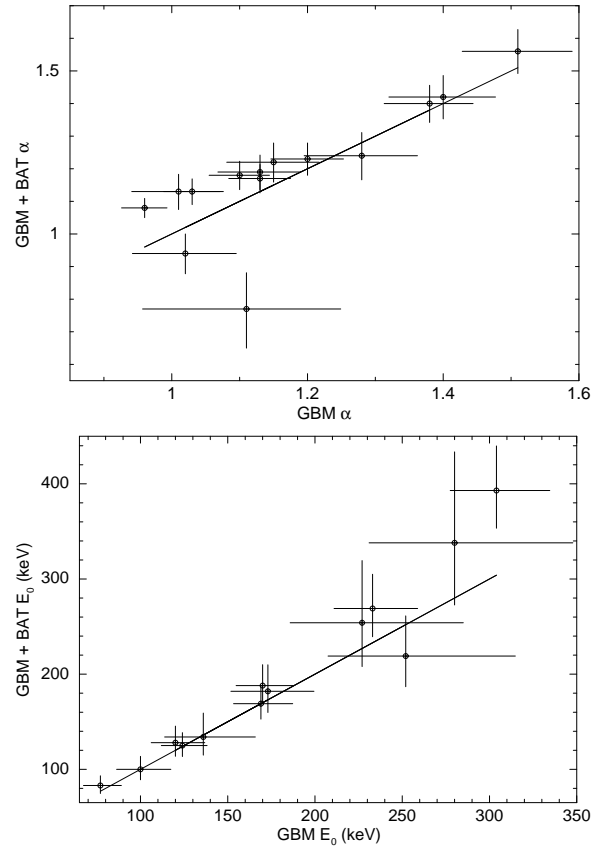}
\caption[Parameters for joint spectral fitting as compared to fitting only the GBM data]
{Parameters for joint spectral fitting as compared to fitting only the GBM data. \textit{Upper panel:} parameter
$\alpha$, \textit{Lower panel:} parameter $E_0$. Source: \cite{Basak_Rao_2012_090618}.}
\label{ch2_fig1}
\end{figure}

\begin{figure}\centering
\includegraphics[width=3.0in]{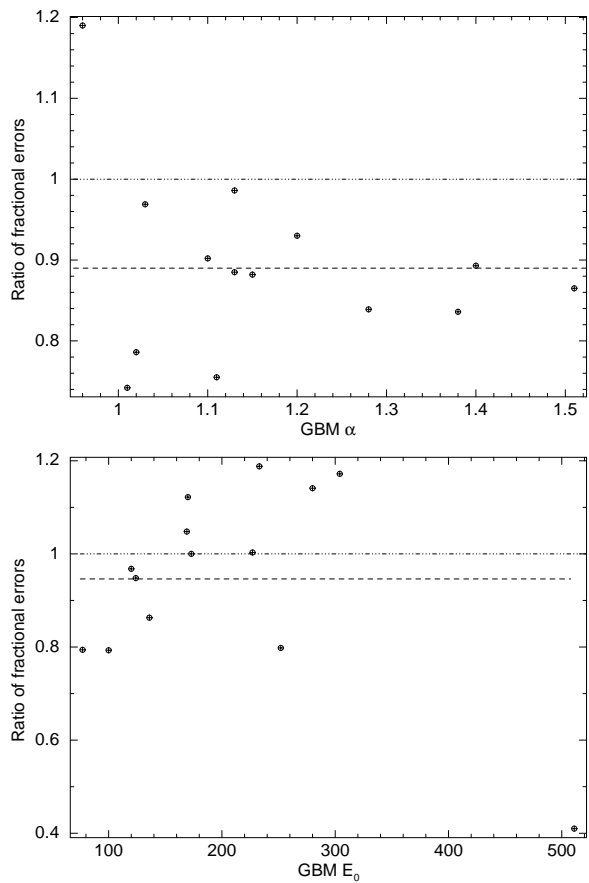}
\caption[Improvement of fractional error by joint spectral fitting over the GBM only fitting]
{Improvement of fractional error by joint spectral fitting over the GBM only fitting. \textit{Upper panel:} parameter
$\alpha$, \textit{Lower panel:} parameter $E_0$. Dot-dashed line shows the equality, while dashed line shows the average 
of the ratio of fractional errors. Source: \cite{Basak_Rao_2012_090618}.}
\label{ch2_fig2}
\end{figure}

The Data Processing Unit (DPU) of the GBM unit generates three type of data files: (i) CSPEC,
(ii) CTIME and (iii) TTE. The data types are summarized in Table~\ref{ch2_t6}.
The data can be accessed either from the FTP site: \textit{http://heasarc.gsfc.nasa.gov/FTP/fermi/data/gbm/triggers},
or Browsed from the \textit{http://fermi.gsfc.nasa.gov/ssc/data/access/}. The analysis softwares can be accessed from
\textit{http://fermi.gsfc.nasa.gov/ssc/data/analysis/user/}. Specifically the IDL-based {\tt rmfit v3.3pr7} software 
developed by user contribution of Fermi Science Support Center (FSSC) is used. We also use {\tt XSPEC v12.6.0}
for x-ray spectral analysis.

The CSPEC and TTE data are used for both time-integrated and time-resolved spectral analysis. The choice 
of detectors are made as follows. In the quicklook folder, the count rate of the 12 NaI detectors are 
shown. The detectors are marked by $nx$, where `$n$' denotes that the detectors are NaI, and `$x$' has 
values from 0 to 11 (in hexadecimal). Generally, we choose 2 NaI detectors with highest count rate. Due 
to the design of the GBM instrument, the BGO detectors can be chosen by the choice of NaI detectors. The 
BGO detectors are denoted by $by$, where `$y$' has the values 0 or 1. We choose the detector $b0$ if $x\leq5$, 
and $b1$ otherwise. If the number `$x$' of the chosen NaI detectors lie in different sectors, and have 
comparable rate, then we use the BGO detector with higher rate. 

For each of the chosen detectors, we extract a background. The background interval is chosen 
both before and after the burst. The chosen background is modelled with a polynomial of degree $n\leq3$.
The region should be large enough to give enough statistics for extrapolation. Also, the selection should avoid 
the burst. The GBM is an open detector, and its spectral capability is limited primarily by the diffused x-ray background.
As the background in the low earth orbit can vary, one should be cautious in choosing not too large a background 
interval. 

The spectrum in a given interval is extracted using {\tt gtbindef} and {\tt gtbin} tasks. The spectra are then 
binned using {\tt grppha} by requiring minimum $\sim40$ counts in the detectors. For spectral fitting, we choose 
8-900 keV of NaI and 200 keV - 30 MeV of BGO detectors. The response matrix for each detector is supplied in the 
downloaded data. There are two types of response files --- rsp and rsp2. A rsp2 file contains response for each 
$5^{\circ}$ slew of the spacecraft. If the GRB occurs for a long time, it is instructive to use the response of the 
corresponding time interval. The spectra are fitted using {\tt XSPEC v12.6.0} software.

\subsection{\textit{Fermi}/LAT}
The LAT data is found in \textit{http://fermi.gsfc.nasa.gov/cgi-bin/ssc/LAT/LATDataQuery.cgi}.
The software used for the LAT analysis is ScienceTools–v9r23p1 package, which can be installed 
from \textit{http://fermi.gsfc.nasa.gov/ssc/data/analysis/software/}.
In the following the major steps are mentioned.

\begin{enumerate}
\item \textit{Earth Limb Emission:} The data file of the LAT contains photon data within $15^{\circ}$ 
of the corresponding the XRT position of the GRB. The earth limb emission is omitted from the data by 
using the task {\tt gtselect} with $105^{\circ}$ zenith angle cut, and assuming ``transient class''
event. For GRB analysis, the background is less significant.

\item \textit{Lightcurve:} The filtered file obtained by omitting earth limb emission is used to 
generate count lightcurve. The task is {\tt gtbin}.

\item \textit{Likelihood Spectral Analysis:} For transients with long exposure, it is recommended to 
use likelihood analysis using a point source (GRB) along with an isotropic component (for the 
extra-galactic diffused background) and a Galactic diffuse component to extract exposure map, 
diffuse response etc. However, GRBs are significantly brighter than the background, and are 
short lived. Hence, it suffices to use {\tt gtbin} to extract the spectrum. The response is 
generated using ``PS'' response calculation by {\tt gtrspgen} task. Background is extracted 
in the off-source region.

\end{enumerate}

\subsection{Joint Analysis And Issues}

The instruments as discussed above are useful to study different aspects of GRB timing and spectral 
data. For example, the \textit{Swift}/BAT and the \textit{Fermi}/GBM are useful for prompt emission analysis, 
while the \textit{Swift}/XRT and the \textit{Fermi}/LAT are used to study x-ray afterglow, and occasional GeV 
emission, respectively. As our goal is to understand the prompt emission, we shall primarily use 
the \textit{Swift}/BAT and the \textit{Fermi}/GBM data. 

It is interesting to employ the BAT and the GBM instruments simultaneously in order to constrain the parameters.
As the BAT energy range is a subset of the the GBM/NaI energy range, the joint fitting can find out extra 
features in the data. Let us illustrate the procedure for one of the brightest GRBs, namely GRB 090618
(\citealt{Ghirlandaetal_2010, Raoetal_2011, Pageetal_2011, Basak_Rao_2012_090618}).
This GRB is the brightest one till August 2009, and remains one of the brightest till date, next to 
two GRBs 130427A, and 090902B (the first one is so bright that it saturates the GBM, and the data is unusable
for a large duration; the second one has a rapidly variable lightcurve). We shall use this GRB for 
developing various fitting schemes in the subsequent chapters.

For joint time-resolved analysis, we use the time divisions as provided by \cite{Ghirlandaetal_2010}. 
As there is a trigger time difference between the two instruments, we correct for this time (see 
chapter 3 for details). We fit the spectra with cutoff power-law (CPL), and using different constant 
multiplier for the different detectors to account for systematic errors of the area calibration 
(e.g., \citealt{Sakamotoetal_2011_cross}).

In Figure~\ref{ch2_fig1}, we have shown the comparison of the spectral parameters as obtained by only
the GBM data, and the joint data. The upper panel shows the comparison of index ($\alpha$; negative value 
shown for convenience), and the lower panel shows the cutoff energy ($E_0$). It is noted from the upper 
panel that the values of $\alpha$ in the first bin (0-3 s) differ from each other. This can be attributed 
to the low count rate in the BAT during this time interval. More important is the fact that the 
value of $\alpha$ (negative value) becomes lower on average for joint analysis. We have also found that 
the value of $E_0$ becomes higher. The constant factor of the BAT is lower by 10-20\%. \cite{Sakamotoetal_2011_cross}
have done a detailed time-resolved spectral fitting with Band function for a set of GRBs using simultaneous 
data from \emph{Konus}/Wind and \emph{Suzaku}/WAM along with the \emph{Swift}/BAT. They have found that the 
constant factor of the BAT is lower by a similar amount (10-20\%), while the $\alpha$ is steeper by 0.1-0.2,
and $E_{\rm peak}$ is higher by 10-20\% due to the inclusion of the BAT. The results presented here are in 
agreement with their findings.

In Figure~\ref{ch2_fig2}, we have plotted the ratio of fractional errors of the two methods of fitting as functions 
of the corresponding parameters, $\alpha$ (upper panel) $E_0$ (lower panel). The average of the ratios are shown by 
dashed lines. Note that the average is lower than the equality (dot-dashed line) for both the parameters. We also 
note that due to the lower energy coverage of the BAT, it has little impact on the measured error of $E_0$, while it 
has significant effect on the measured error of $\alpha$. Joint analysis always gives improvement in the measured 
error of parameters. It also gives confidence that the parameters are essentially unaffected by the systematic 
error of the instruments. However, as noted above, the values of the measured parameters systematically shifts 
for the inclusion of the BAT in the joint analysis. This effect may be a combination of systematic error in both 
the instruments. In fact, \cite{Sakamotoetal_2011_cross} have shown that the systematic error of the BAT in 
the $\sim25-100$ keV is $\sim4\%$, whereas the error can be as large as 20\% in both ends of detector band width.
Hence, in our analysis, we mostly use the GBM data for spectral fitting. In one case, we shall illustrate 
the analysis procedure with the BAT only data. A joint GBM/BAT data is used sometimes to get unambiguous spectral parameters. 

\begin{table*}\centering
\caption{Results of time-integrated spectral analysis of the GRBs.}
\begin{tabular}{c|c|c|c|c}
 \hline
  GRB   & $t_1$, $t_2$ & \multicolumn{2}{c|}{This work} & Nava et al. (2011) \\
\cline{3-4}
(Model) &                & C-stat & $\chi^2$         &                    \\
\hline
\hline

 080904 & -4.096, 21.504    & $\alpha$=-1.22$^{+0.21}_{-0.20}$      & $\alpha$=-1.21$^{+0.20}_{-0.19}$   &  $\alpha$=-1.14$\pm0.05^{\bf(a)}$  \\
 (CPL)  &                   & $E_p=40.1^{+3.92}_{-3.56}$            & $E_p=39.8^{+3.68}_{-3.34}$         &  $E_p=39.24\pm 0.75$ \\
        &                   & $C^{\bf (b)}$=1.08 (597)                  & $\chi^2_{red}$=1.23 (597)          &  $C=1.14(587)$ \\

\hline 

 080925 & -3.840, 32.0      & $\alpha$=-1.06$^{+0.11}_{-0.10}$       & $\alpha$=-1.06$^{+0.11}_{-0.10}$   &  $\alpha$=-1.03$\pm0.03$   \\
 (Band) &                   & $\beta=$=-2.34$^{+0.30}_{-1.13}$       & $\beta=$=-2.24$^{+0.24}_{-0.74}$   &  $\beta=$=-2.29$\pm0.08$   \\
        &                   & $E_p=158.9^{+31.6}_{-24.4}$            & $E_p=157.3^{+33.5}_{-24.9}$        &  $E_p=156.8\pm 7.07$ \\
        &                   & $C=1.17 (712)$                           & $\chi^2_{red}$=1.14 (712)          &  $C=1.13(716)$ \\

\hline 

 081118 & 0.003, 19.968     & $\alpha$=-0.42$^{+0.70}_{-0.48}$       & $\alpha$=-0.37$^{+0.70}_{-0.49}$   &  $\alpha$=-0.46$\pm0.10$   \\
 (Band) &                   & $\beta=$=-2.18$^{+0.16}_{-0.35}$       & $\beta=$=-2.14$^{+0.15}_{-0.19}$   &  $\beta=$=-2.29$\pm0.05$   \\
        &                   & $E_p=55.93^{+22.2}_{-12.5}$            & $E_p=54.0^{+19.7}_{-12.0}$         &  $E_p=56.79\pm 2.77$ \\
        &                   & $C=1.17 (716)$                           & $\chi^2_{red}$=1.02 (716)          &  $C=1.16(601)$ \\

\hline 

 081207 & 0.003, 103.426    & $\alpha$=-0.58$^{+0.10}_{-0.09}$       & $\alpha$=-0.58$^{+0.12}_{-0.11}$   &  $\alpha$=-0.58$\pm0.02$   \\
 (Band) &                   & $\beta=$=-2.15$^{+0.17}_{-0.33}$       & $\beta=$=-2.13$^{+0.20}_{-0.41}$   &  $\beta=$=-2.22$\pm0.7$   \\
        &                   & $E_p=363.4^{+70.7}_{-51.5}$            & $E_p=364.5^{+82.8}_{-59.4}$        &  $E_p=375.1\pm 13.2$ \\
        &                   & $C=1.43 (713)$                           & $\chi^2_{red}$=1.02 (713)          &  $C=1.74(596)$ \\

\hline

 081217 & -28.672, 29.696   & $\alpha$=-1.09$^{+   0.15}_{  -0.14}$  & $\alpha$=-1.10$^{+0.16}_{-0.14}$   &  $\alpha$=-1.05$\pm0.04$  \\
 (CPL)  &                   & $E_p=193.0^{+65.9}_{-37.3}$            & $E_p=200.5^{+77.2}_{-41.7}$        &  $E_p=189.7\pm 11.2$ \\
        &                   & $C=1.19 (715)$                           & $\chi^2_{red}$=1.06 (715)          &  $C=1.46(599)$ \\

\hline 

 081221 & 0.003, 39.425     & $\alpha$=-0.84$^{+0.06}_{-0.05}$       & $\alpha$=-0.84$^{+0.06}_{-0.06}$   &  $\alpha$=-0.82$\pm0.01$   \\
 (Band) &                   & $\beta=$=-4.24$^{+0.93}_{-10.2}$       & $\beta=$=-3.89$^{+0.69}_{-7.1}$    &  $\beta=$=-3.73$\pm0.20$   \\
        &                   & $E_p=85.25^{+2.89}_{-3.08}$            & $E_p=85.09^{+3.23}_{-3.19}$         &  $E_p=85.86\pm 0.74$ \\

        &                   & $C=1.64 (595) $                          & $\chi^2_{red}$=1.49 (595)          &  $C=1.67(600)$ \\

\hline 

 081222 & -0.768, 20.736    & $\alpha$=-0.89$^{+0.14}_{-0.12}$       & $\alpha$=-0.89$^{+0.14}_{-0.12}$   &  $\alpha$=-0.90$\pm0.03$   \\
 (Band) &                   & $\beta=$=-2.46$^{+0.37}_{-1.37}$       & $\beta=$=-2.32$^{+0.31}_{-0.98}$   &  $\beta=$=-2.33$\pm0.10$   \\
        &                   & $E_p=169.2^{+37.3}_{-27.4}$            & $E_p=168.9^{+39.1}_{-29.8}$        &  $E_p=167.2\pm 8.28$ \\
        &                   & $C=1.12 (595)$                           & $\chi^2_{red}$=1.07 (595)          &  $C=1.23(604)$ \\

\hline

 090129 & -0.256, 16.128    & $\alpha$=-1.43$^{+   0.19}_{  -0.16}$  & $\alpha$=-1.46$^{+0.18}_{-0.16}$   &  $\alpha$=-1.46$\pm0.04$  \\
 (CPL)  &                   & $E_p=170.4^{+130.0}_{-48.5}$           & $E_p=195.5^{+212}_{-63.5}$         &  $E_p=166.0\pm 15.1$ \\
        &                   & $C=1.09 (596) $                          & $\chi^2_{red}$=1.03 (596)          & $ C=1.12(602)$ \\

\hline

 090709 & 0.003, 18.432     & $\alpha$=-1.04$^{+   0.38}_{  -0.32}$  & $\alpha$=-1.08$^{+0.37}_{-0.31}$   &  $\alpha$=-0.96$\pm0.08$  \\
 (CPL)  &                   & $E_p=116.7^{+76.9}_{-30.6}$            & $E_p=124.1^{+101}_{-34.7}$         &  $E_p=137.5\pm 12.5$ \\
        &                   & $C=1.05 (596)  $                         & $\chi^2_{red}$=1.01 (596)          & $ C=1.17(602)$ \\

\hline

 091020 & -3.584, 25.088    & $\alpha$=-1.31$^{+   0.29}_{  -0.18}$  & $\alpha$=-1.32$^{+0.22}_{-0.19}$   &  $\alpha$=-1.20$\pm0.06$  \\
 (CPL)$^{(\bf c)}$  &                   & $E_p=255.7^{+332.0}_{-92.0}$           & $E_p=276.4^{+485.0}_{-107.0}$      &  $\beta=$=-2.29$\pm0.18$  \\
        &                   &$ C=1.03 (354)     $                      & $\chi^2_{red}$=0.95 (354)          &  $E_p=186.8\pm 24.8$  \\
        &                   &                                        &                                    &  $C=1.18(354) $\\

\hline 

 091221 & -2.048, 37.889    & $\alpha$=-0.62$^{+0.27}_{-0.21}$       & $\alpha$=-0.62$^{+0.34}_{-0.23}$   &  $\alpha$=-0.57$\pm0.05$   \\
 (Band) &                   & $\beta=$=-2.40$^{+0.50}_{-3.15}$       & $\beta=$=-2.26$^{+0.45}_{-2.80}$   &  $\beta=$=-2.22$\pm0.10$   \\
        &                   & $E_p=191.3^{+67.4}_{-47.5}$            & $E_p=189.5^{+76.8}_{-57.1}$        &  $E_p=194.9\pm 11.6$ \\
        &                   & $C=1.42 (474)   $                        & $\chi^2_{red}$=1.12 (474)          &  $C=1.44(466)$ \\

\hline
\end{tabular}

\begin{footnotesize} 

\vspace{0.1in}
$^{(\bf a)}$ The errors quoted from Nava et al. (2011) are symmetric errors. Errors for this work are 3$\sigma$ errors. \\
$^{(\bf b)}$ $C$ is the reduced C-stat value, the number in the parentheses are dof.\\
$^{(\bf c)}$ The Band spectrum showed unbound 3$\sigma$ errors, we found better fit with CPL for this GRB
\end{footnotesize}
\end{table*}

\begin{table*}[ht]\centering
\caption{Sample of GRBs}

\begin{tabular}{cccc}
\hline
Satellite & Parameter & Number & Source \\
\hline
\hline
\textit{Fermi} & GBM GRB sample (First four years) & 953 & [a], [b] \\ 
               & Redshift sample & 45 & [a], [b] \\
               & LAT GRB sample & 35 & [c] \\
\hline
\textit{Swift} & BAT GRB sample (First five years) & 476 & [d] \\
\hline
     Combined Sample      & Total sample & 1270 & [e] \\
  (till 2014 April 23)    & X-ray afterglow & 854 & [e] \\
                          & Optical afterglow & 538 & [e] \\
                          & Radio afterglow & 95 & [e] \\
                          & Redshift sample & 350 & [e] \\
\hline
\end{tabular}
\vspace{0.1in}

\begin{footnotesize}
[a] \cite{von_Kienlinetal_2014}, [b] \cite{Gruberetal_2014}, [c] \cite{Ackermannetal_2013_LAT}, [d] \cite{Sakamotoetal_2011}, [e] http://www.mpe.mpg.de/~jcg/grbgen.html
 
\end{footnotesize}

\label{ch2_t8}
\end{table*}

\subsection{Use Of Statistics}
We shall use $\chi^2$ minimization for the spectral fitting of both the GBM and the BAT data. It is suggested in the BAT 
analysis guide that the deconvolution technique to extract the background subtracted flux produces gaussian 
errors rather than poissonian errors. Hence, it is recommended that the fitting procedure should use 
$\chi^2$ minimization which assumes a gaussian error in the data. It is customary to use C-statistics 
for the LAT analysis of the \textit{Fermi}. This is due to the low photon flux in the LAT energy band. C-statistics 
method can be used for the GBM data as well. However, as C-statistics is valid for poissonian error, this method 
should be avoided for the BAT data. As we are interested in using both the BAT and the GBM data, we shall generally use 
$\chi^2$ minimization technique and compare between different models using $F$-test. 

The sample we choose for analysis are all bright GRBs. Hence, it is expected that the minimization technique
using $\chi^2$ and C-statistics should give similar parameters. In order to check how the statistics affect the 
parameters of the model fitted to the GBM data, let us use both the methods for a bright sample. We have searched the GRB 
catalogue provided by \citet[][N11 hereafter]{Navaetal_2011}. We use the following criteria for our sample selection: 
(i) fluence $\geq10^{-6}$ erg, (ii) duration (see N11), $\delta t\geq15$ s, (iii) 
single/separable pulse structure. We have found 11 such GRBs. We fit the spectral data of these 
bursts by either Band or CPL model (whichever is preferred), using both $\chi^2$ and C-statistics 
minimization methods. In Table 2.7, we have shown the values of the parameters as obtained by 
these two methods. The reduced $\chi^2$ and reduced C-stat ($C$) values are also given. For reference,
we have shown the parameters and reduced C-stat as provided by N11 for these GRBs.
Note that the errors quoted by N11 are measured with 68\% confidence, while we have 
measured the errors at nominal 90\% confidence level.

The following observations are apparent from Table 2.7. We note that the parameters as obtained by 
$\chi^2$ and C-stat minimization are similar to each other for our analysis. The parameters also agree
with the values quoted by N11. The source of minor deviation of the parameter values obtained by C-stat 
minimization in our analysis and N11 are (i) use of different number of detectors, (ii) the choice of 
different detector band widths, and (iii) the difference in the choice of the background region. 
While the deviations are small, these are higher 
compared to the minor deviations due to the use of different statistics in our analysis. Hence, it 
is apparent that the choice of statistics will have minimal effect on our spectral analysis. This is 
indeed expected as the sample are chosen with a lower limit on the fluence. As we shall frequently use 
the BAT data, we shall generally stick to $\chi^2$ minimization.

Another important tool we shall use is the $F$-test to compare different models. In particular it 
is useful when we try to quantify the significance of adding a new component on the existing model, e.g., 
adding another blackbody with a model having a blackbody plus a power-law. Such models are called 
inclusive. We define $F= \frac{(\chi^2_1-\chi^2_2)/(dof_1-dof_2)}{\chi^2_2/dof_2}$, where the index 
`1' denotes the original hypothesis (null) and `2' denotes the alternative hypothesis. `dof' is the 
degree of freedom. For exclusive models, $F=\frac{\chi^2_1/dof_1}{\chi^2_2/dof_2}$. We compute the 
probably ($p$) for a given F value. This provides the significance (in terms of $\sigma$), and the 
confidence level (\% CL) of the alternative model as preferred over the original (null) model.

\section{Scientific Aspects Of The Satellites}\label{ch2_s5}
The \textit{Swift} and the \textit{Fermi} together has become the primary workhorse of GRB science. Throughout 
the thesis, we shall show their versatile applications for the prompt emission analysis. In the following,
we briefly describe the impact of the two satellites on GRB science. 

\subsection{GRB Science With \textit{Swift}}
The \textit{Swift} has opened up a new era for GRB science. Note that the Interplanetary Network (IPN) used 
to take a few days to months in order to localize a GRB position with arcmin accuracy. This situation was 
improved by Beppo-SAX which could detect the fading x-ray afterglow after a delay of hours. Of course, 
Beppo-SAX was not designed for GRBs. The \textit{Swift} satellite was born out of the requirement of rapid 
localization of GRBs. Due to the fantastic slewing capability, and arcmin position accuracy obtained by the BAT, 
the \textit{Swift} can detect 
the x-ray afterglow just about a minute after the trigger. This provides an unprecedented position accuracy 
(3 arcsec) of a GRB within a few minutes, and facilitates the afterglow observation from ground based 
telescopes. Due to the requirement of quick follow up observation, the early position information for a 
triggered burst is sent through Tracking and Data Relay Satellite System (TDRSS). The GRB Coordinates 
Network (GCN) automatically gets the \textit{Swift} TDRSS messages for GRB, and distributes to the community.
The follow-up mission of early afterglow leads to redshift ($z$) measurement for a large fraction of 
GRB. The impact of the \textit{Swift} on GRB science is summarized in several reviews ({\citealt{Meszaros_2006,
O'Brienetal_2006, Zhang_2007_review, Gehrelsetal_2009_review}).

\subsection{GRB Science With The \textit{Fermi}}
The major contribution of the \textit{Fermi}/LAT is providing an unprecedented effective area in a wide energy band of
20 MeV-300 GeV. In addition, the GBM provides a good spectral data in 8 keV-30 MeV. Together these 
detectors cover seven decades of energy band, which is very useful to study different spectral models, 
and finding additional spectral components. The results of the \textit{Fermi}/GBM can be found in 
\cite{Paciesasetal_2012, Goldsteinetal_2012, von_Kienlinetal_2014, Gruberetal_2014}.
The LAT source catalogue is \cite{Nolanetal_2012}, and the GRB catalogue is \cite{Ackermannetal_2013_LAT}.

In the subsequent chapters, we shall discuss the scientific aspects of the data provided by the \textit{Swift}
and the \textit{Fermi} satellites. The current sample as obtained from different sources are listed in Table~\ref{ch2_t8}.
While the GBM provides the maximum number of GRBs, the \textit{Swift} provides the major fraction with afterglow 
study. It is important to note that while the \textit{Swift}/BAT sees a much lower fraction of the sky (1.4 sr) 
compared to that of the \textit{Fermi}/GBM (9.5 sr), the BAT can see fainter sources. Also, most of the redshift 
measurements are provided due to the follow up observation facilitated by the \textit{Swift} satellite. However, 
due to a limited band width, the spectral data often cannot constrain the value of $E_{\rm peak}$. 
Hence, GRBs with redshift measurement and good energy coverage are quite rare.

\chapter{A New Description Of GRB Pulses} \label{ch3}

\section{Overview}

In this chapter, we shall develop a technique to analyze GRB data simultaneously in time 
and energy domain. This model will be applied for the individual pulses within a GRB.
This simultaneous pulse description is motivated by the following reasons.

\begin{itemize}
 \item (i) As discussed in the first chapter (section 1.4.1), a GRB exhibits pulses in its
lightcurve (LC). For most of the cases one can define broad pulses, and assume 
the rapid variability timescales as ``weather'' on top of the broad variations. With this assumption,
we shall try to obtain a description of these broad pulses. Once the individual pulses 
are generated one can add them to describe the full GRB, except for the rapid variability. 
Essentially, the analysis of a GRB thus reduces to the description of the individual pulses. 
As our model has a handle on the pulses, and as it is a simultaneous timing and spectral 
description, it has versatile applicability, e.g., studying GRB properties within the pulses, and
deriving pulse properties e.g., pulse width, spectral lag etc. using a single description.
 
\item (ii) We know that various timing and spectral parameters of GRBs correlate with the 
energy related physical parameters e.g., the peak energy ($E_{\rm peak}$) correlates with
the isotropic energy ($E_{\rm \gamma, iso}$), known as Amati correlation (\citealt{Amatietal_2002}),
the spectral lag ($\tau_{\rm lag}$) correlates with the isotropic peak luminosity ($L_{\rm iso}$; \citealt{Norrisetal_2000}) etc.
The correlations are important to constrain a given model as well as to use GRBs as high-$z$ 
luminosity indicators (\citealt{Schaefer_2003_HD, Schaefer_2007_HD}). However, these correlations are independently 
studied either in energy, or in time domain. For example, the Amati correlation uses time-integrated 
$E_{\rm peak}$ (i.e., averaged over the burst duration), ignoring the spectral evolution in a 
burst, and within the pulses of a burst. A pulse-average (rather than a burst-average) correlation 
is a reasonable first step to get a physical meaning of a correlation, if one assumes that 
the pulses are independent entities. A pulse-wise study has another important consequence. If 
a burst-average correlation holds within the pulses it shows that the correlation is unbiased by 
the selection effect of the instrument. Pulse-wise correlations are found to hold similar or sometimes 
even better as compared to the average GRB correlations. For example, \cite{Krimmetal_2009} have studied
Amati correlation within the pulses, and have found a good pulse-wise correlation. \cite{Hakkilaetal_2008},
on the other hand, have studied the lag-luminosity correlation, and they have conclusively 
shown that this correlation is a pulse property rather than a burst property. Further 
improvement is expected by incorporating the temporal information in the energy related 
correlations, and vice versa. Some attempts have been made to obtain an empirical description of 
the $E_{\rm peak}$ evolution in a set of GRBs with single pulses (e.g., \citealt{LK_1996, Kocevski_Liang_2003}). 
However, a simultaneous model which preserves both the time and energy
information of a GRB pulse is lacking. This is possibly due to the forward convolution procedure followed in x-ray spectral analysis.
One assumes a model for the spectral data, convolves the model with detector response, 
and fits with the data to determine the spectral parameters (\citealt{Arnaud_1996}). Hence, it is
difficult to incorporate the time evolution of the spectral parameters in the scheme.
The fact that most GRBs also come with multiple overlapping pulses further complicates the 
unification scheme of temporal and spectral description.

\end{itemize}

For our purpose, we shall target the long GRBs (LGRBs), as they have usually higher flux and longer duration.
\cite{Hakkila_Preece_2011} have shown that typical pulses of short GRBs (SGRBs) exhibit 
similar properties as the pulses of LGRBs in terms of correlations among the 
pulse properties e.g.,  duration, luminosity at the peak, fluence, spectral hardness, spectral lag, and asymmetry.
Hence, our results are possibly meaningful even when one deals with SGRB pulses.

We shall explicitly develop the simultaneous timing and spectral model using one of the brightest 
GRBs in the \textit{Fermi} era --- GRB 090618 (\citealt{Basak_Rao_2012_090618}). 
Apart from the high flux, this GRB also displays well-defined separable pulses. Later, we shall use this model 
for a set of GRBs (\citealt{Basak_Rao_2012_correlation, Basak_Rao_2012_germany}) to study 
the Amati correlation (correlation between $E_{\rm peak}$ and $E_{\rm \gamma, iso}$). Our 
particular aim will be to study this correlation within the individual pulses. In addition to the 
pulse-wise Amati correlation, we shall also propose a new correlation which is obtained as 
a by-product of our simultaneous pulse description. Finally, we shall present an updated pulse-wise Amati 
correlation for a larger sample. We shall study the evolution of the correlation with 
redshift ($z$), and discuss about the possible bias in the correlation (\citealt{Basak_Rao_2013_MNRAS}).

\section{GRB 090618}

\begin{figure}\centering
\includegraphics[width=3.4in]{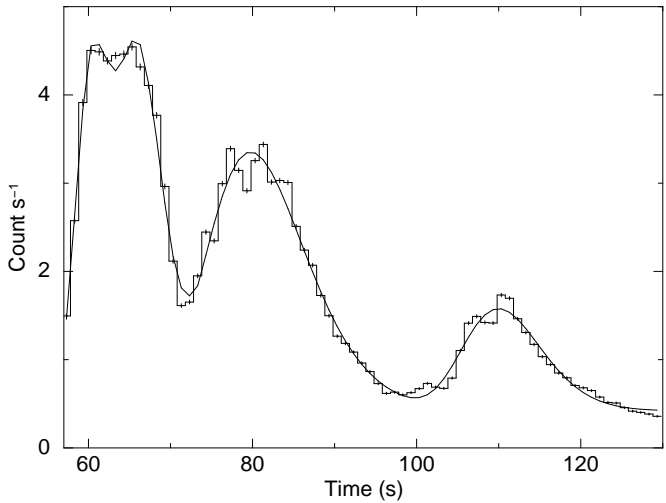}
\caption[The observed BAT LC of GRB 090618 in 15-200 keV, fitted with the Norris model]
{The observed BAT LC of GRB 090618 in 15-200 keV, fitted with the Norris model (Source: \citealt{Basak_Rao_2012_090618}).}
\label{ch3_fig1}
\end{figure}

We shall use GRB 090618 to develop the method of simultaneous timing and spectral description of GRB pulses. 
GRB 090618 is one of the brightest GRBs in the \textit{Fermi} era. This GRB was detected on 18th June, 2009
by many satellites --- \textit{Fermi}/GBM (\citealt{McBreenetal_2009GCN}), 
\textit{Swift}/BAT (\citealt{Schadyetal_2009GCN1, Schadyetal_2009GCN2, Schadyetal_2009GCN3}),
\textit{Suzaku}/WAM (\citealt{Konoetal_2009GCN}), 
\textit{ AGILE} (\citealt{Longoetal_2009GCN}), \textit{Coronas-Photon}/Konus-RF (\citealt{Golenetskiietal_2009GCN}),
\textit{Wind}/Konus-Wind, \textit{Coronas-Photon}/RT-2 (\citealt{Raoetal_2009GCN, Raoetal_2011}). From the duration of this GRB
(T$_{90}\sim113$ s), it is classified as a LGRB. The time integrated flux (i.e., fluence) is $3398.1\pm62.0\times10^{-7}$
erg/cm$^{2}$ (flux integrated over $\triangle t$=182.27 s), which was the highest till GRB~090902B, and remains 
one of the highest till date.
The \textit{Swift}/BAT reports a detection time as 2009 June, 18 at 08:28:29.85 UT (\citealt{Schadyetal_2009GCN2}). 
The \textit{Fermi}/GBM gives a detection time 08:28:26.66
UT (\citealt{McBreenetal_2009GCN}). This GRB has one precursor burst, followed by a flattening, and then 
four pulses. Though the first two pulses have heavy overlap, the other pulses including the 
precursor are well-defined and separable. The spectrum of the GRB shows rapid 
time evolution within the pulses. The precursor shows a clear ``hard-to-soft'' (HTS)
characteristics, while the other pulses show ``intensity tracking'' (IT), or a ``soft-to-hard-to-soft''(SHS)
evolution. However, as discussed in chapter 1 (section 1.4.1.C), IT feature 
can be a overlapping effect (cf. \citealt{Hakkila_Preece_2011}).

The afterglow monitoring in x-ray band (WT mode) started $\sim$ 125 s after the \textit{Swift}/BAT trigger
(\citealt{Schadyetal_2009GCN3}). Initially the burst was very bright in the x-ray, then the flux started to decay very fast
with a slope $\sim-6$ ($F_{\rm X}(\rm t)\propto t^{-6}$) till 310 s post-trigger. At this 
point, the x-ray flux entered the shallower decay phase (with slope -0.71$\pm$0.02 -- \citealt{Beardmore_Schady_2009GCN}). 
The optical afterglow was observed by 60-inch Palomar telescope (\citealt{Cenko_2009GCN}), and Katzman Automatic Imaging
Telescope (\citealt{Perley_2009GCN}). The burst was subsequently followed by various other 
ground based optical and near-infrared (NIR) telescopes. A redshift of $z=0.54$  
was found by Lick observatory using 3-m Shane telescope (\citealt{Cenkoetal_2009GCN}).
See \cite{Raoetal_2011} for the details of the afterglow evolution.

\cite{Raoetal_2011}, using the simultaneous prompt emission data of the \textit{Swift}/BAT and the \textit{Coronas-Photon}/RT-2
find four pulses. They fit the LC with four fast rise exponential decay (FRED) 
profile, and the time-integrated as well as the pulse-wise spectrum with Band function. 
The parameters of Band function, fitted to the time-integrated data are as follows: 
low energy photon index $\left(\alpha\right)=-1.40\pm 0.02$, high energy photon index 
$\left(\beta\right)=-2.50_{-0.5}^{+0.3}$, and peak energy $\left(E_{\rm peak}\right)=164\pm24$ keV.
\cite{Ghirlandaetal_2010} have selected a set of \textit{Fermi} GRBs with known $z$, including GRB 090618.
They perform a detailed time-resolved analysis of this GRB, and obtain the time-resolved peak 
energy-isotropic luminosity $\left(E_{\rm peak}^{t}-L_{\rm iso}^{t}\right)$ relation. They have shown that 
the time-resolved correlation is consistent with the time-integrated $\left(E_{\rm peak}-L_{\rm iso}\right)$
correlation (Yonetoku correlation; \citealt{Yonetokuetal_2004}). The fact that time-integrated correlation holds 
for the time-resolved study strongly indicates that the correlation is real, i.e., devoid of 
selection bias.

\section{Spectral And Timing Data Analysis Of GRB 090618}
In this section, we shall describe the average spectral and timing features of GRB 090618.
Through timing analysis, we shall identify the pulses, and derive various pulse properties, 
e.g., pulse width ($w$), and their variation with time. The spectral analysis of the individual 
pulses give the average spectral properties, e.g., average peak energy. These quantities are required 
for our later purpose. The analyses are done by using both the \textit{Swift}/BAT and the \textit{Fermi}/GBM data.
We essentially follow the data analysis procedure as described in chapter 2.
We use the \textit{Swift}/BAT detector, two NaI detectors, namely, n4 and n7, and one BGO detector, either b0, or b1, one at a time.
To determine the average values of the parameters, we use both the instruments. The parameters 
are in good agreement with those obtained by a single detector. 
However, as pointed out in chapter 2, the joint \textit{Swift}/BAT and \textit{Fermi}/GBM
fitting have some unexplained issues. Hence, for developing our model, we generally 
prefer to use data from a single satellite (i.e., either the \textit{Swift}/BAT or the \textit{Fermi}/GBM), 
rather than both. Note that the \textit{Fermi}/GBM has larger energy coverage, hence it is preferred over the \textit{Swift}/BAT.
However, one can also use the \textit{Swift}/BAT detector, if the peak energy is not too
high. GRB 090618 has $E_{\rm peak}=164\pm24$ keV. Hence, for illustration purpose, 
we use the \textit{Swift}/BAT detector to develop the method. However, in the next section, 
We shall apply this model for a larger set of GRBs, some of which have large $E_{\rm peak}$. 
Hence, for uniformity, we shall exclusively use the \textit{Fermi}/GBM for the data analysis in a global sense.

The trigger time of a GRB is generally expressed in terms of Mission Elapsed Time (MET).
The MET of \textit{Fermi} is measured in seconds from 2001.0 UT, not including leap seconds, 
while for \emph{Swift}, the leap second is added. The \emph{Swift}/BAT trigger time for 
the GRB is, $T_{0}$(BAT) = 2009-06-18 at 08:28:29.851 UT, which is equivalent to a MET 267006514.688 
MET (s). The \emph{Fermi}/GBM trigger time is $T_{0}$(GBM) = 2009-06-18 at 08:28:26.659 UT, 
or 267006508.659 MET (s). Comparison of the UT trigger time shows a delay between the BAT data 
and the GBM data. This must be subtracted if one wants to do a joint analysis. This time delay 
is 3.192 s. When we convert the subtracted time into MET of \emph{Swift}/BAT, this gives 
267006511.496 MET (s). In all joint analysis, we have used this subtracted time, whenever 
required.

\subsection{Timing Analysis}
Other than the precursor, we identify four pulses  in the $\sim$60
s to $\sim$130 s interval after the trigger. We use both the \emph{Swift}/BAT and
the \emph{Fermi}/GBM (n4) for the timing analysis. The background is subtracted, 
and the LC is shifted to the trigger time of the respective detectors.
We extract the LCs in different energies for the BAT and the GBM (n4) detectors. For the \emph{Swift}/BAT,
we choose 15-25 keV, 25-50 keV, 50-100 keV and 100-200 keV. For the \emph{Fermi}/GBM,
we essentially choose the same energy ranges, with two additional bands --- 8-15 keV, and
200-500 keV. We exclude the $>500$ keV band, because, at these high energies, the third and fourth pulses have 
too low counts to correctly model them.

We fit the LCs in various energy bands with Norris exponential model (\citealt{Norrisetal_2005}, chapter 1, section 1.4.1).
The Norris model consists of two exponential functions with time constants $\tau_{1}$, $\tau_{2}$, which 
characterize the rising and falling part of a pulse, respectively. The model can be written as follows.

\begin{equation}
F(t)=A_n\lambda {\rm exp}\{-\tau_{1}/(t-t_{s})-(t-t_{s})/\tau_{2}\}=A_n f_n(t, t_s, \tau_1, \tau_2)
\label{n1}
\end{equation}
\vspace{0.1in}

for $t>t_{s}$. Here, $\mu=\left(\tau_{1}/\tau_{2}\right)^{\frac{1}{2}}$, and
$\lambda=exp\left(2\mu\right)$. $A_n$ is defined as the pulse amplitude, $t_{s}$
is the start time. Following \cite{Norrisetal_2005} we can determine the 
pulse width ($w$) and asymmetry ($\kappa$). The errors in the model parameters
are determined at nominal 90\% confidence level ($\triangle\chi^{2}$=2.7), and these 
errors are propagated in the derived parameters. Figure~\ref{ch3_fig1} shows the 
fitting of the four pulses with the Norris model.

\begin{figure}\centering
\includegraphics[width=3.4in]{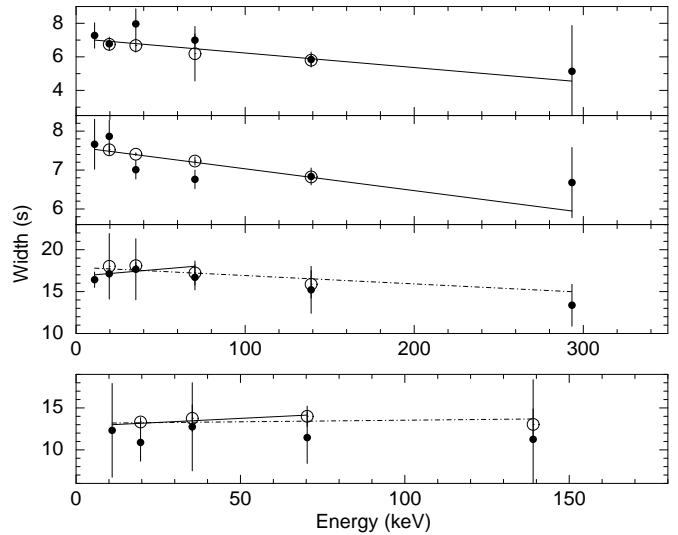}
\caption[Pulse width ($w$) variation as a function of energy ($E$)]
{Pulse width ($w$) variation as a function of energy ($E$) is shown for the four pulses (pulse 1 to pulse 
4 from top to bottom). As the fourth pulse is insignificant in higher energies, the scale shown is different.
We have used open circles to show the values derived from the \textit{Swift}/BAT. The \textit{Fermi}/GBM data points 
are shown by filled circles. We note that width generally increases with decreasing energy (top two panels). 
This trend is the normal width broadening feature (\citealt{Norrisetal_1996}). For the third and fourth pulse, 
there is a tentative evidence of anomalous width broadening i.e., width decreases with decreasing 
energy. This feature is seen in the $\sim$ 10-70 keV band (see text). Source: \cite{Basak_Rao_2012_090618}}
\label{ch3_fig2}
\end{figure}

One of the important derived parameters in our analysis is pulse width ($w$) in various energy bands.
For later purpose, we have shown in Figure~\ref{ch3_fig2} the variation of $w$ with energy for the 
individual pulses. It is suggested that GRB pulses should be broader at lower energies 
(\citealt{Norrisetal_1996, Hakkila_Preece_2011}). Figure~\ref{ch3_fig2} shows that in general $w$ at lower 
energies is indeed higher than that in higher energies. However, in some pulses (namely third
and fourth), we observe a reverse variation at very low energy bands --- $w$ decreases (or 
remains constant within error; see the lower panels of Figure~\ref{ch3_fig2}). In order to 
prove this reverse variation we perform the following simple tests. We fit the $w-E$ data 
with a (i) constant function ($w=c$), and (ii) linear function ($w=mE+c$). For the first two pulses,
we note that $\chi^2$ of the linear fit gives much improvement from a constant fit.
The changes are respectively, from 16.2(9) to 5.1(8), and from 140.9(9) to 7.9(8). The 
numbers in the parentheses are degrees of freedom (dof). We obtain the slope of linear fit as 
$(-7.4\pm2.6)\times10^{-3}$ s $\rm keV^{-1}$ and $(-5.5\pm0.4)\times10^{-3}$ s $\rm keV^{-1}$, respectively.
The negative values of the slope show the normal width broadening for the first two pulses. 
However, when we apply the same method for 
the third and fourth pulse, we find a negligible improvement of $\chi^2$ --- from 8.7(9)
to 6.0(8), and from 2.1(8) to 2.1(7), respectively. Interestingly, when we use the data 
below 70 keV for these pulses, we get marginal improvement in $\chi^2$ --- from 8.7(9) to 
3.8(5) and 2.1(8) to  1.5(5). Though these improvements are not significant given the 
number of dof, but, with the poor quality of data, we can at least claim that the width 
does not broaden in very low energies. For later use, this result is important.
We shall show that a reverse width variation can indeed occur, given some particular 
combination of model parameters, rather than being at the edge of the energy sensitivity
of an instrument. For the third and fourth pulses, we find that the slope of $w-E$
variation are $(15\pm14)\times10^{-3}$ s $\rm keV^{-1}$ and $(10\pm21)\times10^{-3}$ s $\rm keV^{-1}$, respectively. 
Thus we find positive slopes with large errors. Hence, the evidence of the reverse width 
variation is only \emph{tentative} at this moment. In the lower two panels of Figure~\ref{ch3_fig2}, 
we have shown the linear fit to the $w-E$ data by solid lines (in the lower energies), and 
by dot-dashed lines (in the full energy band). The 200-500 keV band of the fourth pulse 
is ignored, because, this pulse was barely visible in that \textit{Fermi} energy band.

\subsection{Pulse-wise Spectral Analysis}

\begin{table*}\centering
\caption{Spectral parameters of Band function fitted to the pulses and time-averaged data of GRB 090618}

\begin{tabular}{cccccc}
\hline 
Part & $\alpha$ & $\beta$ & $E_{peak}$(keV) & $\chi^{2}_{\rm red}$(dof) & $E_{0}$ (keV) \\
\hline
\hline 
Full & -1.35$\pm0.02$ & -2.5 (fixed) & 154$\pm7$ & 1.43(299) & 236$\pm18$ \\
Part 1 (Precursor) & -1.25$\pm0.05$ & -2.5(fixed) & $166_{-14}^{+18}$ & 1.23(299) & $221_{-33}^{+39}$ \\
Part 2 (Pulse 1 \& 2) & -1.11$\pm0.02$ & -2.5(fixed) & 212$\pm9$ & 1.38(299) & 238$\pm15$ \\
Part 3 (Pulse 3) & -1.15$\pm0.03$ & -2.5(fixed) & 114$\pm4$ & 1.29(259) & 134$\pm9$ \\
Part 4 (Pulse 4) & $-1.65_{-0.08}^{+0.11}$ & -2.5(fixed) & 33$\pm3$ & 1.44(299) & $94_{-30}^{+38}$ \\
Full (free $\beta$) & -1.36$\pm0.02$ & -2.96$\pm$0.48 & $160_{-8}^{+9}$ & 1.45(288) & $250_{-20}^{+22}$ \\
\hline
\end{tabular}
\label{pulse_spectrum}
\end{table*}

From the timing analysis, we have identified four pulses, apart from the precursor. In order to
obtain the pulse spectral property, we divide the LC into four bins. As we have seen that 
the first two pulses have large overlap, we cannot disentangle their contribution in the 
spectrum. However, we can generate these pulses separately by a simultaneous timing and spectral 
description, as we shall show in Section 3.5. The time interval we choose are as follows:                   
Part 1 ($T_{0}$ to $T_{0}+50$), Part 2 ($T_{0}+50$ to $T_{0}+77$), Part 3 ($T_{0}+77$
to $T_{0}+100$) and Part 4 ($T_{0}+100$ to $T_{0}+180$). This choice of time interval is subjective, 
and we consider only broad pulses as identified in Section 3.3.1.
Though this is a subjective choice, this will not affect our simultaneous model.

The pulse-wise spectra are fitted with Band function (\citealt{Bandetal_1993}) in the $\sim$ 8 keV to $\sim$ 1 MeV
energy range. As discussed in chapter 1 (Section 1.4.1), Band function is the most acceptable model
for GRB spectrum (\citealt{Kanekoetal_2006}). 
Band function has four parameters: two photon indices in low ($\alpha$), and high energies ($\beta$), 
normalization ($A_b$) and the peak energy ($E_{\rm peak}$). The function
shown in equation~\ref{ch1_band1} represents the photon spectrum --- $N(E)$. One can obtain the $F(E)$ spectrum 
by multiplying the photon spectrum with photon energy. In detectors, a binned 
data of spectrum is obtained, hence, one should multiply by some mean energy of the 
corresponding energy bin, which is automatically done by {\tt XSPEC}. Equation~\ref{ch1_band1} can 
also be written as in $F(E)$ representation as

\begin{equation}
 F(E)=A_b f_b \left(E, \alpha, \beta,E_{\rm peak} \right)
\label{band2}
\end{equation}
\vspace{0.05in}

In Table~\ref{pulse_spectrum}, we have shown the spectral parameters of Band function fitted to the 
pulses, and time-integrated data. The corresponding errors in the parameters are determined at
nominal 90\% confidence level ($\triangle\chi^{2}$=2.7). We have used the joint data of the BAT,
NaI --- n4, n7, and BGO --- b1 for the fitting. If we replace b1 by b0, the results remain unchanged.
The high energy index ($\beta$) could not be constrained for a few cases. 
But we note that $\beta \approx -2.25$ to $-2.5$ in general for \textit{Fermi} GRBs. Hence, to get 
precise values for the other parameters, we freeze the value of $\beta$ to -2.5, and perform the fitting.
For one case, namely, time-integrated spectrum, we set $\beta$ free and obtain a reasonable value of  $-2.96\pm0.48$.

We see from Table~\ref{pulse_spectrum} that though the $\chi^{2}_{\rm red}$ of these fits are 
acceptable, they are in general high, particularly for the time-integrated 
fit. An inspection of the residual reveals that the main contribution comes from 
disagreement between the fit of the BAT and the NaI data in the 15-150 keV energies. However, the parameters
obtained here are comparable to other studies found in the literature 
(\citealt{Ghirlandaetal_2010, Navaetal_2011, Raoetal_2011}). Also, the errors and $\chi^{2}_{\rm red}$
are comparable to the GBM only fit (\citealt{Navaetal_2011}), and better than a BAT/RT-2 joint
fit (\citealt{Raoetal_2011}). Since a joint analysis, performed using the data from different detectors,
can pin down the systematic errors of each instrument, we strongly believe that the spectral parameters
which are determined from a joint spectral fit are unbiased, and devoid of any instrumental artifacts.

\section{A Simultaneous Description of GRB Pulses}
Aided with the average timing and spectral behaviours, we now proceed to obtain a simultaneous 
description. We assume that a pulse LC can be represented by the Norris model (equation~\ref{n1}) 
in a given energy band, and the instantaneous spectrum can be described by the Band function 
(equation~\ref{ch1_band1} and equation~\ref{band2}). In addition, there exists a certain function to describe 
the spectral evolution. Such a functional form can be used to relate one of the spectral parameters 
with the timing properties at a given instance, and the resulting description carry an implicit 
temporal information.

\begin{figure*}\centering
\includegraphics[width=5.9in]{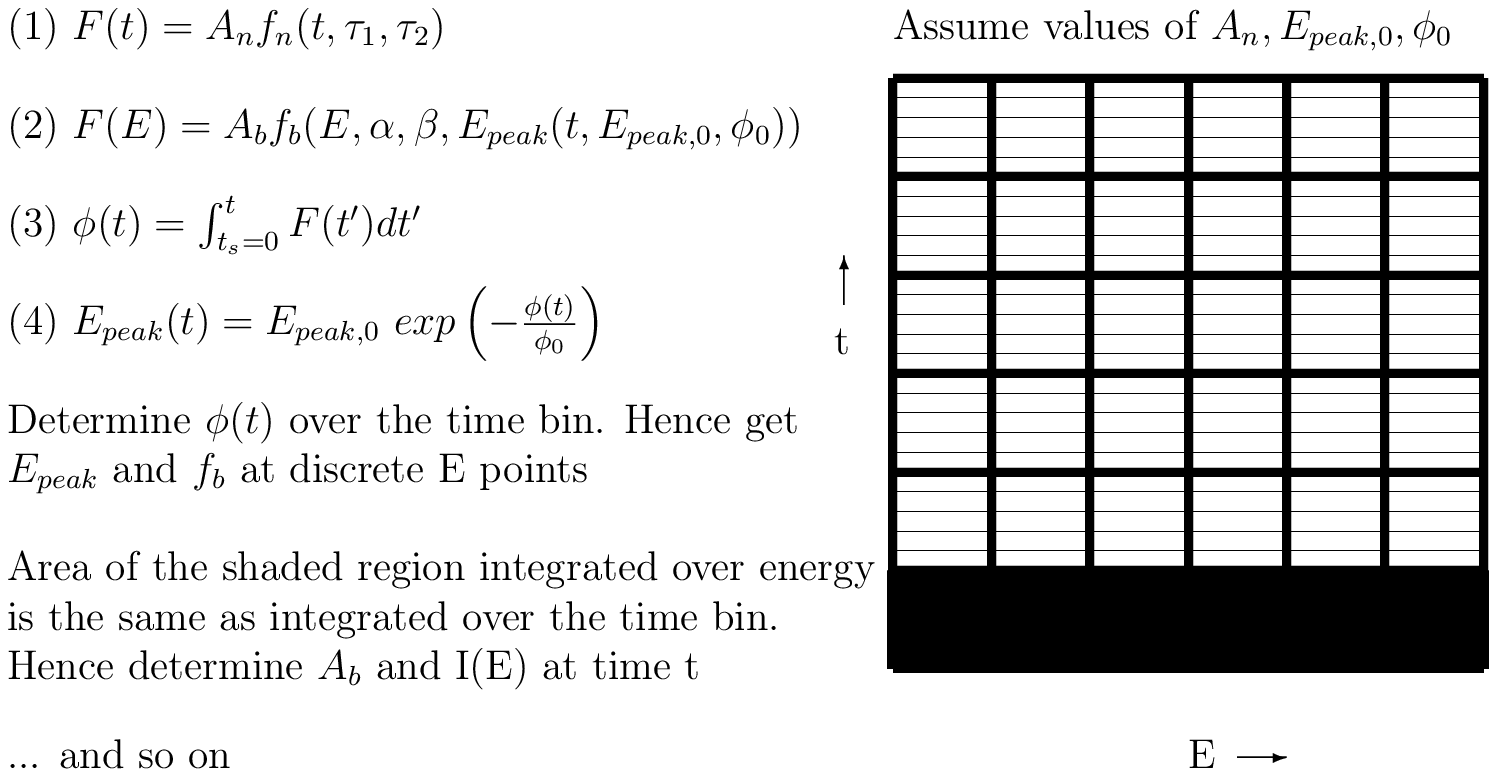}
\caption{Illustration of the simultaneous model for a set of $E_{\rm peak,0}$, $\phi_0$ and the
normalization of Norris model ($A_{n}$).}
\label{ch3_fig3}
\end{figure*}

\subsection{Assumptions: Pulse-wise $\bf E_{\rm \bf peak}$ Evolution}
The essential idea is to first segregate the global pulse parameters from the 
evolving parameters. First, we assume that the peak energy ($E_{\rm peak}$) in a pulse
shows a HTS evolution, which is characterized as follows (\citealt{LK_1996}; LK96 hereafter).

\begin{equation}
 E_{\rm peak} (t)=E_{\rm peak, 0}{\rm exp \left(-\frac{\phi(t)}{\phi_0} \right)}
\label{lk96}
\end{equation}
\vspace{0.05in}

Here, $\phi(t)=\int_{t_s}^{t} {F(t') dt'}$ is defined as the integrated flux up to time, $t$
(i.e., ``running'' fluence) from the start of the pulse, $t_s$. For our convenience, let us 
shift the time coordinate to get $t_s=0$. The value of $t_s$ for each pulse is known from
the Norris model fit, hence, we can shift each pulse, and co-add them to generate the full GRB. 
We further assume that certain spectral parameters do not have appreciable time evolution in a pulse. 
That is, their values can be deemed constant at the pulse-averaged values. These parameters are 
the photon indices ($\alpha$, and $\beta$). Similarly, we assume that certain 
timing parameters can be represented by the corresponding energy-integrated values.
These are the time constants of Norris model ($\tau_1$, and $\tau_2$).

\cite{Kocevski_Liang_2003} have performed time-resolved analysis to 
obtain the running fluence, $\phi(t)$ of LK96 model at different time bins.
The values of $E_{\rm peak}(t)$ at these times are found by spectral fit with Band function.
A linear fit to the $\phi(t)-E_{\rm peak}(t)$ plot gives the model parameters. This procedure 
is applied for GRBs with single FRED pulses. The fact that each FRED pulse has a 
characteristic slope of $\phi(t)-E_{\rm peak}(t)$ plot, supports the HTS spectral evolution.
Our aim is to use this spectral evolution, and generate a synthetic model pulse with a simultaneous 
timing and spectral information. Note that as time-resolved 
description essentially reduces the statistics, one can use the broad 
pulses rather than using intensity guided time-resolved bins. But, a simultaneous description 
is always better, as one can generate the synthetic pulse with any desired resolution, and then 
use it to describe the pulse properties. An added benefit is that the individual synthetic
pulses are essentially independent, with no overlapping effect. Hence, they can be used for GRBs with 
multiple pulses. If the derived properties of the individual pulses conform with the data, then 
  this model also favours HTS evolution.

\subsection{A 3-Dimensional Pulse Description}

With the assumptions of Section 3.4.1, we now have 4 global parameters --- 2 time indices ($\tau_1$, $\tau_2$),
and 2 spectral indices ($\alpha$, $\beta$). We also have 2 LK96 model parameters --- $E_{\rm peak, 0}$
and $\phi_0$. The $E_{peak}$ is dependent on these parameters as well as the timing model (equation~\ref{n1}).
The other parameters are the normalizations of Norris model and Band model ($A_n$ and $A_b$).
These will be determined in the process of developing the model. 

We determine the model as follows (see Figure~\ref{ch3_fig3}). 

\begin{itemize}
 \item (i) Take a grid of energy ($E$) and time ($t$) (thicker lines in Figure~\ref{ch3_fig3}). 
Take a finer resolution of each time bin. Also, assign some values to $E_{\rm peak,0}$, $\phi_0$ and the
normalization of Norris model ($A_{n}$). From the assumption of Section 3.4.1, we have the 
values of $\tau_1$ and $\tau_2$ as the energy-integrated values. Hence, at 
time $t$ (the average of 0 and the first time grid), one can find $\phi(t)$, 
integrating equation~\ref{n1} from 0 to $t$. We use the finer time resolution for this purpose.

\item (ii) Now for the derived value of $\phi(t)$, we use equation~\ref{lk96} to get $E_{\rm peak}$
at time $t$. Using the value of $E_{\rm peak}$, and the pulse-average values of $\alpha$
and $\beta$ in equation~\ref{band2}, we find $f_b$ at the average of each energy grid. 

\item (iii) We now want to express the value of $A_b$ in terms of $A_n$ so that the derived 
model has only one normalization parameter. Note that
the area integrated over energy for the first time grid (shaded region in Figure~\ref{ch3_fig3})
is equal to the time-integrated $F(t)$ i.e., $\phi(t)$. Hence, one gets $A_b=\phi(t)/\int f_b dE$,
where the integration is done over the entire energy range.

\end{itemize}

We perform the procedure described above for all the time grids. Thus we get $F(t, E)$ at each of these 
grid points $F(t, E)= A_b f_b (E, \alpha, \beta, E_{\rm peak}(t, E_{\rm peak, 0}, \phi_0))$, where 
$A_b$ can be determined from $A_n$. This is a three-dimensional representation of the pulse (3D pulse model),
which contains both the time and energy information simultaneously.

\subsection{XSPEC Table Model}

\begin{table*}\centering
\caption[The best-fit values of the 3D pulse model parameters]
{The best-fit values of the 3D pulse model parameters, $E_{peak,0}$, and $\phi_{0}$ obtained by $\chi^{2}$ minimization
in {\tt XSPEC}. The quoted errors are determined at nominal 90\% confidence. The average pulse properties ($\tau_{1}$,
$\tau_{2}$, and $t_{s}$) are also reported}

\begin{tabular}{cccccccc}
\hline 
Pulse & $E_{peak,0}$ & $\phi_{0}$ & Norm & $\chi^{2}$ (dof) & $\tau_{1}$(s) & $\tau_{2}$(s) & $t_{s}$(s)\\
\hline
\hline 
1 & $359_{-92}^{+65}$ & $12.2_{-1.3}^{+2.5}$ & $0.74\pm0.06$ & 40.12 (75) & $795.4_{-7.1}^{+7.2}$ & $0.54_{-0.005}^{+0.005}$ & $40.1_{-0.2}^{+0.2}$\\
\hline 
2 &  &  &  &  & $858.2_{-7.7}^{+6.4}$ & $0.58_{-0.005}^{+0.004}$ & $44.2_{-0.2}^{+0.2}$\\
\hline 
3 & $324_{-83}^{+82}$ & $18.8_{-6.2}^{+9.6}$ & $0.54\pm0.04$ & 31.41 (75) & $353.4_{-50}^{+43}$ & $2.47_{-0.09}^{+0.12}$ & $50.3_{-1.2}^{+1.5}$\\
\hline 
4 & $307_{-99}^{+41}$ & $12.0_{-2.1}^{+2.0}$ & $0.19\pm0.04$ & 56.42 (75) & $532.0_{-132}^{+209}$ & $1.58_{-0.16}^{+0.15}$ & $81.0_{-3.4}^{+2.6}$\\
\hline
\end{tabular}
\label{3Dmodel}
\end{table*}

The synthetic 3D pulse model is naturally dependent on 3 model parameters --- $E_{\rm peak,0}$, $\phi_0$ and the
normalization of Norris model ($A_{n}$). In order to find these parameters for a particular 
pulse, we take a grid of $E_{\rm peak,0}$ and $\phi_0$. At each grid point we generate 
the 3D pulse model. These synthetic pulse models can be used in two ways: (i) if we integrate
the model over time, we shall get spectrum at each of the $E_{\rm peak,0}-\phi_0$ grid points. (ii) an 
integration over a given energy band, on the other hand, will give a pulse LC. Note the immense 
flexibility of the 3D model --- we can opt for the finest desired resolution, and we can generate 
the LCs in any desired energy bin. 

In our analysis, we first integrate the 3D pulse over time to get spectrum at each $E_{\rm peak,0}$, $\phi_0$ grid point.
This is called a {\tt XSPEC} table model. The model falls under the additive category of {\tt XSPEC} models.
This model can be used to determine the best-fit values of 
$E_{\rm peak,0}$ and $\phi_0$ by $\chi^{2}$ minimization. {\tt XSPEC} adds normalization 
as a third parameter during the fitting. This way we get the third variable ($A_n$) of our model.

\subsubsection{Specification Of The Model And Best-fit Values}
For the pulses we analyze, we use a time grid ($t$-grid) resolution 
of 0.5 s, while a energy grid ($E$-grid) has a resolution of 2.0 keV. 
As each pulse has a characteristic set of timing and spectral parameters,
we generate a separate {\tt XSPEC} table model for each of them.
As we use the \textit{Swift}/BAT for this analysis, we generate the 
spectrum in 2-200 keV energy band. The time axis is obtained in 
excess of the total pulse duration. A typical table model contains 
200 values of $E_{peak,0}$ in the range $100-1100$ keV, and 50 values
of $\phi_{0}$ in the range $2 - 77$ photon $\rm cm^{-2}$. In addition,
it requires some specifications ---
the energy binning information, the range of parameter values, and the 
header keywords as required for {\tt XSPEC} table model specification 
(OGIP Memo OGIP/92-009).

The table model is used in {\tt XSPEC} to determine the model parameters.
In Table~\ref{3Dmodel}, we have shown the best-fit values of $E_{\rm peak,0}$ 
and $\phi_0$ as obtained by $\chi^{2}$ minimization. The errors in the parameters 
are determined with nominal 90\% confidence. We have considered the first two pulses 
(pulse 1 and pulse 2) together, as they have a large overlap. We also give the
average Norris model parameters ($\tau_1$ and $\tau_2$), and the start time ($t_s$)
of each pulse. 

\subsubsection{Three Dimensional Pulse Model For The Best Fit Values}

\begin{figure}\centering
\includegraphics[width=3.4in]{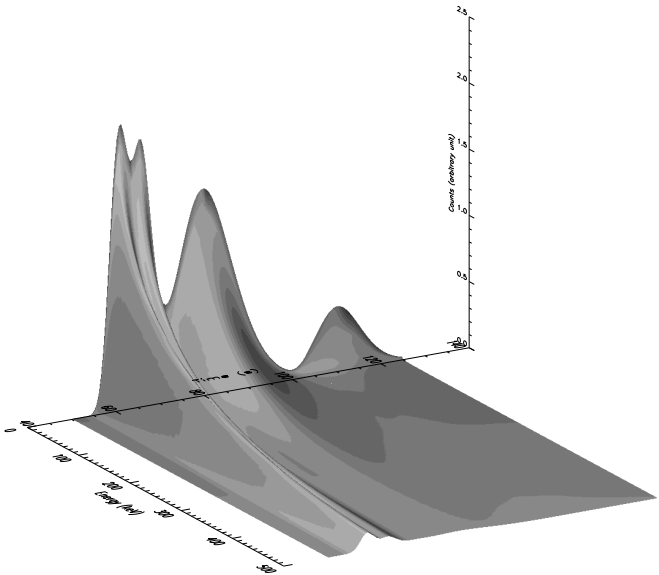}
\caption[Synthetic 3D model of GRB 090618]{The synthetic 3D model of GRB 090618. Each pulse is generated separately 
with their respective best-fit parameters. The pulses are then shifted to their respective
start time, and co-added to generate the full GRB (Source: \citealt{Basak_Rao_2012_090618})}
\label{ch3_fig4}
\end{figure}

As we have all the required parameters for the 3D pulse description, we regenerate the pulses, shift 
them to the respective start time, and co-add them to generate the full GRB. For illustration purpose, 
we have shown the 3D view of GRB 090618 in Figure~\ref{ch3_fig4}. The four pulses of the GRB are clearly 
visible. The count is shown (in arbitrary units) as a function of time and energy. The photon spectrum 
at any instance is a Band function, and the LC in any energy band is a Norris model. Note that this 
figure is used to display the 3D structure of the pulses up to normalization factors. When we use 
this 3D model to obtain various pulse parameters, we normalize the pulses and then derive the 
parameters. However, as pulse width ($w$) is defined as the time interval between the points with 
1/e flux values relative to the peak, and spectral lag ($\tau$) is defined as the difference in the 
peak position of the lightcurve in two energy bands, they do not depend on the normalization.

\section{Timing Analysis Using the 3D Model}
A 3D pulse model can be used in many ways to derive various pulse properties. In our study, we have done 
the following analyses.

\subsection{Synthetic Lightcurves}
For the best-fit values of $E_{\rm peak,0}$ and $\phi_0$, we generate the 3D model of each pulse, and
integrate them over some desired energy band ($E_1$ to $E_2$). This gives LC of each pulse. We shift the LCs
to the respective start time to generate the full LC. We choose the same energy bands which we used earlier 
for Norris model fit to the LCs (section 3.3.1). These are 15-25, 25-50, 50-100, and 100-200 keV. 

\begin{figure}\centering
\includegraphics[width=3.4in]{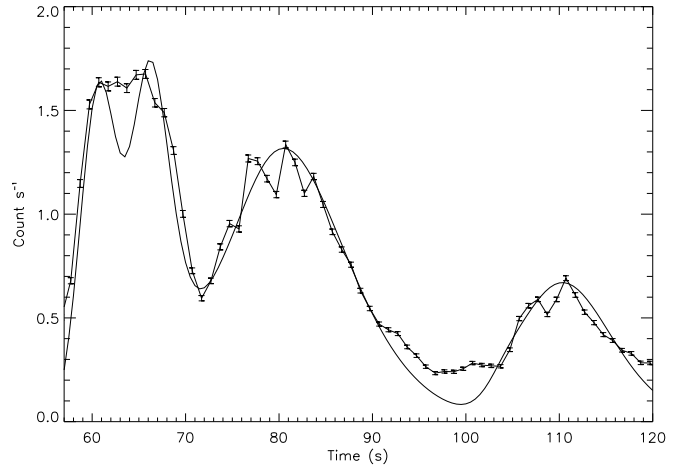}
\caption [Synthetic LC of GRB 090618 in 25-50 keV energy band, over-plotted on the BAT LC in the same energy band]
{The synthetic LC of GRB 090618 in 25-50 keV energy band, over-plotted on the BAT LC in the same energy band.
Note that, except for the rapid variability, the synthetic LC captures the observed LC quite well (Source: \citealt{Basak_Rao_2012_090618}).}
\label{ch3_fig5}
\end{figure}

\begin{figure*}\centering
{
\begin{tabular}{ll}
\includegraphics[width=2.8in]{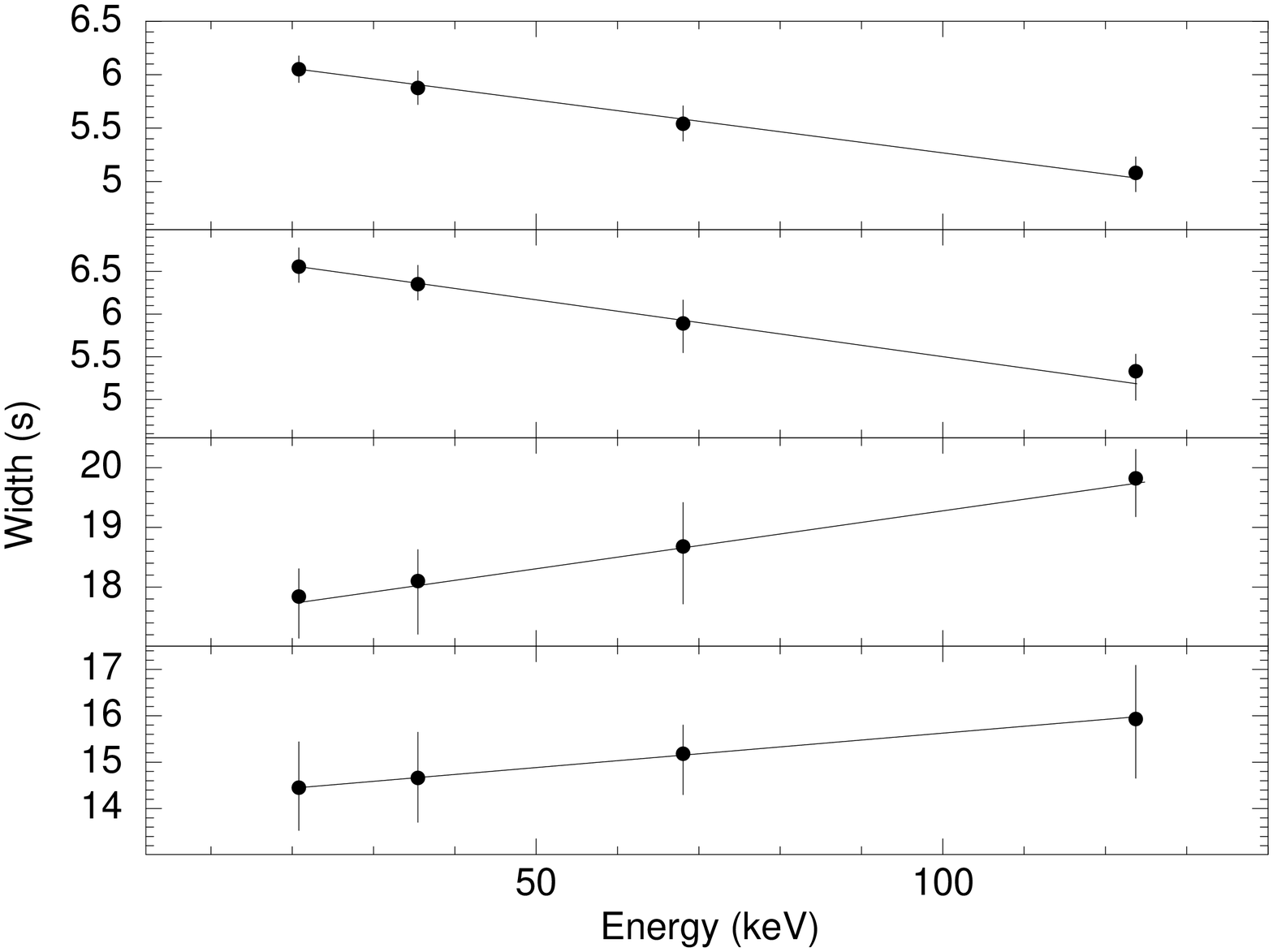} &
\includegraphics[width=2.8in]{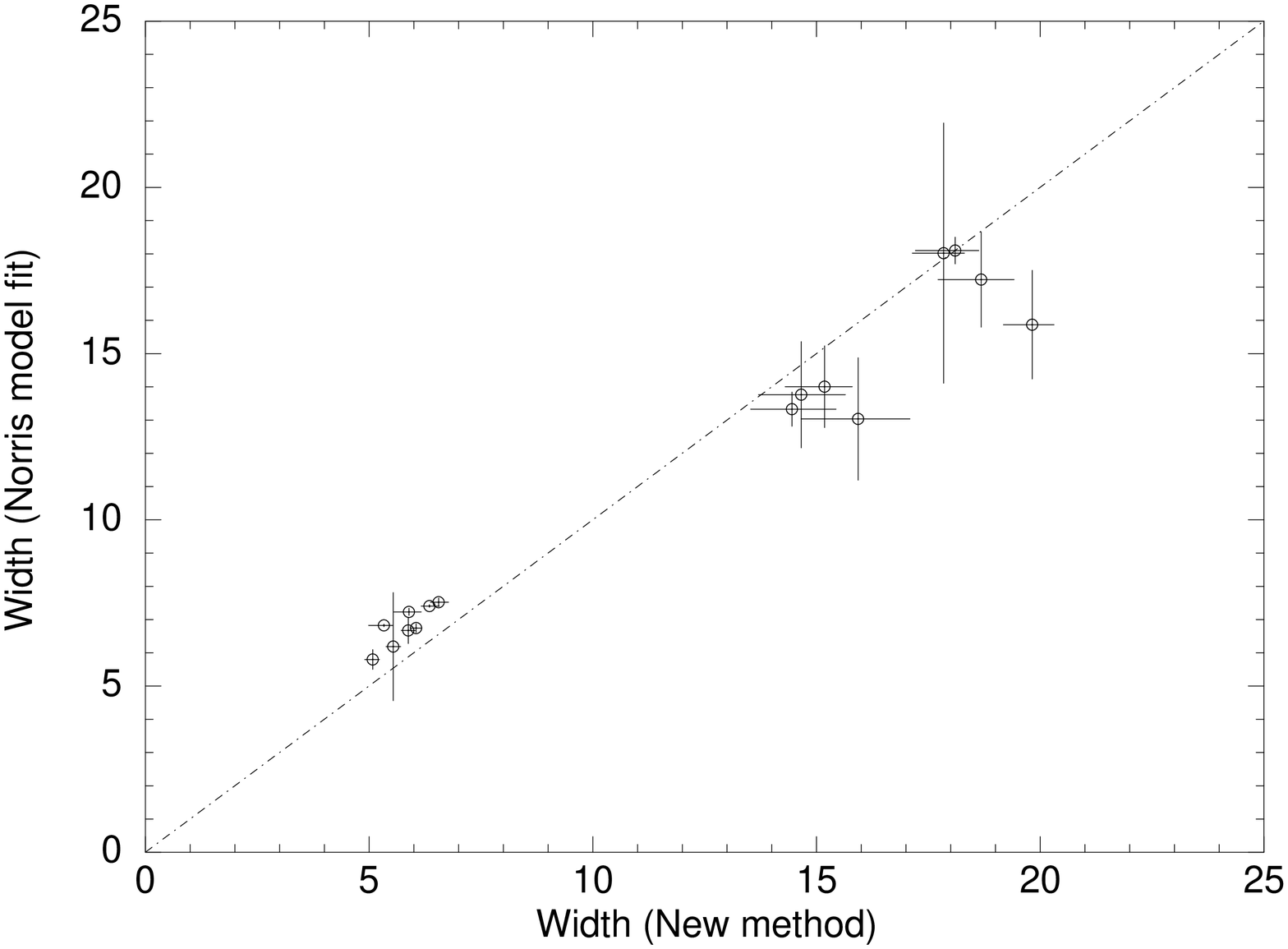} 
\end{tabular}
}
\caption[Variation of $w$ in different energy bands as obtained by using the 3D pulse model,
and a comparison with the direct model fitting] {\textit{Left Panel:} Variation of $w$ in different energy bands 
for the four pulses (upper to lower --- pulse 1-4) as obtained by using the 3D pulse model. Note that the first two pulses 
have \textit{normal width broadening}, while the last two pulses show \textit{anomalous width broadening}. 
\textit{Right Panel:} A comparison between the width ($w$) of the four pulses 
obtained from directly fitting the LCs (Norris model fit), and those obtained from the 
synthetic pulse LCs. The dot-dashed line is the line of equality. (Source: \citealt{Basak_Rao_2012_090618}).
}
\label{ch3_fig6}
\end{figure*}

In Figure~\ref{ch3_fig5}, we have shown the synthetic LC in 25-50 keV energy band, over-plotted with the 
BAT LC in the same energies. The count rate is the BAT mask weighted count rate. Note that the LC
has rapid time variability superimposed on the the broad pulse structure. In our pulse description,
we consider only the smooth and broad time variability. The rapid variability timescales generally
give large $\chi^2$ when we use the smooth synthetic LC for fitting. Hence, the normalization 
factors cannot be determined by using $\chi^2$ minimization technique. Instead, we estimate the 
factors by physically inspecting the LCs. It is worthwhile to mention again that due to the large 
overlap, the first two pulses have been considered as a single entity. Hence, the derived 
model parameters ($E_{\rm peak,0}$ and $\phi_0$) are average quantities. This averaging effect
essentially shows up in the derived LC. However, note that except for the rapid variability, the synthetic
LC is in good agreement with the data. Hence, we conclude that the assumption of $E_{\rm peak}$ 
evolution (equation~\ref{lk96}), along with the the global parameters of our model correctly reproduce 
the energy-resolved LCs. This finding also validates the fact that each pulse can be considered as a
HTS pulse. The IT behaviour is possibly a superposition effect. 

\subsection{Deriving The Timing Parameters}

Using the synthetic light curves in different energy bands, one can derive various parameters characterizing 
a pulse. We shall derive the following pulse properties --- pulse width ($w$), and spectral lag ($\tau$).
The derived parameters, and their energy evolution will be checked with   the values as derived from the
direct observation. We can derive the pulse width of the individual pulses by directly fitting 
the LCs (equation~\ref{n1}), and then using the best-fit model parameters ($\tau_1$, $\tau_2$)
as done in section 3.3.1. The spectral lag ($\tau$) between two energy bands can be calculated by 
cross-correlating the LCs in those energy bands (\citealt{Raoetal_2011}). Note that 
due to the overlapping effect, the direct measurement of $w$ and $\tau$ can be erroneous. This 
effect will be less severe for the $w$ measurement. Of course, the rising part of a pulse can be 
affected by the falling part of the preceding pulse, broadening the width. But, the fact that 
all the pulses are simultaneously fitted with Norris model, takes care of the overlapping effect.
However, $\tau$ would be affected by the overlapping effect in a similar way as two overlapping 
HTS pulses leads to a IT (or rather soft-to-hard-to-soft --- SHS) behaviour in the overlapping 
region. In these cases, the 3D pulse model is very useful. As the pulses are independently 
generated, the derived $w$ and $\tau$ will be unaffected by the overlapping effect.

\begin{table*}\centering
\caption[Pulse width ($w$) variation of the four pulses with the energy bands]
{Pulse width ($w$) variation of the four pulses with the energy bands. The numbers in parentheses are $w$ measured 
by the Norris model fit}

\begin{tabular}{ccccc}
\hline 
Pulse & 15-25 keV & 25-50 keV & 50-100 keV & 100-200 keV \\
\hline
\hline 
1 & $6.05_{-0.12}^{+0.12}$  & $5.88_{-0.16}^{+0.16}$  & $5.54_{-0.17}^{+0.16}$  & $5.08_{-0.18}^{+0.15}$ \\
  & ($6.74_{-0.13}^{+0.13}$) & ($6.67_{-0.39}^{+0.39}$) & ($6.18_{-1.62}^{+1.62}$) & ($5.79_{-0.30}^{+0.30}$)\\
\hline 
2 & $6.56_{-0.19}^{+0.22}$  & $6.35_{-0.19}^{+0.22}$  & $5.89_{-0.34}^{+0.27}$  & $5.33_{-0.34}^{+0.20}$  \\
  & ($7.52_{-0.13}^{+0.13}$) & ($7.40_{-0.04}^{+0.04}$) & ($7.23_{-0.11}^{+0.11}$) & ($6.82_{-0.04}^{+0.04}$) \\
\hline 
3 & $17.84_{-0.70}^{+0.47}$  & $18.10_{-0.89}^{+0.53}$  & $18.68_{-0.96}^{+0.74}$  & $19.82_{-0.64}^{+0.49}$  \\
  & ($18.01_{-3.91}^{+3.91}$) & ($18.09_{-0.40}^{+0.40}$) & ($17.22_{-1.43}^{+1.43}$) & ($15.87_{-1.63}^{+1.63}$) \\
\hline 
4 & $14.45_{-0.92}^{+0.99}$  & $14.66_{-0.96}^{+0.98}$  & $15.18_{-0.88}^{+0.62}$  & $15.93_{-1.27}^{+1.16}$  \\
  & ($13.32_{-0.51}^{+0.51}$) & ($13.76_{-1.60}^{+1.60}$) & ($14.00_{-1.23}^{+1.23}$) & ($13.03_{-1.84}^{+1.84}$) \\
\hline
\end{tabular}
\label{w_model}
\end{table*}

\subsubsection{A. Pulse Width Variation With Energy}
We generate the LCs in various energy bands, measure the peak position, and calculate the width 
at the two points where intensity is $\rm exp(-1)$ factor of the peak intensity. This is essentially
the same definition used for $w$ calculation using Norris exponential model.
We define $\tau$ of an energy band as the peak position of the LC with respect to 
that of the lowest energy band (15-25 keV).
Note that our model assumes global values of certain parameters ($\tau_1$, $\tau_2$,
$\alpha$ and $\beta$) characterized by the pulse-averaged values. The $E_{peak}(t)$
is the only time variable parameter which depends on LK96 model parameters --- $E_{peak,0}$, $\phi_{0}$.
Hence, all the timing properties of a pulse (i.e., $w$ and $\tau$ at various energy bands)
are essentially determined by the evolution law of $E_{peak}(t)$. 

In Table~\ref{w_model} and \ref{lag_model}, we have shown the $w$ and $\tau$ as derived 
from the synthetic LCs of the pulses. In Table~\ref{w_model}, the numbers shown in parentheses 
are those obtained by directly fitting the pulse LCs with Norris model. In Figure~\ref{ch3_fig6} (left panel),
we compare $w$ as derived from these very different methods. We note that 
the observations and predictions match quite well. From Table~\ref{w_model}, we notice for the first two pulses that 
$w$, as derived by our model have systematically lower values than the directly measured values.
This is due to the fact that these two pulses are considered together in our model.
But, as we have developed the 3D model of these pulses separately, we believe that 
the derived widths are devoid of overlapping effect. A complete disentanglement of 
heavily overlapping pulses can give more secured values, however, it is difficult to 
achieve in our model.

\begin{figure}\centering
\includegraphics[width=3.4in]{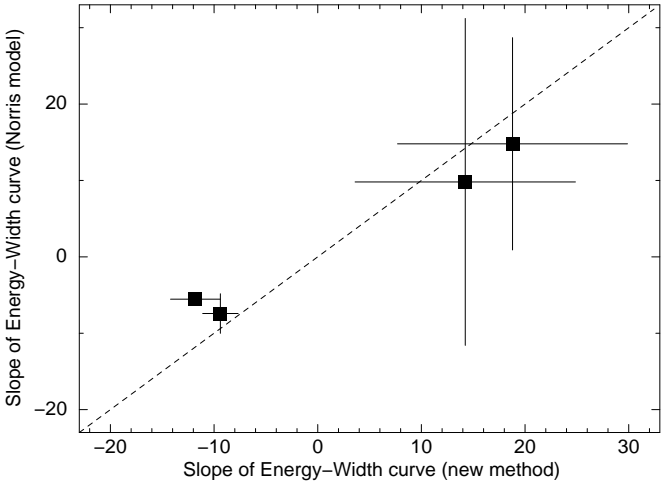}
\caption[Comparison of the slopes of linear fit to the $E-w$ data as 
found by directly fitting the data with Norris model, and by using our 3D pulse model]
{A comparison of the slopes of linear fit to the $E-w$ data as 
found by directly fitting the data with Norris model, and by using our 3D pulse model
(Source: \citealt{Basak_Rao_2012_090618}).}
\label{ch3_fig8}
\end{figure}

We now study the variation of $w$ with energy. As discussed in section 3.3.1 pulse widths 
are expected to be broader at lower energies (\citealt{Norrisetal_1996, Hakkila_Preece_2011}). We call this 
\textit{normal width broadening}. However, as already pointed out (Figure~\ref{ch3_fig2}),
we also have \textit{anomalous width broadening} (i.e., width decreases with decreasing energy
at lower energy bands) for the last two pulses. In Figure~\ref{ch3_fig6} (left panel),
we have shown the variation of $w$ as a function of energy for the four pulses. 
It is clear that the first two pulses follow the normal trend, while the last two 
pulses have a reverse $E-w$ variation. This reverse variation cannot arise 
from any contamination effect as could have been an argument for direct LC fitting.
For the individual pulses, we fit the $E-w$ data with linear function. We obtain 
$(-9.4\pm1.7)\times10^{-3}$ s $\rm keV^{-1}$, and $(-11.8\pm2.4)\times10^{-3}$ s $\rm keV^{-1}$
for the first two pulses with $\chi^2$ (dof) = 0.13 (2) and 0.11 (2), respectively.
These slopes are convincingly negative. However, the $E-w$ slope of the last 
pulses are positive ($(18.8\pm11.1)\times10^{-3}$ s $\rm keV^{-1}$,
and $(14.2\pm10.6)\times10^{-3}$ s $\rm keV^{-1}$ with $\chi^2$ (dof) = 
0.0046 (2) and 0.0021 (2), respectively). Note that the errors in the $E-w$
slope has improved from our previous study of LC with direct Norris model fit.
The improvement is apparent for the third pulse ($(18.8\pm11.1)\times10^{-3}$ s $\rm keV^{-1}$
compared to $(15\pm14)\times10^{-3}$ s $\rm keV^{-1}$). Hence, at least for this 
pulse the evidence of anomalous width broadening is convincing. In the right 
panel of Figure~\ref{ch3_fig6}, we have compared the values of $w$ as obtained 
by the new pulse model and Norris model fitting. We note that the derived values 
are similar to each other. In Figure~\ref{ch3_fig8}, we have compared the slope of 
the $E-w$ data of the four pulses as obtained from the two methods. We again note
a general agreement between the data.

\begin{figure}\centering
\includegraphics[width=3.4in]{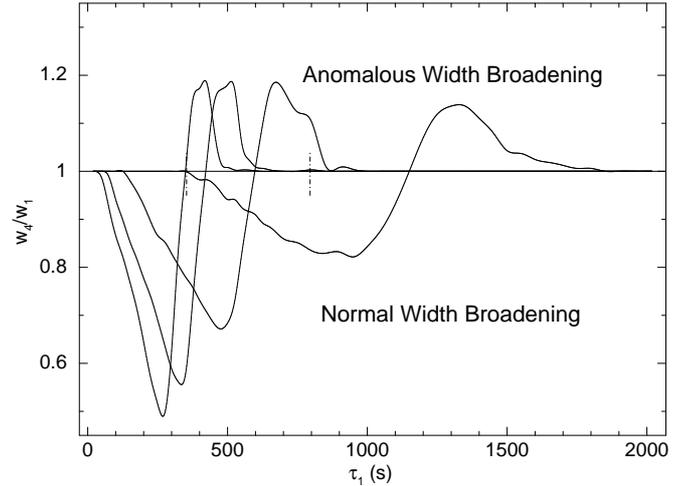}
\caption[The regions of \textit{normal} and \textit{anomalous} width broadening in the parameter space]
{The \textit{normal} and \textit{anomalous} width broadening region as found by 
the ratio plot of $w_4/w_1$ as a function of $\tau_1$. The two regions are separated 
by the equality line $w_4/w_1=1$. Different curves are obtained for 
different values of $\tau_2$, which are varied in the range (0.54 to 2.47) from 0.54 
(right most), 1.20, 1.90 and 2.47 (left most). The values of $\tau_1$ for the first 
and thirst pulse (795.4 and 353.4, respectively) are shown by dot-dashed line.
For example, the set of values for pulse 1 ($\tau_1=795.4$, $\tau_2=0.54$) is covered 
by the right most curve, and it falls in normal width broadening region
(Source: \citealt{Basak_Rao_2012_090618}).}
\label{ch3_fig9}
\end{figure}

\begin{table*}\centering
\caption[The model predicted spectral delay of higher energy photons with respect to the photons in
15-25 keV energy band]
{The model predicted spectral delay of higher energy photons with respect to the photons in
15-25 keV energy band (mean energy = 20.82 keV). Values are calculated for the individual pulses}

\begin{tabular}{cccc}
\hline 
Pulse & Energy Channel & Mean Energy (keV) & Delay (s)\\
\hline
\hline 
1 & 15-25 vs. 25-50 keV & 35.45 & $-0.135_{-0.022}^{+0.011}$\\
 & 15-25 vs. 50-100 keV & 68.07 & $-0.375_{-0.039}^{+0.023}$\\
 & 15-25 vs. 100-200 keV & 123.73 & $-0.730_{-0.045}^{+0.043}$\\
\hline 
2 & 15-25 vs. 25-50 keV & 35.45 & $-0.195_{-0.005}^{+0.005}$\\
 & 15-25 vs. 50-100 keV & 68.07 & $-0.555_{-0.033}^{+0.033}$\\
 & 15-25 vs. 100-200 keV & 123.73 & $-1.070_{-0.037}^{+0.023}$\\
\hline 
3 & 15-25 vs. 25-50 keV & 35.45 & $-0.015_{0.040}^{+0.041}$\\
 & 15-25 vs. 50-100 keV & 68.07 & $-0.040_{-0.030}^{+0.012}$\\
 & 15-25 vs. 100-200 keV & 123.73 & $-0.085_{-0.005}^{+0.005}$\\
\hline 
4 & 15-25 vs. 25-50 keV & 35.45 & $-0.020_{-0.010}^{+0.010}$\\
 & 15-25 vs. 50-100 keV & 68.07 & $-0.065_{-0.040}^{+0.035}$\\
 & 15-25 vs. 100-200 keV & 123.73 & $-0.095_{-0.005}^{+0.005}$\\
\hline
\end{tabular}
\label{lag_model}
\end{table*}

The anomalous width broadening as observed for the last two pulses, may 
have a direct physical reason. However, in our analysis, we have used only 
empirical models for a pulse description. Hence, we shall try to give a phenomenological 
reasoning for reverse width variation. Our motivation for such study is to show that a reverse width 
variation can indeed occur due to a particular combination of model parameters instead of a sensitivity 
limit of the detector at the lower energy band. First, we consider pulse 1 and pulse 3, 
which are the two most convincing cases of normal and anomalous width variation, respectively.
From Table~\ref{3Dmodel}, let us compare their parameters. In the parentheses, we have shown the 
values for pulse 3. These are (in usual units): LK96 parameters --- $E_{peak,0}=359 ~(324)$, $\phi_{0}=12.2 ~(18.8)$,
pulse-average photon indices --- $\alpha= -1.11 ~(-1.15)$, $\beta= -2.5~ (-2.5)$, and time constants of
Norris model --- $\tau_{1}=795.4~ (353.4)$, $\tau_{2}=0.54~ (2.47)$. It is evident that the two 
pulses have all the parameters nearly similar, except for $\tau_{1}$ and $\tau_{2}$. Hence,
these two global parameters are the main contributors for the difference in $E-w$ relation 
of these two pulses. To test how these parameters affect the width variation, let us 
explore the parameter space of $\tau_{1}$ and $\tau_{2}$ with all other parameters 
fixed to the values of the first pulse. Let us define broadening as the ratio of width 
in 100-200 keV band ($w_4$) to that in 15-25 keV band ($w_1$). We say that a pulse 
follows a normal broadening law if $\frac{w_{4}}{w_{1}}>1$, and anomalous, otherwise. 
The procedure is as follows. We shall assume a few values of $\tau_{2}$ in the 
range $0.54 - 2.47$ (the values of $\tau_{2}$ for pulse 1 and 3, respectively), and vary $\tau_1$.
This is because $\tau_{2}$ has a relatively small range. For these values of $\tau_{2}$,
we shall essentially generate many synthetic LCs with a range of $\tau_1$. For each realization,
we calculate $\frac{w_{4}}{w_{1}}$. A plot of $\frac{w_{4}}{w_{1}}$ with $\tau_1$
for a particular value of $\tau_{2}$ gives the region of normal and anomalous width broadening
characterized by $\tau_{2}$.

In Figure~\ref{ch3_fig9}, we have shown the regions of parameter space where a normal and 
anomalous width broadening can occur. The whole plot is divided into two regions by $w_{4}$/$w_{1}=1$ 
line --- (i) $w_{4}<w_{1}$ (lower region), and (ii) $w_{4}>w_{1}$ (upper region) for which we 
get a normal and anomalous broadening, respectively. Different curves represent different 
values of $\tau_2$, from right to left 0.54, 1.20, 1.90 and 2.47. For a given value of 
$\tau_2$, a pulse tends to show anomalous broadening for higher values of $\tau_1$.
Also, the lower the value of $\tau_2$, the higher the allowed value 
of $\tau_1$ for a normal width broadening. 
The set of values for the first pulse ($\tau_1=795.4$, $\tau_2=0.54$) is securely 
positioned in the normal broadening region of the right most curve. The values of 
the third pulse ($\tau_1=353.4$, $\tau_2=2.47$) marginally appears in the anomalous
region of the left most curve. Hence, the third pulse is likely to show a 
reverse width broadening. Note that the parameters used for the third pulse
are those of the first pulse, except for the values of $\tau_1$ and $\tau_2$.
Still we get anomalous broadening for this pulse. This shows that the combination 
of $\tau_1$ and $\tau_2$ of this pulse is prone to the reverse width variation. 
It also signifies that the $E-w$ data is really insensitive to the other parameters.

\subsubsection{B. Spectral Lag of Different Energy Bands}

We now study the spectral lag of different energy bands with respect to the 
lowest energy band (15-25 keV). As we have assumed a HTS evolution in each of the pulses, 
we expect the soft x-rays to lag behind the hard x-rays.
In Figure~\ref{ch3_fig10}, we have shown the spectral lag of the four 
pulses in three different energy bands. It is clear that the lags are negative, 
showing the soft delay. The model predicted spectral delay can be compared with 
the observed delay (\citealt{Raoetal_2011}). These are shown in Figure~\ref{ch3_fig11} .
The time bins used by \cite{Raoetal_2011} are $0-50$ s, $50-77$ s, $77-100$ s
$100-180$ s (post-trigger). The first bin is the precursor, hence neglected for the
lag calculation. In the second time bin, pulse 1 and 2 appear together. 
The third bin captures the third pulse, while we neglect the fourth bin owing 
to the large difference of time cut. In Figure~\ref{ch3_fig11},
we have shown the data of pulse 1 and 2 both for a single delay quoted for the 
second time interval, $50-77$ (the lowest 6 data points, with left points for pulse 2). 
We note that the data in general agrees with the model predicted delay, though with a
deviation for the first pulse. This is, of course, due to the fact that the 
two pulses are combined for the calculation of the observed delay. Note that the 
second pulse is closer to the observed value. Another important point is that 
we can find the spectral lag even within the overlapping pulses by the simultaneous 
timing and spectral description. Also, one has to consider the fact that
we are using only the peak position rather than performing a cross-correlation
as was done  by \cite{Raoetal_2011}.

\subsection{Summary Of 3D Pulse Description}

Before going to the next section, let me summarize the essential points that we have discussed.
We have attempted to describe the individual pulses of a GRB simultaneously in time and energy domain.
Our main assumptions are 
(i) $E_{\rm peak}$ has a HTS evolution (equation~\ref{lk96}) within a pulse, (ii) an instantaneous spectrum 
can be described by Band function (equation~\ref{band2}) with the spectral 
indices represented by the pulse-average value, and $E_{\rm peak}(t)$ is given by equation~\ref{lk96}, and 
(iii) the pulse LC has a Norris model shape with the time constants determined by the average values.
With these simple assumptions, we have developed a table model in {\tt XSPEC} to derive the model parameters.
We have successfully generated the model with best-fit parameters, and derived various pulse properties 
(LC, width, lag), and matched with the data. We found a tentative evidence of an \textit{anomalous} width 
broadening in the data, which is phenomenologically explained by a combination of model parameters.

One of the most important applications of the pulse-wise description is that it can be used for GRBs with
multiple pulses. As these pulses can be generated separately, the parameters have little overlapping effect.
In the next section, we shall apply this model for a set of GRBs, most of which have multiple pulses.

\begin{figure}\centering
\includegraphics[width=3.4in]{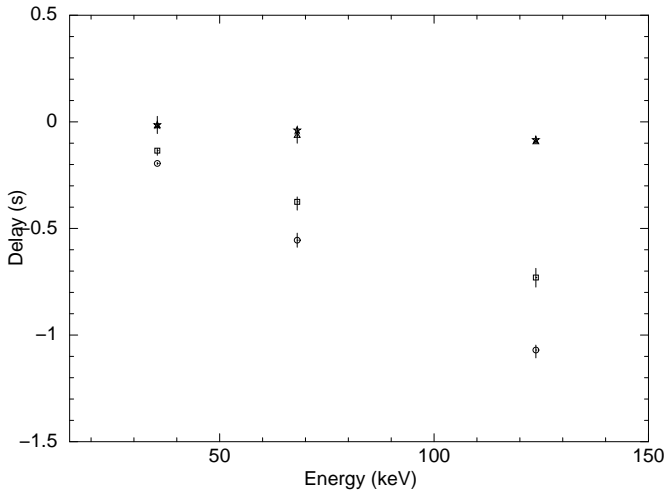} 
\caption[Spectral lag calculated by 3D pulse model]{The model predicted spectral lag for the 
four pulses (squares, circles, stars and triangle, respectively) are shown. The values 
are calculated using 15-25 keV as the reference band.
Hence, a negative value denotes a soft lag. In fact soft lag is expected 
due to the assumption of HTS evolution (Source: \citealt{Basak_Rao_2012_090618}).
}
\label{ch3_fig10}
\end{figure}

\begin{figure}\centering
\includegraphics[width=3.4in]{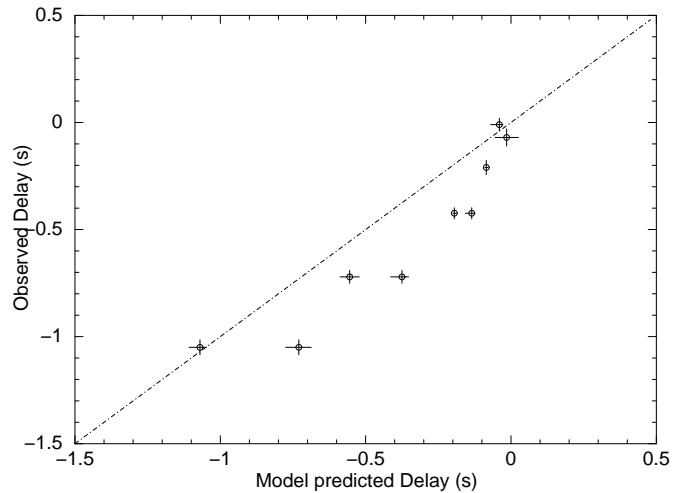} 
\caption[Comparison of spectral lag obtained by two different methods]
{A comparison between the spectral lag derived from 
the new model, and the observed lag calculated by cross correlation (\citealt{Raoetal_2011}).
the dot-dashed line shows the equality.
The lower 6 data points are model predicted lag for pulse 1 (the ones in the right) 
and pulse 2 (the ones in the left), as compared to the 50-77 s data of \citet[][see text]{Raoetal_2011}.
Source: \cite{Basak_Rao_2012_090618}.
}
\label{ch3_fig11}
\end{figure}

\section{A New Pulse-wise GRB Correlation}

Aided with a simultaneous timing and spectral description of the individual pulses of a GRB,
we now apply this model on a set of GRBs (\citealt{Basak_Rao_2012_correlation}). 
In this section, we shall study Amati correlation (\citealt{Amatietal_2002}),
which essentially says that a GRB with high peak energy ($E_{\rm peak}$) has high isotropic energy 
($E_{\gamma, \rm iso}$). This statement can be reverted as --- a GRB with high energy ($E_{\gamma, \rm iso}$)
produces higher energy photons. The physical reason of this correlation is unknown, but 
\cite{Amatietal_2002}, using 9 BATSE GRBs found a significant correlation (Spearman rank, $\rho = 0.92$,
corresponding to a chance probability, $P_{\rho}=5.0\times 10^{-4}$). In the source frame this correlation is approximately
$E_{\rm peak}\sim E_{\gamma, \rm iso}^{0.52\pm0.06}$. With a larger set of GRBs (41), \cite{Amatietal_2006}
have shown that the correlation still holds ($\rho = 0.89$, $P_{\rho}=7.0\times 10^{-15}$).
Similar correlation between $E_{\rm peak}$ and isotropic luminosity ($L_{\rm iso}$)
was later found by \cite{Yonetokuetal_2004} for a set of 16 GRBs (Pearson correlation, 
$r=0.958$, $P_{\rm r}=5.31\times 10^{-9}$). \cite{Ghirlandaetal_2004}, on the other hand, replaced  
$E_{\gamma, \rm iso}$ by the collimation corrected energy ($E_{\gamma}$) and found 
reasonable correlation for 27 GRBs ($\rho = 0.80$, and $P_{\rho}=7.6\times 10^{-7}$).
For a selected sample of 24 GRBs, this correlation improves ($\rho = 0.88$).

However, all these correlations are empirical, and can as well arise due 
to the selection bias of the instrument (\citealt{Band_Preece_2005, Nakar_Piran_2005,
Schaefer_Collazzi_2007_HD, Collazzietal_2012}). One way to argue against the 
selection bias is to prove a correlation within the time-resolved data. For example, 
\cite{Ghirlandaetal_2010} have studied 9 GRBs with known $z$. They have studied time-resolved 
$E_{\rm peak}-L_{\rm iso}$ correlation, and established its reliability. In this regard,
it is important to check the time-resolved Amati correlation. Hence, we take the 
sample used by \cite{Ghirlandaetal_2010} and study the Amati correlation.

In the following analysis, a $\Lambda$-CDM cosmology is assumed. Hence, we use the following parameters of cosmology. 
The Hubble parameter, $H_{\rm 0}=\rm 71~km ~s^{-1}~Mpc^{-1}$, the dark energy density, $\Omega_{\rm \Lambda}=0.73 $, 
total density of baryonic and dark matter, $\Omega_{\rm m}=0.27$, and a spatially flat universe. The values used here 
are determined by combining 7-year data of Wilkinson Microwave Anisotropy Probe (WMAP; \citealt{Jarosiketal_2011}), data from
Baryon Acoustic Oscillations (\citealt{Percivaletal_2010}), and the data from Type Ia supernova observation 
(\citealt{Riessetal_2011}). The recent measurements done by Planck mission indicates slightly different values of 
the parameters, namely, $H_{\rm 0}=\rm 67.3~km ~s^{-1}~Mpc^{-1}$, $\Omega_{\rm m}=0.315$ (\citealt{Planck_2013}).
The new values primarily alters the values of $E_{\rm \gamma, iso}$ at low $z$, e.g., at $z=1$, 3 and 10,
the changes are $\approx$ 7.0\%, 3.6\% and 1.7\%. As the $E_{\rm peak}$ values remain unaltered,
the slope and normalization of the relations will slightly change.

\subsection{Time-integrated, Time-resolved And Pulse-wise Amati Correlation}
We use the sample of \cite{Ghirlandaetal_2010} for our correlation study. They have selected
GRBs detected by the \textit{Fermi}/GBM, and having secured redshift ($z$) measurement.
A \textit{Swift}/BAT sample can provide a larger set. However, due to a limited 
energy coverage (15-150 keV), the \textit{Swift}/BAT often cannot give reliable value of $E_{\rm peak}$.
A joint \textit{Fermi}/GBM-\textit{Swift}/BAT analysis gives $\sim 10\%$ higher $E_{\rm peak}$ 
(chapter 2). For this reason, we shall also re-analyze the \textit{Fermi}/GBM data of
GRB 090618. To remind, the \textit{Swift}/BAT data of this GRB was used to demonstrate 
the 3D pulse model. The original \cite{Ghirlandaetal_2010} sample contains 12 GRBs. Among these
three GRBs (GRB 080905, GRB 080928, and GRB 081007) have very low flux.
\cite{Ghirlandaetal_2010} could fit only a single power-law with 
unconstrained peak energy for these bursts. Hence, these bursts are not suitable 
for a detailed time-resolved study. Hence, we are left with 9 GRBs (after sample selection).
In Figure~\ref{ch3_fig12}, we have shown the time-integrated data of the 9 GRBs by filled 
boxes. We obtain a reasonable correlation (Pearson, $r=0.80$, with $P_{\rm r}=0.0096$).
For the time-resolved study, we use the same bins of \cite{Ghirlandaetal_2010}.
When we study the time-resolved data, we see the correlation is poor (small circles
in Figure~\ref{ch3_fig12}). The correlation coefficient is only $r=0.37$, with $P_{\rm r}=0.0095$.
The reason that this correlation is poor can be attributed to the HTS evolution of the pulses. As a HTS evolution,
by definition, has high $E_{\rm peak}$ values even when the flux is low, it fills the 
upper left region of the Amati correlation (see Figure~\ref{ch3_fig12}).

\begin{table*}\centering
\caption[The isotropic energy ($E_{\gamma, \rm iso}$), observer frame values of peak energy ($E_{peak}$) 
and peak energy at zero fluence ($E_{\rm peak, 0}$)]
{The isotropic energy ($E_{\gamma, \rm iso}$), observer frame values of peak energy ($E_{peak}$) 
and peak energy at zero fluence ($E_{\rm peak, 0}$) are shown for the pulses of GRBs with known redshift ($z$).
The GRBs are taken from \cite{Ghirlandaetal_2010}.}

\begin{tabular}{cccccccccc}
\hline 
GRB & z & Pulse & $t_{1}$ & $t_{2}$ & $E_{peak}$ & $\chi_{red}^{2}$ & $E_{peak,0}$ & $\chi_{red}^{2}$ & $E_{\gamma,iso}$\tabularnewline
   & & & (s) & (s) & (keV) & & (keV) & & ($10^{52}$erg)\\
\hline
\hline 
080810 & 3.35 & 1 & 20.0 & 28.0 & $354_{-61}^{+188}$ & 0.95 & $875_{-180}^{+155}$ & 0.99 & 7.7\tabularnewline
\hline 
080916C & 4.35 & 1 & 0.0 & 13.0 & $430_{-67}^{+87}$ & 1.17 & $2420_{-397}^{+523}$ & 1.21 & 158.9\tabularnewline
 &  & 2 & 16.0 & 43.0 & $477_{-82}^{+108}$ & 1.11 & $1575_{-150}^{+170}$ & 1.38 & 130.1\tabularnewline
\hline 
080916 & 0.689 & 1 & -1.0 & 10.0 & $155_{-19}^{+23}$ & 1.12 & $519_{-59}^{+44}$ & 1.36 & 0.78\tabularnewline
 &  & 2 & 13.0 & 25.0 & $70_{-9}^{+13}$ & 1.0 & $226_{-28}^{+285}$ & 1.00 & 0.25\tabularnewline
 &  & 3 & 28.0 & 39.0 & $39.7_{-7}^{+9}$ & 1.09 & $70_{-23}^{+12}$ & 1.10 & 0.05\tabularnewline
\hline 
081222 & 2.77 & 1 & -2.0 & 20.0 & $159_{-17}^{+22}$ & 1.42 & $488_{-156}^{+173}$ & 1.07 & 23.7\tabularnewline
\hline 
090323 & 3.57 & 1 & -2.0 & 30.0 & $697_{-51}^{+51}$ & 1.33 & $2247_{-298}^{+392}$ & 1.46 & 127.9\tabularnewline
 &  & 2 & 59.0 & 74.0 & $476_{-47}^{+57}$ & 1.38 & $1600_{-94}^{+35}$ & 1.95 & 90.3\tabularnewline
 &  & 3 & 137.0 & 150.0 & $117_{-28}^{+31}$ & 1.37 & $211_{-63}^{+54}$ & 1.17 & 20.9\tabularnewline
\hline 
090328 & 0.736 & 1 & 3.0 & 9.0 & $648_{-124}^{170}$ & 0.93 & $1234_{-146}^{+174}$ & 0.92 & 2.8\tabularnewline
 &  & 2 & 9.0 & 20.0 & $659_{-106}^{+115}$ & 1.25 & $1726_{-122}^{+221}$ & 1.41 & 4.4\tabularnewline
 &  & 3 & 55.0 & 68.0 & $89_{-20}^{+41}$ & 1.16 & $180_{-96}^{+267}$ & 1.03 & 0.36\tabularnewline
\hline 
090423 & 8.2 & 1 & -11.0 & 13.0 & $76.9_{-26}^{+56}$ & 1.10 & $131_{-43}^{+99}$ & 1.04 & 20.3\tabularnewline
\hline 
090424 & 0.544 & 1 & -0.5 & 3.0 & $153_{-5}^{+6}$ & 1.73 & $184.5_{-19}^{+38}$ & 1.43 & 2.0\tabularnewline
 &  & 2 & 3.0 & 6.0 & $148_{-7}^{+8}$ & 1.41 & $162_{-9.2}^{+61}$ & 1.32 & 1.4\tabularnewline
 &  & 3 & 6.5 & 13.0 & $39.1_{-8.4}^{+0.2}$ & 1.32 & $104.8_{-16}^{+17}$ & 1.27 & 0.18\tabularnewline
 &  & 4 & 13.5 & 20.0 & $19.6_{-14.8}^{+6.4}$ & 0.98 & $75_{-34}^{+23}$ & 0.95 & 0.10\tabularnewline
\hline 
090618 & 0.54 & 1 & -1.0 & 41.0 & $185_{-25}^{+26}$ & 1.24 & $415_{-28}^{+37}$ & 1.19 & 3.5\tabularnewline
 &  & 2 & 61.0 & 76.0 & $226_{-9}^{+10}$ & 1.25 & $382_{-30}^{+106}$ & 1.33 & 9.8\tabularnewline
 &  & 3 & 76.0 & 95.0 & $128_{-5}^{+6}$ & 1.15 & $218_{-5.8}^{+6.8}$ & 1.09 & 5.4\tabularnewline
 &  & 4 & 106.0 & 126.0 & $57.7_{-3.3}^{+3.5}$ & 1.08 & $205_{-12}^{+14}$ & 1.19 & 1.5\tabularnewline
\hline
\end{tabular}

\label{9grbs}
\end{table*}

\begin{figure}\centering
\includegraphics[scale=0.34]{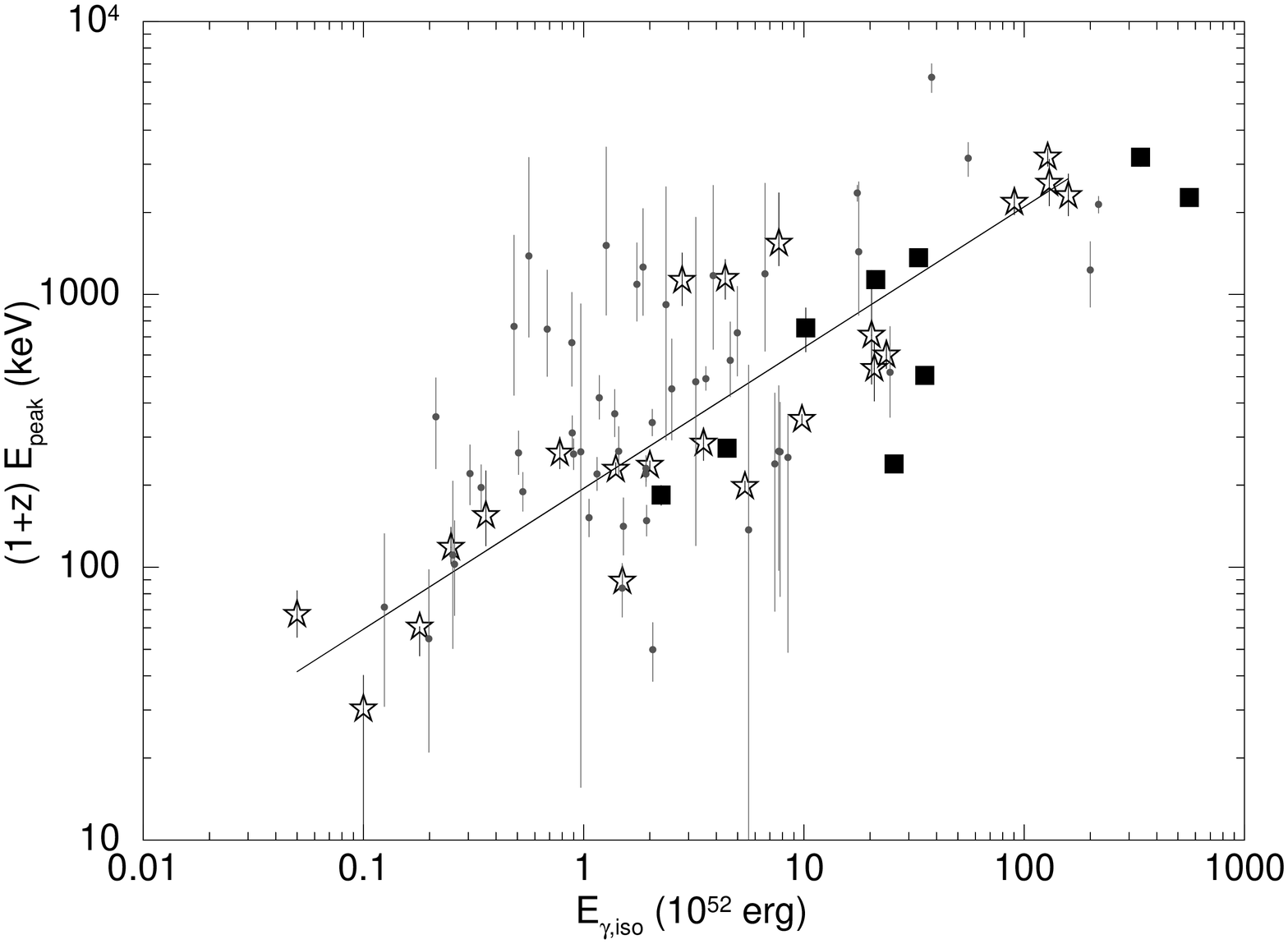}
\caption[Correlation between the isotropic energy ($E_{\gamma, \rm iso}$) and peak energy ($E_{\rm peak}$)
in the source frame for three different methods: time-integrated, time-resolved and pulse-wise]
{Correlation between the isotropic energy ($E_{\gamma, \rm iso}$) and peak energy ($E_{\rm peak}$)
in the source frame, known as the Amati correlation (\citealt{Amatietal_2002}) is shown. The correlation 
is studied for time-integrated (filled boxes), time-resolved (smallest circles) and pulse-wise
(stars) data. The time-resolved data is obtained by requiring equal integrated flux in each time bin
(\citealt{Ghirlandaetal_2010}), and it does not consider the broad pulse shape. The linear fit to the 
pulse-wise data is also shown (Source: \citealt{Basak_Rao_2012_correlation}).}
\label{ch3_fig12}
\end{figure}

\begin{table*}\centering

\caption[Results of the statistical analysis of the correlations]
{Results of the statistical analysis: $E_{\rm peak}$-$E_{\gamma, \rm iso}$(I to IV),
and $E_{\rm peak, 0}$-$E_{\gamma, \rm iso}$ (V) correlations. Parameters of the linear functional 
fit to the individual data are also shown.}

\begin{tabular}{ccccccc}
\hline 
Method & r & $P_{\rm r}$ & K & $\delta$ & $\sigma_{int}$ & $\chi_{red}^{2}$(dof)\tabularnewline
       & ($\rho$) & ($\rm P_{\rho}$) & & & & \tabularnewline
\hline
\hline 
I & 0.80 & 0.0096 & 0.166$\pm$0.080 & 0.473$\pm$0.048 & 0.225$\pm$0.067 & 0.64 (7)\tabularnewline
  & (0.75) & 0.0199 & & & & \tabularnewline

II & --- & 0.004 & 0.162$\pm$0.085 & 0.476$\pm$0.079 & --- & 0.47 (8)\tabularnewline

III & 0.37 & 0.0095 & --- & --- & --- & ---\tabularnewline
    & (0.486) & (0.0003) &  & & & \tabularnewline

IV & 0.89 & 2.95$\times10^{-8}$ & 0.289$\pm$0.055 & 0.516$\pm$0.049 & 0.244$\pm$0.048 & 0.56 (20)\tabularnewline
   & (0.88) & (4.57$\times10^{-8}$) & & & & \tabularnewline

V & 0.96 & 1.60$\times10^{-12}$ & 0.640$\pm$0.050 & 0.555$\pm$0.050 & 0.291$\pm$0.039 & 1.04 (20)\tabularnewline
  & (0.87) & (1.43$\times10^{-7}$) & & & & \tabularnewline
\hline
\end{tabular}
\vspace{0.1in}

\textbf {Notes.} Here, the methods are --- I: Time-integrated correlation study for 9 GRBs (our work), 
II: Time-integrated correlation study for 10 GRBs (quoted from \citealt{Ghirlandaetal_2010}. Note that they have used 
9 GRBs for time-resolved study).
III: Time-resolved correlation study (calculated from \citealt{Ghirlandaetal_2010}). Time-resolved 
analysis does not consider broad pulses.
IV: Pulse-wise correlation study (our work). 
V: Pulse-wise $E_{peak,0}$ - $E_{\gamma,iso}$ correlation study. The other parameters are: 
Pearson correlation coefficient --- $r$, Spearman rank correlation coefficient --- $\rho$, 
chance probability --- P) (both the correlations), and the best-fit parameters (K, $\delta$ and $\sigma_{int}$).
\label{stat}
\end{table*}

In order to restore the correlation, we now try pulse-wise analysis. Note that time-resolved 
analysis does not consider the broad pulses in a GRB. Rather it is guided by the requirement
of equal time-integrated flux in a time bin. Hence, we are bound to loosen the correlation.
A pulse-wise analysis, on the other hand, takes the broad pulses into account. Hence,
it is likely that a pulse-wise analysis is better suited for such a correlation study.
We select 22 pulses in these GRBs, and study the pulse-wise correlation. The pulse selection is 
subjective. Following \cite{Fishman_Meegan_1995}, we categorized the pulses as (i) single, (ii) smooth, 
multiple, (iii) separate episodic, and (iv) rapid, or chaotic. In selecting pulses, we neglect 
the fourth category of pulses/ a portion of a pulse. This way we get clean broad pulses. It is worthwhile
to mention that the selected pulses almost always cover the full GRB, except for a few very rapid portions.
In Table~\ref{9grbs}, we have shown the GRB sample, their $z$, and the pulse start and stop times. 
We neglect some portions of a few bursts, because, either these portions have very low count to 
be properly used for pulse-wise analysis, or these have rapid variability without any 
proper pulse structure. GRB 080810 and GRB 080916C have very low count rates after 30 s and 55 s, respectively. 
GRB 090323 has multiple spiky events in the range $30 - 59$ s, and $75-135$ s. Also, it has low count
rate in $75-135$ s region. GRB 090328 contains two overlapping spiky pulses in the range $20-26$ s 
(which is divided into two region by \citealt{Ghirlandaetal_2010} --- $20-24$ s, and $24-26$ s).
Except for these few regions, we incorporate essentially all the pulses in our analysis.

We have fitted the spectrum of each pulse with a Band function. The $E_{peak}$
and the $\chi_{red}^{2}$ of these fits are reported in the table.
In Figure~\ref{ch3_fig12}, we have shown the data points of the pulse-wise 
analysis by stars. In order to obtain the relation between the 
$E_{peak}$ (both time-integrated and pulse-wise), and $E_{\gamma, \rm iso}$ we apply the following technique.
We fit the data by a linear function which can be written as 

\begin{equation}
 \rm log(\frac{E_{\rm peak}}{100 \rm keV})=K+\delta \rm log(\frac{E_{\rm \gamma, iso}}{10^{52} \rm erg})
\end{equation}

We also assume an intrinsic data scatter ($\sigma_{\rm int}$) in $E_{peak}$ (\citealt{D'Agostini_2005}). 
An intrinsic scatter in the dependent variable denotes our limited knowledge of its relation with the independent
variable (or, variables). It admits that the dependent variable can be a function of some 
extra \textit{``hidden''} parameter (or, parameters), which we are not aware of. This is a 
general practice in GRB correlation study (e.g., see \citealt{Ghirlandaetal_2010, Wangetal_2011}).
With these parameters, we maximize the joint likelihood function ($L$) (\citealt{Wangetal_2011}) of the form

\begin{multline}
L(K, \delta, \sigma_{\rm int}) \propto \prod_i \frac{1}{\sqrt{\sigma_{\rm int}^2+\sigma_{y_i^2}+\delta^2\sigma_{x_i^2}}}\\
\times{\rm exp}\left[-\frac{(y_i-K-\delta x_i)^2}{2(\sigma_{\rm int}^2+\sigma_{y_i^2}+\delta^2\sigma_{x_i^2})} \right]
\end{multline}
\vspace{0.1in}

Using equation 24 of \cite{Wangetal_2011}, we can find the $\chi^2$ as $\chi^2$=-2ln$L$.
In Table~\ref{stat}, we have shown the correlation coefficient, chance probability as well 
as the best-fit values of the linear fit parameters ($K$, $\delta$). The methods are 
summarized in the note of this table. Method I and Method II are essentially the time-integrated 
correlation. Method III is the time-resolved correlation, which we found very poor.
 Method IV is the pulse-wise Amati correlation as found in our study. We can immediately
see a much better correlation compared to the I-III correlations. This correlation has 
a coefficient of $r=0.89$ with a chance probability, $P=2.95\times10^{-8}$. The value of $\delta$, 
which is the power-law of the relation is 0.516$\pm$0.049. Hence, we obtain a similar 
index as the original Amati correlation ($0.52\pm0.06$). This relation can be written as follows.

\begin{equation}
 \left[\frac{E_{\rm peak, 0}}{100 \rm keV}\right]=(0.289\pm 0.55)\times \left[\frac{E_{\gamma, \rm iso}}{10^{52}~\rm erg} \right]^{(0.516\pm0.049)}
\end{equation}
\vspace{0.1in}

Another important fact we notice from Table~\ref{stat} by comparing $\sigma_{\rm int}$
of method I-III with method IV is that the ratio of $\sigma_{\rm int}$ to number of data 
is reduced. For example, in method I, this ratio is 0.028, while it is 0.011 for method IV.

\subsection{Correlation of $\bf E_{\gamma, \bf \rm iso}$ with $\bf E_{\rm \bf peak, 0}$}
The pulse-wise correlation not only restores the Amati correlation, it improves the correlation.
For example, \cite{Krimmetal_2009} have used a sample of 22 GRBs from \textit{Swift} and
\textit{Suzaku} satellites to study both time-integrated and pulse-wise Amati correlation.
They have obtained a coefficient, $\rho=0.74$, with $P_{\rho}=7.58\times 10^{-5}$ for time-integrated data.
The pulse-wise data of 59 pulses shows an improvement $\rho=0.80$, with $P_{\rho}=5.32\times 10^{-14}$. 
A pulse-wise correlation also suggests that pulses are (possibly) independent entities.

Though a pulse-wise correlation is more meaningful than a time-integrated correlation study, the fact that 
$E_{\rm peak}$ is still an average quantity demands an alternative parameter to be used for such analysis.
We have one such parameter, namely the peak energy at zero fluence ($E_{\rm peak, 0}$). As this parameter
is a constant for a given pulse, it is more fundamental. Also, it denotes the initial $E_{\rm peak}$
of a burst, and hence, it carries the initial time information rather than pulse-average $E_{\rm peak}$, where
the time information is lost due to the averaging. Some GRB correlations are studied by 
incorporating the time information e.g., $L_{\rm iso}-E_{\rm peak}-T_{0.45}$ correlation 
(\citealt{Firmanietal_2005, Firmanietal_2006, Firmanietal_2007}). However, in our analysis, 
time information is implicit, and hence, requires one less parameter. 

For the set of 22 pulses of the 9 GRBs, we generate their 3D pulse models, and determine the $E_{\rm peak, 0}$.
We now replace the source frame $E_{\rm peak}$ with the source frame $E_{\rm peak, 0}$ and study the 
pulse-wise correlation. In Table~\ref{stat},
we have shown the correlation coefficient (method V). We immediately see an improvement of the 
correlation ($r=0.96$, $P_{\rm r}=1.60\times 10^{-12}$). The Spearman rank correlation,
however, is similar to the pulse-wise Amati correlation ($\rho=0.87$, $P_{\rho}=1.43\times 10^{-7}$;
method IV: $\rho=0.88$, $P_{\rho}=4.57\times 10^{-8}$). This is also reflected in the 
ratio of $\sigma_{\rm int}$ to number of data, which is 0.013 (compare with method IV: 0.011). Hence, we 
conclude that $E_{\rm peak, 0}$-$E_{\gamma,\rm iso}$ correlation is either comparable (for 
Spearman correlation), or better (for Pearson correlation) than the pulse-wise Amati
correlation. However, note that both the correlations are pulse-wise. Hence, we conclude
that pulse-wise correlation is better than time-integrated and time-resolved 
correlation (that does not account for the broad pulses) in a general sense.

In Figure~\ref{ch3_fig13}, we have shown the correlation. The data is fitted with a straight line to 
derive the relation, which is

\begin{equation}
 \left[\frac{E_{\rm peak, 0}}{100 \rm keV}\right]=(0.64\pm 0.05)\times \left[\frac{E_{\gamma, \rm iso}}{10^{52}~\rm erg} \right]^{(0.56\pm0.05)}
\end{equation}
\vspace{0.1in}

The dot-dashed lines denote the $3\sigma_{\rm int}$ data scatter. In this figure, we have 
marked a few GRBs which have highest number of pulses. As the correlation holds within $3\sigma_{\rm int}$
between the individual pulses, this correlation, like the pulse-wise Amati correlation, is also free from selection bias.

\cite{Ghirlandaetal_2004} have suggested that if GRBs are indeed jetted events,
a collimation corrected energy ($E_{\gamma}$) should be better correlated. Though for a set of 
27 GRBs they found reasonable correlation ($\rho=0.80$, which improves to $\rho=0.88$, for a 
selected sample of 24 GRBs), it is only comparable to $E_{\rm peak}-E_{\gamma,\rm iso}$ correlation.
A close inspection of Figure~\ref{ch3_fig13} reveals that the pulses of a given burst are 
scattered on the same direction of the correlation line. This may point towards the fact the 
actual energy is not $E_{\gamma,\rm iso}$, but $E_{\gamma}$. As the collimation correction is different 
for different burst, but same for all the pulses of a given burst, it is possible that all the pulses of a GRB
scatters away in the same direction from the correlation line. Hence, one can expect to 
have better correlation by the replacement of $E_{\gamma,\rm iso}$ with $E_{\gamma}$. However, in practice, the value of the jet 
opening angle ($\theta_j$) is either impossible to measure, or has large errors (\citealt{Goldsteinetal_2011}).
For the available values of $\theta_j$ (6 cases), we replace $E_{\gamma,\rm iso}$ by
$E_{\gamma}$, and found a reasonable correlation ($r=0.91$). 
We believe that the parameters $E_{\rm peak, 0}$ and  $E_{\gamma}$ can give a
much improved correlation provided  good measurements of $\theta_j$ are available.

\begin{figure}\centering
\includegraphics[scale=0.34]{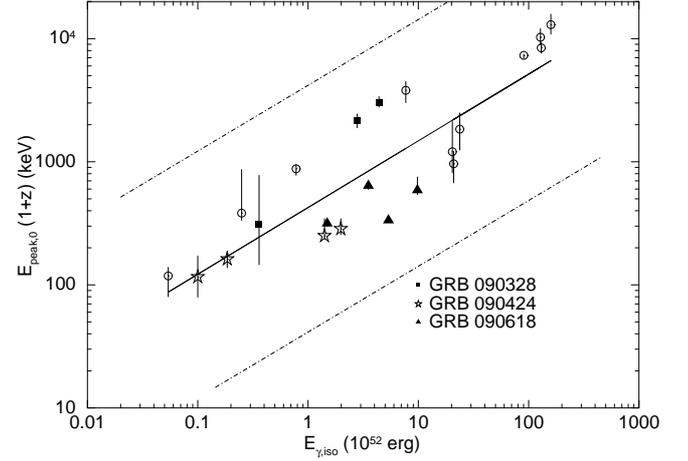}
\caption[A new correlation obtained between isotropic energy ($E_{\gamma, \rm iso}$) and peak energy at zero fluence ($E_{\rm peak, 0}$)
in the source frame]{Correlation between isotropic energy ($E_{\gamma, \rm iso}$) and peak energy at zero fluence ($E_{\rm peak, 0}$)
in the source frame is shown. Different pulses of three GRBs are marked, and explained in the legend.
The solid line shows the linear fit to the data, and the dot-dashed lines denote the 3$\sigma$ data scatter
(Source: \citealt{Basak_Rao_2012_correlation}).
}
\label{ch3_fig13}
\end{figure}

\begin{table*}\centering

\caption[Additional sample of GRBs with known $z$]
{The additional sample of GRBs with known $z$, detected by the \textit{Fermi}/GBM. We have reported here 
the start and stop time of the pulses, the corresponding best-fit parameters for a Band functional fit 
($\alpha$, $\beta$, $E_{\rm peak}$), and the isotropic energy, $E_{\rm \gamma,iso}$ (\citealt{Basak_Rao_2013_MNRAS}).}


\begin{tabular}{cccccccccc}
\hline 

GRB & z & Pulse & $t_{\rm 1}$ & $t_{\rm 2}$ & $\alpha$ & $\beta$ & $E_{\rm peak,obs}$ & $\chi_{red}^{2}$ & $E_{\rm \gamma,iso}$ \\
    &   &       &     (s)     &   (s)       &          &         & (keV)      & (dof)             &  ($10^{52}$erg) \\
\hline
\hline 
090902B & 1.822 & 1 & 5.0 & 13.0 & $-0.23_{-0.13}^{+0.13}$ & $-3.56_{-0.56}^{+0.22}$ & $828.9_{-28.7}^{+31.6}$ & 1.03 (243) & 178.24 \\
        &       & 2 & 12.0 & 18.0 & $-0.76_{-0.04}^{+0.07}$ & $-3.21_{-0.38}^{+0.23}$ & $537.3_{-23.7}^{+23.5}$ & 1.34 (290) & 111.81 \\
        &       & 3 & 18.0 & 23.0 & $-0.76_{-0.03}^{+0.04}$ & $-2.44_{-0.11}^{+0.08}$ & $285.4_{-17.0}^{+15.6}$ & 1.34 (275) & 62.14 \\
\hline
090926A & 2.1062 & 1 & 0.0 & 8.0 & $-0.55_{-0.02}^{+0.02}$ & $-2.44_{-0.05}^{+0.05}$ & $332.1_{-9.6}^{+9.7}$ & 1.59 (492) & 116.83 \\
        &       & 2 & 8.0 & 15.0 & $-0.80_{-0.02}^{+0.02}$ & $-2.90_{-0.18}^{+0.13}$ & $241.3_{-7.3}^{+7.6}$ & 1.55 (466) & 61.15 \\
\hline
090926B & 1.24 & 1 & 12.0 & 65.0 & $0.13_{-0.40}^{+0.45}$ & $-10.0$ & $73.8_{-6.1}^{+7.7}$ & 0.65 (42) & 2.54 \\
\hline
091003A & 0.8969 & 1 & 13.95 & 26.24 & $-0.953_{-0.06}^{+0.07}$ & $-2.38_{-0.51}^{+0.20}$ & $299.4_{-41.0}^{+48.2}$ & 1.10 (293) & 6.05 \\
\hline
091020  & 1.71 & 1 & -2.0 & 15.0 & $-1.16_{-0.15}^{+0.22}$ & $-2.07_{-0.50}^{+0.24}$ & $197.3_{-75.6}^{+115.9}$ & 1.07 (134) & 8.0 \\
\hline
091024  & 1.092 & 1 & -7.94 & 33.02 & $-0.95_{-0.14}^{+0.22}$ & $-2.08_{-\infty}^{+0.47}$ & $725.0_{-162.8}^{+226.7}$ & 0.82 (119) & 9.10 \\
        &       & 2 & 200.71 & 249.86 & $-0.81_{-0.26}^{+0.40}$ & $-9.37$ & $112.7_{-14.6}^{+14.4}$ & 1.43 (81) & 3.74 \\
        &       & 3 & 313.35 & 346.12 & $-1.18_{-0.07}^{+0.10}$ & $-9.36$ & $225.7_{-27.5}^{+19.4}$ & 1.60 (335) & 9.14 \\
        &       & 4 & 622.7 & 664.7 & $-1.17_{-0.07}^{+0.07}$ & $-2.15$ & $371.0_{-71.0}^{+111.0}$ & 1.09 (473) & 2.45 \\
\hline
091127  & 0.490 & 1 & -2.0 & 4.0 & $-0.92_{-0.16}^{+0.19}$ & $-2.20_{-0.14}^{+0.08}$ & $65.8_{-9.3}^{+12.1}$ & 1.35 (151) & 1.06 \\
        &       & 2 & 5.0 & 14.0 & $-1.34_{-0.33}^{+0.77}$ & $-2.88_{-0.17}^{+0.17}$ & $14.6_{-3.5}^{+1.7}$  & 1.00 (140) & 0.44 \\
\hline
091208B  & 1.063 & 1 & -1.0 & 5.0  & $-1.36_{-0.24}^{+1.08}$ & $-2.30$ & $74.2_{-34.9}^{+41.2}$ & 1.19 (154) & 0.55 \\
         &       & 2 &  6.0 & 13.0 & $-1.25_{-0.13}^{+0.13}$ & $-2.84_{-\infty}^{+0.48}$ & $113.7_{-15.8}^{+30.8}$  & 1.19 (223) & 1.31 \\
\hline
100414A & 1.368 & 1 & 1.0  & 13.0 & $-0.14_{-0.07}^{+0.08}$ & $-4.90_{-\infty}^{+1.47}$ & $557.5_{-28.5}^{+31.1}$ & 1.10 (275) & 22.73 \\
        &       & 2 & 14.0 & 20.0 & $-0.56_{-0.06}^{+0.06}$ & $-3.52_{-\infty}^{+0.71}$ & $599.4_{-44.3}^{+49.7}$ & 1.01 (238) & 14.26 \\
        &       & 3 & 21.0 & 28.0 & $-0.91_{-0.05}^{+0.06}$ & $-2.76_{-2.42}^{+0.39}$ & $635.1_{-78.3}^{+93.5}$ & 1.17 (240) & 12.00 \\
\hline
100814A  & 1.44  & 1 & -3.0 & 5.0  & $1.04_{-0.50}^{+0.65}$ & $-3.00_{-2.55}^{+0.71}$ & $168.6_{-22.1}^{+25.8}$ & 0.92 (176) & 2.14 \\
         &       & 2 &  4.0 & 14.0 & $0.84_{-0.36}^{+0.55}$ & $-3.43_{-\infty}^{+0.95}$ & $133.5_{-16.42}^{+13.8}$  & 0.79 (130) & 2.32 \\
\hline

\end{tabular}

\label{complete_sample}
\end{table*}

\section{Pulse-wise Amati Correlation Revisited}
The new correlation based on our 3D pulse description is marginally better, or comparable to the 
pulse-wise Amati correlation. However, the ratio of intrinsic data scatter to the number of 
points does not improve. The reason for our limited success is two-fold. (i) The models 
which are used to develop the 3D pulse model are all empirical. We note that the pulse-wise analysis gives 
a better result than the time-integrated and time-resolved analysis. This is because pulses are possibly physical 
(e.g., multiple episodes of central engine activity on a longer time scale). On the other hand, the spectral 
model, and its evolution used to develop our pulse model is empirical. A more physical model of a pulse may 
help in identifying the correct parameter/parameters for a correlation study. (ii) We have assumed that the 
pulses are all HTS, and the IT behaviour is a superposition effect. Of course, we have validated 
this assumption essentially by reproducing LC, width variation and spectral lag. But, as 
shown in the next chapter (\citealt{Basak_Rao_2014_MNRAS}), IT behaviour is found even for a few GRBs with single pulses.
Hence, our assumption may not be applicable in a global sense. However, we can still use the 
pulses and study pulse-wise analysis. 

In this section, we shall study pulse-wise Amati correlation for a larger set of GRBs. The \cite{Ghirlandaetal_2010}
sample contained all $z$-measured GRBs up to June, 2009. We complement the sample by adding 
all $z$-measured GRBs detected by the \textit{Fermi}/GBM during June, 2009 to August, 2010. With these 
additional GRBs (10) our sample contains a total of 19 GRBs. In Table~\ref{complete_sample}, we 
show the additional sample. The start and stop time of the 21 pulses are shown. We fit the 
GBM data with Band function. For GRB 090902B, a Band model is unacceptable. Hence, we 
fit the pulses of this GRB with Band+PL model (\citealt{Rydeetal_2010_090902B}). The best-fit 
parameters ($\alpha$, $\beta$ and $E_{\rm peak}$) along with $\chi_{red}^{2}$ (dof),
and $E_{\rm \gamma,iso}$ are shown in the same table. In Figure~\ref{ch3_fig14},
we show the pulse-wise Amati correlation for our complete sample. We find a 
good correlation ($r=0.86$, $P_{\rm r}=1.50 \times 10^{-13}$; $\rho=0.86$, 
$P_{\rho}=7.47\times 10^{-14}$). A Pearson correlation without using the logarithmic
values is $r=0.80$. Due to the larger sample size, the significance has increased i.e., chance
probability has decreased (previous value was $P_{\rho}=4.57\times 10^{-8}$).
The coefficient of the time-integrated Amati correlation for these GRBs is $r=0.83$ with 
$P_{\rm r}=1.10 \times 10^{-5}$; or, $\rho=0.85$ with $P_{\rho}=4.27\times 10^{-6}$.
The logarithmic data is fitted with a linear function as before. The corresponding 
values are shown in Table~\ref{complete_sample_stat}. Note that the slope is close to $\sim0.5$, 
which gives the empirical relation $E_{\rm peak}\propto E_{\gamma, \rm iso}^{1/2}$.
The values of $K$, $\delta$, and $\sigma_{\rm int}$ are within $0.36\sigma$, $0.36\sigma$, and $0.25\sigma$
of the previous values. The ratio of $\sigma_{\rm int}$ to the number of data points has decreased (0.006).

\begin{figure}\centering
\includegraphics[scale=0.34]{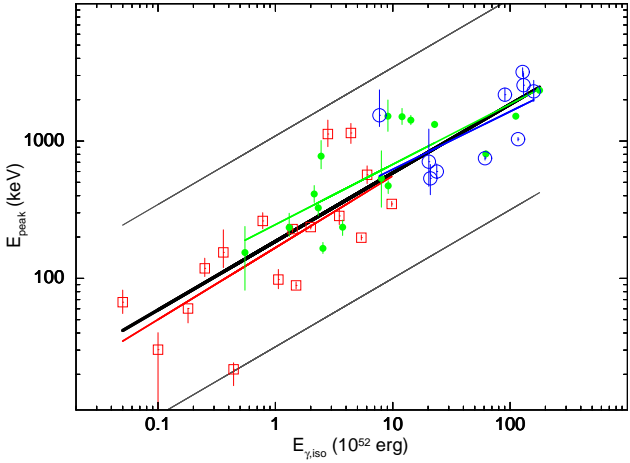}
\caption[Pulse-wise Amati correlation studied for the 43 pulses of 19 GRBs]
{Pulse-wise Amati correlation studied for the 43 pulses of 19 GRBs (Table~\ref{9grbs}, and \ref{complete_sample}). 
The log($E_{\rm peak}$)-log($E_{\rm \gamma, iso}$) data is fitted with a straight line assuming intrinsic
data scatter ($\sigma_{\rm int}$), and shown by thick black line. Thin black lines show the $3\sigma_{\rm int}$ 
data scatter. Markers are: red boxes (z\textless1), green filled circles (1\textless z\textless2), 
and blue open circles (z\textgreater2). Each of these binned data is fitted by a straight line (same colour as the 
markers), as done for the complete set (Source: \citealt{Basak_Rao_2013_MNRAS}).}
\label{ch3_fig14}
\end{figure}

\begin{table*}\centering
\caption[Linear fit results for the complete sample]
{Linear fit results for the complete sample is shown. The fitting is performed on the 
logarithmic values of $E_{peak}$, normalized by 100 keV, and $E_{\gamma,iso}$, normalized by $10^{52}$ erg. 
The parameters of the linear fit are: intercept ($K$), slope ($\delta$), and intrinsic data scatter ($\sigma_{int}$).
}

 \begin{tabular}{c|c|c|c|c}
\hline
 Data set & K & $\delta$ & $\sigma_{int}$ & $\chi^2_{red}$ (dof)\\
\hline
\hline 
All & $0.269\pm0.041$ & $0.499\pm0.035$ & $0.256\pm0.0344$ & 1.04 (41) \\
z\textless1 bin & $0.223\pm0.077$ & $0.523\pm0.113$ & $0.306\pm0.065$ & 1.13 (15) \\
1\textless z\textless2 bin & $0.391\pm0.056$ & $0.439\pm0.048$ & $0.208\pm0.047$ & 1.10 (14) \\
z\textgreater2 bin & $0.373\pm0.068$ & $0.421\pm0.038$ & $0.200\pm0.063$ & 1.24 (8) \\
\hline
\hline
First/Single pulses & $0.352\pm0.055$ & $0.465\pm0.044$ & $0.221\pm0.048$ & 1.10 (17) \\
Rest of the pulses & $0.228\pm0.057$ & $0.503\pm0.050$ & $0.273\pm0.048$ & 1.09 (22) \\

\hline
\end{tabular}
\label{complete_sample_stat}
\end{table*}

\begin{figure}\centering
\includegraphics[scale=0.34]{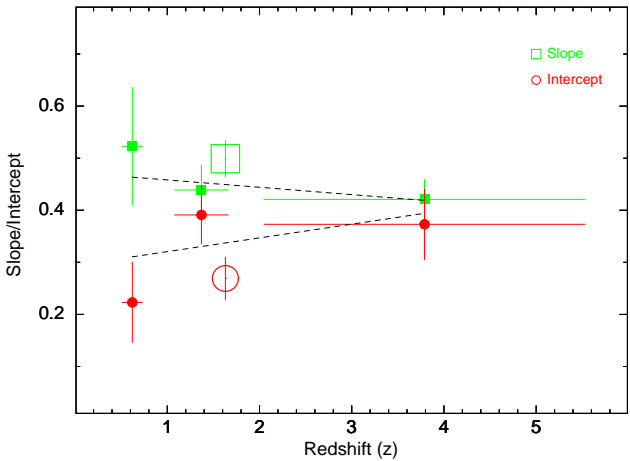}
\caption[Redshift ($z$) evolution of the pulse-wise Amati correlation]
{Redshift ($z$) evolution of the pulse-wise Amati correlation is shown. Green filled boxes are the values of
slope ($\delta$), and red filled circles are the values of intercept (K) in different redshift bins. The average
value of these parameters ($\delta$ and K) are shown by similar open symbols. We fit each of the parameter evolutions 
with a straight line (Source: \citealt{Basak_Rao_2013_MNRAS}).
}
\label{ch3_fig15}
\end{figure}

\begin{figure}\centering
\includegraphics[scale=0.62]{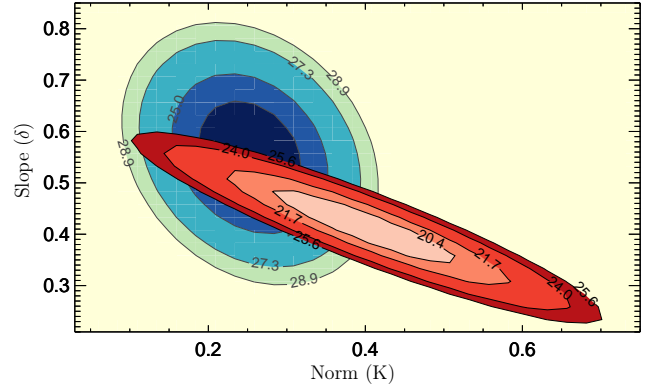}
\caption[$\chi^2$ contour for the parameters $K$ and $\delta$ for GRBs in two redshift bins]
{$\chi^2$ contour for the parameters $K$ and $\delta$. The parameters are studied in 
two redshifts --- z$\leqslant 1.092$ (shown by blue contours on the left hand side), and z\textgreater1.092 
(shown by red contours on the right side). In both of the cases, contour levels are --- 
$\Delta \chi^2=1.0, 2.3,4.61,6.17$ which correspond to 1$\sigma$ (1 parameter), 1$\sigma$, 2$\sigma$, and 3$\sigma$ 
(2 parameters). We note that the parameters agree within $\Delta \chi^2=1.0$ (Source: \citealt{Basak_Rao_2013_MNRAS}).
}
\label{ch3_fig16}
\end{figure}

\subsection{Redshift Evolution Of The Pulse-wise Amati Correlation}
As we have a larger sample compared to the previous study, we can perform some critical 
tests on the correlation. The first test is to check any possible redshift evolution 
of the correlation. If we want to use the correlation for cosmology, it is necessary to 
check the cosmological evolution. For this purpose, we divide our sample into three 
$z$ bins: (i) $z<1$ (this contains 17 data points), (ii) $1<z<2$ (contains 16 data points),
and (iii) $z>2$ (contains 10 data points). In Figure~\ref{ch3_fig14}, these sets are 
shown by different markers (red boxes for z\textless1, green filled circles for 1\textless z\textless2, 
and blue open circles for z\textgreater2). Each of the data sets is fitted with a 
straight line as before. The best-fit values of the linear fits are shown in Table~\ref{complete_sample_stat}.
In Figure~\ref{ch3_fig15}, we have shown these parameters (green filled boxes for $K$,
and red filled circles for $\delta$) as a function of the weighted mean $z$ of the corresponding bin. 
The average values of these parameters are shown by similar open symbols. As we have only three bins, we can at most 
fit a straight line to quantify the evolution. We find that the evolution of the 
parameters are consistent with zero. The slopes of the evolution are ---
$\frac{d\delta}{dz}=(-1.41\pm3.74)\times10^{-2}$, and $\frac{dK}{dz}=(2.64\pm4.88)\times10^{-2}$.
The errors are estimated at nominal 90\% confidence level.

The evolution of $K$ and $\delta$ with $z$ is not statistically significant, as the errors are 
quite large. The evolution of Amati correlation, and other GRB correlation with redshift
has been investigated by many authors (e.g., \citealt{Li_2007, Ghirlandaetal_2008_correlation, 
Tsutsuietal_2008, Azzam_2012}). In all the study, the evolution is found to be insignificant. Note that 
the evolution as reported by \cite{Tsutsuietal_2008} is monotonic. But, in our analysis we 
do not find any such evidence. Rather the evolution saturates at higher $z$ (see high $z$ 
data points in Figure~\ref{ch3_fig15}). To further investigate whether their is any significant 
evolution, we divide the sample into two redshift bins. This way we get larger numbers  in each bin 
though we cannot fit the evolution. As the number of pulses in a given GRB is arbitrary,
we cannot get an equal division of sample. The closest we find is 23 pulses in the 
lower $z$ bin (up to $z=1.096$), and 20 in the higher bin. We fit the individual set of data 
with straight line, and obtain the parameters. In Figure~\ref{ch3_fig16}, we have shown the 
$\chi^2$ contour of $K$ and $\delta$. The levels shown are: single parameter $1\sigma$
($\triangle \chi^2=1.0$), two-parameter $1\sigma$ ($\triangle \chi^2=2.3$), $2\sigma$ ($\triangle \chi^2=4.61$),
and $3\sigma$ ($\triangle \chi^2=6.17$). We note that the two sets agree within $\triangle \chi^2=1.0$.
Hence, the evolution is not significant. \cite{Yonetokuetal_2010} have studied the evolution of
both Amati and Yonetoku correlation in the time-integrated data. They have found only $1\sigma$
and $2\sigma$ evolution, respectively.

\begin{figure}\centering
\includegraphics[scale=0.34]{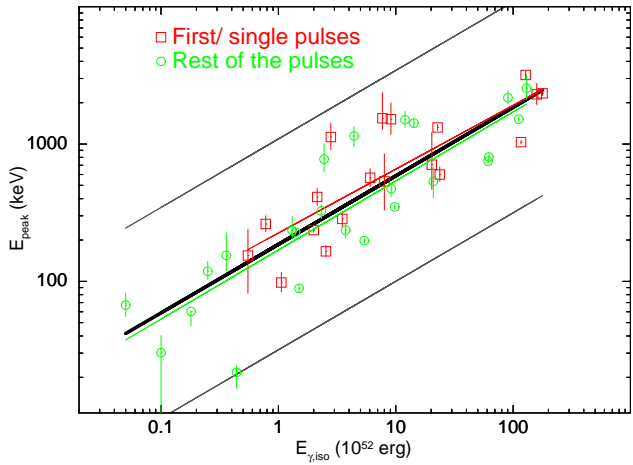}
\caption[Pulse-wise Amati correlation studied for first/single pulses and rest of the pulses]
{Same as Figure~\ref{ch3_fig14}, with the set divided into 19 first/single pulses (red open boxes),
and 24 rest of the pulses (green open circles). The straight line fitting these sets are shown by the same colour 
as the markers (Source: \citealt{Basak_Rao_2013_MNRAS}).
}
\label{ch3_fig17}
\end{figure}

\begin{figure}\centering
\includegraphics[scale=0.62]{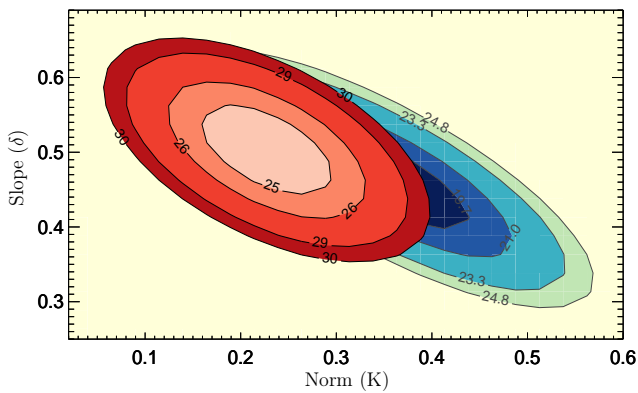}
\caption[$\chi^2$ contour for the parameters $K$ and $\delta$ for GRBs in first/single and rest of the pulses]
{$\chi^2$ contours for the parameters $K$ and $\delta$. Contour levels are 
similar as in Figure~\ref{ch3_fig15}. Blue contours on the right hand side shows the first/single pulses,
while the red contours on the left show the rest of the pulses (Source: \citealt{Basak_Rao_2013_MNRAS}).
}
\label{ch3_fig18}
\end{figure}

\subsection{Possible Bias For Harder First/Single Pulse}
Another possible bias in the pulse-wise correlation can arise from 
the dependence of spectral hardness of pulses on their time sequence. 
It is suggested that during the initial phase, GRBs generally tend to be harder  
than the rest of the prompt emission phase (\citealt{Crideretal_1997, Ghirlandaetal_2003,
Kanekoetal_2003, Ryde_Pe'er_2009}). A spectral hardness naturally 
tends to make higher $E_{\rm peak}$. Hence, it is expected that the first pulse of a GRB should 
be biased towards higher $E_{\rm peak}$, giving a systematic shift in the pulse-wise 
correlation.

In order to estimate this bias, we divide our sample in first/single pulses (19 pulses) and the 
rest of the pulses (24 pulses). We fit the logarithmic values of $E_{\rm peak}-E_{\gamma, \rm iso}$
data by a straight line, and obtained the individual best-fit parameters. In Figure~\ref{ch3_fig17},
we have shown the two sets of data by red open boxes (first/single pulses) and green open circles 
(rest of the pulses). The linear fits are shown by the same colours. A comparison with the linear fit 
to the complete sample (black line) immediately shows that the bias is insignificant. In Table~\ref{complete_sample_stat},
we have reported the fit parameters of these sets. While we see that the intercept ($K$)
of the first/single pulse fitting is higher ($0.352\pm0.055$) than that of the rest of the pulse
($0.228\pm0.057$), the slope ($\delta$) is lower ($0.465\pm0.055$ compared to $0.503\pm0.057$).
As the relation between $E_{\rm peak}$ and $E_{\gamma, \rm iso}$ is $E_{\rm peak} \propto 10^K E_{\gamma, \rm iso}^{\delta}$,
the bias due to higher intercept is compensated by the lower slope. 

To further investigate the matter, we plot the $\chi^2$ contour of the two sets as before.
In Figure~\ref{ch3_fig18}, the contours are shown. The blue contours on the right side
denote the parameter range for the first/single pulse data fitting. The red contours on the 
left side are those for the rest of the pulses. Again we see that the parameters agree within
$\triangle \chi^2=1.0$. Hence, we conclude that the bias due to harder first/single pulse 
is not statistically significant for a pulse-wise Amati correlation.

\subsection{Comparison With The Time-integrated Correlation}
It is interesting to compare the pulse-wise Amati correlation with a time-integrated Amati correlation.
For this purpose, let us use the recent updated Amati correlation as studied by \cite{Navaetal_2012}.
Let us denote the values quoted by \cite{Navaetal_2012} by subscript `1'.
For a complete sample of 46 GRBs, they have reported the a correlation coefficient, $\rho_1=0.76$.
The slope is found to be $\delta_1=0.61\pm0.04$, while the intercept is $K_1=-29.60\pm2.23$, and $\sigma_{int}=0.25$.
They have also studied a larger complementary sample of 90 GRBs with $\rho_1=0.78$, $\delta_1=0.531\pm0.02$,
$K_1=-25.63\pm1.35$, $\sigma_{int}=0.25$. Together the set contains 136 GRBs with $\rho_1=0.77$, $\delta_1=0.55\pm0.02$,
$K_1=-26.74\pm1.13$, $\sigma_{int}=0.23$.

In Figure~\ref{ch3_fig19}, we have shown the time-integrated Amati correlation line of the complete sample (\citealt{Navaetal_2012})
by red dashed line. The shaded region is the 3$\sigma_{int}$ data scatter of the correlation. In the same figure, 
we have shown the data points obtained for our pulse-wise analysis (blue filled circles). We have over-plotted the time-integrated 
data (taken from David Gruber --- private communication) of our sample (marked by open circles). The linear fit
to the time-integrated data is shown by a dot-dashed line. For comparison, we have also shown the line fitted 
to the pulse-wise data (thick black line). In our analysis, $E_{\rm peak}$ is normalized by 100 keV, while
$E_{\rm \gamma,iso}$ is normalized by $10^{52}$ erg. Hence, in order to compare the best-fit parameters we 
convert these values in the same units of \cite{Navaetal_2012}. We obtain $K=-23.68 \pm 1.82$. This value is 
comparable to the values obtained for the total sample ($K_1=-26.74\pm1.13$). The values of $\delta$
are also comparable ($0.499\pm0.035$ for pulse-wise analysis, and $0.55\pm0.02$ for time-integrated analysis
of the total sample). Figure~\ref{ch3_fig19} clearly shows that all these correlations are comparable 
within $3\sigma_{\rm int}$ data scatter. A closer look at the plot reveals that the time-integrated 
correlation line of our sample always lies lower than the pulse-wise correlation line. Hence, the 
slopes are similar, but the normalization (i.e., the intercepts) of the correlations are different. The difference
can be explained as follows. If a GRB with two identical pulse (i.e., same energy and $E_{\rm peak}$)
are summed, their total energy will be added up, while the $E_{\rm peak}$ remains the same.
Similarly, for pulses with different energies and $E_{\rm peak}$, the resultant total energy will 
add-up, while the $E_{\rm peak}$ is only an average of them. Hence, we expect a lower 
normalization for a time-integrated correlation. Note that the correlation coefficient of 
pulse-wise analysis is always better than the time-integrated analysis. 

\begin{figure}\centering
\includegraphics[scale=0.35, angle=-90]{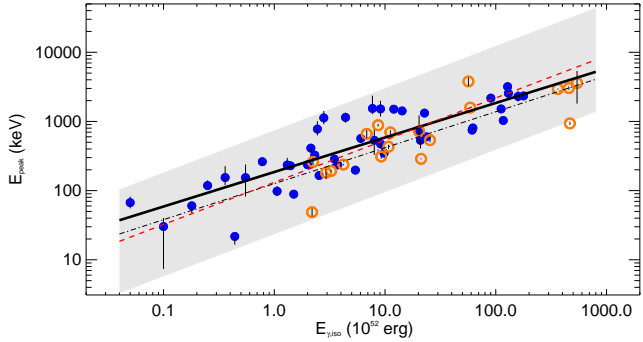}
\caption[Comparison of pulse-wise correlation with time-integrated Amati correlation]
{Data set of the complete sample of our study (blue filled circle) as plotted on the plane of
time-integrated $E_{\rm peak}$-$E_{\rm \gamma, iso}$ correlation (for the complete sample from \citealt{Navaetal_2012}). 
The data for the time-integrated study of 19 GRBs (present analysis) are also shown (orange open circle).
The straight line fitted to the pulse-wise log($E_{\rm peak}$)-log($E_{\rm \gamma, iso}$) data is shown 
by thick black line. The red dashed line denotes the relation between time-integrated $E_{\rm peak}$ 
with $E_{\rm \gamma, iso}$ (\citealt{Navaetal_2012}). The blue dot-dashed line shows the similar relation 
in this study (Source: \citealt{Basak_Rao_2013_MNRAS}).}
\label{ch3_fig19}
\end{figure}

\section{Summary}
In this chapter, we have developed a simultaneous timing and spectral description ($F(t, E)$) of a GRB pulse.
Using the synthetic 3D pulse, we have derived various pulse properties and have found consistent results 
with the data. The agreement between the model and the data extends even to the minute details of the 
parameters e.g., \textit{reverse} width variation in some pulses. Such a finding explicitly shows the 
immense applicability provided by simple assumptions of spectral evolution, even if they are empirical.

We have applied the model to a set of GRBs with known $z$ and found that one of the spectral parameters 
($E_{\rm peak, 0}$) correlates with the isotropic energy ($E_{\gamma, \rm iso}$) of the pulse.
We have found that this correlation is marginally better, or comparable with a pulse-wise 
$E_{\rm peak}-E_{\gamma, \rm iso}$ correlation. Also, these pulse-wise correlations are 
always better than time-integrated, or time-resolved (not accounting for broad pulses) correlation.
The pulse-wise analysis seems to be more physical, and it is more useful as it 
favours the reliability of the correlation. Note that the ultimate aim of GRB correlation study
is to find luminosity distance e.g., measuring the redshift by an independent way. Using 
the pulses for such study is quite useful. The pulses, despite having different 
$E_{\rm peak}$ and $E_{\gamma, \rm iso}$, have the same redshift, and follow the same pulse-wise correlation.
Hence, a redshift measured by multiple pulses using the pulse-wise correlation should give 
a better constraint. 

In spite of these versatile applicability, the pulse model is limited by the assumption
of spectral model, and its evolution. The assumption of broad pulses seems to be reasonable, as 
we always get better correlation. Also, the lag-luminosity correlation is found to be a pulse
property (\citealt{Hakkilaetal_2008}). However, the other assumptions e.g., instantaneous spectrum 
is a Band function, and the spectral evolution is always HTS should be critically checked.
In the next chapters, we shall scrutinize these assumptions, and explore various other options.
We shall find that indeed the assumptions are not valid in a global sense (In fact,
there is no physical reason to expect these to be global properties). However, the model 
is still useful as an empirical description of GRB pulses. Moreover, the pulse model developed 
here is quite generic in a sense that any empirical timing and spectral function, and spectral 
evolution can be incorporated, and checked against the data. This is the first attempt to 
describe GRB pulses in a different approach, and predict various pulse properties from a 
single description (\citealt{Basak_Rao_2012_090618}). We again emphasize that a simultaneous description
has an immense flexibility to check the validity of a model against the data from various directions.

\chapter{Alternative Spectral Models of GRB Pulses} \label{ch4}

\section{Motivation}
In this chapter (\citealt{Basak_Rao_2014_MNRAS}), we shall explore various prompt emission models, 
and compare them with the Band function. The motivation to look for alternative models in the first 
place is described below. However, note that the Band function till date is the most 
appropriate function to describe both the time-integrated and time-resolved prompt emission spectrum 
of GRBs (e.g., \citealt{Kanekoetal_2006, Navaetal_2011, Zhangetal_2011}). Hence, it is unlikely that 
we shall get any order of magnitude improvement in the spectral fitting. In fact, the alternative models 
in many cases will be just comparable to the Band function. Also, we do not expect all the GRBs to show any 
immediate improvement, simply because the data may be flux limited. Hence, we shall apply these models 
on a selected sample. 

\begin{itemize}
\item (i) The main reason to study alternative spectral models is to understand the radiation mechanism.
Though the Band function is undoubtedly an acceptable function, the physical origin of this function 
is still debated. In the internal shock model (\citealt{Rees_and_meszaros_1994_prompt}), the prompt emission 
is produced via synchrotron radiation of electrons accelerated by ``internal shock'' 
in an optically thin region ($r_{\rm IS} \sim 3 \times 10^{13}$ cm $\gg r_{\rm ph}\sim10^{11}$ cm). 
In certain circumstances, the spectrum can have a synchrotron self-compton (SSC) component, but at a 
fairly higher energy ($\sim10$ MeV; section 1.5.2; \citealt{Sarietal_1996}). Hence, a Band function, in principle, 
can be a phenomenological representation of an optically thin synchrotron emission. As discussed
in section 1.5.2, the electrons emitting synchrotron radiation are expected to cool fast 
($\gamma_{\rm e, c}<\gamma_{\rm e, min}$; \citealt{Piran_1999_review}). Hence, the spectral index is expected 
to be $-\frac{1}{2}$ (\citealt{Cohenetal_1997}), i.e., a photon index, $\alpha=-\frac{3}{2}$. Even if the 
electrons are cooling slowly, the $\alpha$ is expected to be at most $-\frac{2}{3}$. These two 
limits are referred to as ``synchrotron lines of death''. A GRB photon index cannot be 
greater than -3/2 (or, -2/3, for slow cooling). However, the value of $\alpha$ is
found to be greater than -2/3 in many cases (\citealt{Crideretal_1997, Crideretal_1999, Preeceetal_1998}).  
In addition, the synchrotron radiation is supposed to be produced by electrons accelerated 
in IS. As IS requires interactions of relativistic shells with low relative velocity, 
the efficiency of energy conversion is low ($\lesssim20\%$; \citealt{Piran_1999_review}). 
Hence, both the IS model and synchrotron emission are inadequate to fully explain the spectrum.

\item (ii) It is possible that Band is actually an average function which is produced by the evolution 
of a different, more physical function (\citealt{Ghirlandaetal_2003, Ryde_2004, Ryde_2005, Ryde_Pe'er_2009, Pe'er_Ryde_2011}). 
For example, \cite{Ryde_2004} has shown that the instantaneous spectrum of the BATSE GRBs with single pulses
can be modelled by a blackbody (BB), or a BB along with a power-law (PL). Note that while Band function 
has a single non-thermal spectral shape, a BBPL model segregates the spectrum into 
a thermal and a non-thermal component. In a BBPL model, the $\nu F_{\nu}$ peak of 
the spectrum is represented by the peak of the photospheric temperature
as seen by an observer ($\sim 3kT$). Also, as the PL index becomes shallower 
than $\alpha$ of the Band function, it can be accommodated in the 
synchrotron interpretation. In a few \textit{Fermi} GRBs, this PL appears to have a second slope, or 
a cut-off at high energies. This non-thermal component is modelled with either a cut-off PL 
(e.g., \citealt{Ackermannetal_2011_090926A}), or a Band function (\citealt{Guiriecetal_2011, Axelssonetal_2012, Guiriecetal_2013}).
In one case (GRB 090902B), the time-resolved spectrum is interpreted as a multicolour BB with 
a PL (mBBPL; \citealt{Rydeetal_2010_090902B}), where mBB represents a departure from the BB shape due to dissipative processes inside 
the photosphere, and a geometric broadening due to the finite size of the photosphere (\citealt{Pe'er_2008}).

\item (iii) The Band function was actually derived by an extensive study of the \emph{CGRO}/BATSE data in a limited band width.
Extension of the spectrum to the lower energies ($<50$ keV, i.e., the BATSE lower range),
sometimes require additional features, e.g., time-resolved spectral study of GRB 041006
using \textit{HETE-2} data (2-400 keV) indicates multiple components with diverse time evoltuion
(\citealt{Shirasakietal_2008}). Additional high energy component ($>8$ MeV) is found in the 
high energy detector, EGRET of CGRO (\citealt{Gonzalezetal_2003}). With \textit{Fermi}/LAT,
additional high energy components are found in several GRBs (\citealt{Abdoetal_2009_090902B, Ackermannetal_2010_090510,
Ackermannetal_2011_090926A}). However, note that the exceptions are found in a limited cases.
Hence, the search for alternative models is motivated mainly by our inability to associate
Band function with a physical process.

\end{itemize}

\section{Alternative Models and Data Selection}
In the original photon-lepton fireball model, the spectrum was predicted as a BB (\citealt{Goodman_1986}).
Due to the finite size of the photosphere, the BB can be broader than a Planck model (though 
still far from a typical GRB spectrum). As a BB shows up in the time-resolved spectrum
(\citealt{Ryde_2004, Ryde_2005}), it is possible to associate the BB with the photospheric emission.
It is shown that the temperature evolution of a BB follows the fireball prediction (i.e.,
an adiabatic cooling). This evolution also shows a break, which can be interpreted as 
the saturation break. The photosphere ($r_{\rm ph}$), in this case, occurs below the 
saturation radius ($r_{\rm s}$; see discussions in section 1.5.2). Note that a 
kT evolution with a break disfavours a HTS evolution. For a baryonic fireball,
\cite{rees_and_meszaros_2005_prompt} have argued that dissipative processes, e.g., magnetic reconnection, or 
IS can enhance the photospheric flux by comptonization. For fast dissipation, an effective 
pair photosphere ($r_{\rm pair}$) is generated above the baryonic photosphere ($r_{\rm baryon}$), 
and the comptonization occurs both below and above $r_{\rm baryon}$. For slow dissipation,
comptonization is effective only below $r_{\rm baryon}$. In both cases, a re-energized 
thermal emission (a ``grey body'') along with a synchrotron emission is predicted. The 
$\nu F_{\nu}$ peak is represented by the comptonized BB. Though the detail of the 
spectrum depends on unknown dissipative processes, the emergent spectrum can be effectively
modelled with a BBPL, or a mBBPL function. In addition to these models we shall introduce
a new model, namely, two BBs with a PL function (2BBPL) to incorporate multiple peaks 
in the spectrum (e.g., \citealt{Shirasakietal_2008}). The two 
BBs here represent two bumps in the spectrum. Though this model is phenomenological, 
we shall show some convincing evidence, and its strong predictive power (chapter 5 and 6). 

In summary, we shall use four models for spectral fitting: (i) Band, (ii) BBPL, (iii) mBBPL and
(iv) 2BBPL. Except for the first function, all the other models have a thermal and a non-thermal 
component. Recently, \citet[][Lu12 hereafter]{Luetal_2012} have studied $E_{\rm peak}$ evolution in the 
time-resolved data of a set of a \textit{Fermi}/GBM sample (51 LGRBs, 11 SGRBs). In this context, 
they have specifically discussed the GRBs with single pulses (8 GRBs) in their sample. Note that 
the BATSE sample of \cite{Ryde_2004} also consists of single pulse GRBs. The reason to give special 
attention to single pulses are --- firstly, we expect broad single 
pulse for a single episode of emission, and secondly, a single pulse is free from an overlapping
effect. Hence, one can meaningfully investigate the evolution within a pulse. If the 
spectral evolution is a pulse property (rather than a burst property), the knowledge gained 
from single pulse events can be used for complex GRBs. The Lu12 sample is selected based on the following
criteria: (i) fluence in 8-900 keV band $\geq10^{-5}$ erg cm$^{-2}$ (LGRBs), and $\geq8\times10^{-7}$ erg cm$^{-2}$
(SGRBs), (ii) at least 5 time-resolved bins are obtained for a signal-to-noise ratio of $35\sigma$.
We select all the single pulse GRBs of Lu12 sample. In addition, we use GRB 110721A, because, this is 
an approximately single pulse GRB, with HTS spectral evolution, and we could obtain 15 time-resolved 
bins owing to the high peak flux. This GRB is extensively studied for multiple spectral components
by \cite{Axelssonetal_2012}.

\begin{figure}\centering
{

\includegraphics[width=3.4in]{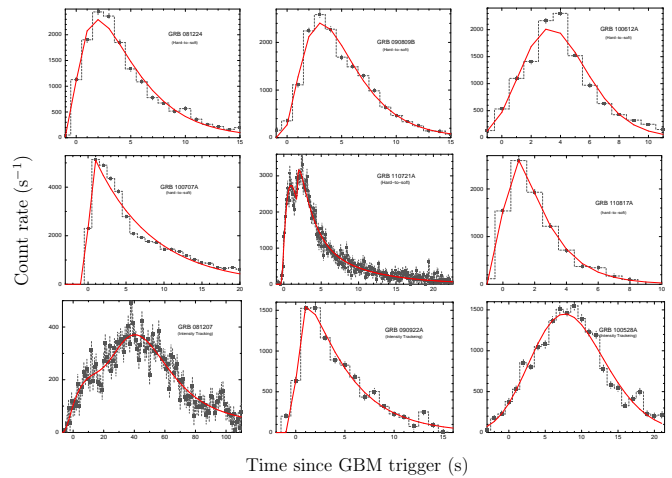} 

}
\caption[Background subtracted LCs of the GRBs in the energy range of NaI detector]
{Background subtracted LCs of the GRBs in the energy range of NaI detector. We use the detector which has the highest
count rate. \textit{Upper 6 panels}: ``hard-to-soft'' (HTS) pulses, and \textit{lower 3 panels}: ``intensity tracking'' 
(IT) pulses. LCs are fitted with exponential model (\citealt{Norrisetal_2005}). The values of the corresponding best-fit parameters
are given in Table~\ref{ch4_t1}. Source: \cite{Basak_Rao_2014_MNRAS}.
}
\label{lc}
\end{figure}

\section{Results of Timing Analysis}
We generate the light curve (LC) of each GRB in the 8-900 keV band of one NaI detector having maximum count rate
(Figure~\ref{lc}). The LCs are fitted with Norris model. The best-fit values of the model 
parameters are shown in Table
\ref{ch4_t1}. We note that the pulse profiles of all the GRBs are adequately captured by Norris model (Figure~\ref{lc}). 
However, we note that $\chi^2_{red}$ of these fits are generally high. This is expected as the Norris model
only captures the broad pulses, and does not account for the finer variability (see \citealt{Raoetal_2011}). 
Our motivation to fit the LCs is to quantify the pulse width and asymmetry of the global pulse 
structure (see below). Hence, Norris model is adequate for this purpose.  

In Table~\ref{ch4_t1}, we have also shown the derived parameters --- the pulse peak position ($p$), 
width ($w$) and asymmetry ($\kappa$). These quantities are derived following \cite{Norrisetal_2005}. 
The errors in the derived parameters 
are found by propagating the errors of the model parameters. However, error in $p$ is obtained by 
$\triangle \chi^2=2.7$ by directly tabulating $\chi^2$ for different values of $p$. This is because, the 
the correlation between $\tau_1$ and $\tau_2$ leads to large errors in $p$, particularly for the symmetric 
pulses. The derived errors in $w$ and $\kappa$, however, are small. From Figure~\ref{lc}, we see that
though in general the GRBs have single broad pulse structures, in two cases we find evidence of 
multiple pulses. For GRB 081207, which is a very long GRB in comparison to the other GRBs, we find 
two pulses in the main bursting period. We also find two pulses for GRB 110721A ($\Delta \chi^2 = 205.1 $
is found for an additional pulse). Hence, one has to be cautious while interpreting the results obtained 
for these two GRBs. 

In Table~\ref{ch4_t1} and Figure~\ref{lc}, the first 6 GRBs are categorized as HTS, and the rest as IT GRBs
(section 4.3.2). By comparing the LCs and the values of asymmetry ($\kappa$) of these two classes, we find
no trend in the asymmetry that can characterize any particular class. We find that two HTS GRBs (GRB~100707A, 
and the first pulse of GRB~110721A) have the highest asymmetry ($0.72\pm0.01$, and $0.83 \pm 0.05$, respectively). 
However, GRB~100612A (a HTS GRB) is very symmetric ($\kappa=0.14\pm0.02$ only). Moreover, GRB~090922A, and the first pulse
of GRB~081207 (IT GRBs) are very asymmetric ($0.62\pm0.05$, and $0.57\pm0.06$, respectively). Though
we should be cautious about GRB~081207, it is clear that HTS and IT GRBs do not show any preference 
for asymmetry in their respective LCs.

\section{Spectral Analysis}

\begin{figure}\centering
{

\includegraphics[width=3.4in]{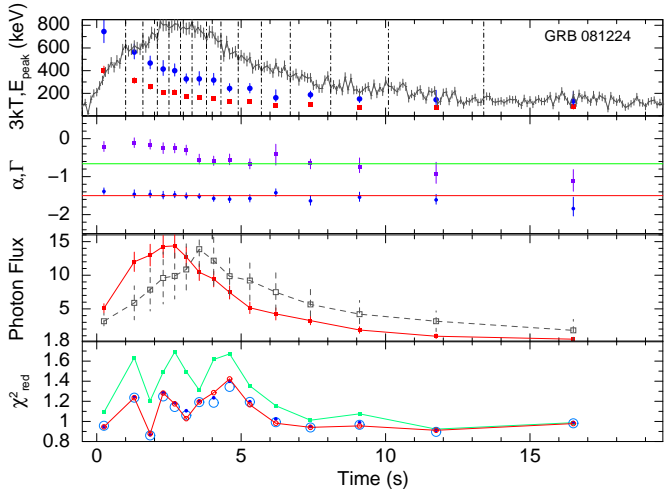} 

}
\caption[Analysis of GRB 081224]
{Analysis of GRB 081224. \textit{Panel 1}: Evolution of $E_{\rm peak}$ (blue filled circles), and
3kT (red filled boxes) with time are shown. The corresponding LCs are shown in background. The LCs are
plotted with the same scale as in Figure~\ref{lc}. The time intervals chosen for spectral analysis
are shown by dot-dashed lines. We show the parameters at the mean time of each of the intervals.
\textit{Panel 2}: Evolution of $\Gamma$ (blue circles) of BBPL, and $\alpha$ (violet boxes) of 
Band function are shown. The ``synchrotron lines of death'' are shown at -3/2 and -2/3 values.
\textit{Panel 3}: Flux (photons cm$^{-2}$ s$^{-1}$) evolution of different components of BBPL model is shown.
BB flux evolution is shown by the filled boxes (joined by continuous line), and PL flux, scaled with the 
total BB flux is shown by open boxes (joined by dashed line).
\textit{Panel 4}: $\chi^{\rm2}_{\rm red}$ shown for Band (red open circles), BBPL (green filled boxes),
mBBPL (blue filled circles), and 2BBPL (light blue open circles)
}
\label{grb1}
\end{figure}

\subsection{Choice Of Time-resolved Spectral Bins}
We now turn to the time-resolved spectral analysis of the GRBs. For time-resolved study, 
the LC of a GRB is divided by requiring minimum counts $C_{\rm min}=1200$ i.e., $\sim35\sigma$
(background subtracted) per time bin. We use one of the NaI detectors (section 4.3.1) having 
the highest count rate as the reference LC. The time divisions are achieved by integrating 
the LC from a start time ($T_{\rm s}$) till this required $C_{\rm min}$ is reached. Note 
that we do not require a definite $T_{\rm s}$ for integration. All we need is equal statistics
in each bin. In all cases, we have chosen $T_{\rm s}\leq T_0$ (trigger time) to use as much 
data as possible. The choice of $C_{\rm min}$ depends on two competitive requirements:
(i) The bins must be wide enough to give enough count, and (ii) small enough to 
capture the evolution. Note that the errors in the model parameters are dependent 
on the total flux in a bin. An approximate choice of total flux in a bin can be made such that 
the error in the evolving parameter (e.g., $E_{\rm peak}$) should be comparable to the 
variation. We have found that this is the case for a $\sim0.5$ s bin size at the peak. As all the 
GRBs have $\sim 2500$ count rate at the peak (except GRB 100707A), we choose 1200 
counts/bin, which is essentially the same as in Lu12. For GRB 100707A, which has $\sim5000$ counts 
at the peak, we set $C_{\rm min}=2500$ count/bin (i.e., 50$\sigma$).  
We use $\chi^2$ minimization technique to estimate the model parameters. Note that
the choice of $C_{\rm min}$ is also reasonable for $\chi^2$ method, as the estimated 
parameters have a maximum deviation of 10\% from those determined by C-stat for a 
count of 1000 (\citealt{Nousek_Shue_1989}).

\begin{figure}\centering
{

\includegraphics[width=3.4in]{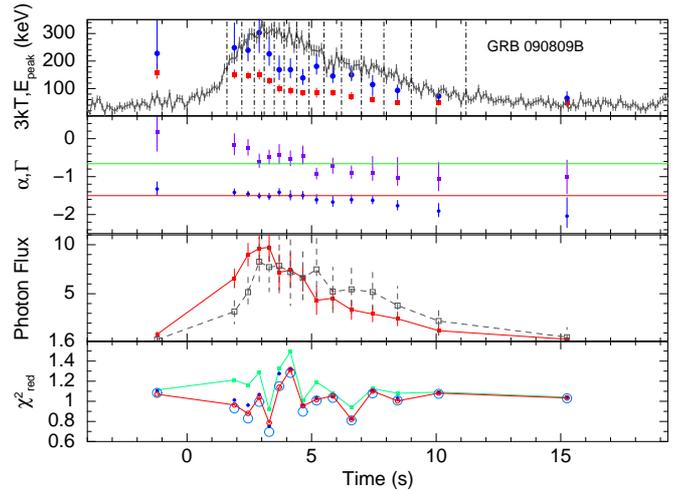} 

}
\caption[Analysis of GRB 090809B]{Analysis of GRB 090809B. The symbols used here are the same as Figure~\ref{grb1}
}
\label{grb2}
\end{figure}

For spectral analysis, we choose three NaI detectors, and one BGO detector. The detectors used for 
the individual cases are shown in Table~\ref{ch4_t1}. Note that due to the lower effective area,
BGO detectors are mostly redundant for time-resolved study (e.g., \citealt{Ghirlandaetal_2010}). 
We found that the count rate in the BGO energy bins are often less than 2$\sigma$, particularly
at energies above $\sim 500$ keV. However, to constrain the high energy portion of the spectral 
function, we use the BGO detector with heavy binning (5 to 7 logarithmic bins). The spectrum of 
the NaI, on the other hand, is binned by requiring 40 counts per bin. 
In Table~\ref{ch4_t1}, we have shown the start ($t_1$) and stop ($t_2$) time, and 
the number of bins ($n$) for the time-resolved spectral analysis. 
For GRB~081207, as we note another pulse at $>80$ s, we use the data till 76 s.
However, this selection will not alter any conclusion we make for this GRB. 

\begin{figure}\centering
{

\includegraphics[width=3.4in]{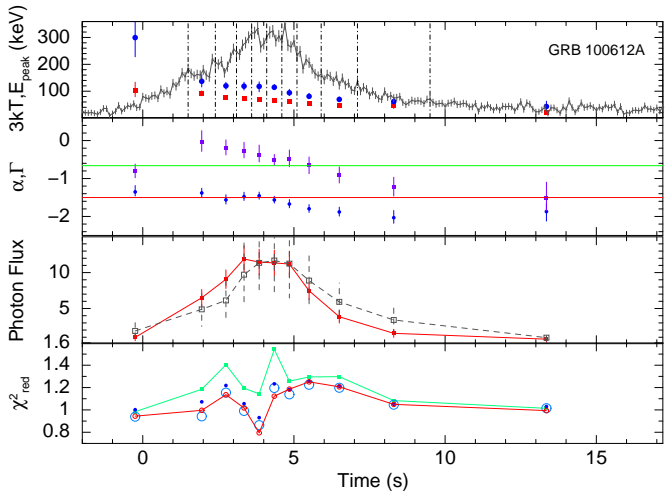} 

}
\caption[Analysis of GRB 100612A]{Analysis of GRB 100612A. The symbols used here are the same as Figure~\ref{grb1}
}
\label{grb3}
\end{figure}

\begin{figure}\centering
{

\includegraphics[width=3.4in]{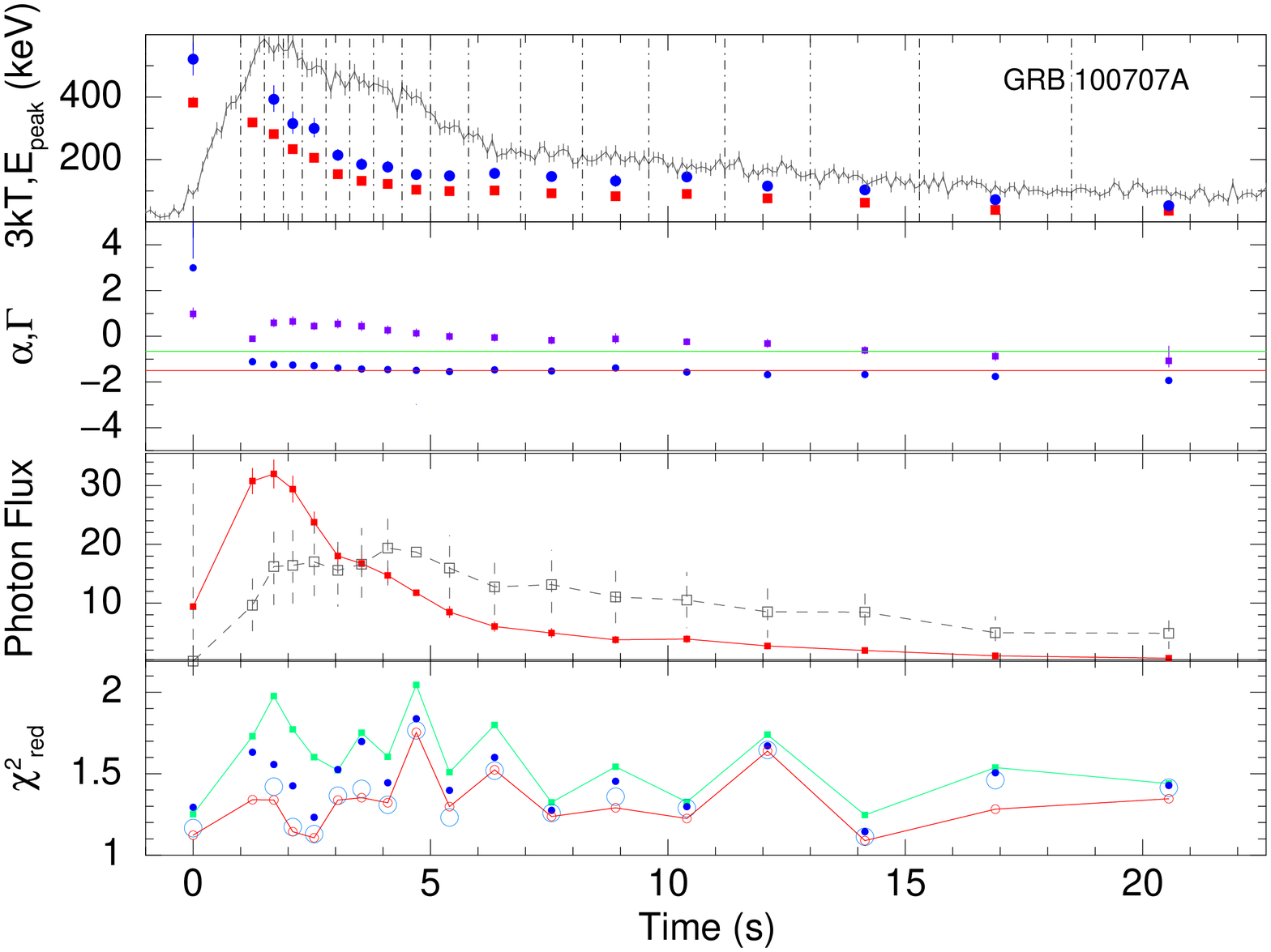} 

}
\caption[Analysis of GRB 100707A]{Analysis of GRB 100707A. The symbols used here are the same as Figure~\ref{grb1}
}
\label{grb4}
\end{figure}

\begin{figure}\centering
{

\includegraphics[width=3.4in]{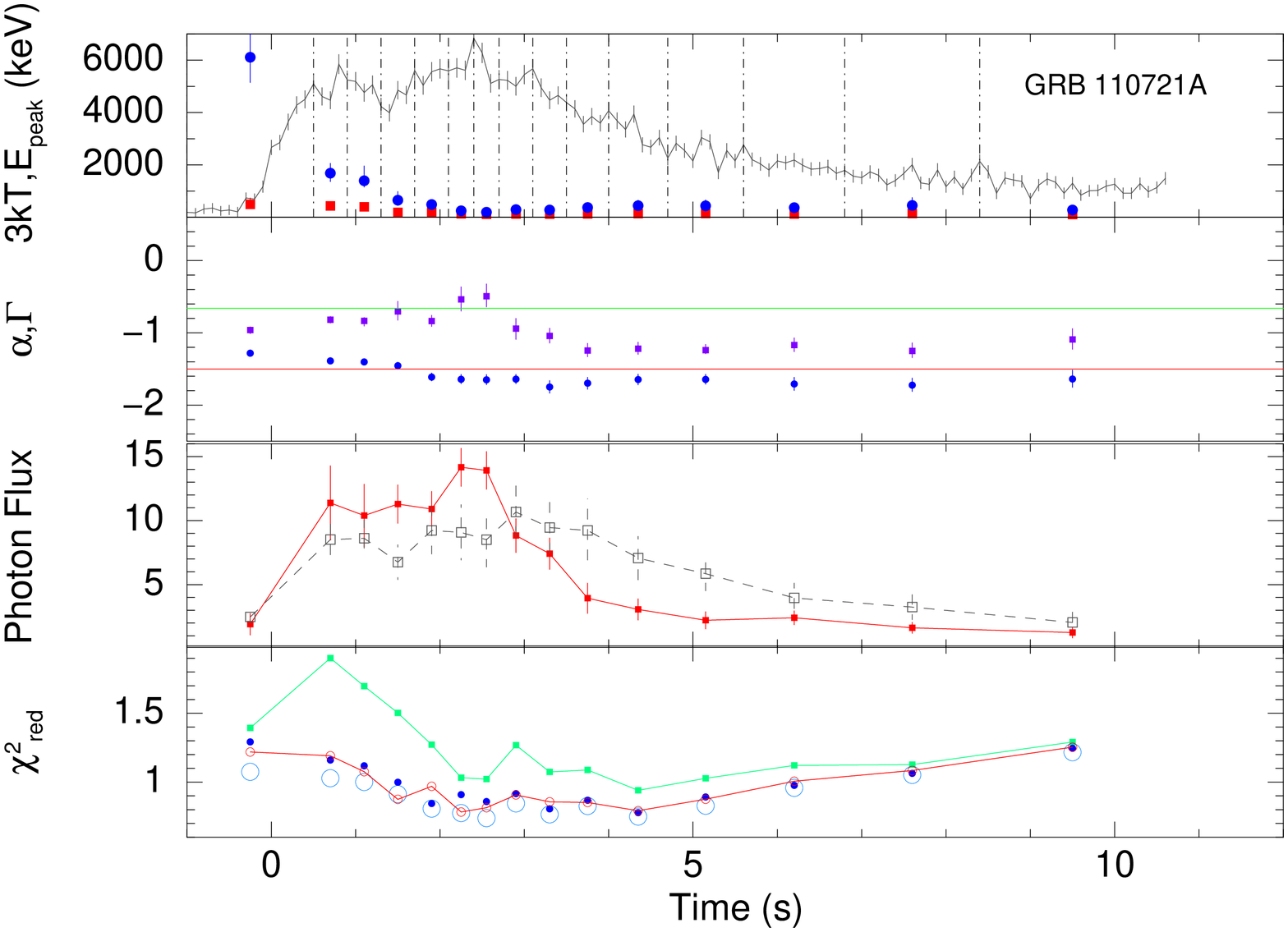} 

}
\caption[Analysis of GRB 110721A]{Analysis of GRB 110721A. The symbols used here are the same as Figure~\ref{grb1}
}
\label{grb5}
\end{figure}

\begin{figure}\centering
{

\includegraphics[width=3.4in]{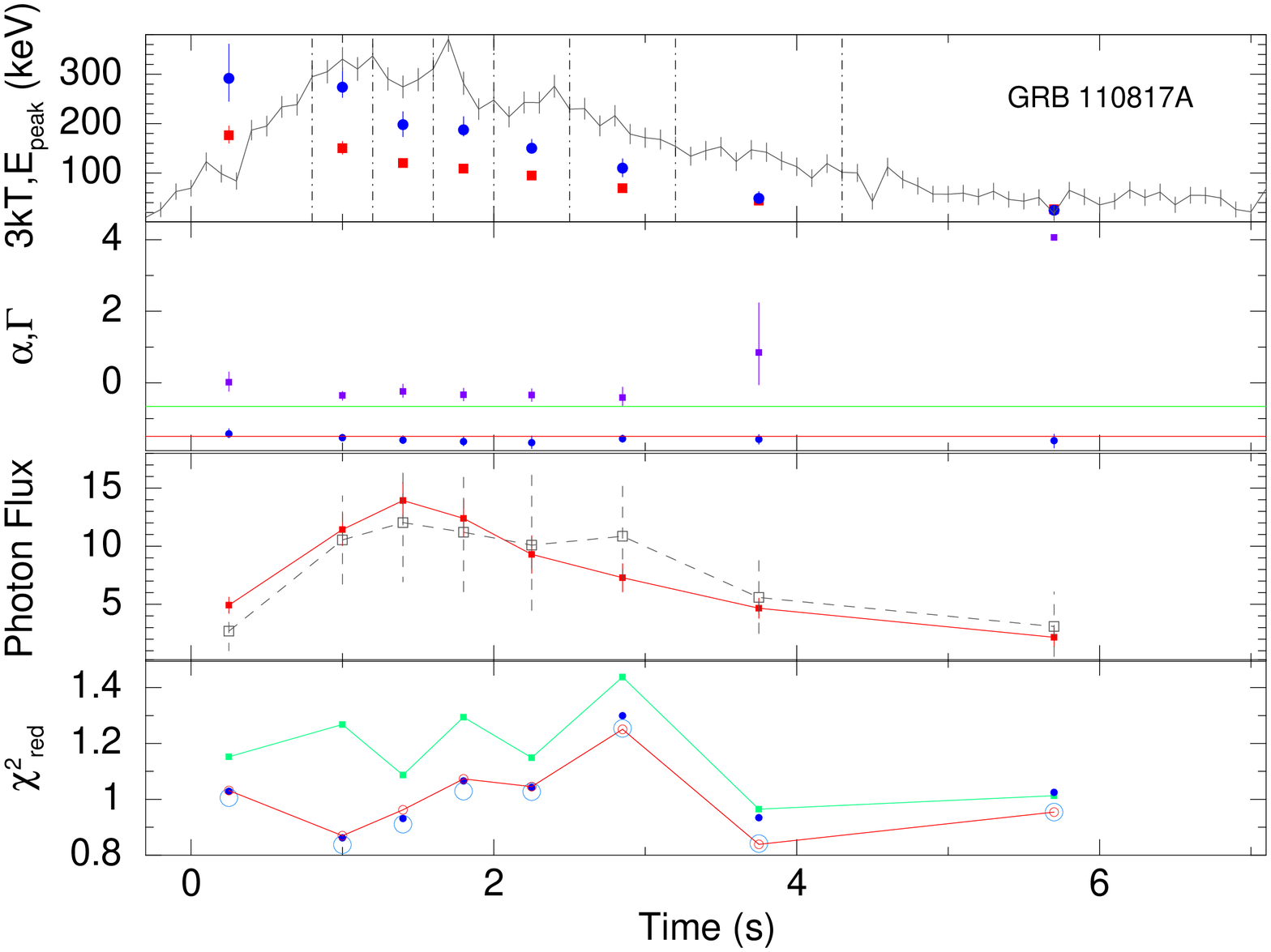} 

}
\caption[Analysis of GRB 110817A]{Analysis of GRB 110817A. The symbols used here are the same as Figure~\ref{grb1}
}
\label{grb6}
\end{figure}

\begin{figure}\centering
{

\includegraphics[width=3.4in]{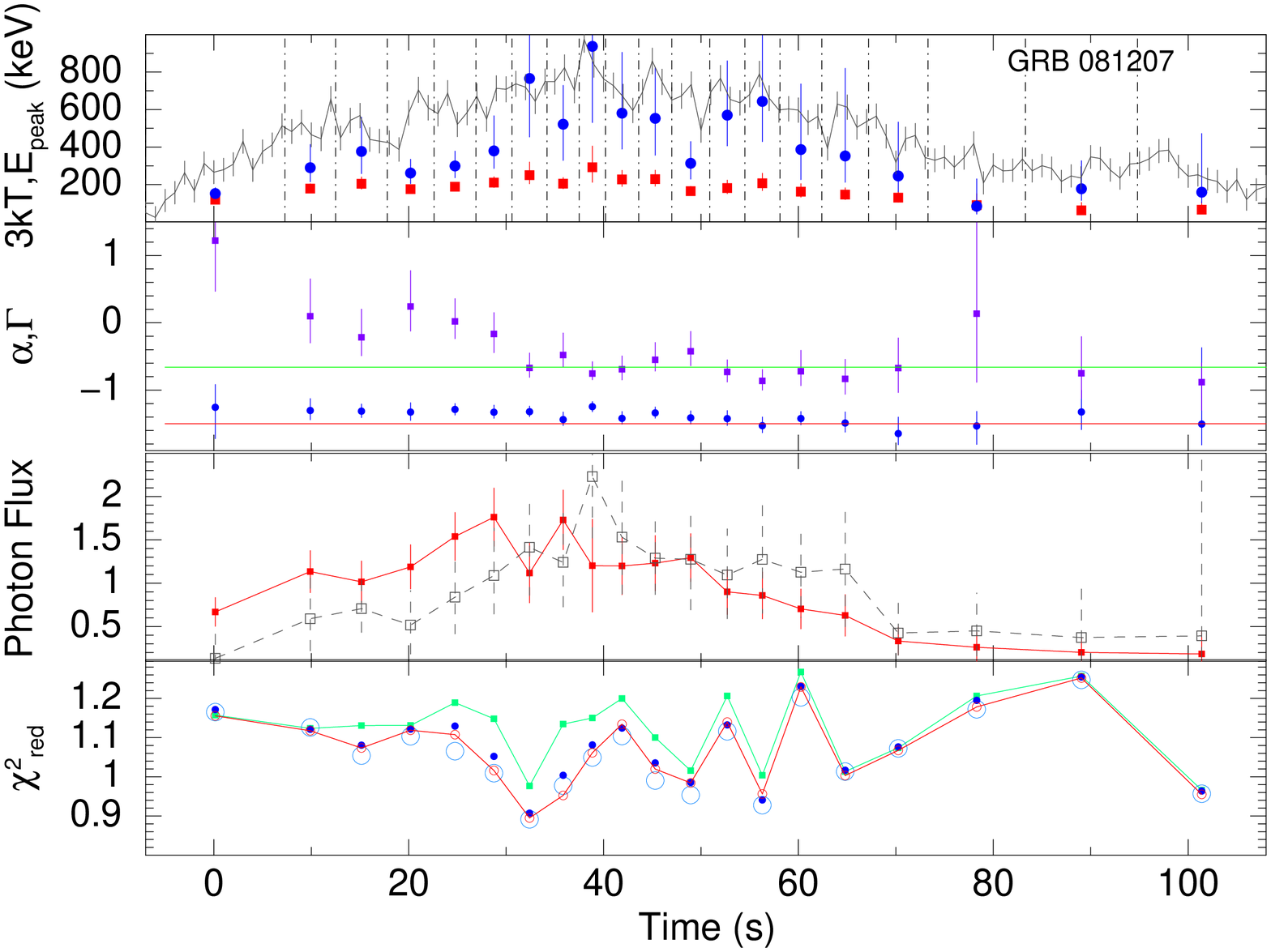} 

}
\caption[Analysis of GRB 081207]{Analysis of GRB 081207. The symbols used here are the same as Figure~\ref{grb1}
}
\label{grb7}
\end{figure}

\begin{figure}\centering
{

\includegraphics[width=3.4in]{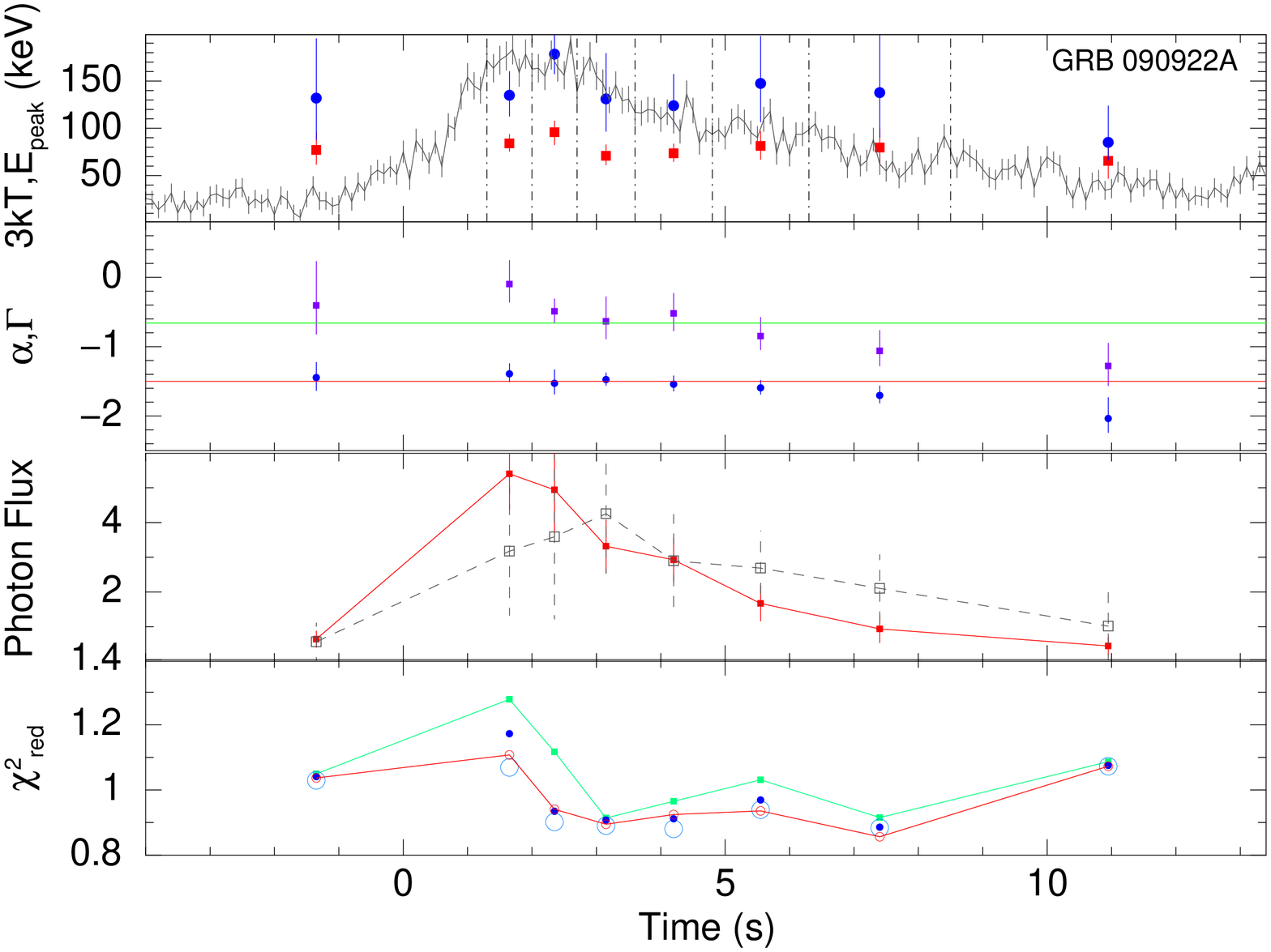} 

}
\caption[Analysis of GRB 090922A]{Analysis of GRB 090922A. The symbols used here are the same as Figure~\ref{grb1}
}
\label{grb8}
\end{figure}

\begin{figure}\centering
{

\includegraphics[width=3.4in]{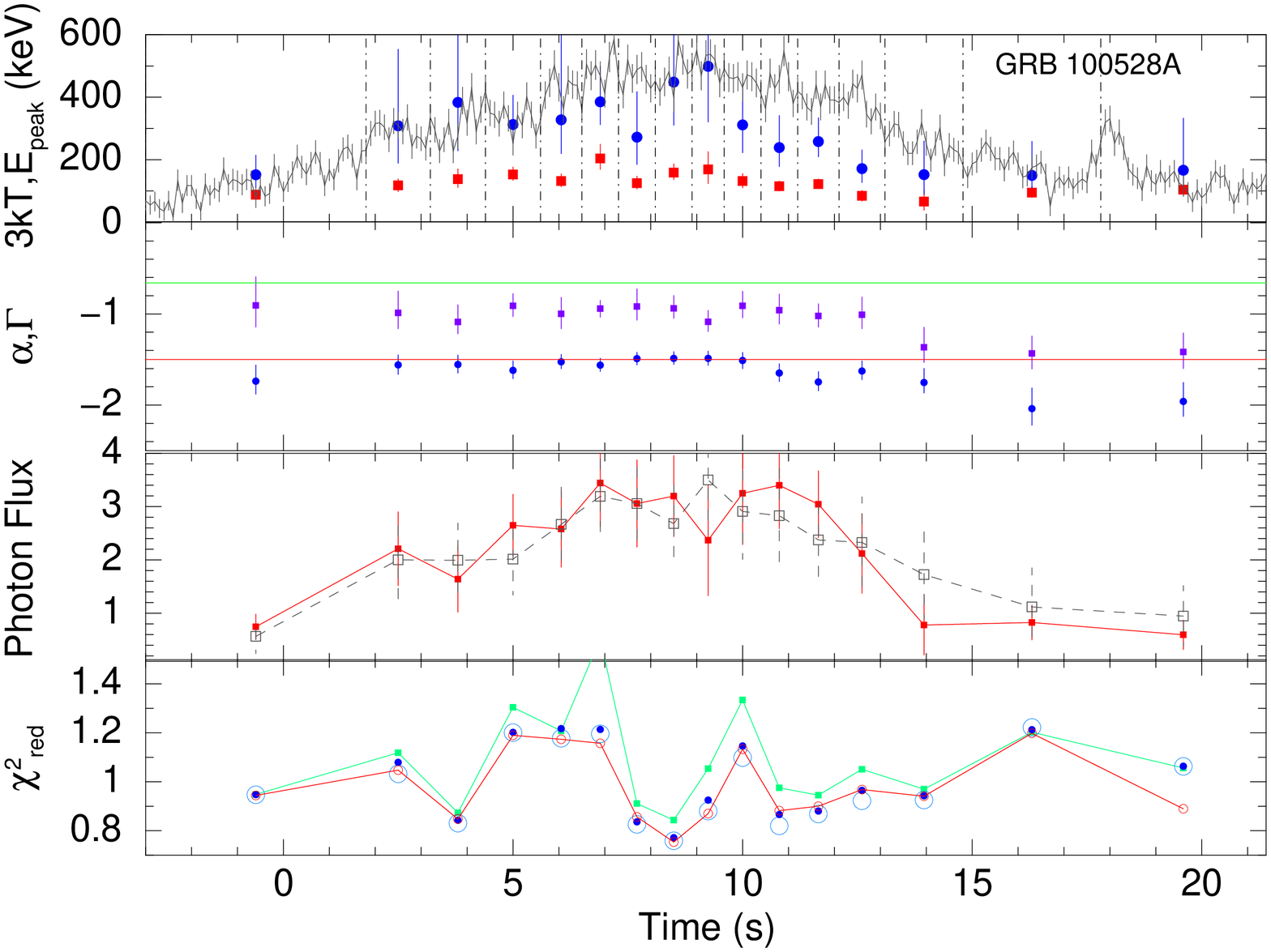} 

}
\caption[Analysis of GRB 100528A]{Analysis of GRB 100528A. The symbols used here are the same as Figure~\ref{grb1}
}
\label{grb9}
\end{figure}

\subsection{Results Of Time-resolved Spectroscopy}

\subsubsection{A. Evolution Of $\bf E_{\rm \bf peak}$ And HTS-IT Classification}
In Figure~\ref{grb1} through Figure~\ref{grb9}, we have shown the results of our time-resolved spectral analysis 
of all the GRBs. The upper panels show the evolution of $E_{\rm peak}$ of the Band function, and $kT$
of the BBPL model as a function of time. The LCs of the individual GRBs are shown in grey colour to clearly show the 
evolution. The dot-dashed lines show the time divisions of the time-resolved 
spectral study. The values of the parameters are shown at the mean of $t_1$ and $t_2$.
From the evolution of $E_{\rm peak}$, we see that the first 6 
GRBs belong to HTS class, i.e., $E_{\rm peak}$ starts with a high value and decreases monotonically.
The next 3 GRBs are IT GRBs, with $E_{\rm peak}$ tracking the pulse LC. Note that as the pulse 
LC is low-to-high-to-low, $E_{\rm peak}$ evolution can as well be said a soft-to-hard-to-soft 
(SHS) evolution (as discussed in chapter 3). Another important fact we find by comparing the 
$E_{\rm peak}$ and $kT$ evolution is as follows. In this panel, we have plotted the $3kT$ rather than kT to 
show the peak of the BBPL. Though the evolution of $E_{\rm peak}$ and $kT$ generally agrees,
the $\nu F_{\nu}$ peak of the Band function is always higher than the corresponding 
peak of the BBPL model ($3kT$).

\subsubsection{B. Evolution Of The Spectral Index}
The evolution of the index are compared in the second panels from the top. We have shown the low energy index
$\alpha$ of the Band function (violet boxes), and PL index ($\Gamma$) of the BBPL model 
(Please do not confuse $\Gamma$ of BBPL with bulk \textit{Lorentz} factor). The two 
lines of death of synchrotron emission are shown by green line ($\alpha=-2/3$, i.e., slow cooling regime),
and red line ($\alpha=-3/2$, i.e., fast cooling regime). We first note that the general trend 
of $\alpha$ evolution is high values to low values. Also, the value of $\alpha$ is always found greater than 
-3/2. In order to quantify the significance of the deviation from the predicted spectrum of fast cooling 
electrons, we find the following quantities. (i) We find the mean value of $\alpha$,
and the deviation of the mean from $\alpha=-3/2$ in the units of $\sigma$. (ii)
We assume $\alpha=-3/2$ as the model spectrum, and find the $\chi^2$ of the fit.
The mean values of $\alpha$ are shown in Table~\ref{ch4_t2} (4th column). The 
calculated errors in the means are 1$\sigma$, and these are calculated by 
using the two tailed nominal 90\% errors of $\alpha$. Column 5 shows the 
first quantity, i.e., the deviation of mean $\alpha$ from $\alpha=-3/2$ in the units of $\sigma$.
The significance of the trend as denoted by the $\chi^2_{\rm red}$ is shown in parenthesis.
In each GRB, we see that the deviation is quite significant. In some GRBs, 
we see that the mean $\alpha$ is even greater than -2/3, the slow cooling line.
However, the deviation from $\alpha=-2/3$ is not always significant. For some of the 
HTS GRBs, we found a significant deviation. For example, GRB 100707A (40$\sigma$), GRB~081224 (5.8$\sigma$),
and GRB~110817A (5.0$\sigma$). For IT GRBs, we found no significant deviation from $\alpha=-2/3$ 
(all deviation are within 1$\sigma$). For IT GRB~100528A, we found a mean $\alpha=-1.02\pm0.04$,
which is well within the slow cooling line of death (at 8.8$\sigma$). In fact, we found that 
the values of $\alpha$ is always less than -2/3 for this GRB. We find another extreme case ---
GRB 110817A, which is a HTS GRB, the value of $\alpha$ is always greater than -2/3.
From this discussion, we see that the HTS GRBs have a general preference of higher 
(harder) value of $\alpha$. Another important trend is shown by the PL index ($\Gamma$) of the BBPL model.
This can be seen from the second panels of Figure~\ref{grb1} through Figure~\ref{grb9}. 
It is apparent that $\Gamma$ has a preference for 
the value -3/2, which is exactly the index of synchrotron emission predicted for the fast cooling electrons. 
In Table~\ref{gamma_32}, we have shown the deviation of mean value of $\Gamma$ from -3/2, and the 
$\chi^2_{\rm red}$ of the $\Gamma=-3/2$ fit. Note that the value of $\Gamma$ generally clusters 
near -3/2 almost in all cases (very low $\sigma$ deviation). The $\chi^2_{\rm red}$, which 
shows the significance of the deviations are small compared to the deviation of $\alpha$.
\vspace{0.1in}

\begin{table*}\centering
 \caption{Deviation of $\Gamma$ from -3/2 line}

\begin{tabular}{cccc}
\hline
GRB & $\chi^2_{\rm red}$ (dof) for $\Gamma=-\frac{3}{2}$ fit & Mean $\Gamma$ & Deviation from $\Gamma=-\frac{3}{2}$ (in $\sigma$)\\
\hline
\hline
081224 & 0.49 (15) & $-1.52\pm0.03$ & $0.67\sigma(\downarrow)^{(a)}$ \\
090809B & 1.03 (15) & $-1.53\pm0.03$ & $1.0\sigma(\downarrow)$ \\
100612A & 2.86 (11) & $-1.60\pm0.04$ & $2.5\sigma(\downarrow)$ \\
100707A & 4.58 (18) & $-1.41\pm0.02$ & $4.5\sigma(\uparrow)$ \\
110721A & 6.86 (15) & $-1.50\pm0.01$ & $<0.4\sigma(\uparrow)$ \\
110817A & 0.54 (8) & $-1.57\pm0.04$ & $1.75\sigma(\downarrow)$ \\
\hline
081207 & 1.86 (20) & $-1.36\pm0.03$ & $3.5\sigma(\uparrow)$ \\
090922A & 0.87 (8) & $-1.55\pm0.05$ & $1.0\sigma(\downarrow)$ \\
090922A & 1.63 (16) & $-1.57\pm0.02$ & $3.5\sigma(\downarrow)$ \\
\hline

\end{tabular}
\vspace{0.1in}

\begin{footnotesize}
 $^{(a)}$ $\uparrow$ (or $\downarrow$) denotes the value of $\Gamma>$ (or $<$) -3/2 
\end{footnotesize}
\label{gamma_32}
\end{table*}

\subsubsection{C. Flux Evolution}
In the third panels of the figures, we have plotted the flux evolution of the BB and PL component.
Interestingly, for three GRBs, namely GRB~081224, GRB~100707A and GRB~110721A, we see a distinct 
behaviour of the PL flux from the BB flux. The PL flux is evidently delayed from the BB. Also,
the PL tends to linger at the late phase of the prompt emission. In chapter 6, we shall discuss 
the delay of the PL component of 2BBPL model, and show that the delayed and lingering behaviour 
of the PL component has a remarkable similarity with the high energy (GeV) evolution. As for the 
current study, we find reported \textit{Fermi}/LAT detection (LLE data) for these three GRBs 
in our sample, i.e., these GRBs accompany high energy (GeV) emission (see chapter 6 for extensive discussion). 
\cite{Ackermannetal_2013_LAT} report the following LAT detection levels --- 3.1$\sigma$ (GRB~081224), 3.7$\sigma$ (GRB~100707A), 
and 30.0$\sigma$ (GRB~110721A)

\subsubsection{D. A Comparative Study}
Finally, the $\chi^2_{\rm red}$ of all the models are shown in panel 4. As we have seen interesting 
results by applying BBPL model for the spectral fitting, it is interesting to compare this model with Band function.
However, we note that $\chi^2_{\rm red}$ of the BBPL model (green boxes) are generally 
worse than those of the Band function (blue filled circles).
To quantify the superiority of the Band function over BBPL, we perform $F$-test assuming BBPL as the original (null) hypothesis, 
and Band function as the alternative hypothesis. Note that as these models are not nested, the F-test is different 
from the general F-test procedure used for nested models. For each time-resolved bin of a GRB,
we calculate the confidence level (CL) of rejecting the original hypothesis as compared to
the alternative model (Band). We compute the mean ($\langle \rm CL \rangle$), and standard deviation (SD) 
of the CLs (see Table~\ref{ch4_t2}). It is clear from the comparison of HTS and IT classes that BBPL model 
better fits the spectrum of IT GRBs (also see Figure~\ref{grb7}, ~\ref{grb8}, and ~\ref{grb9}).
But for HTS class, in general, Band shows much better $\chi^2_{\rm red}$. Hence, BBPL model may be 
more physical than the Band function, but the BB component is probably an \textit{approximation}
of a more physical function. The obvious choice are multi-colour BB (mBB), or two BBs (2BB).
Note that both these models have comparable $\chi^2_{\rm red}$ (in the range 0.8-1.2) as the Band function.
Hence, one of these models is preferred for spectral study. Note that both of these models are 
extension of the simple BBPL model, hence, all the interesting results of BBPL model, e.g., -3/2 
PL index, delayed PL evolution etc., should hold for both of them (see chapter 6).

The results of our detailed analysis are summarized in Table~\ref{ch4_t2}. Based on the above discussions, we can 
draw some approximate, but important conclusions regarding HTS and IT classes. We note the following differences.
 
\begin{itemize}

\item (i) The HTS GRBs have generally higher values of $\alpha$ than the IT GRBs. Note that the significance of 
deviation of $\alpha$ from the -3/2 line is quite high  for HTS GRBs. In 3 cases, we find significant 
deviation even from -2/3 line. In 63.4\% cases of HTS pulses, $\alpha$ is greater than -2/3. The only exception
of this trend is GRB 110721A. Exclusion of this GRB leads to 74.6\% such cases. The IT GRBs, on the 
other hand, shows generally lower values of $\alpha$. Though the $\alpha$ of IT GRBs also show high 
deviation from -3/2 line, the $\chi^2_{\rm red}$ are generally lower than those of HTS GRBs
(see Table~\ref{ch4_t2}). In only 44\% cases, the value of $\alpha$ is found to be greater than -2/3. 
Also, as discussed, the mean value of $\alpha$ has an insignificant deviation (within 1$\sigma$)
from -2/3 in all IT GRBs. Finally, the HTS GRB 110817A has $\alpha$ always greater than 
-2/3, with mean deviation of $5\sigma$, while the IT GRB 100528A has $\alpha$ always less than 
-2/3, with mean deviation of $8.8\sigma$.

\item (ii) We also study the trend of $\alpha$ evolution with time. In Figure~\ref{alpha_hts_it} (left panel),
the values of $\alpha$ are plotted in the x-axis, with time sequence in the y-axis.
We have also indicated the -3/2 and -2/3 death lines by red and purple solid lines. 
We note that the value of $\alpha$ generally becomes lower (softer) at the later part (right to left 
transition). This possibly indicates that the synchrotron emission, if present, dominates 
at the later phase. We have marked the HTS (red circles) and IT (black open boxes) GRBs to 
show the evolution of each class. We note that the evolution of $\alpha$ is generally 
high-to-low values. In the right panel of the figure, we have plotted the 
$\alpha$ values in ascending order to visualize the deviations from the death lines.
It is clear from this panel that the $\alpha<-2/3$ region (left of the purple line)
is mainly populated by IT GRBs, while majority of HTS GRBs have $\alpha>-2/3$ (right of the 
purple line).

\item (iii) For HTS GRBs, the mean value of CL that Band function is preferred over BBPL fitting
is 79.3\%. For IT GRBs, the mean value is 67\%.

\item (iv) Note that the LAT detection is found only for 3 HTS GRBs, and none of the IT GRBs.

\end{itemize}

\begin{figure}\centering
{

\includegraphics[width=3.4in]{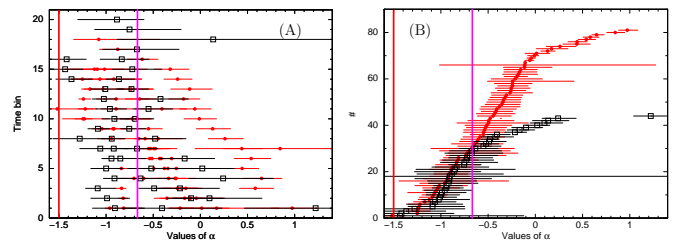} 

}
\caption[Low energy index ($\alpha$) for the GRBs with single pulses]
{Low energy index ($\alpha$) for the GRBs in our sample --- HTS (filled circles), and IT (open boxes).
The errors are measured at nominal 90\% confidence level. \textit{(A)} For each GRB, the distribution of 
$\alpha$ (x-axis) of each time-resolved bin is plotted with the time sequence (y-axis). The values of $\alpha$
show higher deviation from the synchrotron death lines ($\alpha=-3/2$: red line, $\alpha=-2/3$: purple line)
at the earlier times. \textit{(B)} $\alpha$ values,
sorted in ascending order to show the deviations. Its value is always greater than -3/2 (for significance, 
see Table~\ref{ch4_t2}). The region where $\alpha<-2/3$ (i.e., within the slow cooling regime of synchrotron
emission) is significantly populated by IT GRBs (see the left side of the purple line), while $\alpha>-2/3$ 
region is generally acquired by HTS pulses. The mean value of $\alpha$ for HTS and IT GRBs are -0.42 and -0.68, respectively
}
\label{alpha_hts_it}
\end{figure}

\begin{figure}\centering
{

\includegraphics[width=3.4in]{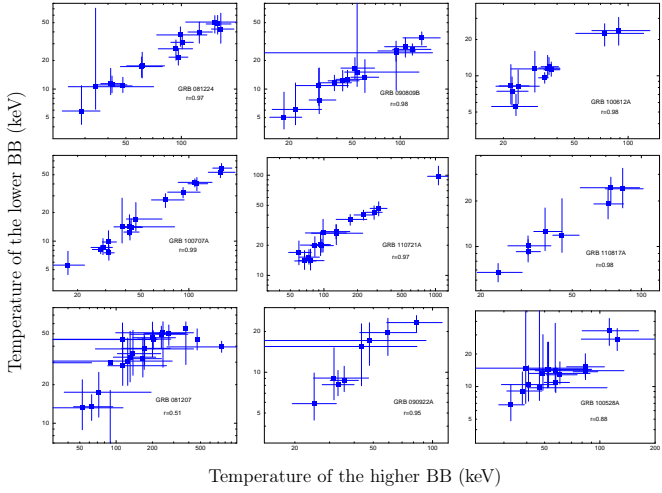} 

}
\caption[Correlation between the temperature (kT) of the two BBs of 2BBPL model]
{The correlation between the temperature (kT) of the two BBs of 2BBPL model are shown. 
A significant correlation is found in each case. \textit{Upper 6 panels}: HTS GRBs. \textit{Lower 3 panels}: IT GRBs
}
\label{kt_corr}
\end{figure}

\section{Summary And Discussion}
To summarize, we have investigated the timing and spectral properties of GRBs with single pulses. We have 
applied four models, namely Band, BBPL, mBBPL and 2BBPL for time-resolved spectroscopy of the pulses.
We found that the value of $\alpha$ of the Band function is significantly higher than the fast cooling 
line $\alpha=-3/2$, and in many cases, even higher than $\alpha=-2/3$ line with reasonable significance.
Hence, a synchrotron interpretation is unacceptable. At least the spectrum should have other contributors e.g.,
the emission from the photosphere. In this spirit, we have applied the BBPL model (following earlier works),
and have found interesting results e.g., the PL index of the BBPL model ($\Gamma$) has a preference for -3/2 
value, showing a remarkable consistency with the synchrotron origin of the PL component. However, 
we have found that the $\chi^2_{red}$ of BBPL model is generally worse than that of the Band function.
The other two functions, namely mBBPL and 2BBPL models have comparable $\chi^2_{red}$ as the Band function.
Hence, mBBPL and 2BBPL are the preferred models for our set of GRBs. As these models are modification 
of the thermal component of the BBPL model, we expect the PL component to show similar behaviour (see 
the next chapters).

Based on the peak energy evolution, we have found that the pulses can belong to either HTS or IT (or, SHS) 
class. As we have chosen only single pulse GRBs, the IT evolution cannot have any overlapping effect. 
By performing a detailed spectral analysis, we have found interesting differences of these two classes, 
though the origin is not clear at the moment. We have also found delayed and lingering behaviour of 
the PL component for 3 GRBs, and interestingly, these GRBs are accompanied by high energy (GeV) emission. 

Finally, we have used a new model, namely 2BBPL which gives comparable fit as mBBPL model. 
Though the physical mechanism of this model is not known at this moment, in the next chapter, we 
shall try to give some important evidences in support of this model. We shall also show the 
predictive power of this model in chapter 6. Before finishing this chapter, let us show one of the 
most important feature of 2BBPL model, which we shall use in subsequent analyses. 
In Figure~\ref{kt_corr}, we have shown the temperature (kT) of the two BBs of 2BBPL model.
It is evident from the figure that the two BB temperatures are highly correlated.
The correlation coefficient ($r$) are generally high (0.51-0.99). There can be two reasons
for such correlation, first, the 2BBPL may be an approximation of a more fundamental
function. In this case, the correlation is a natural consequence of the approximation. 
Second, the 2BBPL is indeed a physical model. It may be an approximate function, but 
all the components have distinct source of emission. However, the sources of the 
2BBs must have some common feature which leads to the observed correlation. In the 
subsequent chapters, we shall try to give evidences in favour a physical origin of the
2BBPL model.

\begin{sidewaystable}\centering
\begin{footnotesize}

\caption[Parameters of Norris model fit to the LCs of the GRBs with single pulses]
{Parameters of Norris model fit to the LCs of the GRBs. We also show the time interval 
($t_1$ to $t_2$), total number of time-resolved bins ($n$), and the detectors which are used for 
the time-resolved spectroscopy.}

\begin{tabular}{c|ccccccc|ccc}
\hline 

GRB & \multicolumn{7}{c|}{Norris Model parameters} & \multicolumn{3}{c}{Specification of time-resolved analysis}\\

\cline{2-11}

    & $t_s$ (s) & $\tau_1$ (s) & $\tau_2$ (s) & $\chi^2_{red} (dof)$ & p (s) & w (s) & $\kappa$ & $t_1$,$t_2$ (s) & $n$ & Detectors used \\
\hline
\hline 
081224 & $-1.53_{-0.26}^{+0.34}$ & $3.60_{-1.07}^{+0.99}$ & $3.30_{-0.21}^{+0.27}$ & 6.80 (13) & $1.90\pm 0.10$ & $7.5\pm 0.7$& $0.44\pm 0.03$& -0.5, 19.6 & 15 & n6, n7, n9, b1\\
090809B & $-1.74_{-0.38}^{+0.31}$ & $9.35_{-1.85}^{+2.59}$ & $2.50_{-0.17}^{+0.17}$ & 5.88 (13) & $3.10\pm 0.09$& $7.4\pm 0.6$& $0.34\pm 0.02$& -4.0, 19.3 & 15 & n3, n4, n5, b0\\
100612A & $-8.43_{-2.65}^{+1.61}$ & $156.4_{-57.4}^{+137.6}$ & $0.88_{-0.17}^{+0.14}$ & 16.8 (9) & $3.30\pm 0.06$ & $6.5\pm 1.5$& $0.14\pm 0.02$& -2.0, 17.2 & 11 & n3, n4, n8, b0\\
100707A & $-2.5_{-2.6}^{+3.2}\times 10^{-2}$ & $0.37_{-0.05}^{+0.04}$ & $7.19_{-0.16}^{+0.18}$ & 4.01 (233) & $1.60\pm0.07$ & $9.9\pm0.28$& $0.72\pm0.01$&  -1.0, 22.6 & 18 & n4, n7, n8, b1 \\
110721A & $-0.62_{-0.15}^{+0.08}$ & $1.58_{-0.46}^{+0.67}$ & $1.59_{-0.23}^{+0.15}$ & 1.32 (229) & $0.96\pm0.10$ & $3.55\pm0.62$& $0.45\pm0.03$&  -1.0, 10.6 & 15 & n6, n7, n9, b1 \\
        & $1.75_{-0.07}^{+0.10}$ & $0.079_{-0.053}^{+0.074}$ & $6.75_{-0.66}^{+0.86}$ &  & $2.48\pm0.12$ & $8.08\pm1.04$& $0.83\pm0.05$&  &  &  \\
110817A & $-0.62_{-0.42}^{+0.30}$ & $1.11_{-0.66}^{+1.31}$ & $1.76_{-0.24}^{+0.24}$ & 1.85 (8) & $0.78\pm 0.12$ & $3.6\pm 0.8$& $0.49\pm 0.08$& -0.3, 7.1 & 8 & n6, n7, n9, b1\\
081207 & $-8.55_{-2.21}^{+3.30}$ & $14.14_{-6.10}^{+9.10}$ & $50.79_{-10.22}^{+11.10}$ & 1.36 (114) & $18.25\pm 3.36$ & $90\pm 21$& $0.57\pm 0.06$& -7.0, 76.0 & 20 & n1, n9, na, b1\\
       & $-9.64$ & $304.7$ & $9.24_{-2.20}^{+3.50}$ & & $43.42 \pm 1.19$ & $45$& $0.20$&  & &\\
090922A & $-0.39_{-0.19}^{+0.15}$ & $0.62_{-0.27}^{+0.38}$ & $4.00_{-0.30}^{+0.30}$ & 6.23 (15) & $1.18\pm 0.18$ & $6.4\pm 0.7$& $0.62\pm 0.05$& -4.0, 13.4 & 8 & n0, n6, n9, b1\\
100528A & $-32.6_{-2.36}^{+2.36}$ & $1262.0_{-61.1}^{+60.9}$ & $1.29_{-0.38}^{+0.14}$ & 6.56 (21) & $7.70\pm 0.06$ & $14.5\pm 3.0$& $0.09\pm 0.004$& -3.0, 21.4 & 16 & n6, n7, n9, b1\\

\hline
\end{tabular}
\label{ch4_t1}
\end{footnotesize}

\end{sidewaystable}

\begin{sidewaystable}\centering

\caption{Classification of the GRBs based on the spectral analysis: ``hard-to-soft'' (HTS), and ``intensity tracking'' (IT) }

\begin{tabular}{cccccccc}
\hline 
GRB & Type & Behaviour of  & Mean $\alpha$ & Deviation of $\alpha$ $^{(a)}$  & $\alpha$ crossing -2/3  & Band/BBPL$^{(b)}$ & LAT detection\\
    &              &   the PL Flux &  & from -3/2 ($\chi^2_{\rm red}$)& ``line of death''  &          &              \\
\hline
\hline
081224	& HTS & Clear delay and lingering & $-0.43 \pm 0.04 $ & 26.7$\sigma$ (53.5) & 11/15 & 0.80 (0.14) & 3.1$\sigma$\\
090809B & HTS & Mild delay and lingering & $-0.64 \pm 0.06 $ & 14.3$\sigma$ (13.5) & 8/15 & 0.74 (0.15) & No\\
100612A & HTS & Very mild lingering & $-0.58 \pm 0.07 $ & 13.1$\sigma$ (19.1) & 7/11 & 0.73 (0.15) & No\\
100707A & HTS & Clear delay and lingering & $0.013 \pm 0.017 $ & 89.0$\sigma$ (71.6) & 16/18 & 0.74 (0.23) & 3.7$\sigma$\\
110721A & HTS & Clear delay and lingering & $-0.95 \pm 0.02 $ & 27.5$\sigma$ (56.2) & 2/15 & 0.94 (0.10) & 30.0$\sigma$\\
110817A & HTS & Mild lingering & $-0.31 \pm 0.07 $ & 16.9$\sigma$ (34.1) & 8/8 & 0.81 (0.09) & No\\
\hline
081207	& IT & Mild delay & $-0.63 \pm 0.06 $ & 14.5$\sigma$ (10.5) & 10/20 & 0.66 (0.12) & No\\
090922A & IT & Mild delay and lingering & $-0.66 \pm 0.10 $ & 8.4$\sigma$ (9.7) & 5/8 & 0.67 (0.13) & No\\
100528A & IT & Mild lingering & $-1.02 \pm 0.04 $ & 12.0$\sigma$ (9.8) & 0/16 & 0.68 (0.14) & No\\
\hline
\end{tabular}
\label{ch4_t2}
\begin{flushleft}
\begin{footnotesize}

$^{(a)}$ Here deviation is quantified as the difference of the mean value of $\alpha$ 
($\langle \alpha \rangle$) from the fast cooling ``line of death'' (i.e., $\alpha=-3/2$ line) 
in the units of $\sigma$. The $\chi^2_{red}$ is calculated by fitting $\alpha$ values 
assuming the model $\alpha=-3/2$. Hence, higher the value, higher is the deviation from
the fast cooling synchrotron emission.
\\
\vspace{0.1in}
$^{(b)}$ Band function is compared to the BBPL model. We have performed F-test for all 
time-resolved spectrum to find the confidence level (CL) of alternative model (i.e., Band function) over 
the original model (i.e., BBPL model). The quantity shown here is the mean (and standard deviation)
of CL of the F-test for each GRB.
\end{footnotesize}
\end{flushleft}

\end{sidewaystable}

\chapter{Parametrized Joint Fit} \label{ch5}

\section{Overview}
In the previous chapter, we have discussed the alternative models of the prompt 
emission spectrum of GRBs. We specifically chose GRBs with single pulses and high flux
for our study. The primary motivation of selecting single pulses was to get an idea about the parameter 
variations within a pulse which is essentially unaffected by the overlapping effect. We also
wanted to know whether the $E_{\rm peak}$ evolution is always hard-to-soft (HTS), or there are 
intensity-tracking (IT) behaviours as well. We have found that the $E_{\rm peak}$ evolution 
within some of the pulses are indeed HTS. However, we have also found pulses which rather 
show the IT evolution. The finding of such a spectral evolution in a single pulse undoubtedly tells us that 
the IT evolution is real, and cannot be always a superposition effect of two HTS pulses. 
Though the physical reason for these two distinct behaviours are not known at present, 
but we can use these evolution properties to get a better handle on the time-resolved 
spectral study, which is the subject of this chapter. Aided with the knowledge of single pulses,
we now turn our attention to GRBs with multiple (but separable) pulses (\citealt{Basak_Rao_2013_parametrized}). 
We shall also study some GRBs having high flux, but with more rapid variability (\citealt{Raoetal_2014}).

\subsection{Time-resolved Study With Parametrization}
Time-resolved spectroscopy is a natural choice to study the time evolution of a spectrum. It is possible that 
a time-integrated spectrum appears completely different from the actual snap-shots of the 
spectrum. For example, \cite{Ryde_2004} have shown that an instantaneous spectrum is consistent 
with a blackbody and a power-law (BBPL) model, whereas the integrated spectrum appears 
as a Band function, and a BBPL model does not fit a time-integrated spectrum at all.
Hence, one has to investigate the spectral evolution within as small time bin as possible.
But, as discussed in the previous chapter, the choice of bin size is limited by the statistics, 
meaning, with a low flux data, one can fit any model. For example, \cite{Ghirlandaetal_2010}
have studied the time-resolved data of 9 GRBs detected by the \textit{Fermi}/GBM. Due to 
the requirement of fine bin size, they could use only a cut-off power-law (CPL), which is a 
three-parameter model. The parameters of a more complex function like Band cannot 
be constrained with such data quality. On the other hand, we cannot possibly afford to 
make a large bin size if we want to follow the time evolution.

The solution to our dilemma lies in the realization that the spectral evolution 
need not be too drastic. In fact, one can suitably parametrize the evolution in order to 
reduce the number of free parameters ($P_f$) of the description. For example, if we have $n$ 
time-resolved bins, a 4-parameter function, such as Band, requires $4n$ parameters in total to 
describe the time evolution. If we believe e.g., that $E_{\rm peak}$ evolution is HTS, and 
assume certain functional form of the time evolution, then effectively it reduces 
$P_f$ to $3n+1$. For a GRB with 20\,s duration and 1\,s uniform bin, we have $n=20$. Hence, such a scheme reduces the 
total number of parameters from 80 to 61. Note that such a strategy was indeed applied when we 
developed the simultaneous timing and spectral model of GRB pulses (chapter 3). Of course, we 
know that GRB pulses can also have IT evolution, and hence, we shall not assume a HTS
evolution. In section 5.3, we shall describe our assumptions in detail. We shall 
specifically develop a new technique, named ``parametrized joint fit'' for the spectral analysis.
The primary motivation of the method is quite apparent --- we are aiming to reduce 
the number of model parameters in order to study the spectral evolution, and compare 
various models. 

\subsection{The Spectral Models}
The models we use for the spectral study (as in the previous chapter) are: (i) Band, (ii) BBPL, (iii) mBBPL, and 
(iv) 2BBPL. Except for the Band function, which has a non-thermal spectral shape, all other models 
have a thermal and a non-thermal component. We have often seen that the Band function is unacceptable based on 
the fact that the low energy index ($\alpha$) has higher value than -3/2, and sometime the value 
even exceeds the slow cooling limit at -2/3. However, the goodness of a fit 
(in terms of $\chi^2_{\rm red}$) using a Band function is comparable to those of mBBPL and 2BBPL models. 
From a purely phenomenological point of view, a general spectrum has two regions 
of interest --- the peak of the $\nu F_{\nu}$ spectrum, 
and the ``wings'' (i.e., the low and high energy parts). While a Band function has two slopes joining at 
the peak, each of the alternative functions have a thermal peak, and a single slope describing 
both the low and high energy parts. This slope is lower than the low energy slope ($\alpha$), 
and higher than the high energy slope ($\beta$) of the Band function. Hence, the PL 
component of these alternative models ``hold'' the spectrum at the wings. If the difference 
occurs in the peak then we immediately see the effect in the value of $\chi^2$. For example, BBPL model 
has a narrower peak than all other models. If the spectrum is not as narrow as a BB spectrum 
at the peak, the BBPL model should give an inferior fit. However, the peak of both mBBPL and 2BBPL are 
as broad as that of a Band function; the only difference occurs in the wing. As the statistics is low at 
the two ends of the detector's energy band, we do not expect an order of magnitude improvement 
if the differences occur at the wings. This was precisely the reason that all these very different 
functions, namely, Band, mBBPL, and 2BBPL, gave similar values of $\chi^2$ while fitting the time-resolved data
(see chapter 4). Hence, it is interesting to investigate how the models fit the data when we demand 
certain extra conditions on their individual parameter evolution. For example, we have seen that the 
two temperatures of the 2BBPL model are highly correlated. Hence, we shall assume a fixed factor 
for the two temperatures, and tie the temperature ratio in all time bins. If the temperatures are indeed correlated, we 
should still get good $\chi^2$ values. As a benefit, we shall reduce the number of free parameters ($P_f$), 
because determining the factor is enough to infer the temperature of one of the BBs, given the 
temperature of the other one.

\begin{figure}\centering
{

\includegraphics[width=3.4in]{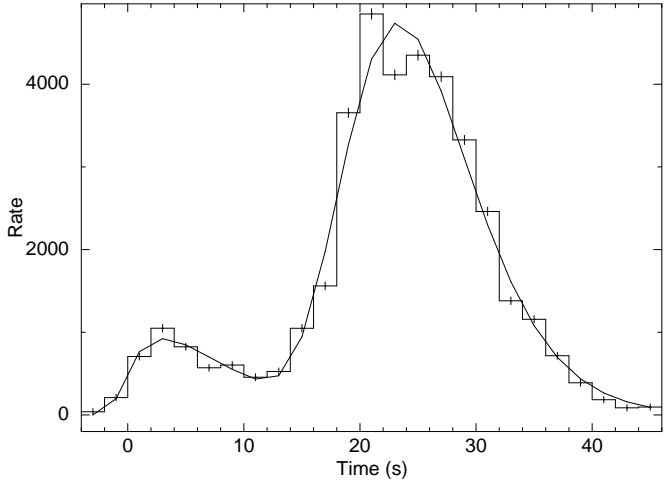} 

}
\caption[Norris model fitted to the background subtracted LC of GRB 081221]
{The LC of GRB 081221 (background subtracted), fitted with Norris model (\citealt{Norrisetal_2005}).
The LC is generated by adding the counts of two NaI ($n1$ and $n2$) and one BGO ($b0$) detector.
Source: \cite{Basak_Rao_2013_parametrized}.}
\label{ch5_f1}
\end{figure}

\section{Sample Selection}
The requirements of our study are GRBs with high background subtracted counts and long enough duration for time-resolved 
study. In addition, we want GRBs with separable pulses. We use \cite{Navaetal_2011} catalogue for selecting bright 
(fluence $\geq10^{-6}$ erg), and long ($\delta t\geq15$ s) GRBs with single/separable pulse (s). We find 11 such 
GRBs (see chapter 2, section 2.4.6). As GRB 081221 is the brightest GRB (fluence $=3.7\times10^{-5}$ erg) in the sample, 
we use it for the time-resolved spectral study using our new technique. In Figure~\ref{ch5_f1}, we have shown the LC of the GRB, 
with the two pulses fitted by Norris model (\citealt{Norrisetal_2005}). In addition to this GRB, we shall use GRB 090618, which has 
high fluence ($=3.4\times10^{-4}$ erg), and broad pulse structure. Apart from studying GRBs with clean pulses, it is 
also important to study GRBs with variable LC. In this regard, we shall discuss 3 GRBs, namely, GRB 080916C, 
GRB 090902B, and GRB 090926A. As these GRBs have highly variable time profile, some modification of our technique 
is required for them.
\vspace{1.0in}

\section{Assumptions Of The New Technique}
We have the following parameters of various models for the spectral fitting.
\begin{itemize}
 \item (i) Band function: normalization ($A_b$), photon indices ($\alpha$, and $\beta$), and peak energy ($E_{\rm peak}$)
 \item (ii) BBPL: temperature ($kT$) and normalization ($K_1$) of BB, and index ($\Gamma$) and normalization ($K_2$) of PL.
 \item (iii) mBBPL: the local disc temperature $kT(r)\propto r^{-p}$, where r is the distance from the 
centre of the disc. Hence, the parameters are inner temperature ($T_{in}$), index $p$, normalization of mBB ($K_{mbb}$), and
PL index and normalization
\item (iv) 2BBPL: two BB temperatures ($kT_h$, $kT_l$), normalizations ($N_h$, $N_l$), and PL index and normalization.
\end{itemize}

First, we divide the LC of a GRB into the constituent pulses e.g., GRB 081221 is divided into two pulses: -1.0 to 12.05 s, 
and 17.0 to 40.55 s. Note that the overlapping region is neglected to facilitate parametrization. 
We further divide each pulse into a rising and a falling \textit{sector}. This is based on the fact that 
$kT$ evolution of BBPL model generally has a break near the peak flux. \cite{Ryde_Pe'er_2009} have studied a 
set of bright GBM GRBs, and have found that the time evolution of both flux and $kT$ have similar break 
time (within errors). Following the procedure of the previous chapter, we choose the start time of each sector
and integrate the LC till we get a minimum count per bin $C_{\rm min}$.

\subsection{Assumptions For The Band Function}
For the Band function, we assume that the photon indices have little variation within a sector. 
Hence, the value of these parameters can be determined by tying the parameters in all 
time bins of a sector. Note that this tying will not only help us reducing $P_f$, but we 
can also study the differences in the spectral slopes in the rising and falling sectors.
For example, we have seen that the value of $\alpha$ tend to be high at the beginning, and evolves 
to a lower value. Hence, by tying this parameter in each sector, we can find the contrast of the spectral 
slope, and can possibly comment more on the origin of the spectrum.

We further assume that $E_{\rm peak}$ evolution is a power-law function of time ($E_{\rm peak}\propto t^{\mu}$).
As the evolution can be either HTS or IT type, we shall obtain the parameter $\mu$ independently in each sector.
Note that ideally one should also put a start time for the parametrization. In our case, we 
have used this time either as zero, or -10 (for negative start time). As the parametrization is not 
corrected for the start time, the $\mu$ values in different pulses should not be compared. Finally, 
the normalization of Band function ($A_b$) is assumed as a free parameter of our model. With these 
assumptions we reduce $P_f$ from $4n$ to $n+4$, where $n$ is the number of time bins in a sector. 
The free parameters are $n$ normalizations, $\alpha$, $\beta$, $\mu$ and the peak energy at the starting 
time bin ($E_{\rm peak}(t_0)$). The peak energy at any time can be found by $E_{\rm peak}(t_0)\times (t/t_0)^{\mu}$.
Note that the choice of a time bin to specify the value of $E_{\rm peak}$ is arbitrary. All we require is the 
$E_{\rm peak}$ at any time bin, and $\mu$ to specify the evolution.

\subsection{Assumptions For The BBPL Model}
Following the scheme used for the Band function, we assume that the $kT$ of BBPL model has a power-law 
time evolution ($kT\propto t^{\mu}$) in each sector. We tie the index of PL spectral component 
in all bins of a sector. The parametrization of the normalization is more complicated than that 
of the Band function, as we have two normalizations ($K_1$ and $K_2$). We can either assume an 
overall free normalization, and parametrize $K_1$ and $K_2$, or we can parametrize one of the 
normalizations and treat the other as a free parameter. Note that in both of the cases, we get an 
equal number of parameters as the Band function. \cite{Ryde_Pe'er_2009} have investigated the parameter 
${\cal{R}} = (F_{\rm BB}/\sigma T^4)^{1/2}$,
which either remain constant or increases with time, where $F_{\rm BB}$ is the observed 
BB flux. Now, if we assume that $F_{\rm BB}$ evolves as a simple function of time as $F_{\rm BB}\propto t^{\zeta}$,
then ${\cal{R}} \propto t^{\zeta/2-2\mu}$. For $\mu\leq \zeta/4$, we expect the observed 
time evolution of ${\cal{R}}$. Hence, the BB normalization ($K_1$) can be chosen as 
a power-law function of time ($K_1\propto t^{\nu_1}$). Now, we can assume this parametrization 
and use the PL normalization ($K_2$) as a free parameter, and this will give us the same 
number of free parameters as the Band function. However, note that this scheme will put 
more constraints on the BBPL parametrization, as the overall norm is not a free parameter
as the Band function. Hence, to overcome this situation, we also assume that the PL component 
has as smooth a time evolution as the BB component i.e., $K_2\propto t^{\nu_2}$. Now, we 
can make the overall normalization ($K$) as a free parameter by assuming the time evolution 
of the ratio of the normalization of BB and PL component as 
$K_1/K_2\propto t^{\nu_1/\nu_2}\propto t^{\nu}$. Thus the parametrized joint fit 
of the BBPL model has equal number of parameters as that of the Band function ($n+4$).

\subsection{Assumptions For mBBPL And 2BBPL Models}
For mBBPL we choose similar parametrization as the BBPL model. In addition, we
tie the parameter $p$ in all the time bins of a sector. For 2BBPL model, we 
assume that the two temperatures ($kT_h$, $kT_l$), and the normalizations ($N_h$, $N_l$) 
are highly correlated. Hence, we tie their ratio in each sector. Compared to Band and
BBPL, these models have one more, and two more parameters, respectively. For example,
if we have 25 time-resolved bins, then a Band model without parametrization would require
$25\times 4=100$ free parameters. Using the technique as described above, this number 
is respectively reduced to 29 (for Band and BBPL), 30 (for mBBPL) and 31 (for 2BBPL).
In Figure~\ref{ch5_f1a}, we have shown $P_f$ as a function of the number of time 
bins. As the number of bins increases, we gain in terms of $P_f$. The models using the 
new technique give similar $P_f$.

\begin{figure}\centering
{

\includegraphics[width=3.4in]{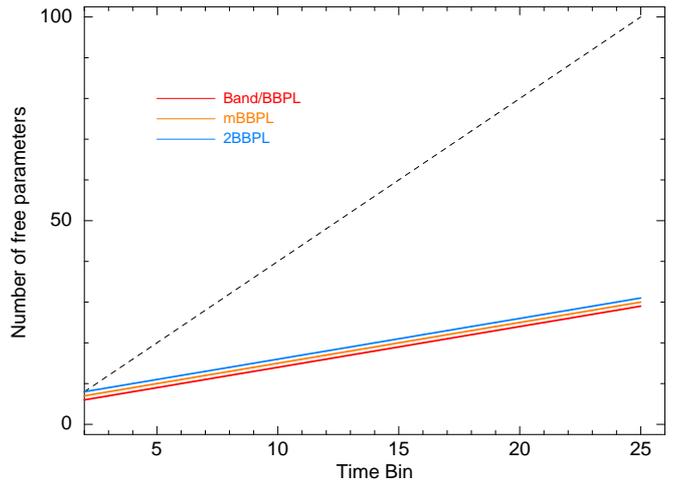} 

}
\caption[Number of free parameters ($P_f$) as a function of number of time bins ($n$)]
{Number of free parameters ($P_f$) as a function of number of time bins ($n$) is shown. The dashed line shows the 
number of parameters ($4n$) required for a normal time-resolved spectroscopy using Band/BBPL model. The solid 
lines show that required by the parametrized joint fit technique for all the models. $P_f$ is comparable
for all the models using the new technique. 
}
\label{ch5_f1a}
\end{figure}

\section{Time-resolved Spectral Analysis of GRB 081221}
We shall start with a time-resolved spectral study of GRB 081221 (see Figure~\ref{ch5_f1}) to validate the 
assumptions of the parametrized joint fit technique. In order to check whether the choice of 
bin size makes any effect on the analysis, we try various schemes to extract the time-resolved data
(i) We choose 3 s uniform time bin to extract the time-resolved data. (ii) Later, we shall also use a 
finer bin size of 1 s uniform bin to check whether finer bin changes any conclusion we make. 
(iii) Finally we shall choose unequal bins by requiring minimum count per bin ($C_{\rm min}$) 
for the parametrized joint fit technique. 

\begin{figure}\centering
{

\includegraphics[width=3.4in]{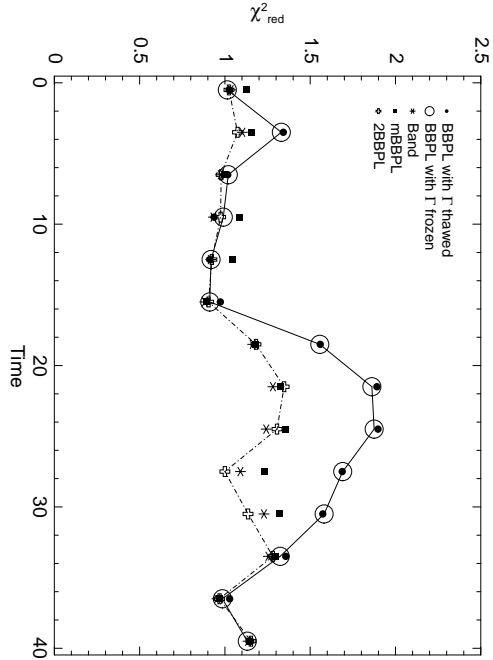} 

}
\caption[Comparison between $\chi^2_{\rm red}$ obtained by fitting Band, BBPL ($\Gamma$ thawed and frozen cases), mBBPL, and
2BBPL models to the time-resolved spectra of GRB 081221]
{Comparison between $\chi^2_{\rm red}$ obtained by fitting Band, BBPL ($\Gamma$ thawed and frozen cases), mBBPL, and
2BBPL models to the time-resolved spectra of GRB 081221. The markers are explained in the inset. 
We draw lines to join $\chi^2_{\rm red}$ points obtained by BBPL ($\Gamma$ free case) fitting (solid line), and those obtained 
by 2BBPL fitting (dot-dashed line) to the spectral data. Source \cite{Basak_Rao_2013_parametrized}.
}
\label{ch5_f2}
\end{figure}

\subsection{Case I: Uniform Bin of 3.0 s}
First, we choose 3.0\,s uniform bin size, and obtain a total of 14 time-resolved bins starting with -1.0\,s.
Among these time bins the first pulse contains approximately the first four bins, while 
the second pulse contains the last 8 bins. The two intermediate bins belong to
the overlapping region of the two pulses. In Table~\ref{081221_a}, we have shown the 
time bins by numbering them 0-13. We fit the time-resolved data with Band and BBPL model. The 
corresponding model parameters and $\chi^2_{\rm red}$ are reported in the table.
As we want to tie the PL index ($\Gamma$) in each sector (rising and falling part)
of a pulse for the parametrized joint fit, let us investigate the effect of freezing 
this parameter in each pulse. From the table, we first note that the value of $\Gamma$ 
remains almost constant in the major portion of each pulse (note that we have shown -$\Gamma$ 
for convenience). We calculate the average value of $\Gamma$ in each pulse and find 
$\Gamma=-1.83$ in each case, with a dispersion of 0.14 and 0.10 in the respective pulses.
The average value of $\Gamma$ is determined by using the values of 0-2 bins and 6-11 bins
of the first and second pulse, respectively. We have neglected the overlapping regions 
and the low flux bins for calculating the average. We now freeze the value of $\Gamma$
at -1.83 and perform the spectral analysis again. The corresponding values are also shown in 
the table. By comparing the values of the parameters, we note that the assumption of 
constant $\Gamma$ does not change the spectral parameters significantly.  

In Figure~\ref{ch5_f2}, we plot $\chi_{red}^{2}$ obtained by fitting a BBPL with free $\Gamma$ 
(filled circles), BBPL with frozen $\Gamma$ (open circles), and Band function (stars). We note that 
the two cases of BBPL fitting have remarkably similar $\chi_{red}^{2}$. Hence, freezing 
$\Gamma$ essentially does not affect the statistics. In fact, it points towards the fact that 
the data is consistent with a BBPL model with the PL index constant throughout a pulse.
We also note that the BBPL model is inferior to the Band function specially in the second pulse.
We now fit the time-resolved spectra with mBBPL and 2BBPL models. A comparison of the
mBBPL (filled boxes) and 2BBPL (pluses) models with the Band function clearly shows that
all these models have similar $\chi_{red}^{2}$ throughout (Figure~\ref{ch5_f2}).

\begin{sidewaystable}\centering
\begin{scriptsize}
\caption[Results of fitting the time-resolved spectra of GRB 081221 with Band and BBPL function]
{Results of fitting the time-resolved spectra of GRB 081221 with Band and BBPL function. For BBPL fitting, we 
study two cases, namely, PL index ($\Gamma$) free, and frozen to the mean value -1.83 which is obtained 
during the highest count rate region.
}
\label{081221_a}

\begin{tabular}{c|ccccc|cccc|cccc}

\hline

Bin$^{(a)}$ & \multicolumn{5}{c|}{BBPL ($\Gamma$ free)} & \multicolumn{4}{c|}{BBPL ($\Gamma$ frozen to 1.83)} & \multicolumn{4}{c}{Band} \\
\cline{2-14}
&&&&&&&&&&&&&\\
\# & $kT$ & $K_1^{(b)}$ & $-\Gamma^{(c)}$ & $K_2$ & $\chi^{2}_{red}(dof)$ & $kT$ & $K_1$ & $K_2$ & $\chi^{2}_{red}(dof)$ & $\alpha$ & $\beta$ & $E_{peak}$ & $\chi^{2}_{red}(dof)$\\
&&&&&&&&&&&&&\\
\hline
\hline
&&&&&&&&&&&&&\\
0 & $38.03_{-4.28}^{+4.91}$ & $2.99_{-0.68}^{+0.67}$ & $1.73_{-0.25}^{+0.42}$ & $5.31_{-3.15}^{+12.72}$ & $1.03(67)$ & $38.28_{-3.97}^{+4.78}$ & $3.14_{-0.46}^{+0.49}$ & $7.18_{-2.16}^{+2.25}$ & $1.01(68)$ & $-0.28_{-0.30}^{+0.36}$ & $-10.0$ & $178.06_{-27.57}^{+42.39}$ & $1.03(67)$ \\
1 & $16.26_{-2.21}^{+2.45}$ & $1.76_{-0.36}^{+0.39}$ & $1.77_{-0.13}^{+0.15}$ & $14.27_{-5.98}^{+9.51}$ & $1.34(76)$ & $16.92_{-1.63}^{+1.87}$ & $1.83_{-0.33}^{+0.34}$ & $17.33_{-2.99}^{+3.08}$ & $1.33(77)$ & $-0.69_{-0.22}^{+0.39}$ & $-3.76_{-\infty}^{+1.30}$ & $77.92_{-15.58}^{+11.08}$ & $1.10(76)$ \\
2 & $10.14_{-2.07}^{+3.12}$ & $0.78_{-0.27}^{+0.28}$ & $1.99_{-0.23}^{+0.27}$ & $20.66_{-13.36}^{+28.87}$ & $1.02(67)$ & $9.07_{-1.33}^{+1.49}$ & $0.84_{-0.25}^{+0.26}$ & $10.94_{-3.10}^{+3.28}$ & $1.02(68)$ & $-0.24_{-1.10}^{+1.64}$ & $-2.55_{-\infty}^{+0.29}$ & $35.41_{-8.53}^{+18.41}$ & $0.98(67)$ \\
3 & $10.98_{-2.14}^{+2.56}$ & $0.74_{-0.27}^{+0.31}$ & $2.16_{-0.25}^{+0.40}$ & $32.99_{-20.23}^{+65.24}$ & $0.94(69)$ & $9.07_{-1.32}^{+1.39}$ & $0.80_{-0.24}^{+0.26}$ & $9.21_{-2.87}^{+3.02}$ & $0.99(70)$ & $-0.85_{-0.54}^{+0.80}$ & $-3.07_{-\infty}^{+0.62}$ & $39.74_{-7.77}^{+8.60}$ & $0.93(69)$ \\
4 & $6.82_{-1.13}^{+1.69}$ & $0.61_{-0.26}^{+0.28}$ & $1.85_{-0.34}^{+0.27}$ & $9.05_{-7.42}^{+17.99}$ & $0.92(134)$ & $6.76_{-1.08}^{+1.17}$ & $0.63_{-0.20}^{+0.21}$ & $8.13_{-2.71}^{+2.89}$ & $0.91(135)$ & $0.52_{-1.60}^{+3.16}$ & $-2.47_{-0.54}^{+0.25}$ & $24.48_{-5.39}^{+9.05}$ & $0.92(134)$ \\
5 & $11.36_{-1.74}^{+2.01}$ & $1.12_{-0.29}^{+0.30}$ & $2.08_{-0.13}^{+0.16}$ & $49.09_{-19.27}^{+31.10}$ & $0.91(147)$ & $9.18_{-1.02}^{+1.06}$ & $1.19_{-0.26}^{+0.27}$ & $18.14_{-3.07}^{+3.19}$ & $0.97(148)$ & $-1.06_{-0.32}^{+0.46}$ & $-2.92_{-\infty}^{+0.45}$ & $43.19_{-7.51}^{+7.34}$ & $0.88(147)$ \\
6 & $22.61_{-0.99}^{+1.00}$ & $8.52_{-0.65}^{+0.68}$ & $1.77_{-0.06}^{+0.07}$ & $35.88_{-7.30}^{+9.67}$ & $1.56(178)$ & $23.10_{-0.78}^{+0.82}$ & $8.92_{-0.49}^{+0.50}$ & $43.47_{-3.33}^{+3.37}$ & $1.56(179)$ & $-0.45_{-0.09}^{+0.09}$ & $-10.0$ & $102.66_{-4.34}^{+4.77}$ & $1.16(178)$ \\
7 & $23.69_{-0.73}^{+0.74}$ & $13.67_{-0.74}^{+0.77}$ & $1.73_{-0.05}^{+0.06}$ & $41.65_{-6.89}^{+8.52}$ & $1.86(182)$ & $24.34_{-0.59}^{+0.61}$ & $14.53_{-0.58}^{+0.58}$ & $55.80_{-3.53}^{+3.57}$ & $1.89(183)$ & $-0.31_{-0.08}^{+0.08}$ & $-3.82_{-\infty}^{+0.52}$ & $105.94_{-3.95}^{+4.20}$ & $1.28(182)$ \\
8 & $19.77_{-0.82}^{+0.84}$ & $8.93_{-0.58}^{+0.60}$ & $1.76_{-0.04}^{+0.05}$ & $53.13_{-7.89}^{+9.22}$ & $1.87(180)$ & $20.51_{-0.65}^{+0.67}$ & $9.46_{-0.49}^{+0.50}$ & $66.35_{-3.90}^{+3.95}$ & $1.90(181)$ & $-0.61_{-0.08}^{+0.09}$ & $-3.30_{-1.09}^{+0.41}$ & $91.76_{-4.95}^{+4.85}$ & $1.24(180)$ \\
9 & $14.41_{-0.77}^{+0.81}$ & $5.93_{-0.44}^{+0.44}$ & $1.86_{-0.04}^{+0.05}$ & $75.42_{-12.23}^{+14.09}$ & $1.69(175)$ & $14.06_{-0.54}^{+0.57}$ & $5.87_{-0.42}^{+0.43}$ & $67.70_{-4.14}^{+4.20}$ & $1.69(176)$ & $-0.88_{-0.08}^{+0.14}$ & $-9.37_{-\infty}^{+19.37}$ & $70.82_{-2.87}^{+3.00}$ & $1.09(175)$ \\
10 & $12.64_{-1.05}^{+1.17}$ & $3.36_{-0.37}^{+0.38}$ & $1.84_{-0.06}^{+0.06}$ & $54.40_{-11.52}^{+13.64}$ & $1.58(164)$ & $12.41_{-0.72}^{+0.78}$ & $3.35_{-0.37}^{+0.37}$ & $51.03_{-3.86}^{+3.92}$ & $1.57(165)$ & $-1.02_{-0.10}^{+0.11}$ & $-9.37_{-\infty}^{+19.37}$ & $66.62_{-3.22}^{+4.25}$ & $1.23(164)$ \\
11 & $11.33_{-1.06}^{+1.19}$ & $1.97_{-0.30}^{+0.31}$ & $2.01_{-0.12}^{+0.14}$ & $40.27_{-15.24}^{+22.81}$ & $1.32(149)$ & $10.27_{-0.70}^{+0.73}$ & $2.03_{-0.29}^{+0.30}$ & $19.73_{-3.23}^{+3.33}$ & $1.36(150)$ & $-0.61_{-0.31}^{+0.45}$ & $-3.11_{-\infty}^{+0.44}$ & $45.39_{-5.74}^{+5.17}$ & $1.26(149)$ \\
12 & $10.03_{-1.07}^{+1.19}$ & $1.42_{-0.27}^{+0.29}$ & $2.21_{-0.24}^{+0.36}$ & $37.04_{-21.85}^{+59.08}$ & $0.99(141)$ & $8.96_{-0.72}^{+0.74}$ & $1.50_{-0.25}^{+0.26}$ & $8.31_{-2.79}^{+2.91}$ & $1.03(142)$ & $-0.51_{-0.37}^{+0.43}$ & $-10.0$ & $39.19_{-3.41}^{+3.82}$ & $0.95(141)$ \\
13 & $7.61_{-1.11}^{+1.35}$ & $0.65_{-0.19}^{+0.17}$ & $3.65_{-1.58}^{+3.10}$ & $164.92$ & $1.13(127)$ & $7.29_{-0.98}^{+1.02}$ & $0.65_{-0.20}^{+0.20}$ & $0.44_{-0.44}^{+2.39}$ & $1.15(128)$ & $1.18_{-1.59}^{+1.72}$ & $-4.48_{-\infty}^{+1.25}$ & $27.93_{-3.58}^{+4.48}$ & $1.14(127)$ \\
&&&&&&&&&&&&&\\
\hline
\end{tabular}
\end{scriptsize}
\vspace{0.1in}

\begin{footnotesize}
\textbf{Note:} $^{(a)}$Bin numbers are 0 to 13 starting with -1 s, and a uniform bin size 
of 3 s. $^{(b)}K_1$ is the BB normalization, $K_2$ is PL normalization. As detector effective area is applied 
for the fitting, only the relative ratio of the normalizations should be used. $^{(c)}$ Note that negative 
value of $\Gamma$ is shown for convenience.

\end{footnotesize}

\end{sidewaystable}

\subsection{Case II: Uniform Bin of 1.0 s}
From the analysis of the time-resolved data of 3.0\,s bin size, we have found that the $\chi_{red}^{2}$ of 
the BBPL model is inferior to the other models. However, this can be due to our inability to capture the 
evolution in a 3.0\,s time bin. It is pointed out that the spectral evolution can effectively make a narrow 
BB peak much broader in the time-integration of a broad bin. For example, \cite{Zhangetal_2011} have 
studied the spectrum of GRB 090902B with various bin size, and have found that the peak becomes narrower 
with finer bin tending to a BB spectrum. We investigate the effect of smaller bin size on the 
$\chi_{red}^{2}$ of the models by choosing a uniform bin size of 1.0\,s. The result of our revised 
analysis is shown in Table~\ref{av_chi}. In this table, we have shown the average value of $\chi_{red}^{2}$ 
for the full GRB and the second pulse (which is brighter) obtained for 3.0\,s and 1.0\,s bin size.
We note that $\chi_{red}^{2}$ of the BBPL model apparently improves with lower bin size. However, 
the amount of improvement is roughly of the same order as that of all the other models we use. Hence, 
we require an alternative model other than BBPL to describe the spectrum. In Figure~\ref{ch5_f2},
we note that the BBPL model gives comparable fit with respect to the other models in the first pulse,
except for 2.0-5.0\,s bin. Hence, the spectrum is likely to have thermal origin in the first pulse.
However, the first pulse has lower flux, which makes this claim inconclusive. In the major portion of 
the second pulse, where flux is high, we clearly require either a Band function or a more complex model 
than BBPL to describe the spectrum. Hence, it is important to identify whether a model fit is 
limited by the flux or actual physical mechanism. In the parametrized joint fit technique, as 
we reduce the number of free parameters by a large factor, we can hope to resolve some of these issues.

\begin{table*}\centering
 \caption{$\chi_{red}^2$ obtained by fitting the time-resolved spectra of GRB 081221 with various models}
\begin{footnotesize}
 \begin{tabular}{c|c|c|c|c}
\hline
 Method & \multicolumn{2}{c|}{3 Second time bins} & \multicolumn{2}{c}{1 Second time bins} \\
\cline{2-5}

       & $\langle \chi_{red}^2 \rangle$ of full GRB &  $\langle \chi_{red}^2 \rangle$ (2nd pulse)  & $\langle \chi_{red}^2 \rangle$ of full GRB &  $\langle \chi_{red}^2 \rangle$ (2nd pulse)\\
\hline
\hline 
BBPL ($\Gamma$ free) & $1.31 \pm 0.35$ & $1.52 \pm 0.32$ & $1.11 \pm 0.26$ & $1.21 \pm 0.28$\\
BBPL ($\Gamma$ frozen) & $1.30 \pm 0.35$ & $1.50 \pm 0.33$ & $1.16 \pm 0.27$ & $1.22 \pm 0.29$\\
Band & $1.09 \pm 0.14$ & $1.17 \pm 0.11$ & $1.00 \pm 0.16$ & $1.04 \pm 0.18$ \\
mBBPL & $1.15 \pm 0.14$ & $1.23 \pm 0.13$ & $1.07 \pm 0.17$ & $1.06 \pm 0.19$ \\
2BBPL & $1.09 \pm 0.15$ & $1.17 \pm 0.13$ & $1.02 \pm 0.16$ & $1.05 \pm 0.17$ \\
\hline
\end{tabular}
\end{footnotesize}
\label{av_chi}
\end{table*}

\subsection{Parameter Evolution}
We now focus our attention on the evolution of the spectral parameters. We shall show that the spectral 
evolution is not arbitrary, and we can indeed parametrize, or tie certain parameters in each sector
i.e., the rising and falling part of a pulse. In Figure~\ref{ch5_f3}, we have shown the evolution of 
$kT$ of the BBPL model (filled circles for $\Gamma$ free case, and open circles for $\Gamma$ frozen case),
and $E_{\rm peak}$ of Band function (pluses). We have shown the LC of the GRB as a histogram in the background.
First, notice that both the $E_{\rm peak}$ and $kT$ have smooth time evolution. \cite{Ryde_Pe'er_2009}
have studied the $kT$ evolution in single pulses and have found that the evolution can indeed be 
described as a power-law. This evolution has a break time which is consistent with the 
flux peak time (within errors). The power-law index, averaged over all the analyzed GRBs, below the break is 
found to be $\langle a_{\rm T}\rangle=-0.07$ (with a dispersion of $\sigma(a_{\rm T})=0.19$), while that after 
the break is $\langle b_{\rm T}\rangle=-0.68$ (with $\sigma(b_{\rm T})=0.24$). In the second pulse 
(Figure~\ref{ch5_f3}), we see similar break in the evolution of $kT$ as well as $E_{\rm peak}$.
Hence, we can parametrize the evolution as $\propto t^{\mu}$. The index ($\mu$), in principle
can have two values in the two sectors i.e., in the rising and falling sector of a pulse, $\mu$ 
can be same or different.

\begin{figure}\centering
{

\includegraphics[width=3.4in]{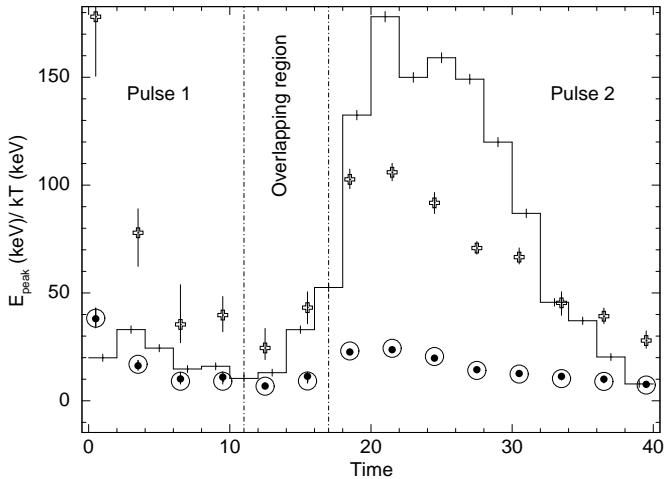} 

}
\caption[Time evolution of the $E_{\rm peak}$ of the Band function and $kT$ of the BBPL model]
{Time evolution of the $E_{\rm peak}$ of the Band function and $kT$ of the BBPL model.
The markers are: crosses ($E_{\rm peak}$), filled circles ($kT$, for $\Gamma$ free case),
open circles ($kT$, for $\Gamma$ frozen case). The LC is shown in the background to track 
the evolution. The evolution in the first pulse is clearly HTS, while in the overlapping 
region, the evolution is rather soft-to-hard. This can be an effect due to a superposition of 
two HTS pulses, or it can be a genuine IT behaviour. Source: \cite{Basak_Rao_2013_parametrized}.
}
\label{ch5_f3}
\end{figure}

From the discussion in chapter 3 (section 3.4.1), we know that the $E_{\rm peak}$ evolution within 
a single pulse can be described by the \citet[][LK96]{LK_1996} law. As both the 
LK96 evolution and the present evolution are empirical we can use either of them for the parametrization.
However, note that while the power-law evolution is a simple function, the LK96 formula 
is crucially dependent on the calculated fluence. Specifically we calculated the running
fluence using a Band function to characterize the evolution (see equation 3.1.4):

\begin{equation}
 E_{\rm peak} (t)= E_{\rm peak, 0}~{\rm exp \left(- \frac{\phi_{\rm Band}}{\phi_{\rm Band, 0}} \right)} 
\end{equation} 
\vspace{0.1in}

Also note that the function generally assumes a HTS evolution. Hence, it is crucial to determine 
the start point while using this formula for the falling part of IT pulses. Note that as $\phi_{\rm Band}$
is a monotonically increasing function of time (which is indeed a different way of saying a HTS evolution), 
we require negative $\phi_{\rm Band,0}$ for the rising part of a IT pulse. Hence, we can use either a 
modified LK96 function, or a power-law function with positive and negative index. For simplicity, we 
have chosen a power-law function following \cite{Ryde_Pe'er_2009}. In the following, we shall validate 
the LK96 model by studying the evolution of $E_{\rm peak}$ and $kT$. The assumption of a power-law 
evolution for the parametrized joint fitting will be justified by noting the $\chi^2_{\rm red}$ and
the smooth transition of $E_{\rm peak}$ and $kT$ at the peak flux.

\begin{figure}\centering
{

\includegraphics[width=3.4in]{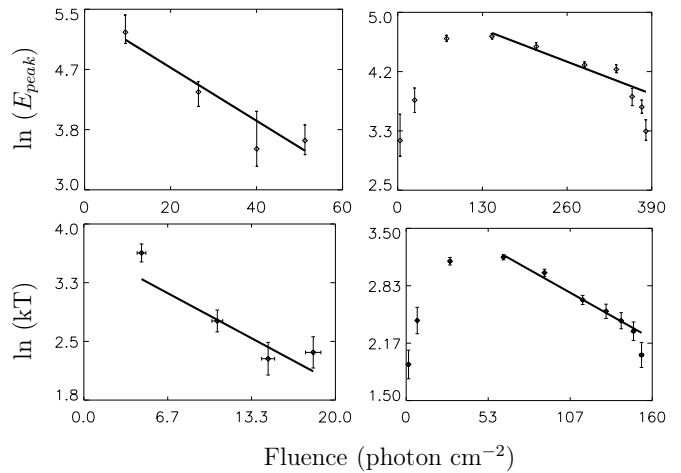} 

}
\caption[Verification of LK96 description of spectral evolution]
{Verification of LK96 description of spectral evolution. 
The evolution of $E_{\rm peak}$ and $kT$ is shown as a function of ``running fluence''. 
In case of BBPL model, we have used the BB fluence. The left panels show the evolution 
in the first pulse. Here we note that both $E_{\rm peak}$ and $kT$ show a HTS (or, a hot-to-cold) 
evolution (see text). The evolution in the second pulse is an IT type. However, the later part 
of the evolution follows a LK96 evolution. Source: \cite{Basak_Rao_2013_parametrized}.
}
\label{ch5_f4}
\end{figure}

As the $kT$ evolution shows similar behaviour as the $E_{\rm peak}$ evolution it is interesting to check the 
applicability of LK96 formula for the $kT$ evolution. In fact, it is obvious that $kT$ evolution should 
follow the $E_{\rm peak}$ evolution as the BB has a peak at $\sim3kT$. However, in chapter 4 (section 4.4.2), 
we found that though the evolutions are similar, the values of $\sim3kT$ are always lower than the values of
$E_{\rm peak}$. Hence, we would prefer to study the evolution of these parameters independently. In Figure~\ref{ch5_f4},
we have plotted the logarithmic values of $E_{\rm peak}$ (upper panels) and $kT$ (lower panels) as 
functions of the running fluence. Note that the fluence of the BBPL model is calculated for the BB component 
only. In the left panels, we have shown the evolutions of these parameters for the first pulse, while 
the right panels show those for the second pulse. Note that the evolution in the first pulse is always 
HTS (or hot-to-cold), while only the later portion of the second pulse shows this behaviour.
From the plot, at least in the falling sector of a pulse, we can describe the $kT$ evolution as 
a function of BB running fluence, $\phi_{\rm BB}=\int_{t_s}^{t}F_{\rm BB}(t') dt'$ as follows.

\begin{equation}
 kT (t)= kT_{0}~{\rm exp \left(- \frac{\phi_{\rm BB}}{\phi_{\rm BB, 0}} \right)} 
\end{equation} 
\vspace{0.1in}

In addition to the $kT$ evolution, we also check the flux evolution of the individual components of the BBPL model.
In Figure~\ref{ch5_f5}, we have shown the evolution of the energy flux (upper panels) and the photon flux (lower panels) 
of each component calculated in 8-900 keV energy band. We study the evolution both for the free $\Gamma$ 
(left panels) and frozen $\Gamma$ (right panels) cases. The data points of the individual components 
are shown by crosses (BB), triangles (PL), and diamonds (total). We note that the evolutions are similar for 
both $\Gamma$ free and frozen cases. Interestingly, we see that the flux evolution of each component
smoothly varies with time. Hence, we can indeed assume a flux parametrization as described in section 5.3.2.
In fact, we shall check the $\chi^2$ for every possible combination (i.e., with and without parametrization),
and convince ourselves that the parametrization works. 

Finally, we check the evolution of the spectral indices with time. In Figure~\ref{ch5_f7}, we have shown the
evolution of $\alpha$ (triangles) and $\beta$ (stars) of the Band function, and $\Gamma$ (pluses) of the BBPL 
model. The parameter $\beta$ is either unconstrained, or has large errors. Sometimes the value pegs at -10. 
This is expected for the poor statistics of time-resolved bins. Note that the variation of the 
parameters are small. If we consider each sector, we expect the parameters to remain effectively 
constant during that time span. Hence, we can tie these parameters in each sector, and determine their 
value with better accuracy. Note that the ultimate proof of all these assumptions is getting an acceptable 
$\chi^2_{\rm red}$. If we get similar $\chi^2$ with a minimal set of free parameters then the result confirms
such evolution.

\begin{figure}\centering
{

\includegraphics[width=3.4in]{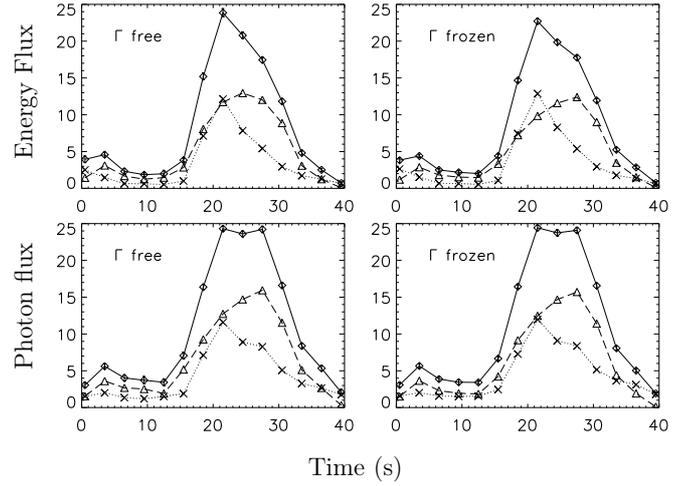} 

}
\caption[Flux evolution of the individual components of the BBPL model]
{Flux evolution of the individual components of the BBPL model: crosses (BB), triangles (PL),
and open diamonds (total). The energy flux (in the units of $10^{-7}$ erg cm$^{-2}$ s$^{-1}$), and the photon 
flux (photons cm$^{-2}$ s$^{-1}$) are shown in the upper and lower panels, respectively. We show both the 
$\Gamma$ free and $\Gamma$ frozen cases. Source: \cite{Basak_Rao_2013_parametrized}.
}
\label{ch5_f5}
\end{figure}

\begin{figure}\centering
{

\includegraphics[width=3.4in]{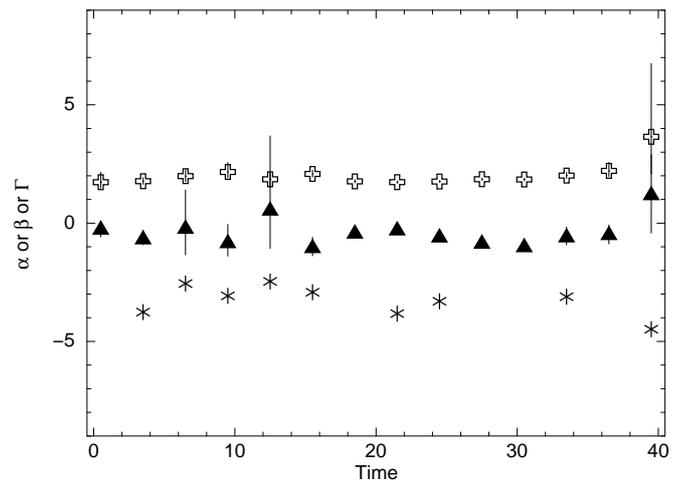} 

}
\caption[Evolution of the spectral indices: $\alpha$ and $\beta$ of the Band function, and
$\Gamma$ of the BBPL model]
{Evolution of the spectral indices: $\alpha$ (triangles) and $\beta$ (stars) of the Band function, and
$\Gamma$ (pluses) of the BBPL model. The errors in $\beta$ are not shown, because either they are large, or 
unconstrained. The parameters vary little in a sector (see text). The negative values of $\Gamma$ are shown 
for convenience. Source: \cite{Basak_Rao_2013_parametrized}.
}
\label{ch5_f6}
\end{figure}

\section{Results Of Parametrized Joint Fit (GRB 081221)}
The fact that the model parameters are smoothly varying functions of time makes the analysis 
of the time-resolved data more tractable. Following the parametrization and tying scheme
(as discussed in section 5.3), we perform the analysis of the individual sectors of each pulse 
of GRB 081221. Firstly, as the detector normalization should not vary with time during a given burst, we 
freeze the normalization as obtained by the time-integrated analysis. The values are 2.25, 2.32, 2.34, and 3.24
for NaI, $n0$, $n1$, $n2$ and BGO, $b0$ detectors, respectively. In addition, we make some changes with 
respect to the time-resolved analysis. As discussed, we obtain the time bins of our study by requiring 
a minimum count per bin, $C_{\rm min}$. Further we regroup the NaI and BGO spectral bins to get 
better S/N at the edge of the detector's energy band. We also note that the 30-40 keV region of the 
spectrum of the GRB has known calibration issue due to the presence of NaI $K$-edge (e.g., see 
\citealt{Guiriecetal_2011}). This will not matter for estimating the parameters, but as we shall compare the $\chi^2$
of different models, it is important to check the $\chi^2$ by omitting this energy band. In the following,
we shall show the $\chi^2$ for the spectral fit with 30-40 keV band omitted. We have verified the results 
by incorporating these bins as well.

\begin{figure}\centering
{

\includegraphics[width=3.4in]{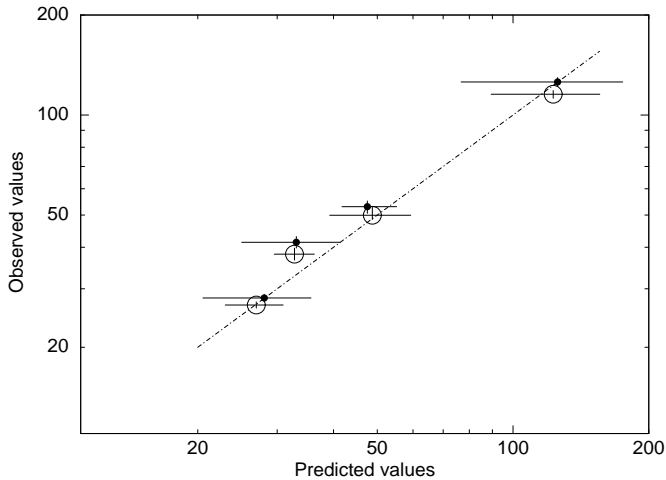} 

}
\caption[Comparison of the predicted and observed values of $E_{\rm peak}$ (Band function), $kT$ (BBPL),
$kT_{in}$ (mBBPL) and $kT_h$ (2BBPL)]
{Comparison of the predicted and observed values of $E_{\rm peak}$ (Band function), $kT$ (BBPL),
$kT_in$ (mBBPL) and $kT_h$ (2BBPL). Values are compared both for wider bin size (excess of 3000 counts per bin;
open circles), and smaller bin size (excess of 3000 counts per bin; filled circles).
The dot-dashed line shows the line of equality. Source: \cite{Basak_Rao_2013_parametrized}.
}
\label{ch5_f7}
\end{figure}

\subsection{Analysis Of The Second Pulse}
As this pulse contains the major portion of the burst, we first analyze this pulse. Note that the count 
rate of this pulse is $\gtrsim 3$ times higher than the first pulse. Hence, we shall use two values of 
$C_{\rm min}$ --- 3000 and 1000. For the first pulse, we shall use only $C_{\rm min}=1000$ counts per 
time bin. The analysis for the second pulse for these two cases are as follows.

\subsubsection{A. Case I: Analysis For $\bf C_{\rm \bf min}=\bf 3000$}
We consider the second pulse from 17.0 s onwards. The pulse is divided into the rising and falling 
sectors. We determine the peak position of the pulse (21.55 s) to define the dividing line. We integrate the 
LC from 17.0 s with a $C_{\rm min}=3000$, till the dividing line. We obtain 3 time-resolved bins 
in the range 17.0-21.45 s. We obtain the time bins of the falling sector in a similar way by integrating 
from 21.55 s onwards. We obtain 9 bins in the range 21.55-40.45 s. We check the counts in each spectral bin,
and find that at energies \textgreater 100 keV, the count is less than 2$\sigma$, while at energies 
\textless 15 keV the count is less than 3$\sigma$. Hence, we merge 8-15 keV bins in a single spectral bin.
The 100-900 keV bins of NaI are re-binned to get 7 logarithmic bins. Similarly, we re-bin the full
energy range of the BGO detector to get 5 logarithmic bins. Note that re-binning of spectral bins increases 
the S/N, and gives better constraints on the derived parameters, while it affects the $\chi^2$. 
Hence, the spectral binning is done by examining the count per bin to achieve at least 3$\sigma$ count
whenever possible. We have listed all the best-fit spectral parameters in Table~\ref{p2}.

\begin{itemize}
 \item (i) Rising Sector: In Table~\ref{p2} (see the first four rows), we have shown the $\chi^2_{\rm red}$ 
of each model fit. Note that the BBPL model is inferior to all other models with $\chi^2$ (dof) = 455.92 (354).
This value is obtained by considering the overall normalization as the free parameter. Instead, if we parametrize
the BB norm, and consider the PL norm as the free parameter, we obtain a worse value: 487.68 (354).
A spectral fit with BBPL having no parametrization and tying gives $\chi^2$ (dof) = 451.09 (348).
Note that while our technique reduces the number of parameters by 6, the $\chi^2$ of the fit 
remains similar. This confirms that the parametrization-joint scheme works. In other words, a BBPL with a
parametrized $kT$ evolution, tied PL index and parametrized flux ratio of the two components 
conforms with the data with similar statistics as a BBPL model without any such constraints.

A comparison of the values of $\chi^2$ of the BBPL model with those of the other models show that the
other models provide better descriptions.
For example, mBBPL gives a better $\chi^2$ (dof) = 355.68 (353) with a significance of $2.55\sigma$
(98.93\% confidence); with $\chi^2$ (dof) = 364.41 (354), Band is better than BBPL at $2.37\sigma$.
This suggests that the radiation mechanism of the rising part of the second pulse may have a 
photospheric origin, but this photosphere has a broader peak e.g., a mBB shape.
A comparison of the BBPL model with 2BBPL model gives a very high significance. 
This is partly because the BBPL model is inclusive of the 2BBPL model. We get a 
$\chi^2$ (dof) = 351.78 (352) for a 2BBPL model fitting. The significance 
of adding another BB on the BBPL model is $9.29\sigma$ ($p=1.5\times 10^{-20}$). 
Note that the 2BBPL model is comparable with the mBBPL and Band function.

\item (ii) Falling Sector: In the falling sector (the next four rows in Table~\ref{p2}), 
the Band function is preferred compared to the BBPL and mBBPL model. Band function has 
35.22 less $\chi^2$ with an additional dof compared to mBBPL. However, if we compare the 
2BBPL model with Band function, we see that the 2BBPL model has 40.64 less $\chi^2$ with 
two less dof than the Band function. The significance of $F$-test with the Band function (exclusive model)
gives $1.03\sigma$ ($\sim70\%$ confidence level). Hence, we see that 2BBPL model is marginally 
better than the Band function. As discussed earlier, we do not expect an order of magnitude 
improvement in terms of $\chi^2$. Compared to mBBPL, we gain 75.86 in $\chi^2$, with one additional
parameter, with a significance of $1.31\sigma$ (81.1\% confidence level). Hence, in this sector 
either a Band function or a 2BBPL model is preferred.

\end{itemize}

\begin{sidewaystable*}\centering
\caption{Results of parametrized joint fit: GRB 081221, pulse 2 (17.0 - 40.55 s)
}
\begin{tabular}{ccccccccccc}

\hline

\hline
Model & $\chi^2$ (dof) & $\mu$ & $\nu$ & $\alpha$ & $\beta$ & $E_{peak}$ $^a$ & p & $-\Gamma$ & $kT_h $/$kT_{in} $/$kT$ $ ^a$  & $kT_l$ $^a$\\
\hline
\hline
\multicolumn{9}{l}{Case I: Count per time bin $\gtrsim$ 3000 --- Rising part (17.0 to 21.45 s 3 bins)}\\
\hline
&&&&&&&&&&\\
Band & 364.41 (354) & $1.0 \pm0.3$ & --- & $-0.44_{-0.03}^{+0.06}$ & $-7.61_{-\infty}^{+2.14}$ & $98.59_{-3.02}^{+2.75}$ & --- & --- & --- & --- \\
BBPL & 455.92 (354) & $0.5 \pm0.3$ & $3.2 \pm1.0$ & ---      & ---  &   --- & --- &  $1.89_{-0.04}^{+0.05}$ &   $24.16_{-0.63}^{+0.64}$ & --- \\
mBBPL & 355.68 (353) & $0.9 \pm 0.2$ & $0.8 \pm 1.0$ & --- & --- & --- &  $0.81_{-0.04}^{+0.08}$ & $1.81_{-0.11}^{+0.20}$ & $40.02_{-2.04}^{+2.87}$ & --- \\
2BBPL & 351.78 (352) & $0.6 \pm0.1$ & $4.7\pm 1.0$ & --- & --- & --- & --- & $1.94_{-0.11}^{+0.16}$ & $28.73_{-1.44}^{+1.67}$ & $9.75_{-1.03}^{+1.14}$ \\
&&&&&&&&&&\\
\hline
\multicolumn{9}{l}{Case I: Count per time bin $\gtrsim$ 3000 --- Falling part (21.55 to 40.55 s 9 bins)}\\
\hline
&&&&&&&&&&\\
Band & 1188.17 (993) & $-2.1 \pm0.1$ & --- & $-0.68 \pm 0.05$ & $-3.55_{-0.44}^{+0.26}$ & $115.5_{-3.2}^{+3.0}$ & --- & --- & --- & --- \\
BBPL & 1557.25 (993) & $-1.9 \pm0.1$ & $-3.2 \pm0.4$ & ---      & ---  &   --- & --- &  $2.02_{-0.02}^{+0.03}$ &   $26.81_{-0.57}^{+0.58}$ & --- \\
mBBPL & 1223.39 (992) & $-2.0 \pm 0.2$ & $3.5 \pm 0.3$ & --- & --- & --- &  $0.74_{-0.03}^{+0.02}$ & $2.03_{-0.06}^{+0.09}$ & $49.93_{-1.44}^{+2.82}$ & --- \\
2BBPL & 1147.53 (991) & $-1.9 \pm0.1$ & $-3.1\pm 0.4$ & --- & --- & --- & --- & $2.15_{-0.08}^{+0.09}$ & $38.13_{-1.52}^{+1.63}$ & $13.33_{-0.68}^{+0.71}$ \\
&&&&&&&&&&\\
\hline
\multicolumn{9}{l}{Case II: Count per time bin $\gtrsim$ 1000 --- Rising part (17.0 to 21.45 s 10 bins)}\\

\hline
&&&&&&&&&&\\
Band & 1247.91 (1187) & $1.5 \pm0.3$ & --- & $-0.44_{-0.06}^{+0.05}$ & $-9.15_{-\infty}^{+4.02}$ & $90.35_{-2.48}^{+2.96}$ & --- & --- & --- & --- \\
BBPL & 1328.50 (1187) & $1.0 \pm0.3$ & $4.4 \pm 0.8$ & ---      & ---  &   --- & --- &  $1.89_{-0.04}^{+0.05}$ &   $22.55_{-0.63}^{+0.64}$ & --- \\
mBBPL & 1239.20 (1186) & $0.9 \pm 0.1$ & $1.3 \pm 0.5$ & --- & --- & --- &  $0.87\pm0.09$ & $1.96_{-0.14}^{+0.28}$ & $39.01_{-1.92}^{+3.48}$ & --- \\
2BBPL & 1222.76 (1185) & $0.4 \pm0.6$ & $9.0\pm 2.0$ & --- & --- & --- & --- & $2.07_{-0.16}^{+0.37}$ & $30.30_{-1.54}^{+1.73}$ & $9.96_{-0.96}^{+1.01}$ \\
&&&&&&&&&&\\
\hline

\multicolumn{9}{l}{Case II: Count per time bin $\gtrsim$ 1000 --- Falling part (21.55 to 40.55 s 29 bins)}\\

\hline
&&&&&&&&&&\\
Band & 3743.36 (3448) & $-2.5 \pm0.1$ & --- & $-0.75_{-0.05}^{+0.06}$ & $-3.56_{-0.77}^{+0.31}$ & $125.7\pm3.9$ & --- & --- & --- & --- \\
BBPL & 4133.63 (3448) & $-2.1 \pm0.1$ & $-3.6 \pm0.2$ & ---      & ---  &   --- & --- &  $2.04\pm0.03$ &   $28.15_{-0.67}^{+0.68}$ & --- \\
mBBPL & 3804.05 (3447) & $-2.0 \pm 0.1$ & $3.0 \pm 0.5$ & --- & --- & --- &  $0.72_{-0.02}^{+0.03}$ & $2.22_{-0.08}^{+0.11}$ & $53.00_{-2.39}^{+2.13}$ & --- \\
2BBPL & 3690.21 (3446) & $-2.0 \pm0.2$ & $-3.6\pm 0.4$ & --- & --- & --- & --- & $2.31_{-0.11}^{+0.13}$ & $41.40_{-1.61}^{+1.69}$ & $13.72_{-0.62}^{+0.65}$ \\
&&&&&&&&&&\\
\hline
\end{tabular}
\label{p2}
 
\begin{footnotesize} $^a$ The values are shown for the first time bin. 

\end{footnotesize}
\end{sidewaystable*}

\subsubsection{B. Case II: Analysis For $\bf C_{\rm \bf min}=\bf 1000$}
To check the effect of lowering of the bin size on our analysis, we choose $C_{\rm min}=1000$. We obtain 
10 bins in the rising and 29 bins in the falling sector. The analysis 
follows the same strategy as discussed, the only change being the binning. As we have smaller bin size, we 
re-bin the 100-900 keV band of the NaI detectors into 5 bins rather than 7 bins. The rest of the binning 
remains the same.

\begin{itemize}
\item (i) Rising Sector: The $\chi^2$ (dof) of the Band, BBPL, mBBPL and 2BBPL models are 1247.91 (1187),
1328.50 (1187), 1239.20 (1186) and 1222.76 (1185). We note that the Band and the mBBPL models are preferred over 
the BBPL model at $1.36\sigma$ (82.76\%) and $1.56\sigma$ (88.17\%), while 2BBPL model is preferred at $9.66\sigma$.
Hence, we see that using a finer bin size does not rule out that an additional BB is required to fit the 
spectrum. In fact, a finer bin has increased the significance of 2BBPL compared to a BBPL model. Hence, the 
conclusions remain unchanged.

\item (ii) Falling Sector: In the falling part (last four rows in Table~\ref{p2}), we see that the 
finer bin size has equal impact on each model. Note that the 2BBPL has 113.84 less $\chi^2$ than the 
the mBBPL model with one less dof. Hence, the 2BBPL model is preferred over the mBBPL model at 
$1.40\sigma$ (83.9\% confidence). Compared to the BBPL model, mBBPL, Band and 2BBPL models are 
preferred at $2.67\sigma$, $3.12\sigma$, and $19.61\sigma$, respectively.

\end{itemize}

\subsection{Analysis Of The First Pulse}
As we have extensively discussed the essential points, we shall only highlight the important points for the 
analysis of the first pulse. We use $C_{\rm min}=1000$ for our analysis, and obtain a total of 5 time-resolved 
bins. The results of our analysis are shown in Table~\ref{p1}. In the rising part of the first pulse, all the 
models are comparable, with BBPL marginally better than the Band function (61.46\% confidence). The
2BBPL model is marginally preferred over the Band function at $1.04\sigma$ (70.17\% confidence), and preferred 
over BBPL at $2.19\sigma$ (97.12\% confidence). In the falling sector, BBPL is no longer the preferred mode, while 
all other models have similar $\chi^2$. In Table~\ref{sigma_level}, we have shown the model comparison in all 
sectors of the two pulses. In all cases, we can see that the 2BBPL model is preferred over all the other models.

\begin{sidewaystable}

\caption{Results of parametrized joint fit: GRB 081221, pulse 1 (-1.0 to 12.05 s). The bins are obtained for excess of 1000 counts per bin
}
\begin{tabular}{ccccccccccc}

\hline

\hline
Model & $\chi^2$ (dof) & $\mu$ & $\nu$ & $\alpha$ & $\beta$ & $E_{peak}$ $^a$ & p & $-\Gamma$ & $kT_h $/$kT_{in} $/$kT$ $ ^a$  & $kT_l$ $^a$\\
\hline
\hline
\multicolumn{9}{l}{Rising part (-1 to 2.15 s 1 bin)}\\
\hline
&&&&&&&&&&\\
Band & 115.78 (116) & --- & --- & $-0.55_{-0.22}^{+0.26}$ & $-10.0$ & $170.3_{-22.7}^{+30.7}$ & --- & --- & --- & --- \\
BBPL & 109.67 (116) & --- & --- & ---      & ---  &   --- & --- &  $1.93_{-0.21}^{+0.35}$ &   $38.27_{-3.76}^{+4.08}$ & --- \\
mBBPL & 110.27 (115) & --- & --- & --- & --- & --- &  $0.98_{-0.28}^{+\infty}$ & $2.15_{-4.46}^{+\infty}$ & $62.78_{-7.22}^{+18.14}$ & --- \\
2BBPL & 103.05 (114) & --- & --- & --- & --- & --- & --- & $1.74_{-3.04}^{+\infty}$ & $38.47_{-3.66}^{+4.40}$ & $6.57_{-1.64}^{+3.17}$ \\
&&&&&&&&&&\\
\hline
\multicolumn{9}{l}{Falling part (2.25 to 12.05 s 4 bins)}\\
\hline
&&&&&&&&&&\\
Band & 544.65 (473) & $-0.7 \pm0.1$ & --- & $-0.86_{-0.19}^{+0.22}$ & $-3.61_{-\infty}^{+0.69}$ & $82.02_{-7.43}^{+7.53}$ & --- & --- & --- & --- \\
BBPL & 571.47 (473) & $-0.7 \pm0.1$ & $-0.2 \pm0.4$ & ---      & ---  &   --- & --- &  $2.09_{-0.11}^{+0.13}$ &   $19.62_{-1.77}^{+1.88}$ & --- \\
mBBPL & 548.22 (472) & $-0.7 \pm 0.2$ & $2.5 \pm 0.8$ & --- & --- & --- &  $0.63_{-0.03}^{+0.10}$ & $1.51_{-\infty}^{+0.52}$ & $41.63_{-6.76}^{+6.49}$ & --- \\
2BBPL & 544.79 (471) & $-0.7 \pm0.2$ & $-0.1\pm 0.5$ & --- & --- & --- & --- & $2.04_{-0.14}^{+0.28}$ & $28.59_{-4.45}^{+5.99}$ & $9.95_{-1.89}^{+2.30}$ \\
&&&&&&&&&&\\
\hline
\end{tabular}
\label{p1}
\vspace{0.1in} 
\begin{footnotesize}

 $^a$ The values are shown for the first time bin.

\end{footnotesize}

\end{sidewaystable}
 
\begin{table*}\centering
 \caption{Comparison between the goodness of fits for different models in GRB 081221}
 \begin{tabular}{c|c|c|c|c}
\hline
Region & Model$_2$/Model$_1$ & p & $\sigma$ & C.L.\\
\hline
\hline 
Pulse 1, Rising part & BBPL/Band & 0.385 & 0.87 & 61.46\% \\
\cline{2-5}
 & mBBPL/Band & 0.415 & 0.81 & 58.50\% \\
\cline{2-5}
 & 2BBPL/Band & 0.298 & 1.04 & 70.17\% \\
\cline{2-5}
 & 2BBPL/BBPL & 0.029 & 2.19 & 97.12\% \\
\hline

Pulse 1, Falling part & Band/BBPL & 0.301 & 1.03 & 69.93\% \\
\cline{2-5}
 & mBBPL/BBPL & 0.334 & 0.965 & 66.57\% \\
\cline{2-5}
 & 2BBPL/BBPL & 1.29$\times10^{-5}$ & 4.36 & 99.99\% \\
\cline{2-5}
 & Band/2BBPL & 0.480 & 0.705 & 51.95\% \\
\hline

Pulse 2, Rising part & Band/BBPL & 0.018 & 2.37 & 98.23\% \\
\cline{2-5}
 ($\gtrsim$ 3000 counts/bin) & mBBPL/BBPL & 0.011 & 2.55 & 98.93\% \\
\cline{2-5}
 & 2BBPL/BBPL & 1.5$\times10^{-20}$ & 9.29 & 100\% \\
\cline{2-5}
 & 2BBPL/Band & 0.390 & 0.86 & 60.94\% \\
\hline

Pulse 2, Falling part & Band/BBPL & 1.04$\times10^{-5}$ & 4.41 & 99.99\% \\
\cline{2-5}
 ($\gtrsim$ 3000 counts/bin) & mBBPL/BBPL & 7.86$\times10^{-5}$ & 3.95 & 99.99\% \\
\cline{2-5}
 & 2BBPL/BBPL & 1.99$\times10^{-66}$ & 17.22 & 100\% \\
\cline{2-5}
 & 2BBPL/Band & 0.303 & 1.03 & 69.71\% \\
\cline{2-5}
 & 2BBPL/mBBPL & 0.188 & 1.31 & 81.1\% \\
\hline

Pulse 2, Rising part & Band/BBPL & 0.172 & 1.36 & 82.76\% \\
\cline{2-5}
 ($\gtrsim$ 1000 counts/bin) & mBBPL/BBPL & 0.118 & 1.56 & 88.17\% \\
\cline{2-5}
 & 2BBPL/BBPL & 4.55$\times10^{-22}$ & 9.66 & 100\% \\
\cline{2-5}
 & 2BBPL/Band & 0.374 & 0.89 & 62.60\% \\
\hline

Pulse 2, Falling part & Band/BBPL & 0.0018 & 3.12 & 99.82\% \\
\cline{2-5}
 ($\gtrsim$ 1000 counts/bin) & mBBPL/BBPL & 0.0075 & 2.67 & 99.24\% \\
\cline{2-5}
 & 2BBPL/BBPL & 1.23$\times10^{-85}$ & 19.61 & 100\% \\
\cline{2-5}
 & 2BBPL/Band & 0.343 & 0.95 & 65.64\% \\
\cline{2-5}
 & 2BBPL/mBBPL & 0.160 & 1.40 & 83.9\% \\
\hline
\end{tabular}
\label{sigma_level}
\end{table*}

\subsection{Connecting The Rising And Falling Part}
The smooth evolution of the peak of the spectrum demands that both $E_{\rm peak}$ (of Band function), and 
$kT$ (of BBPL, mBBPL, or 2BBPL) should be continuous during the break at the peak flux.
In other words, the values of these parameters should agree (within error) with the observed values
at the break. For mBBPL and 2BBPL model we use the parameter $kT_{in}$, and $kT_h$, respectively.
From the tabulated values of pulse 2 (Table~\ref{p2}), we use the evolution formula of each model
for the rising part, and predict the corresponding value of $E_{\rm peak}$ (or, $kT$, $kT_{in}$, $kT_h$)
at the first bin of the falling part. We then compare the value of each parameter with the corresponding
observed value. In Figure~\ref{ch5_f7}, we have compared these values. Note that the predicted values 
have larger uncertainty than the observed values. The major source of error in the predicted values 
comes from the large uncertainty of the time evolution parameter, $\mu$. The dot-dashed line in this figure 
shows the equality of the predicted and observed value. The values agree quite well within errors. Hence, 
the parametrization smoothly joins the two evolution at the peak. 
 
\subsection{Thermal And Synchrotron Origin}
Based on the parameters of the new fitting technique let us examine the possible radiation mechanism
during different phases of GRB 081221. Let us reserve 2BBPL for a later discussion, and compare the Band 
function with BBPL and mBBPL model. 
\begin{itemize}
\item First notice that the spectrum is consistent with a photospheric 
model in the rising part of each pulse. For the first pulse, the rising part could be fitted even with a simple 
BBPL model, while for the second pulse a mBBPL model gives comparable fit as the Band function. Let us check 
the tabulated values of the low energy photon index ($\alpha$) in the rising part. For the second pulse these 
are (see Table~\ref{p2}) $\alpha=-0.44_{-0.03}^{+0.06}$ (for $C_{\rm min}=3000$), and $\alpha=-0.44_{-0.06}^{+0.05}$ 
(for $C_{\rm min}=1000$). For the first pulse (see Table~\ref{p1}), $\alpha=-0.55_{-0.22}^{+0.26}$. 
Clearly, the values are greater than the slow cooling limit of synchrotron radiation (-2/3). 
Hence, the spectrum cannot have a solely synchrotron origin in the rising part of any pulse. 
On the other hand, a model with a thermal and a non-thermal component gives a better physical meaning 
of the PL index, e.g., a mBBPL model fit gives $\Gamma = -1.81_{-0.11}^{+0.20}$ for pulse 2, and a BBPL model 
fit gives $\Gamma = -1.93_{-0.21}^{+0.35}$ for pulse 1. Hence, we conclude that the rising part of each of the 
pulses has a major photospheric component, and the spectrum does not have a fully synchrotron origin.

\item Let us check the values of $\alpha$ in the falling sector: for pulse 2 (see Table~\ref{p2}), 
$\alpha=-0.68_{-0.05}^{+0.05}$ (for $C_{\rm min}=3000$), $\alpha=-0.75_{-0.05}^{+0.06}$ (for $C_{\rm min}=1000$), 
and for pulse 1 (see Table~\ref{p1}), $\alpha=-0.86_{-0.19}^{+0.22}$. These values are within the synchrotron line 
of death for slow cooling electrons. Also, based on the $\chi^2$ values, Band function is 
marginally better than the photospheric models in the falling sector. Hence, it is possible that the 
spectrum gradually becomes synchrotron dominated at a later phase. Note that this behaviour is not 
arbitrary --- the transition is always from a photospheric emission to a synchrotron emission.  
Though our result is purely phenomenological, it is possibly pointing towards some basic radiation 
mechanism of the prompt emission. For standard values of coasting bulk \textit{Lorentz} factor ($\eta$) and 
variability ($\delta t_{\rm var}$), we know that the IS develops at $\sim 100$ times higher radius 
than the baryonic photosphere. Hence, it is rather likely that the synchrotron dominates at a later phase.
However, as the observer sees a ``compressed'' time scaled by $\frac{1}{\eta^2}$, the transition time 
largely depends on the value of $\eta$. Note that the transition need not be abrupt. For our benefit, 
we have tied $\alpha$ in all time bins of a sector. However, in chapter 4, we have seen that 
the value changes gradually. Hence, the transition from a photosphere dominated to a synchrotron dominated emission 
should be a smooth function of time. A suitable parametrization of the evolution of $\alpha$, in principle, 
can show this gradual transition. However, we have seen that the evolution is not always monotonic, and hence, 
the presumption of a functional form may lead to wrong conclusions. In our model, we have instead found the 
contrast of the spectral shape, which shows that there is indeed a transition. 

\item A solely internal shock-external shock (IS-ES) origin of both the pulses is unlikely for the following reason.
We know that ES predicts a HTS evolution, while IS does not predict about the spectral evolution. The 
finding of HTS and IT evolution in the first and second pulse, respectively, disfavours any 
possible combination of the IS and ES origin for both of the pulses. Hence, we require a phtospheric 
component. 

\item We can draw another important conclusion from our discussion. As the second pulse comes after 
the synchrotron dominated falling part of the first pulse, and it also has a photospheric origin in the 
rising part, this pulse should have been independently generated. Hence, broad pulses of GRBs are possibly 
multiple episodes of the central engine activity. 

\end{itemize}

\section{2BBPL Model: A Comparative Study}
In the previous chapter, we introduced 2BBPL model and showed its interesting features e.g., correlated 
temperature of the two BBs. In this chapter, we have used the fact that the temperature and the normalization
of the two BBs are correlated, and hence we have tied their ratios in all bins of a sector. As we find 
acceptable fits even after putting these constraints, the parametrization of 2BBPL model is validated.
A comparison of 2BBPL model with all other models immediately shows that this model is either 
comparable or marginally better than all other models in all episodes. In chapter 6, we shall show 
some convincing proof, and discuss about a possible simplistic physical model of this function. 
Here, we shall discuss this model from a phenomenological 
point of view. For later purposes, we note the following characteristics of this model: 
(i) The value of $\Gamma$ is within the synchrotron regime of fast cooling electron. 
(ii) The ratio of the BB temperature ($kT_h/kT_l$) is $3-6$.
(iii) We have also found that the normalizations of the BBs have a similar ratio. 
Note that the 2BBPL model is a simple extension of the BBPL model, with an additional BB.
We have found in our analysis that the addition of a second BB always gives a better $\chi^2$
with high significance. Note from Table~\ref{sigma_level} that the significance of the other 
models, namely Band and mBBPL with respect to the BBPL model are not as significant.
For example, if we believe that mBBPL model is the correct model in the rising part, then 
the significance of mBBPL over BBPL is only $2.55\sigma$ for pulse 2 in the rising part.
During the same phase, 2BBPL is preferred over the BBPL model at $9.29\sigma$. Of course, 
the $F$-test of mBBPL/BBPL and 2BBPL/BBPL are different as the BBPL and 2BBPL are inclusive models, 
while mBBPL and BBPL are exclusive. However, we also note that with a reduced bin size ($C_{\rm min}$), the significance 
of mBBPL/BBPL becomes lower --- $2.55\sigma$ to $1.56\sigma$. The significance of 2BBPL/BBPL,
on the other hand, does not decrease with a smaller bin size ($9.29\sigma$ to $9.66\sigma$). 
For pulse 1, while the mBBPL model is comparable with the BBPL model, the 2BBPL model 
is still marginally preferred at $2.19\sigma$. Similarly, let us compare the Band 
function with the BBPL model. In the first pulse, the BBPL model is comparable to 
the Band function in the rising part (BBPL/Band$=0.87\sigma$), and the Band function is only marginally 
preferred over the BBPL model in the falling part ($1.03\sigma$). For the second pulse, the significance
decreases with a finer bin: $2.37\sigma$ to $1.36\sigma$ (rising part), and $4.41\sigma$ to
$3.12\sigma$ (falling part). Hence, we conclude that 2BBPL is phenomenologically
preferred over the other models for a time-resolved spectral fitting.

\section{2BBPL Model: Case Studies}
As discussed in section 5.1.2, the major contribution to the residuals comes from the peak position of the 
spectrum. If the models differ in the ``wings'', the contrast is not very apparent. It is only when 
we perform a parametrized joint fit and reduce the number of free parameters of the full description, 
we get a marginal improvement, and can comment on the relative preference of the models. However, we
find some individual cases where we can directly compare the residuals (see below). 

\subsection{I. 2BBPL vs. Band}

\begin{figure}\centering

{

\includegraphics[width=3.4in]{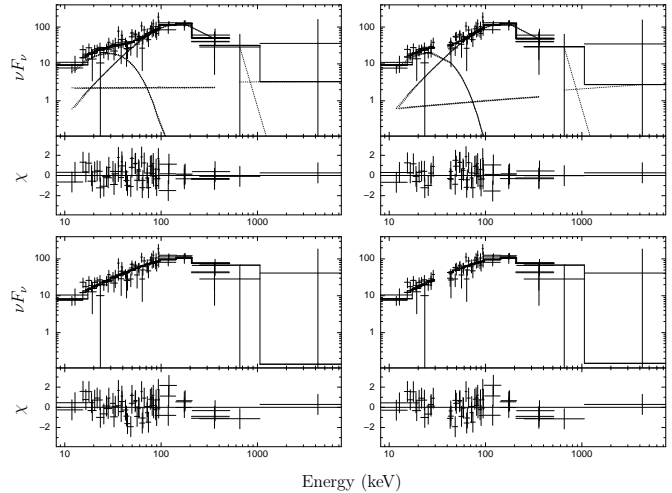} 

}
\caption[Comparison of spectral fitting by Band and 2BBPL model in -1.0-2.15 s bin]
{Comparison of spectral fitting by Band and 2BBPL model in -1.0-2.15 s bin. $\nu F_{\nu}$ is in the units
of keV$^2$ (photons cm$^{-2}$ s$^{-1}$ keV$^{-1}$). The upper panels show the 
2BBPL model fitting with 30-40 keV band included (upper left) and omitted (upper right). The lower 
panels show similar plots for fitting the Band function. Note the structure in the residual 
of Band spectral fitting: positive excess in 15 keV, and 150 keV, and a negative excess in 40-60 keV.
For the case where 30-40 keV bins are neglected, the 2BBPL model is preferred at 1.04$\sigma$ 
(70\% confidence, $p$-value=0.298) over the Band function
based on F-test of two exclusive models. Source: \cite{Basak_Rao_2013_parametrized}.
}
\label{ch5_f8}
\end{figure}

\begin{figure}\centering

{

\includegraphics[width=3.4in]{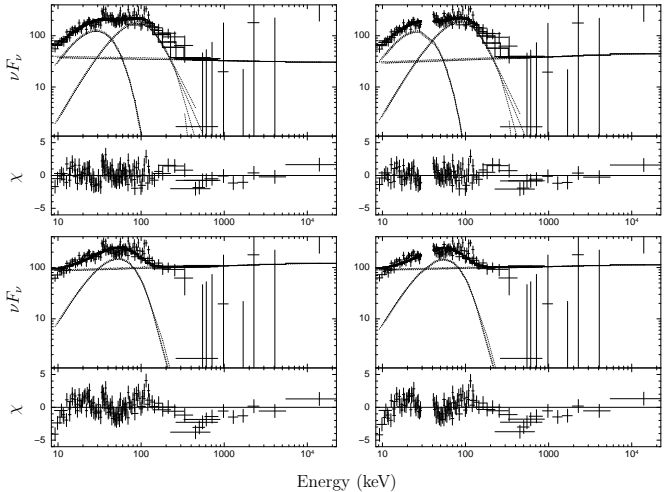} 

}
\caption[Comparison of spectral fitting by BBPL and 2BBPL model in 29.0-32.0 s bin]
{Comparison of spectral fitting by BBPL and 2BBPL model in 29.0-32.0 s bin. $\nu F_{\nu}$ is in the units
of keV$^2$ (photons cm$^{-2}$ s$^{-1}$ keV$^{-1}$). The upper panels show the 
2BBPL model fitting with 30-40 keV band included (upper left) and omitted (upper right). The lower 
panels show the similar plots for BBPL model fitting. Note the double hump structure in the residual
of BBPL fitting which is taken care by the two BBs of 2BBPL model. For the case where 30-40 keV bins are neglected,
the 2BBPL model is preferred at 9.01$\sigma$ (100\% confidence, $p$-value=$2.1\times10^{-19}$) over the BBPL model based on F-test.
Source: \cite{Basak_Rao_2013_parametrized}.
}
\label{ch5_f9}
\end{figure}

In Figure~\ref{ch5_f8}, we have shown the spectral fitting of -1.0-2.15 s data using the 2BBPL model (upper panels) 
and the Band function (lower panels). The left panels show the fit with 30-40 keV data retained, and the right 
panels show the same with 30-40 keV data omitted. We note from the lower panels that the residual of a Band 
functional fit has positive excess at $\sim15$ keV, and $\sim150$, and a negative excess at $\sim40-60$ keV.
These features are absent in the residual of a 2BBPL fitting. Note that the feature does not arise in the 
30-40 keV region where a NaI $K$-edge can give systematic features. In fact, exclusion of the 30-40 keV 
energy band gives a better significance. The 2BBPL is marginally preferred over the Band function at 
1.04$\sigma$ ($p=0.298$, 70.17\% confidence). Hence, we find that the difference between the residual
is not readily visible. In section 5.9.2, we shall study the time-resolved spectra of GRB 090902B, which 
is brighter than GRB 081221 ($\gtrsim 10$ times higher fluence). We shall show some better cases.

\subsection{II. 2BBPL vs. BBPL}
In Figure~\ref{ch5_f9}, we have shown the spectral fitting of 29.0-32.0 s data with a BBPL (lower panels) and 
a 2BBPL (upper panels) model. The left panels show the fit with 30-40 keV band retained, while the right 
panels show the same with the 30-40 keV band excluded. For the BBPL model fit, we see many structures 
in the residual. With 30-40 keV included, we find a $\chi^2$ (dof) = 340.85 (217). The 
residuals are largely minimized by the 2BBPL model, which gives a $\chi^2$ (dof) = 239.82 (215). The $F$-test 
between these two models show that 2BBPL is preferred at $8.42\sigma$ ($p=3.86\times 10^{-17}$). If we 
exclude the 30-40 keV bands, the $\chi^2$ (dof) are 300.26 (197) and 193.17 (195) for BBPL and 2BBPL, 
respectively. Hence, the 2BBPL model is preferred over the BBPL model at $9.01\sigma$ ($p=2.10\times 10^{-19}$).

To visualize the significance of a second BB, and to study its evolution, we do the following.
We first fit the spectrum with a 2BBPL model, freeze the parameters and then omit the lower BB.
The corresponding residual gives an excess at the position of the lower BB, which can be visualized.
This is a well-known method in x-ray spectral analysis, and it is extensively used to find the 
profile of iron line which comes from the inner accretion disc of a blackhole binary (see e.g., 
\citealt{Miller_2007}). In Figure~\ref{ch5_f10}, we have shown this plot for the second, sixth and ninth
time bins of the falling part of the 2nd pulse ($C_{\rm min}=3000$ case). The data of different 
detectors (marked) are plotted in the units of counts s$^{-1}$ keV$^{-1}$. The BB model with different 
detector normalizations are over-plotted to guide the eye. We note that the lower BB temperature has 
a hot-to-cold evolution with time.

\begin{figure}\centering
{

\includegraphics[width=3.2in]{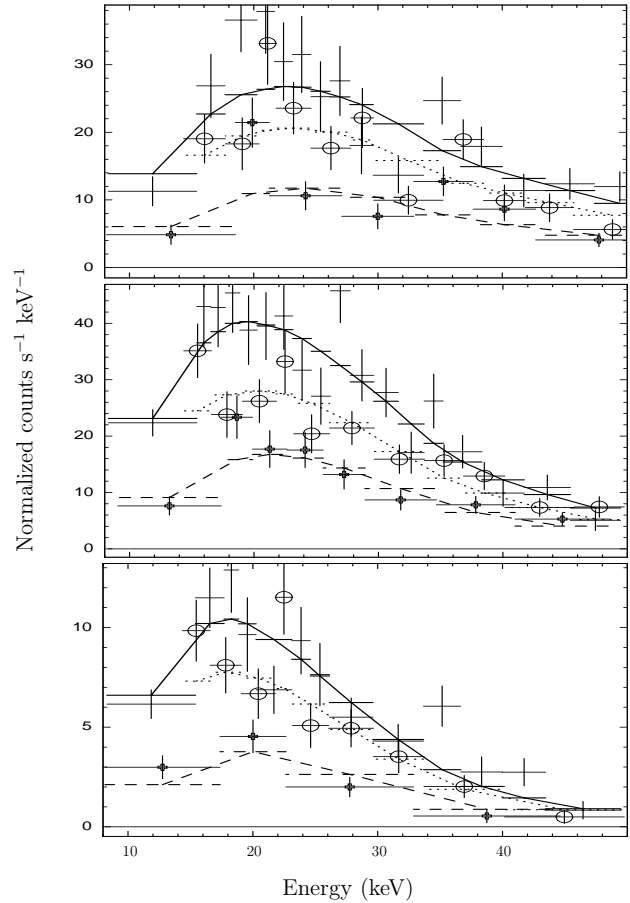} 

}
\caption[Evolution of the lower BB of 2BBPL model with time]
{Evolution of the lower BB of 2BBPL model with time is shown. The plots are showing the residual 
of 2BBPL model fit, with the lower BB omitted. Hence, the residual shows the significance of the lower 
BB in the data. The panels from top to bottom are second, sixth and ninth bins of the falling part of 
the 2nd pulse (count per bin in excess of 3000). The lower BB is over-plotted for the three different 
NaI detectors (shown with different markers) in units of counts s$^{-1}$ keV$^{-1}$. The lower BB 
temperature shifts to lower values with time. Source: \cite{Basak_Rao_2013_parametrized}.
}
\label{ch5_f10}
\end{figure}

\section{Comparison With GRB 090618}
To remind, GRB 090618 was used to develop the simultaneous timing and spectral model of GRB pulses. 
As this GRB also has broad pulse structure, and high fluence ($\sim 10$ times higher than 081221), it is 
interesting to study this GRB using the parametrized joint fit technique. In Figure~\ref{ch5_f4a}, we have 
shown the $kT$ evolution of each pulses of this GRB. Each pulse follows a LK96-type 
evolution (hot-to-cold) at the falling part. This GRB has a precursor pulse in -1 to 40 s, which is a 
separate pulse from the main bursting episodes, and hence, it is ideal for our analysis. Note that this 
pulse always follows hot-to-cold evolution as we have found for the first pulse of GRB 081221.
The major emission of GRB 090618 comes from 50-75 s, where we have two heavily overlapping pulses in 
50-61 s. The other two pulses (75-100 s, and 100-124 s) are less luminous, and are contaminated with a 
long emission (exponential decay) of 50-75 s episode. Hence, we analyze the precursor, and the 61-75 s 
of the data. As the second portion has a large overlap, we shall be cautious in our inferences for this part.

\begin{figure}\centering
{

\includegraphics[width=3.4in]{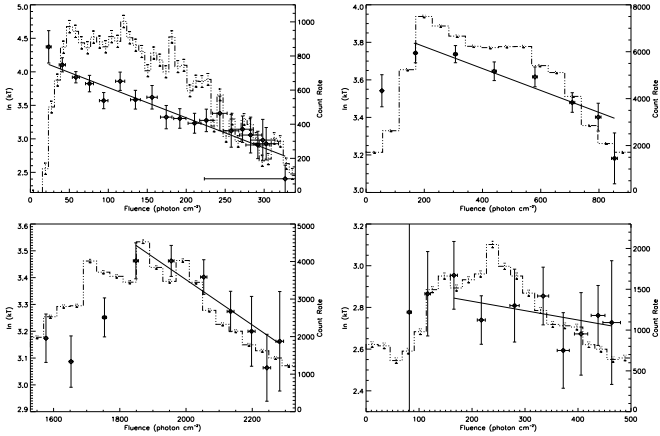} 

}
\caption[Evolution of temperature ($kT$) within the pulses of GRB 090618]
{Evolution of temperature ($kT$) within the pulses of GRB 090618 --- upper left: precursor,
upper right: pulse 1 and 2, lower left: pulse 3, lower right: pulse 4. The falling part of the 
evolution follows LK96 law as we have found for GRB 081221. The precursor pulse follows this 
evolution throughout its duration.
}
\label{ch5_f4a}
\end{figure}

\subsection{Precursor Pulse}
By requiring $C_{\rm min}=1000$, we obtain 10 time bins in the rising sector (-1.0 to 14.15 s), and 
11 bins in the falling sector (14.15 to 40.85 s). In Table~\ref{090618_a}, we have shown the $\chi^2$ 
and the parameters for all the models. We note that the models are all comparable (except BBPL)
in the rising part, and $\alpha$ of Band function has a higher value ($-0.46^{+0.10}_{-0.09}$)
than the synchrotron limit in the slow cool regime (-2/3). The PL index ($\Gamma$) of mBBPL and 
2BBPL are within the fast cooling synchrotron regime. In the falling part, $\alpha=-0.79^{+0.11}_{-0.13}$, within 
the synchrotron limit. 2BBPL is marginally better than the other models.

\subsection{Second Pulse}
Because of the overlap, this pulse is difficult to analyze. We find two structures in 50-61 s of the rising part.
Hence, we ignore this portion first. With the requirement of $C_{\rm min}=2000$, we obtain 1 bin in 61-64.35 s, 
and 24 bins in 64.35-74.95 s. We could have chosen a smaller limit on the counts per bin. However, following 
our experience of GRB 081221, we limit ourselves to a conservative bin size. In Table~\ref{090618_b}, we have shown 
the results of our analysis. It is apparent from the $\chi^2$ values that Band function is better than all the other 
models in the falling part. However, as discussed, there is a large overlap during this period. Hence, we do the analysis
again by neglecting data for the first overlapping pulse. We obtain 11 bins in 69.25-74.95 s, which covers only 
the falling part of the second overlapping pulse. We obtain $\chi^2$ (dof) = 1137.67 (991), 1179.13 (990),
and 1156.25 (989). Hence, we can clearly see that overlapping of pulses can significantly affect our analysis.
Hence, the analysis of this pulse is inconclusive. However, we note that the value of $\alpha$ in the 
rising and falling parts have large contrast even for this pulse ($-0.68^{+0.06}_{-0.05}$ and $-0.88\pm0.03$).

\section{Study Of Brightest Fermi GRBs With Variable LC}
We have studied GRBs with single pulses (chapter 4), and GRBs with multiple, but separable pulses.
In this section, we shall focus on GRBs with highly variable profile. These are GRB 090902B,
090926A, and 080916C. Apart from high keV-MeV flux, these GRBs have high energy (GeV) emission, which 
makes them very special. We shall discuss about the GeV emission characteristics in the next chapter. 
In this section, we shall concentrate on the time-resolved study of the prompt keV-MeV data. 

Among these, GRB 090902B is the brightest (\citealt{Abdoetal_2009_090902B, Ackermannetal_2013_LAT}). 
\cite{Rydeetal_2010_090902B} have found that the time-resolved spectra of this GRB cannot be fitted with a Band function. 
An additional PL is required to fit the broad band data. A Band with a PL is a completely phenomenological 
function, hence, in order to fit a physically realistic model they have proposed the mBBPL model. 
They have found that a mBBPL model gives similar fit as a Band+PL model. \cite{Zhangetal_2011} have found 
that with smaller bin size the peak of the spectrum becomes narrower, and the spectrum in a fine
enough bin is consistent even with a BBPL model. However, for GRB 080916C, the signature of the thermal 
spectrum is not found. It is possible that each of these GRBs has unique property. However, 
it is important to investigate whether a unified description is possible for all the GRBs.
In this spirit, we shall analyze the time-resolved data of the brightest \textit{Fermi} GRBs. 
We shall extensively show the analysis of GRB 090902B, and use our knowledge to describe the other two GRBs.

\begin{sidewaystable}


\caption{Results of parametrized joint fit: GRB 090618, pulse 1 (-1.0 to 40.85 s). The bins are obtained for excess of 1000 counts per bin
}
\begin{tabular}{ccccccccccc}

\hline

\hline
Model & $\chi^2$ (dof) & $\mu$ & $\nu$ & $\alpha$ & $\beta$ & $E_{peak}$ $^a$ & p & $-\Gamma$ & $kT_h $/$kT_{in} $/$kT$ $ ^a$  & $kT_l$ $^a$\\
\hline
\hline
\multicolumn{9}{l}{Rising part (-1 to 14.15 s 10 bins)}\\
\hline
&&&&&&&&&&\\
Band & 623.94 (547) & $-0.8 \pm0.2$ & --- & $-0.46_{-0.09}^{+0.10}$ & $-3.07_{-0.61}^{+0.32}$ & $344.9_{-22.0}^{+22.1}$ & --- & --- & --- & --- \\
BBPL & 661.77 (547) & $-0.6 \pm0.2$ & $-1.4 \pm0.4$ & ---      & ---  &   --- & --- &  $1.71_{-0.04}^{+0.05}$ &   $64.79_{-2.46}^{+2.50}$ & --- \\
mBBPL & 621.93 (546) & $-0.7 \pm0.2$ & $-0.5 \pm2.0$ & --- & --- & --- &  $0.83_{-0.06}^{+0.09}$ & $1.69_{-0.10}^{+0.19}$ & $125.1_{-8.7}^{+12.5}$ & --- \\
2BBPL & 624.55 (545) & $-0.7 \pm0.2$ & $-1.4 \pm0.8$ & --- & --- & --- & --- & $1.72_{-0.08}^{+0.11}$ & $79.69_{-5.39}^{+9.19}$ & $25.88_{-5.08}^{+9.45}$ \\
&&&&&&&&&&\\
\hline
\multicolumn{9}{l}{Falling part (14.15 to 40.85 s 11 bins)}\\
\hline
&&&&&&&&&&\\
Band & 602.53 (571) & $-1.0 \pm0.3$ & --- & $-0.79_{-0.11}^{+0.13}$ & $-3.02_{-1.10}^{+0.38}$ & $186.2_{-19.2}^{+18.3}$ & --- & --- & --- & --- \\
BBPL & 642.37 (571) & $-0.9 \pm0.3$ & $-1.4 \pm0.7$ & ---      & ---  &   --- & --- &  $1.83 \pm 0.05$ &   $37.95_{-2.50}^{+2.67}$ & --- \\
mBBPL & 602.30 (570) & $-1.1 \pm 0.2$ & $1.5 \pm 1.5$ & --- & --- & --- &  $0.69_{-0.03}^{+0.06}$ & $1.70_{-0.19}^{+0.21}$ & $85.81_{-7.67}^{+12.27}$ & --- \\
2BBPL & 593.74 (569) & $-0.3 \pm0.3$ & $-2.9\pm 1.0$ & --- & --- & --- & --- & $2.11_{-0.12}^{+0.10}$ & $50.88_{-4.66}^{+4.88}$ & $15.33_{-1.65}^{+1.93}$ \\
&&&&&&&&&&\\
\hline
\end{tabular}
\label{090618_a}
 \vspace{0.1in} 

\begin{footnotesize}

 $^a$ The values are shown for the first time bin.

\end{footnotesize}

\end{sidewaystable}

\begin{sidewaystable}

\caption{Results of parametrized joint fit: GRB 090618, pulse 2 (61 to 75.0 s) of GRB 090618. The bins are obtained for excess of 2000 counts 
per bin}
\begin{tabular}{ccccccccccc}

\hline

\hline
Model & $\chi^2$ (dof) & $\mu$ & $\nu$ & $\alpha$ & $\beta$ & $E_{peak}$ $^a$ & p & $-\Gamma$ & $kT_h $/$kT_{in} $/$kT$ $ ^a$  & $kT_l$ $^a$\\
\hline
\hline
\multicolumn{9}{l}{Rising part (61 to 64.35 s 1 bin)}\\
\hline
&&&&&&&&&&\\
Band & 2012.93 (1707) & $8.0 \pm3.0$ & --- & $-0.68_{-0.05}^{+0.06}$ & $-2.49_{-0.12}^{+0.09}$ & $192.5_{-10.3}^{+11.3}$ & --- & --- & --- & --- \\
BBPL & 2173.47 (1707) & $5.0 \pm1.0$ & $6.5 \pm5.0$ & ---      & ---  &   --- & --- &  $1.64 \pm 0.02$ &   $42.82_{-1.16}^{+1.17}$ & --- \\
mBBPL & 2099.21 (1706) & $8.0 \pm1.5$ & $10.0 \pm5.0$ & --- & --- & --- &  $0.68 \pm 0.02$ & $1.29 \pm 0.08$ & $97.1_{-4.9}^{+6.4}$ & --- \\
2BBPL & 2030.83 (1705) & $6.0 \pm1.5$ & $3.0 \pm3.0$ & --- & --- & --- & --- & $1.78_{-0.04}^{+0.05}$ & $115.56_{-12.09}^{+13.33}$ & $33.19_{-1.73}^{+1.71}$ \\
&&&&&&&&&&\\
\hline
\multicolumn{9}{l}{Falling part (64.35 to 74.95 s 24 bins)}\\
\hline
&&&&&&&&&&\\
Band & 1574.79 (1317) & $-14.0 \pm 1.0$ & --- & $-0.88 \pm 0.03$ & $-2.74_{-0.10}^{+0.08}$ & $262.8_{-7.9}^{+8.5}$ & --- & --- & --- & --- \\
BBPL & 2402.54 (1317) & $-5.3 \pm0.3$ & $-6.9 \pm0.9$ & ---      & ---  &   --- & --- &  $1.78 \pm 0.01$ &   $53.91_{-1.00}^{+1.01}$ & --- \\
mBBPL & 1794.05 (1316) & $-8.0 \pm 3.0$ & $15.0 \pm 5.0$ & --- & --- & --- &  $0.65_{-0.005}^{+0.006}$ & $1.58_{-0.10}^{+0.03}$ & $157.36_{-5.68}^{+4.18}$ & --- \\
2BBPL & 1794.06 (1315) & $-5.5 \pm0.5$ & $-4.0\pm 0.5$ & --- & --- & --- & --- & $1.78 \pm 0.02$ & $73.08_{-2.57}^{+3.04}$ & $21.72_{-1.44}^{+1.68}$ \\
&&&&&&&&&&\\
\hline
\end{tabular}
\label{090618_b}
 \vspace{0.1in} 

\begin{footnotesize}

 $^a$ The values are shown for the first time bin. 

\end{footnotesize}

\end{sidewaystable}

\subsection{Choice Of Time-resolved Bins}
The LC of GRB 090902B has multiple overlapping peaks. Hence, it is difficult to 
divide the LC into the constituent pulses. More difficult is to use the parametrized 
joint fit technique which we have developed in this chapter. 
\cite{Rydeetal_2010_090902B} have tried the mBBPL model in the initial bins (0.0-12.54) with 22 bins, while 
\cite{Abdoetal_2009_090902B} have chosen 7 time-resolved bins in 0-30.0 s data. 
In order to get equal statistics in each time bin, we extract the time-resolved 
data by requiring $C_{\rm min}=4000$, and $C_{\rm min}=2000$. As we obtain similar 
results for both cases, we show the $C_{\rm min}=2000$ case, for which we obtain 48 
time bins in 0.0-35.2 s. During the time-resolved spectroscopy, we find that Band function 
is particularly unacceptable in 7.2-12.0 s time span. It is reasonable to assume that the 
radiation mechanism changes during this period. Hence, we further divide the LC into three regions: 
Episode 1: $0.0-7.2$ s, Episode 2: $7.2-12.0$ s, and Episode 3: $12.0-35.2$ s.

\begin{figure}\centering

{

\includegraphics[width=3.4in]{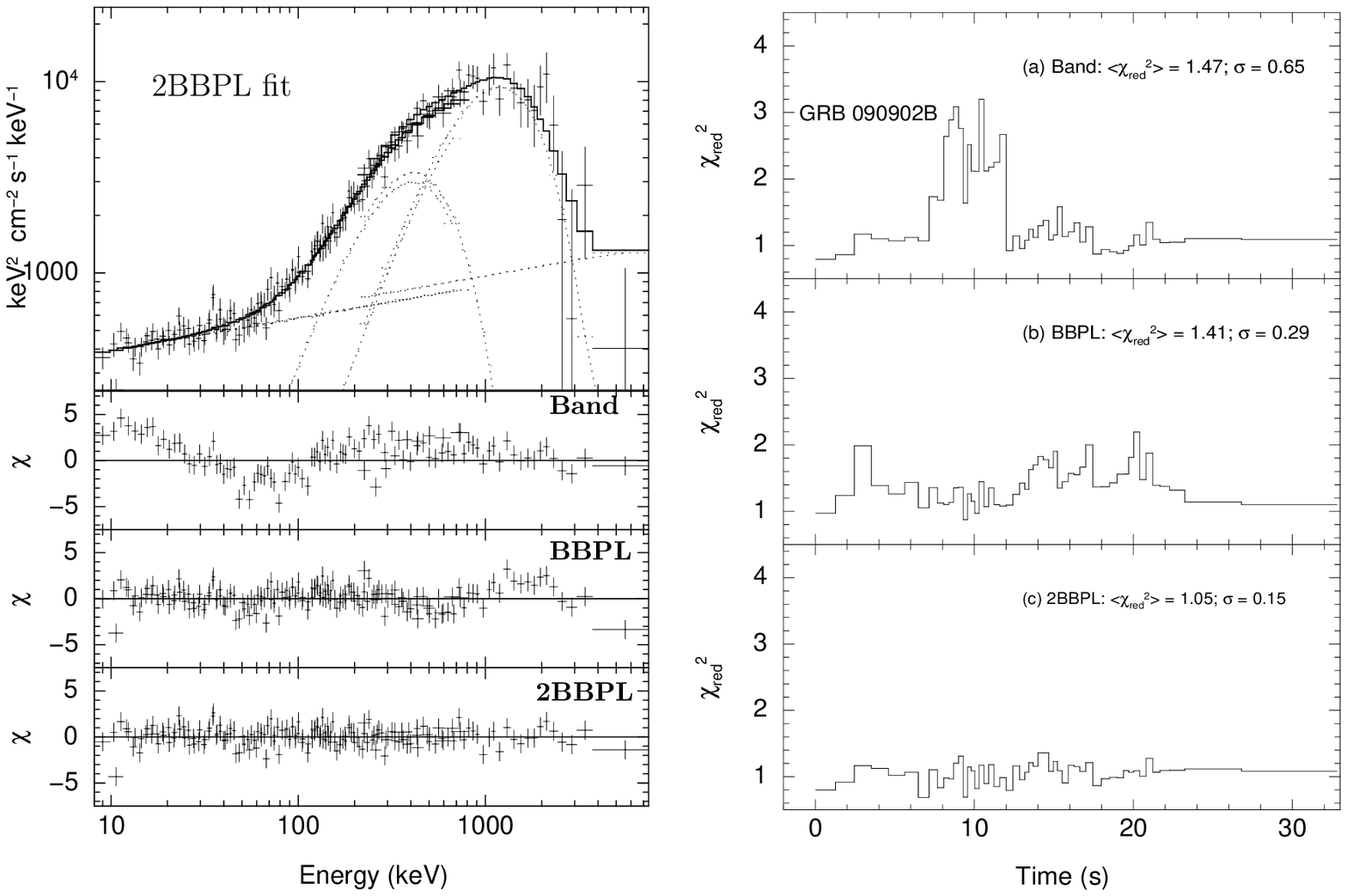} 

}
\caption[Comparison of the spectral fitting to the time-resolved data of GRB 090902B with Band, BBPL 
and 2BBPL model]{Comparison of the spectral fitting to the time-resolved data of GRB 090902B with Band, BBPL 
and 2BBPL model. \textit{Left Panels}: The upper panel shows the unfolded spectrum fitted with 2BBPL model.
The lower panels show the fit residual for the corresponding models. \textit{Right panels}: $\chi^2_{\rm red}$
of the model fits as a function of time bins. Source: \cite{Raoetal_2014}.}
\label{ch5_f11}
\end{figure}

\subsection{Time-resolved Spectral Analysis}
As the LC of these GRBs are variable, we cannot parametrize the spectral evolution. However,
we tie the spectral indices in the time bins to perform a joint fit. For GRB 090902B, we tie the
indices within the time bins of each episode. We also try our analysis with the spectral indices 
as free parameters. 

For GRB 090902B, when we fit the spectrum with a Band function, we obtain unacceptable fits,
specially in Episode 2. In Figure~\ref{ch5_f11} (left panel), we have shown the residual 
of the model fits. In the second left panel, the residual of the Band functional fit is shown and 
it is clearly unacceptable. A spectral fit to the data of the same bin by a BBPL model also gives large structures in 
the residual. Note that the residual is better than the case of GRB 081221 (Figure~\ref{ch5_f8}).
However, the BBPL fit is still not acceptable. When we fit the spectrum with a 2BBPL model we get 
a uniform residual (lowest left panel). In the right panel, we have shown the $\chi^2_{\rm red}$
of the model fits as a function of the time bins. The average and dispersion of $\chi^2_{\rm red}$
of the models are as follows: 1.47, with $\sigma=0.65$ (Band function), 1.41, with $\sigma=0.29$ 
(BBPL model), and 1.05, with $\sigma=0.15$ (2BBPL model). For mBBPL fitting, we get $\chi^2_{\rm red}=1.14$,
with a dispersion of 0.15. We also fit the models by tying the spectral indices in each episode. 
In a tied fitting the mBBPL and 2BBPL models are found to be comparable. For GRB 090926A, and GRB 
080916C we extract the time-resolved data by requiring $C_{\rm min}=2000$, and obtained 37 and 22 bins, 
respectively. We perform both untied and tied spectral fitting for these GRBs. We find that 
the 2BBPL is acceptable in all cases. In Table~\ref{ch5_t1}, we have shown the $\chi^2_{\rm red}$
for tied spectral fitting for all the GRBs

\begin{table}\centering
 \caption{The values of $\chi^2_{\rm red}$ for different model fits}
\begin{small}

\begin{tabular}{ccccc}
\hline
 GRB & Band & BBPL & mBBPL & 2BBPL \\
\hline
\hline
080916C   & 1.05  & 1.14  & 1.07  & 1.04 \\
090902B (0-7.2 s)  & 1.19  & 1.38  & 1.10  & 1.11 \\
090902B (7.2-12.0 s)  & 3.81  & 1.25  & 1.15  & 1.16 \\
090902B (12.0-35.2 s)  &  1.33  &  1.65  &  1.24  &  1.22 \\
090926A   & 1.11  & 1.68  & 1.19  & 1.15  \\
\hline
\end{tabular}
 
\end{small}

\label{ch5_t1}
\end{table}

\begin{figure}\centering
{

\includegraphics[width=3.4in]{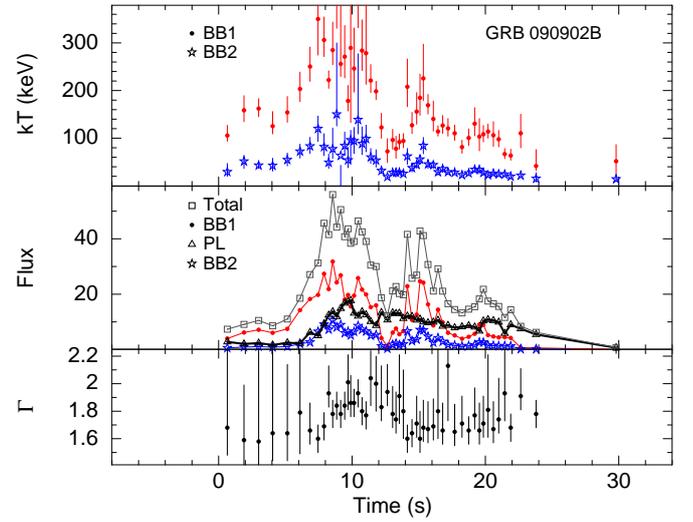} 

}
\caption[The evolution of the parameters as found by time-resolved spectral study of GRB 090902B using 2BBPL model]
{The evolution of the parameters as found by time-resolved spectral study of GRB 090902B using the 2BBPL model.
The parameters are --- \textit{(upper panel):} temperature ($kT$) of the two BBs ; \textit{(second panel):} 
flux (in the units of $10^{-6}$ erg cm$^{-2}$ s$^{-1}$) for the total (open boxes), BB1 (red filled circles), 
BB2 (blue stars), and PL (triangles joined by thick line); \textit{(bottom panel):} power-law index ($\Gamma$).
Source: \cite{Raoetal_2014}.
}
\label{ch5_f12}
\end{figure}

\begin{figure}\centering
{

\includegraphics[width=3.4in]{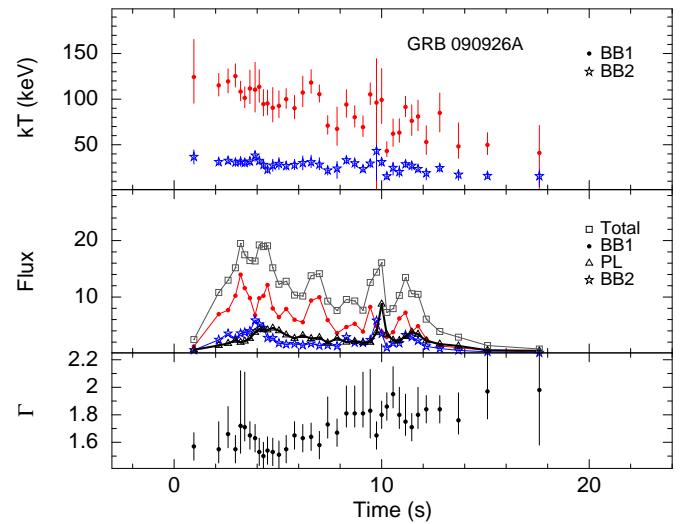} 

}
\caption[The evolution of the parameters as found by time-resolved spectral study of GRB 090926A using 2BBPL model]
{The evolution of the parameters as found by time-resolved spectral study of GRB 090926A using 2BBPL model.
The symbols are the same as in Figure~\ref{ch5_f12}. Source: \cite{Raoetal_2014}.
}
\label{ch5_f13}
\end{figure}

\subsection{Features Of Spectral Evolution}
In Figure~\ref{ch5_f12}, and Figure~\ref{ch5_f13}, we have shown the evolution of various 
parameters for GRB 090902B and GRB 090926A, respectively. In the upper panel, we plot 
the evolution of $kT$ of the two BBs (BB1 and BB2). We note that $kT$ of the two BBs vary in a similar manner.
In the second panel, we show the flux evolution of the total (grey open boxes), 
BB1 (red circles), BB2 (blue stars), and PL component (black triangles joined by thick line).
We note the followings: (i) Evolution of $kT$ of both the BBs track the total flux evolution
(or possibly the evolution of the respective BB flux), (ii) The flux evolution of the two BBs 
are similar to each other, and (iii) the PL component starts with a \textit{delay}, and in GRB 090902B,
the PL component \textit{lingers} at the later phase. All these evolutions will be discussed in detail
when we propose a simplistic model in chapter 6. 

In the lowest panel, we have shown the evolution of $\Gamma$ of 2BBPL model fit. Note that the 
negative values of $\Gamma$ are shown for convenience. We note that the value is always lower 
than -3/2, the line of death of synchrotron emission from fast cooling electrons. 

In Figure~\ref{ch5_f14}, we have studied the correlation between $kT$ (left panels) and normalization 
(right panels) of the two BBs of 2BBPL model. The upper panels show the data obtained for GRB 090902B, 
and the lower panels are those for GRB 090926A. We find high correlation in these parameters.

\begin{figure}\centering
{

\includegraphics[width=3.4in]{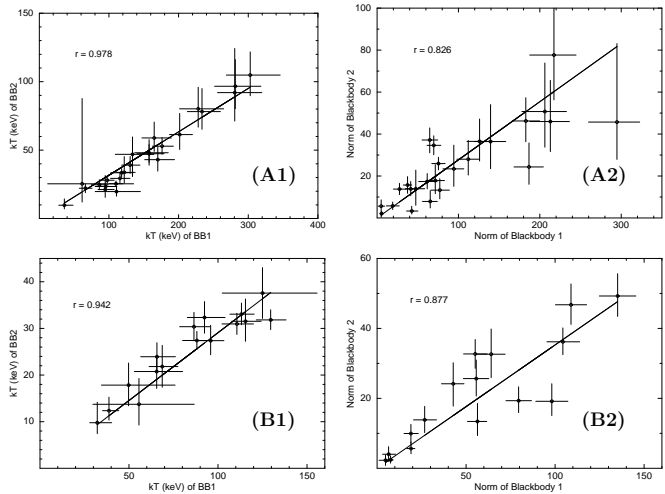} 

}
\caption[Correlation study between different parameters of 2BBPL model]
{Correlation study between different parameters of 2BBPL model. For GRB 090902B, \textit{(A1):} correlation
between temperature ($kT$) of the two BBs, \textit{(A2):} correlation between the flux of the two BBs.
Similar plots for GRB 090926A are shown in \textit{(B1)} and \textit{(B2)}. Source: \cite{Raoetal_2014}.
}
\label{ch5_f14}
\end{figure}

\section{Summary}

Before moving to the next chapter, let us summarize the results of the parametrized joint fit as
applied to the brightest GRBs in the \textit{Fermi} era. For the GRBs with separable pulses,
we have seen that the spectrum has a photospheric origin, which is gradually taken over by a 
synchrotron emission. However, the photon index of the Band function is only within the slow cooling 
regime of the synchrotron emission. As we expect a GRB spectrum in the fast cooling regime, 
it is possible that though the synchrotron becomes important in the later phase, the spectrum 
is still not fully synchrotron. The models like mBBPL and 2BBPL have a more physical PL index 
($\Gamma$). Hence, the spectrum is likely to be a combination of a thermal and a non-thermal 
component in general. Note that the same conclusion was obtained for the GRBs with single pulses. 

By parametrization and tying scheme applied to all the models, we have arrived at the conclusion that 
the 2BBPL model is marginally preferred over all the other models in all episodes. We have found consistent 
results for GRB 081221 and GRB 090618, the two brightest \textit{Fermi} GRBs with multiple broad pulse 
structures. It is unlikely that the results obtained for these two brightest cases can get any better 
for other GRBs. 

For GRBs with high flux, but rapidly varying LC, we have performed our new technique by dropping the parametrization
scheme. We obtain consistent results for three GRBs. Specially for GRB 090902B, we find that the instantaneous
spectrum is far from being a Band function. The residual of a BBPL fit gives a clear indication 
of another BB. A fit with 2BBPL model gives an average $\chi^2_{\rm red}=1.05$, which is even better 
than a mBBPL model fit ($\chi^2_{\rm red}=1.14$). We want to emphasize here that the 2BBPL model is found 
either comparable or preferred than the other models for all cases where we have high signal to noise. 
Though we do not rule out the possibility that each GRB spectrum can be unique, however, the applicability
of a single functional form to a diverse group (single/separable/multiple pulses) possibly indicates 
that a fundamental radiation process prevails in all GRBs in general.

\chapter{Predictions And Physical Picture} \label{ch6}

\section{Overview}

In the last two chapters, we have used various alternative models to describe 
the time evolution of the prompt emission spectrum, which spans a few keV to $\sim 10$ MeV
energy band. We have extensively studied these models for the brightest GRBs detected by 
the \textit{Fermi}/GBM. In addition to the prompt keV-MeV emission, GRBs are also accompanied by high 
energy (GeV) emission. CGRO/EGRET was the first instrument used for detecting GeV 
photons from GRBs. Though the detector severely suffered from backsplash (chapter 2), 
EGRET was quite successful in detecting a few GRBs with GeV emission. In the modern era, 
\textit{Large Area Telescope} (LAT) of \textit{Fermi} is providing a wealth of data for 
this class of GRBs. Through extensive studies by the LAT, it is now established that
compared to the prompt keV-MeV emission, GeV emission has a delayed onset during the 
prompt emission phase, and has a longer lasting emission extending to the early afterglow phase. 
The time evolution of the GeV flux after the prompt emission shows a remarkable similarity with the x-ray 
and optical afterglows. In fact, it is suggested that GeV photons are produced via synchrotron 
emission in the external forward shock (\citealt{Kumar_Duran_2010}). However, this emission 
as early as the prompt emission is rather puzzling. More curious is the fact that the GeV emission 
is prominent in some of the brightest GRBs, while there is apparently very little GeV flux
in others with comparable brightness in keV-MeV energies. In this chapter, we shall attempt 
to address some of the observations of GeV emission based on our knowledge of the 
prompt spectral evolution in the keV-MeV energy band.

Apart from the analysis of high energy spectral data, it is interesting to study the low energy 
x-ray spectrum ($\lesssim 10$ keV) during the early afterglow phase, and connect it with the prompt 
emission. However, it is difficult, as unlike the prompt 
$\gamma$-ray emission which is many orders higher than the background and readily triggers the 
open (or, large field of view) $\gamma$-ray detectors, x-rays are background limited and need to be focused. 
Hence, it takes a considerable time to relocate the burst with a focusing x-ray detector.
The \textit{X-Ray Telescope} (XRT) on-board \textit{Swift} is a dedicated instrument 
for studying x-rays from a very early stage of the GRB afterglow phase (chapter 2). 
The XRT has established a canonical picture of the x-ray afterglow with various breaks in the 
lightcurve. It is suggested that the rapid flux decay before the shallow phase of x-ray 
afterglow is possibly an after-effect of the late phase of the prompt emission. If it is true,
the spectral parameters of the prompt emission should smoothly join those of the early XRT data.
By extending the evolution of the prompt emission model, we shall try to connect these two regimes.

The first topic of this chapter is to obtain some predictions for the emission characteristics 
of the highest energy $\gamma$-rays (GeV, \citealt{Basak_Rao_2013_linger}) as well as the low energy x-rays 
($<10$ keV; Basak et al. in preparation) from the spectral model fitted to the prompt keV-MeV data 
(Part I and II). In the previous chapters, based on detailed analysis of the 
time-resolved spectra, we have found that a new model, viz. two blackbodies along with a 
power-law (2BBPL) is preferred over the other models. In this chapter, we shall assume this 
model for the keV-MeV spectrum and study how the components connect with the emissions in 
other wavelengths. We shall also show that none of the other models (Band, BBPL and mBBPL) 
can explain all the observations. In the third part of our discussion, we shall present 
a simplistic (and tentative) physical model for the origin of the 2BBPL model. 

\section*{Part I: Predicting GeV Emission \newline From MeV Spectrum}
\vspace{0.3in}

\section{Background}

\subsection{Features Of GeV Emission}
Let us begin our discussion with the GeV emission in GRBs. The first long detection of 
GeV emission was found in GRB 940217 (\citealt{Hurleyetal_1994}) detected by CGRO/EGRET. 
The GeV emission continued for 90 minutes after the trigger, which showed the longer lasting 
behaviour of GeV emission. However, with the EGRET and the LAT data it is clear that
GeV emission is not solely an afterglow component,
it appears during the prompt emission phase, sometimes starting with a delay, and showing 
the major bursting features of the prompt keV-MeV emission save for the initial phase in 
some cases. In this regard, the GeV emission behaves like a ``bridge'' between the prompt 
and the afterglow emission. Hence, it is interesting to study the GeV emission in light of 
the evolution of the prompt keV-MeV emission. 

In addition to the unique lightcurve (LC), the spectrum of GeV emission is also interesting, and shows 
diverse features. For example, \cite{Dingusetal_1998} have found a consistent fit for the wide-band 
spectral data using a Band function, while \cite{Gonzalezetal_2003} have found an additional 
PL component which becomes progressively important at a later phase. Recent observations with
the \textit{Fermi}/LAT has validated these observations. With a good sample size (currently 35; 
\citealt{Ackermannetal_2013_LAT}),
the delayed emission, longer lasting behaviour and addition spectral component are found 
in many cases (\citealt{Abdoetal_2009_090902B, Abdoetal_2009_080916C_Sci, Ackermannetal_2010_090217A,
Kumar_Duran_2009, Kumar_Duran_2010, Duran_Kumar_2011, Ackermannetal_2013_LAT}). 
In the following, we shall briefly describe the current understanding 
based on the comparative study of the GBM and the LAT data. For convenience of description, 
we shall interchangably use the ``GeV'' emission with the ``LAT'' emission (and similarly the ``keV-MeV''
emission with the ``GBM'' emission). 

\subsection{GBM-LAT Correlation}
The \textit{Fermi}/LAT is the modern dedicated instrument for studying GeV emission 
(see chapter 2 for details). Recently, Fermi LAT team (\citealt{Ackermannetal_2013_LAT}, 
A13 hereafter) have published the first catalogue of GRBs detected by the LAT (also see 
\citealt{Granotetal_2010, Akerlofetal_2011, Rubtsovetal_2012}). In order to establish a possible 
connection between the prompt keV-MeV and GeV emission, A13 have studied the GBM and the LAT fluence in the 
``GBM time window''. The LAT fluence is calculated by independently fitting the GBM-LAT data and the 
LAT-only data. For bright bursts, the two fits disagree in the LAT energy band due to the presence of 
multiple spectral components in the GBM-LAT joint data. In order to account for 
the longer lasting behaviour of GeV emission, they have also calculated the GeV 
fluence in the ``LAT time window''. They have studied the relative fluence 
in the GBM and the LAT between 19 GRBs (17 LGRBs), and have found only a tentative correlation, 
i.e., GRBs with high keV-MeV emission are likely to produce high GeV emission. 
In fact, for off-axis GRBs from the LAT field of view, the current strategy of the LAT 
observation is to re-point the telescope to a GRB with high GBM flux. 
However, the correlation data of A13 has a large scatter, e.g., GRBs with similar GBM fluence are 
found to have widely different LAT fluence. In fact, A13 have
categorized the GRBs into two classes, namely, hyper-fluent LAT bursts (GRB 090902B, 
GRB 090926A, GRB 080916C, and GRB 090510), and the rest. In an attempt to find a correlation 
between the GBM and the LAT emission components, \citet[][Z12 hereafter]{Zhengetal_2012}
have selected a sample of 22 GRBs 
(17 LGRBs). Since the LAT photons are either simultaneous or delayed than the GBM emission, they 
have selected a uniform 47.5 s time window of match filter technique for both the emissions. 
The LAT emission outside this time window is considered to be much delayed, and unusable for a 
correlation study. For the LGRB sample, Z12 also find a weak correlation with a Pearson 
correlation coefficient, $r=0.537$. 

The lack of a strong connection between the GeV and keV-MeV emission could be a manifestation 
of spectral diversity of the prompt emission. In the standard model, though the basic ingredients 
are roughly known, the subsequent emission strongly depends upon the unknown initial conditions e.g., 
emission region, the amount of energy shared by the magnetic field, and the baryon loading of the ejecta
(\citealt{Meszaros_2006}). For example, \cite{Zhangetal_2011} have analyzed the broadband data of 
a set of 17 GRBs. They have found five possible combination of the spectral models (a combination of
Band, BB, PL etc). Models other than a Band only function is found for the brightest GRBs.
In the following we shall try a single model to describe all the GRBs, and hope for a better GBM-LAT
correlation.

\begin{figure}\centering
{

\includegraphics[width=3.4in]{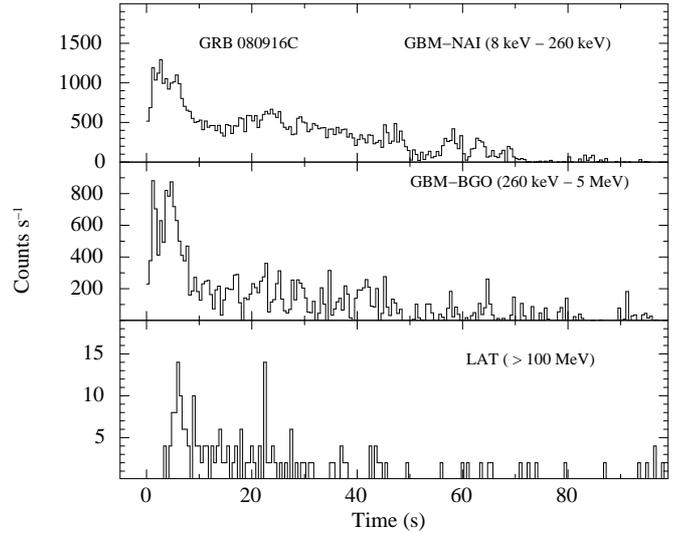} 

}
\caption[Lightcurve of GRB 080916C in various energy bands]
{Lightcurve of GRB 080916C in various energy bands. The GeV emission detected in the LAT 
is delayed by $\sim 4$ s from the keV-MeV emission detected in the GBM.
}
\label{ch6_f1}
\end{figure}

\section{Sample Selection And Analysis Method}
The A13 catalogue has used 17 LGRBs for studying the GBM-LAT fluence correlation. Among these, 
5 GRBs have either much delayed the LAT emission than the GBM emission, or only an upper limit of 
the GBM and/or the LAT fluence. The Z12 catalogue ignores the following GRBs --- 090323, 090328,
090626, 091031, and 100116A, and adds the following GRBs --- 091208B, 100325A,
100724B, 110709A, and 120107A. As we are interested in finding a connection between
the GBM and the LAT emission during the prompt emission, we follow the uniform time selection criteria 
of Z12, and study all the GRBs in their sample. 

Note that our aim is to perform a spectral fit to the GBM data without invoking the LAT data, and 
investigate whether any spectral component can predict the LAT observation. As a Band only function does not give 
a strong correlation, we choose a model having a thermal and non-thermal component, which hopefully 
gives a better understanding of the physical process. As we have found that the 2BBPL model gives either 
comparable or marginally better fit to a variety of data, it is interesting to apply this model.
Hence, we use 2BBPL model to fit the GBM spectrum of all GRBs with GeV emission. In the previous chapter, 
we have developed the ``parametrized-joint fit'' 
technique (\citealt{Basak_Rao_2013_parametrized}) for spectral analysis of GRBs with separable pulses.
However, note that majority of the bursts in our present sample do not have well-defined broad 
pulse structure. Hence, we drop the parametrization scheme in order to perform uniform spectral 
fitting for all the bursts. However, we tie the PL index ($\Gamma$), and the ratio of the temperatures 
($kT$), and normalizations ($N$) of the two BBs. We extract the time-resolved bins by requiring 
$C_{\rm min}$ between 800-1200, which is chosen by considering the peak flux and duration. For 
GRBs with high GBM flux (GRB 090902B, GRB 090926A, and GRB 100724B), we choose a higher value of 
$C_{\rm min}$ ($1800-2000$). For GRB 081006, which has a low GBM count, we use only one bin covering 
the entire burst (-0.26 to 5.9 s; see A13).

\begin{figure}\centering
{

\includegraphics[width=3.4in]{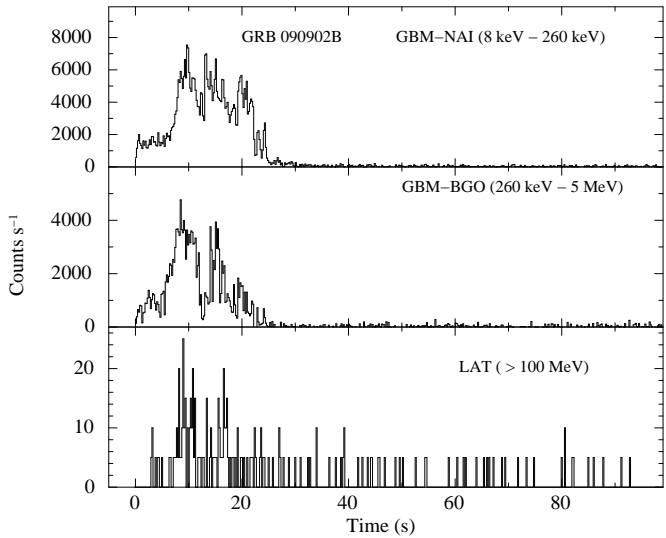} 

}
\caption[Lightcurve of GRB 090902B in various energy bands]
{Lightcurve of GRB 090902B in various energy bands. The GeV emission
is delayed by $\sim 3$ s from the keV-MeV emission.
}
\label{ch6_f2}
\end{figure}

We calculate the flux value of each component of 2BBPL model in each time bin. The error of normalization 
is scaled to calculate the error in flux of each of the components. All these values are used to calculate
the fluence and the corresponding error of each component and the total model. For the LAT analysis, we use
{\tt LAT Science Tools-v9r33p1 package}, choosing ``transient class'' response function (see chapter 2 for detail). 
The LAT fluence (event count) is directly taken from Z12. Note that Z12 provide the LAT fluence in a 
47.5 s time window. We determine the GBM fluence both in $T_{90}$ duration (as provided by A13), and 
within the 47.5 s inetrval. We study both Spearman rank (coefficient $\rho$, and chance probability 
$P_{\rho}$) and Pearson linear correlations (coefficient $r$, and chance probability $P_{\rm r}$).
We study the correlation of the LAT fluence with both the total, and the PL component of the GBM fluence.
In order to determine which among these correlations is more fundamental, we perform Spearman
partial rank correlation test (\citealt{Macklin_1982}). This method finds the correlation of two variables,
say A and X, when a third variable (Y) is present. The confidence level that the A-X correlation 
is unaffected by X-Y correlation is given by the D-parameter, whose value should be at least -1 for 
a significant correlation. The relation between the two fluence quantities are found by fitting a
linear function to the logarithmic values, and assuming a Gaussian noise ($\sigma_{\rm int}$),
denoting an intrinsic scatter of the data (see chapter 3).

\section{Analysis Of GRBs With High GeV Emission}

\subsection{Delayed Onset Of The LAT Lightcurve}
In chapter 5, we have done a detailed time-resolved spectroscopy of three bright GRBs with high 
GeV emission, namely, GRB 080916C, GRB 090902B, and GRB 090926A. These bursts are classified as 
the hyper-fluent LAT GRBs (A13). Here, we shall discuss about the essential features relevant 
to find a relation between the low energy (GBM) and high energy (LAT) emission. In Figure,
\ref{ch6_f1} (GRB 080916C), \ref{ch6_f2} (GRB 090902B), and \ref{ch6_f3} (GRB 090926A), we have 
shown the LC of these GRBs in three energy bands: $8~ {\rm keV} -260 ~{\rm keV}$, 260 keV-5 MeV, and $>100$ MeV.  
Note that in all cases the LAT emission is delayed compared to the prompt emission in the GBM.
The Fermi LAT team reports the following delay of the LAT emission: 4 s (GRB 080916C,
\citealt{Abdoetal_2009_080916C_Sci}), 3 s (GRB 090902B, \citealt{Abdoetal_2009_090902B}), 
and 5 s (GRB 090926A, \citealt{Ackermannetal_2011_090926A}). 
In chapter 5, we have fitted the time-resolved data of GRB 090902B and GRB 090926A
with 2BBPL model. We have found that the two BBs of the 2BBPL model trace each other, 
while the PL component has a delayed onset (see Figue~\ref{ch5_f12}, and \ref{ch5_f13}).
Hence, we suspect that the PL component might be connected with the LAT emission. 
If they are indeed connected the evolution of the PL flux of the keV-MeV data 
should be similar to the LAT flux. Also, the total flux in the PL component, rather
than the total flux of the GBM spectrum should show a better correlation with the total LAT flux.

\begin{figure}\centering
{

\includegraphics[width=3.4in,height=2.75in]{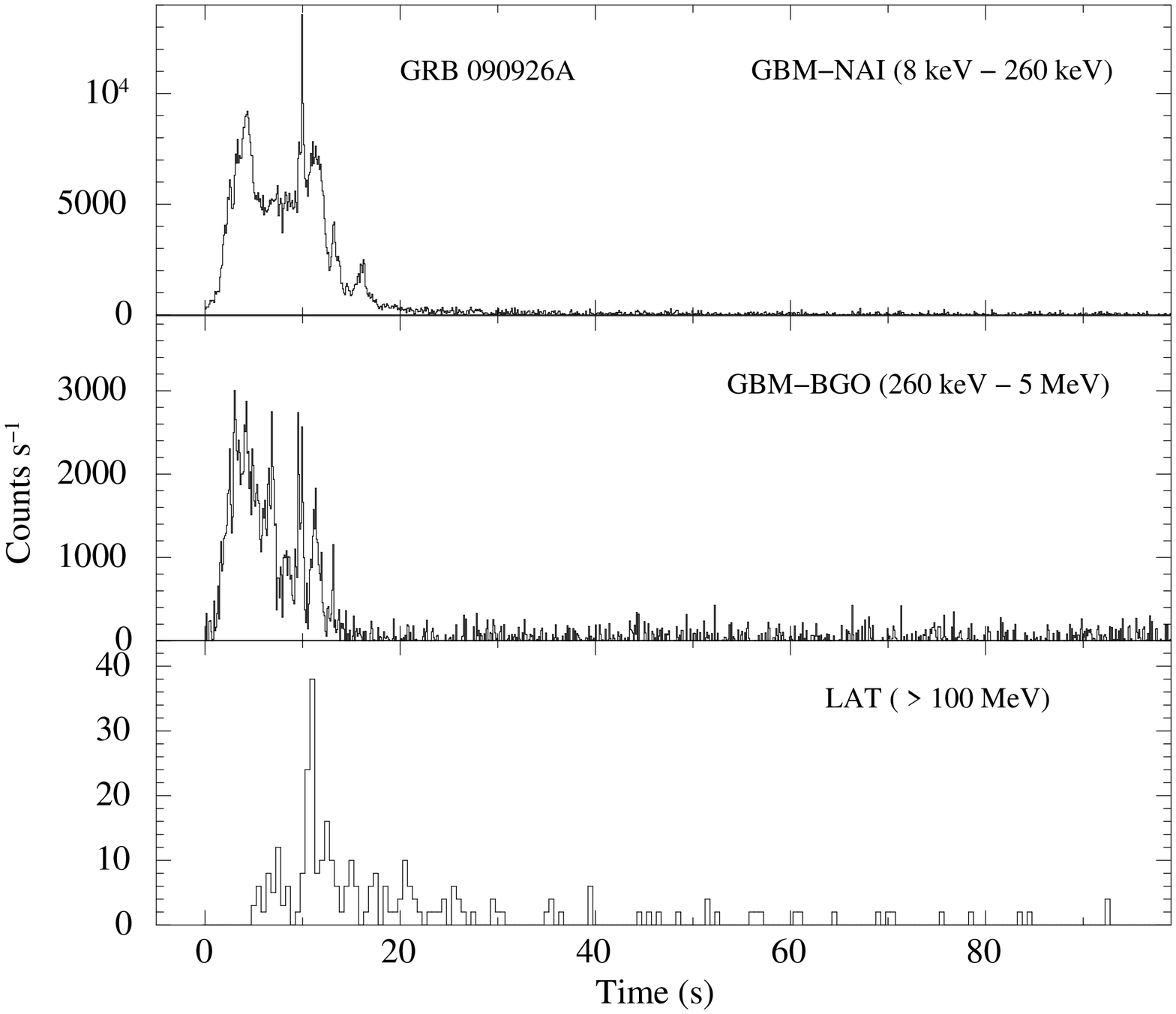} 

}
\caption[Lightcurve of GRB 090926A in various energy bands]
{Lightcurve of GRB 090926A in various energy bands. The GeV emission 
is delayed by $\sim 5$ s from the keV-MeV emission.
}
\label{ch6_f3}
\end{figure}

\subsection{Evolution Of The Power-law Flux}\label{PL_evolution}
We first perform a comparative study of the evolution of the PL flux and the LAT flux. In addition to the time-resolved 
bins in the main bursting phase, we use a few large time bins at the late stage of the prompt emission for our study.
These time bins are: 25-30 s, 30-40 s, 40-60 s, and 60-100 s (for GRB 090902B); 17-30 s, 30-50 s, 50-70 s
(for GRB 090926A), and 64-100 s (for GRB 080916C). We fit a power-law to the broad bin GBM data of each GRB, 
with an index frozen at the average value of that we find during the burst. In the following, we investigate 
whether the PL flux in the GBM spectrum (which is presumably the non-thermal component of the spectrum)
is related with the LAT count (which we assume to have a non-thermal origin).

\begin{figure}\centering
{

\includegraphics[width=3.4in]{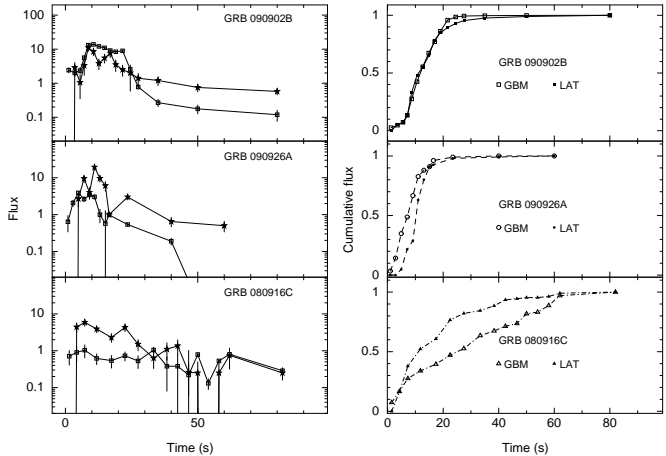} 

}
\caption[Comparison of the power-law (PL) flux with the LAT photon flux]{
Comparison of the power-law (PL) flux with the LAT photon flux. \textit{(Left panels):} The evolution
of PL flux (in units of $10^{-6}$ erg cm$^{-2}$ s$^{-1}$) as calculated by the time-resolved spectroscopy 
with 2BBPL model (open squares). The LAT count flux (in the units of LAT count rate for $>100$ MeV events) is marked
by stars. \textit{(Right panels):} The cumulative integrated flux distribution of the GBM PL flux (open
symbols) compared to that of the LAT count flux (corresponding filled symbols). The GRBs are (top to bottom)
GRB 090902B, GRB 090926A, and GRB 080916C. Source: \cite{Raoetal_2014},
}
\label{ch6_f4}
\end{figure}

Figure~\ref{ch6_f4} (left panel) shows the evolution of the non-thermal GBM flux (open boxes) in the units of 
$10^{-6}$ erg cm$^{-2}$ s$^{-1}$, and the LAT flux (stars) in the units of LAT count rate in the 
$>100$ MeV energies. It is quite fortuitous that our choice of units makes the values of these 
quantities in the same range. As the LAT bore-sight angle is almost similar for these GRBs, namely,
$49^{0}$ (GRB 080916C), $50^{0}$ (GRB 090902B), and $47^{0}$ (GRB 090926A), the observed LAT 
flux can be regarded as the relative flux of a given GRB. It is clear from this figure that 
the flux of the PL component of the GBM and the LAT flux track each other quite well. For GRB 090902B,
for example, we have found that the PL flux is $\approx1$ order of magnitude lower than the 
total flux in the initial $\sim6$ s data (Figure~\ref{ch5_f12}). This evolution coincides with 
the rise of the LAT flux quite smoothly. The peaks near 9-11 s also coincide, though the PL flux 
decays faster than the LAT after 20 s. This probably denotes the ``end'' of the prompt emission.
But, quite remarkably the LAT emission enters into the afterglow phase. For GRB 090926A, the PL
flux is delayed compared to the total flux by $\sim3$ s, which is comparable to the delay of the LAT ($\sim 5$ s).
The sharp drop of the PL flux is noted at the end. For GRB 080916C, the flux evolutions track each other,
including a dip near $\sim55$ s. In the right panel of Figure~\ref{ch6_f4}, we have shown the
cumulative flux distribution which again highlights the similarity between the flux evolution of the LAT 
and the PL component of the GBM spectrum. From the above discussions, we infer that the PL flux 
has a similar behaviour as the LAT flux both in terms of delayed onset and long-lived emission.

\begin{figure}\centering
{

\includegraphics[width=3.4in]{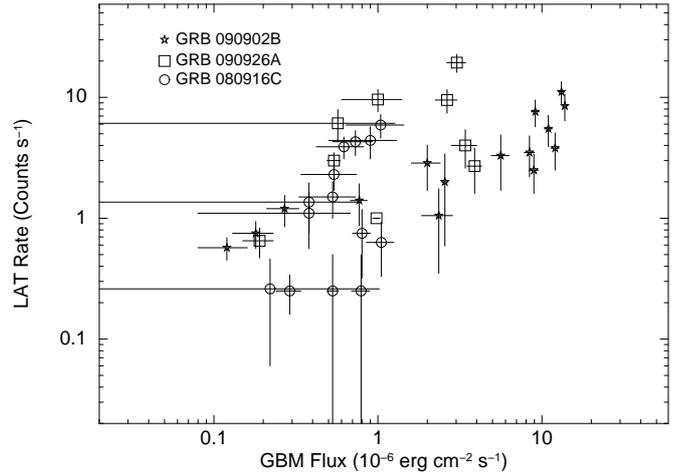} 

}
\caption[A scatter-plot of the PL flux of 2BBPL model with the LAT count rate for the time-resolved spectral
analysis]{A scatter-plot of the PL flux of 2BBPL model with the LAT count rate for the time-resolved spectral
analysis. The GRBs are GRB 090902B (stars), GRB 090926A (open squares), and GRB 080916C (open circles).
Source: \cite{Raoetal_2014}.
}
\label{ch6_f5}
\end{figure}

In Figure~\ref{ch6_f5}, we have shown a scatter-plot of the PL flux of the GBM versus the LAT count rate flux
as obtained in the time-resolved study of these GRBs. Though the correlations obtained 
for GRB 090926A ($r=0.32$), and GRB 080916C ($r=0.36$) are weak, for GRB 090902B, we obtain 
a strong correlation ($r=0.84$) between these two emissions.

\section{Comparison Of Hyper-fluent With Low-LAT Class}\label{lingering}

\begin{figure}\centering
{

\includegraphics[width=3.4in]{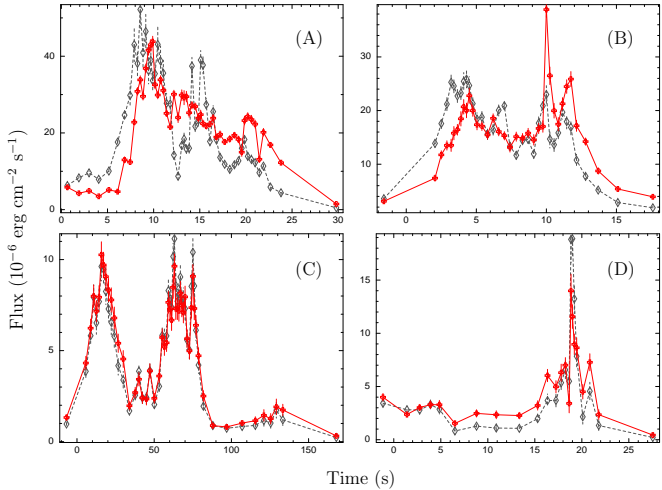} 

}
\caption[The evolution of the total flux and the non-thermal (power-law) flux for two categories of GRBs]
{The evolution of the total flux (diamonds joined by dashed line), and the non-thermal (power-law) flux
(pluses joined by solid line) is shown for two category of GRBs in our sample. \textit{(upper panels):}
GRBs with high GeV emission as detected by the LAT. The GRBs are (A) GRB 090902B with 378.1 photons m$^{-2}$ 
LAT count, and (B) GRB 090926A with 372.2 photons m$^{-2}$ LAT count. \textit{(lower panels):} GRBs with 
similar GBM fluence, but with much lower GeV fluence. The GRBs are (C) GRB 100724B with 23.9 photons 
m$^{-2}$, and (D) GRB 091003 with 14.8 photons m$^{-2}$. The values of fluence are calculated in 47.5 s time window.
Note the delayed onset and lingering behaviour of the PL component for the first category (see text).
Source: \cite{Basak_Rao_2013_linger}.
}
\label{ch6_f6}
\end{figure}

It is evident from our discussion of the hyper-fluent GRBs that their associated 
LAT photons are delayed, and so is the PL component of the prompt keV-MeV emission. 
This behaviour seems to be ubiquitous for the hyper-fluent LAT GRBs. The natural question 
that arises is ``how does the LAT and PL emission evolve for GRBs with low LAT count''?  
In order to address this question let us choose a few sample GRBs. Note that Z12 have found 
a weak but a definite correlation between the GBM and the LAT fluence (Figure 5 of Z12).
That is, we expect high LAT count only for GRBs with high GBM count. Hence, we should choose 
GRBs with comparable GBM fluence. A quick look at Figure 5 of Z12 clearly shows 
two classes of our interest --- (i) GRBs with high GBM fluence as well as high 
LAT fluence (hyper-fluent LAT class), and (ii) GRBs with the same order of magnitude GBM
fluence, but an order of magnitude lower LAT fluence. From set (i), we choose GRB 
090902B, and GRB 090926A. From set (ii) we choose GRB 100724B, and GRB 091003.
Among the class (ii) GRBs, GRB 100724B has a fluence similar to GRB 090902B, and 
even greater than GRB 090926A, but a very low LAT fluence of $23.9\pm7.6$ photons m$^{-2}$
(compared to $378.1\pm29.5$ photons m$^{-2}$ and $372.2\pm28.0$ photons m$^{-2}$ for 
GRB 090902B and GRB 090926A, respectively). GRB 091003 has a factor of 5 lower GBM 
fluence, but a factor of 25 lower LAT fluence ($14.8\pm4.5$ photons m$^{-2}$) than 
GRB 090902B.

In Figure~\ref{ch6_f6}, we have shown the evolution of the total GBM flux (diamonds
joined by dashed line) along with the flux of the PL component (pluses joined by 
solid line). To show both the evolutions in the same scale, we have multiplied the PL 
flux by a constant, which is the ratio of the average GBM flux and average PL flux.
The upper panels show the evolutions for GRBs with high LAT count, while the lower 
panels show those for the low-LAT GRBs. Clearly, the non-thermal component of each 
high-LAT GRB has a \textit{delayed onset}, and this component \textit{lingers at the later phase} 
of the prompt emission. The low-LAT GRBs apparently show no such trend. For this class,
the PL component starts simultaneously, and traces the total GBM flux. 

A quick look at Figure~\ref{ch6_f2} and \ref{ch6_f3} immediately shows that the evolution 
of the LAT flux of the two high-LAT GRBs have a similar characteristic as the PL flux evolution.
The LAT flux also starts with a delay and lingers later (A13). Note that we 
have fitted only the GBM data with a model consisting of a thermal and a non-thermal component, 
without invoking the LAT data. Still the evolution of the PL component of the keV-MeV data 
alone shows this remarkable similarity with the LAT flux evolution (see LAT Low-Energy (LLE: 30-100 MeV) 
lightcurve in Figure 59, 61 of A13 for GRB 090902B and GRB 090926A, respectively). On the other hand, a 
look at the LLE lightcurves in Figure 63 (GRB 091003) and 78 (GRB 100724B) of A13 shows that the 
LAT photons (low energy) do not have a delay. Hence, we conclude that (i) the PL component 
of the 2BBPL model mimics the evolution of the LAT flux in the low energy GBM data, 
and (ii) GRBs with a delayed and lingering PL component are likely to be LAT-bright.

\begin{table*}\centering

 \caption{The GBM and LAT fluence of the 17 GRBs with GeV emission}

 \begin{tabular}{c|cc|cc|c}

\hline
 GRB &  \multicolumn{2}{c|}{GBM $T_{90}$ window$^{(a)}$} & \multicolumn{2}{c|}{47.5 s time window } & LAT fluence\\
 &  \multicolumn{2}{c|}{(Photon cm$^{-2}$)} & \multicolumn{2}{c|}{(Photon cm$^{-2}$)} & (Photon m$^{-2}$)\\

\cline{2-5}
 & Total fluence & PL fluence & Total fluence & PL fluence &  in 47.5 s\\
\hline
\hline 
080825C  & 224.8$\pm$6.2 & 105.2$\pm$5.5 & 245.1$\pm$13.3 & 115.7$\pm$11.9 & 36.6$\pm$11.6\\
080916C  & 369.9$\pm$7.7 & 223.9$\pm$6.4 & 329.3$\pm$5.7 & 196.34$\pm$4.8 & 279.0$\pm$24.9\\
081006A$^{(b)}$ & 6.97$\pm$0.92 & 3.24$\pm$0.62 & 6.97$\pm$0.92 & 3.24$\pm$0.62 & 16.3$\pm$4.7\\
090217 & 124.7$\pm$4.1 & 54.1$\pm$3.4 & 129.0$\pm$4.4 & 55.8$\pm$3.7 & 22.5$\pm$6.0\\
090902B & 1028.4$\pm$18.6 & 498.3$\pm$14.6 & 1102.7$\pm$29.6 & 525.3$\pm$23.2 & 378.1$\pm$29.5\\
090926A & 739.6$\pm$10.8 & 324.9$\pm$8.6 & 785.6$\pm$13.1 & 343.3$\pm$10.4 & 372.2$\pm$28.0\\
091003 & 186.8$\pm$6.3 & 95.9$\pm$4.6 & 210.1$\pm$9.8 & 107.7$\pm$7.2 & 14.8$\pm$4.5\\
091208B & 60.5$\pm$3.5 & 37.2$\pm$3.1 & 82.5$\pm$11.4 & 43.6$\pm$10.0 & 14.6$\pm$6.5\\
100325A & 13.4$\pm$1.7 & 3.3$\pm$0.9 & 13.9$\pm$1.7 & 3.6$\pm$0.9 & 6.7$\pm$3.0\\
100414A & 289.9$\pm$7.6 & 103.4$\pm$6.2 & 384.8$\pm$7.9 & 145.4$\pm$6.4 & 87.5$\pm$33.1\\
100724B & 998.5$\pm$9.5 & 500.6$\pm$7.1 & 396.6$\pm$3.8 & 212.4$\pm$2.9 & 23.9$\pm$7.6\\
110120A & 69.1$\pm$4.7 & 27.5$\pm$2.2 & 77.9$\pm$7.2 & 32.6$\pm$3.4 & 9.5$\pm$3.6\\
110428A & 127.4$\pm$3.5 & 32.4$\pm$2.6 & 147.6$\pm$5.5 & 44.0$\pm$4.0 & 8.0$\pm$3.6\\
110709A & 198.9$\pm$5.6 & 92.2$\pm$5.2 & 212.1$\pm$6.2 & 101.1$\pm$5.7 & 18.7$\pm$7.1\\
110721A & 182.2$\pm$7.2 & 98.8$\pm$4.5 & 192.5$\pm$9.6 & 105.0$\pm$6.0 & 46.4$\pm$9.3\\
110731A & 89.6$\pm$5.7 & 55.5$\pm$2.0 & 102.9$\pm$11.9 & 66.9$\pm$4.1 & 81.5$\pm$10.4\\
120107A & 39.5$\pm$4.1 & 25.8$\pm$4.9 & 39.7$\pm$5.2 & 25.8$\pm$3.9 & 17.6$\pm$7.2\\

\hline
\end{tabular}

\label{ch6_t1}
\vspace{0.1in}

\begin{footnotesize} 

\flushleft $^{(a)}$ $T_{90}$ values from A13 

\flushleft $^{(b)}$ $T_{90}$ value of this GRB is retained for a larger window

\end{footnotesize}

\end{table*}

\section{A Detailed Correlation Analysis}
From the analysis of the previous section (\ref{lingering}), it follows that the PL component of the 2BBPL model 
has some close relation with the LAT emission. In section~\ref{PL_evolution}, we have seen that the 
non-thermal flux of the time-resolved data of hyper-fluent GRBs indicates a correlation with 
the LAT flux. However, a flux correlation suffers from the fact that the LAT flux has a delay compared to
the corresponding GBM flux. Hence, choosing similar bins for the LAT and the data GBM may wipe out the correlation.
Since it is not clear whether the GeV emission is derived from the MeV emission, it may not be possible
to correct for the delay. Hence, a correlation study between the fluence quantities seems more 
reliable. As in this case, we integrate the flux over a chosen time window, it suffers from little 
error (only contribution, if any, comes from the delayed onset). Both A13 and Z12 have studied fluence 
correlation rather than flux correlation. Also note that the total fluence in ``LAT time window''
may not be used for the following reason. We have already seen that the PL flux decreases sharply during 
the late prompt emission phase, while the LAT emission survives. An extended LAT emission is probably 
an afterglow phenomenon, rather than a prompt emission. By choosing a time window in the prompt emission phase,
we are effectively extracting only the prompt contribution of the LAT data. Following Z12, 
we use the 47.5 s data from the trigger time of the GBM. However, as the boundary of the late prompt and 
early afterglow phase is rather arbitrary, the choice of the time window is not unique. 

\begin{figure}\centering
{

\includegraphics[width=3.4in]{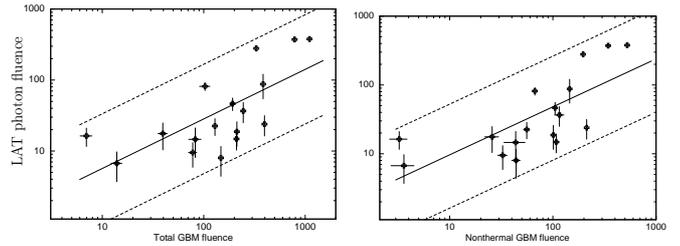} 

}
\caption[Correlations between the GBM photon fluence (photons cm$^{-2}$) and the LAT photon fluence (photons m$^{-2}$)]
{Correlations between the GBM photon fluence (photons cm$^{-2}$) and the LAT photon fluence (photons m$^{-2}$).
The values of fluence are calculated in 47.5 s time window. \textit{(Left panel):} Correlation of the total GBM fluence with the LAT fluence, 
\textit{(Right panel):} correlation of non-thermal GBM fluence with the LAT fluence. Source: \cite{Basak_Rao_2013_linger}.
}
\label{ch6_f7}
\end{figure}

In our analysis, we perform two kinds of correlations: I --- the LAT fluence with the total GBM fluence,
II --- the LAT fluence with the fluence of the non-thermal (PL) component. The GBM fluence is
calculated both in the GBM $T_{90}$, and in the 47.5 s time window (Z12). We designate the former with 
`a', and the latter with `b'. In Table~\ref{ch6_t1}, we have shown the sample of GRBs we have analyzed
(Z12 sample of LGRBs). We have shown the total and the PL fluence of the GBM in the two time windows.  
The values of the LAT fluence is shown in the last column (taken from Z12). 

In Figure~\ref{ch6_f7}, we have shown the data for the LAT-Total GBM fluence correlation (Ib, 
left panel) and the LAT-PL GBM fluence correlation (IIb, right panel) cases, as described above. 
The correlation coefficient of these cases are reported in Table~\ref{ch6_t2} (last two rows).
The $p$-values of the correlations are low, denoting that the correlations are not due to chance.
We note that the correlation using $T_{90}$ time window (first two rows) are inferior to those using
the 47.5 s time window. This suggests that a uniform time window is a better choice. Also, note that the correlation 
of the total GBM fluence (in $T_{90}$) with the LAT fluence (0.68) is better than the Z12 correlation (0.537).
This is possibly due to the different spectral models as well as different values of $T_{90}$.

The correlation of Ib ($r=0.87$) and IIb ($r=0.88$) are comparable in terms of the 
Pearson linear correlation. Using the logarithmic values, the corresponding coefficients 
are 0.68 and 0.72, respectively. We now use the Spearman partial rank correlation to find 
out which is the more fundamental correlation. The coefficient are $\rho=0.75$ and $\rho=0.81$
for Ib and IIb cases, respectively. Note that a Spearman correlation does not assume the 
linearity of data, and hence it is a more robust estimator of the correlation (\citealt{Macklin_1982}). 
Also, the outliers have least effects on the correlation i.e., if some GRBs are really 
exceptional, while others follow a trend, then a Spearman rank is more reliable. 
Note that the correlations using $T_{90}$ time window have comparable $\rho$ to the 
corresponding cases using 47.5 s time  window. We calculate the $D$-parameter of I and II
correlation. For a 47.5 s time window, we see that a correlation of the LAT fluence with 
the total GBM fluence (Ib) has $D=-1.4$, while the correlation of the LAT fluence with the 
non-thermal GBM fluence (IIb) has $D=2.3$. Hence, based on the $D$ value, we conclude that 
the correlation between the LAT and the non-thermal GBM fluence exists in the presence of a third 
parameter, namely the total GBM fluence, and it is the more fundamental correlation.
Using the fluence values in $T_{90}$ instead of 47.5 s window leads to similar conclusions.

If we assume that the fluence are related as a power-law function, then we can fit a 
linear function to the logarithmic data as: ${\rm log}(y)=K+\delta {\rm log} (x)$, where 
$y$ is the LAT fluence, and $x$ is either the total GBM fluence or the PL fluence.
In Table~\ref{ch6_t3}, we have shown the best-fit parameters $K$, $\delta$, and $\sigma_{\rm int}$.
Note that the values of the slopes are remarkably similar, denoting that high PL 
fluence is expected for high total fluence, and either of them acts as a proxy of high
LAT fluence. However, based on the value of $D$-parameter, the PL fluence is a better 
indicator of the LAT fluence. Note that the $\sigma_{\rm int}$ has a lower value for IIb 
case, which again shows that the LAT-PL GBM fluence correlation is ``better understood''
than the LAT-total GBM fluence correlation.

\begin{table}\centering
 \caption{Correlations between (i) the LAT fluence with the GBM fluence, and (ii) the LAT fluence with the GBM PL fluence}
\begin{small}
 \begin{tabular}{c|c|c|c|c|c}
\hline
 Correlation & \multicolumn{2}{c|}{Pearson} & \multicolumn{3}{c}{Spearman} \\
\cline{2-6}

       & r &  $\rm P_{r}$ & $\rho$ &  $\rm P_{\rho}$ & $D$\\
\hline
\hline 
Ia & 0.68 & $2.67\times10^{-7}$ & 0.73 & $8.20\times10^{-4}$ & -0.6\\
IIa & 0.68 & $2.67\times10^{-7}$ & 0.79 & $1.66\times10^{-4}$ & 1.8\\
\hline
Ib & 0.87 & $5.66\times10^{-6}$ & 0.75 & $5.61\times10^{-4}$ & -1.4\\
IIb & 0.88 & $3.20\times10^{-6}$ & 0.81 & $9.23\times10^{-5}$ & 2.3\\

\hline
\end{tabular}

\end{small}

\label{ch6_t2}
\vspace{0.1in}

\begin{footnotesize}

\flushleft \textbf{Note:} I: Correlation between the LAT fluence and the GBM total fluence
II: Correlation between the LAT fluence and the GBM fluence in the non-thermal (PL) component 
a: Fluence measured in $T_{90}$, b: Fluence measured in 47.5 s time window

\end{footnotesize}

\end{table}

\begin{table}\centering
 \caption{Results of linear fit to the correlation data in Figure~\ref{ch6_f7}}
 \begin{tabular}{c|c|c}
\hline
 Correlation$\rightarrow$ & GBM-LAT (Ib) & GBM PL-LAT (IIb) \\ 
\hline
\hline 
$K$ & $0.056\pm0.099$ & $0.288\pm0.096$ \\
$\delta$ & $0.698\pm0.044$ & $0.697\pm0.049$ \\
$\sigma_{int}$ & $0.385\pm0.084$ & $0.368\pm0.081$ \\
$\chi^2_{red}$ (dof) & 1.10 (15) & 1.10 (15) \\

\hline
\end{tabular}
\vspace{0.1in}

\begin{footnotesize}
\flushleft \textbf{Notes:} Function fitted ${\rm log}(y)=K+\delta {\rm log} (x)$.
$\sigma_{int}$ is intrinsic data scatter, $\chi^2_{red}=-2lnL$, where $L$ is the likelihood function (see chapter 1)

\end{footnotesize}

\label{ch6_t3}

\end{table}

\section{Summary And Discussion On GeV Prediction}

\subsection{The Correlation}
To summarize, we have used a model consisting of a thermal and a non-thermal 
component to fit the time-resolved GBM data of GRBs, which have GeV emission 
during the prompt emission phase. The thermal component of our model is represented 
by two correlated BBs, and the non-thermal component is assumed as a PL. 
We have found that a spectral fit using only the GBM data has a predictive power 
for the LAT emission. For example, the fluence of the PL component bears a strong
correlation with the GeV fluence. Previous attempts to find such a connection between keV-MeV 
emission and GeV emission have failed in a sense that (i) no unified spectral model 
is found, and (ii) correlation using Band only function (as used by Z12) is weak. 
The reason behind the success of our study lies in segregating the thermal and 
non-thermal components of the prompt emission data, and using only the non-thermal 
component to investigate the correlation. 

We would like to point out that a mBBPL model is likely to show similar improvement 
over a correlation, which is studied through a Band only fitting. We have seen 
that the data of some GRBs are indeed consistent with a mBBPL model, e.g., GRB 
090902B (\citealt{Rydeetal_2010_090902B}). However, a 2BBPL model also gives a comparable fit.
It is worthwhile to mention that several recent studies indicate that the prompt 
emission data of some GRBs require a separate thermal component, rather than a continuous distribution 
of temperature (\citealt{Guiriecetal_2011, Axelssonetal_2012}). For example, 
\cite{Guiriecetal_2011} find a sub-dominant BB with a temperature, $kT\approx38$ keV, 
along with a Band function with $E_{\rm peak}\approx 350$ keV. Note that the peak 
energy corresponds to a $kT\approx117$ keV, which has a temperature ratio $\sim3$ with 
the lower BB. This ratio is of the same order we find for many GRBs. Can we use BB+Band 
function for the correlation study? A BB+Band model fit to the data cannot give a possible 
connection between the non-thermal (Band function) and GeV photon for the following 
reason. We have explicitly shown for GRBs with high GeV count that the PL of 2BBPL model 
is delayed and mimics the GeV flux evolution, whereas the two BBs are correlated and do 
not have any delay (Figure~\ref{ch5_f11} and \ref{ch5_f12}). Using a BB+Band function 
for such bursts will replace the higher BB peak with a Band peak, and the corresponding 
flux will not show a delay. Incidentally, GRB 100724B, which is fitted with BB+Band, is 
included in our LAT sample, and the data is consistent with a 2BBPL model. It is worthwhile 
to emphasize again that the 2BBPL model stands out as a universal model for time-resolved GRB spectrum, 
and the non-thermal component of this function gives a better insight for GeV emission.

\subsection{Constraining Physical Models}
The origin of the GeV emission is still a matter of intense debate (e.g., \citealt{meszaros_rees_1994_GeV,
Waxman_afterglow, Gupta_Zhang_2007, Panaitescu_2008, Fan_Piran_2008, Zhang_Pe'er_2009, 
meszaros_rees_2011_GeV}). Majority of the mechanisms to produce GeV photons during the prompt and 
afterglow phase involve inverse compton (IC) of some seed photon. \cite{Zhang_2007_review} has listed 
many possible sites of IC including self-Compton in internal and external shock. 
For example, (i) synchrotron self-Compton (SSC) by electrons accelerated in the IS
(e.g., \citealt{Meszarosetal_1994_prompt, Pilla_Loeb_1998_prompt, Razzaqueetal_2004, 
Pe'er_Waxman_2004, Pe'er_Waxman_2005,Pe'eretal_2005, Pe'eretal_2006}), (ii) Synchrotron
emission from accelerated protons or photon-meson interaction in the IS
(\citealt{Totani_1998, Bhattacharjee_Gupta_2003}), (iii) SSC in the ES --- (a) forward 
shock (\citealt{meszaros_rees_1994_GeV, Dermeretal_2000, Panaitescu_Kumar_2000, 
Zhang_Meszaros_2001, Kumar_Duran_2009, Kumar_Duran_2010}),
(b) reverse shock (\citealt{Wangetal_2001, Granot_Guetta2003}), and (c) cross IC of 
photons in either region (\citealt{Wangetal_2001, Wangetal_2001_IC}), (iv) IC of prompt keV-MeV photons in 
the ES (\citealt{Beloborodov_2005, Fanetal_2005}), (v) IC of photons from x-ray flare in ES
(\citealt{Wangetal_2006, Fan_Piran_2006}), or SSC of x-ray flare photons (\citealt{Wangetal_2006}), and so on.
It is important to identify the correct mechanism, as it can give additional constraint
on the unknown fireball parameters. For example, \cite{Gupta_Zhang_2007} have shown that 
for a $\epsilon_{\rm e}$ (fraction of energy carried by electrons) not too low, the 
leptonic models are preferred. Hence, synchrotron emission from protons or photon-meson 
interaction is disfavoured for high energy (GeV) emission. 

The observation of GeV emission can help in constraining the possible models of
prompt emission. If we assume, e.g., an IS origin, and extrapolate the Band function 
fitted to the prompt keV-MeV data, then it generally over-predicts the GeV emission 
in the LAT (\citealt{Le_Dermer_2009}). It is shown that a low detection rate of the LAT is consistent 
with a ratio of GeV to MeV emission $\sim0.1$ (however, see \citealt{Guettaetal_2011}). 
\cite{Beniaminietal_2011}, using 18 bright GRBs with no LAT detection, have found an upper limit of the fluence 
ratio $\sim0.13$ (during the prompt emission phase), $\sim0.45$ (during 600 s time window).
These ratios put a strong constraint on the possible prompt emission model, and particularly rules 
out SSC for both MeV and GeV data. The implication of our finding in the comparative
study of the GBM and the LAT data is that only the PL component is connected with the GeV 
emission. If the photospheric emission is efficient, the 2BBs possibly do not have any 
connection with the GeV emission. This requirement puts more constraints on the GeV 
afterglow model. For example, if we consider SSC as the possible mechanism, then 
the circumburst density required to explain GeV emission for usual values of parameters 
varies as some negative power of the energy (\citealt{Wangetal_2013}). As the energy is channelized 
into two components, the required density increases. For bursts like GRB 090902B, which have 
a low calculated circumburst density from afterglow modelling (e.g., \citealt{Liu_Wang_2011}),
SSC becomes quite impossible. 

\subsection{Spectral Break Or Cut-off}
One of the most important application of GeV emission is to give a lower limit on the 
unknown bulk \textit{Lorentz} factor ($\Gamma$, \citealt{Woods_Loeb_1995}, section 1.5.1). While 
$\Gamma$ can be estimated from the variability time scale, observation of the highest GeV photon
provides an independent measurement. Note that none of these provide an accurate measurement,
and hence, it is important to have such independent estimators. It is expected that the spectrum 
at very high energy should have a cut-off due to the photon-photon interaction. As this cut-off 
directly depends on $\Gamma$, observation of cut-off gives another measurement of $\Gamma$.
It is also interesting to compare the spectral index of the MeV and GeV data to get an 
indication of spectral break or cut-off. In our analysis, we have computed the average 
spectral index of 2BBPL model fit for the hyper-fluent LAT GRBs. For GRB 090902B, the average 
spectral index in the GBM data ($-1.76\pm0.17$) is remarkably consistent with that of the LAT 
data (also $-1.76$, \citealt{Zhangetal_2011}), showing no cut-off. However, for GRB 090926A, the average index 
($-1.65\pm0.35$) is clearly inconsistent with that of the GeV data ($-2.03$). Hence, there 
is possibly a spectral break in the second case. A higher spectral slope can possibly indicate 
that the spectrum approaches a cut-off. \cite{Ackermannetal_2011_090926A} have indeed found a cut-off 
in the spectrum during 9.7-10.5 s of this GRB. 

\subsection{Delayed Onset: Early Indication Of GeV Emission}
Finally, the PL component in the GBM spectrum of the hyper-fluent LAT GRBs have shown 
delayed onset and lingering behaviour. GRBs with similar brightness, but an order of magnitude 
lower LAT count have a coupled PL and total GBM flux variation. In part III of this chapter, 
we shall discuss about the possible reason of this dual behaviour of PL in terms of a tentative model.
We would like to mention that this delayed onset can be used as an early indication 
of high GeV emission. The current strategy for the off-axis LAT events is to target the brightest GRBs. 
Based on our analysis, it is evident that even moderately bright GRBs with a delayed PL 
component may be targeted. A combination of these two criteria can increase the number 
of GRBs detected by the LAT, and it can also shed light on the seemingly different characteristics 
of the LAT emission in GRBs with similar GBM brightness.

\section*{Part II: Predictions For Low Energy Data}
\vspace{0.3in}

We now turn our discussion to the predictions for the low energy data. We shall use the \textit{Swift}/XRT
for our purpose, and illustrate the predictions for two GRBs. 

\section{A Hypothetical Situation} 
Let us first assume a hypothetical situation based on our phenomenological understanding of 2BBPL model.
In Figure~\ref{ch6_f8}, we have shown the evolution of the two BB temperatures ($kT_h$, and $kT_l$)
as functions of ``running fluence''. The evolution of $kT_h$ and $kT_l$ are shown as strictly hot-to-cold 
(LK96-like evolution). We expect such evolution for a GRB with a single smooth pulse. Note that, the 
``running fluence'' is a monotonically increasing function of time. Hence, the evolution will 
be similar if we replace this quantity with time. As we have seen that the ratio of the two BB 
temperatures is $\sim 3-5$, it is possible that the lower BB goes below the lowest sensitive energy 
band of the higher energy detectors (GBM and BAT). Note that though the GBM has a lower energy coverage
($\sim 8$ keV) than the BAT ($\sim15$ keV), we have found that the 8-15 keV band of GBM has $\lesssim 2-3\sigma$
count rate per bin. Hence, a spectral component below the lowest BAT energy is unlikely to be significantly 
detected in the GBM detector. For convenience of description, let us denote the lowest 
sensitive energy of these detectors as $E_{\rm low}$. The transition of $kT_l$ at $E_{\rm low}$ 
will have the following observational effects.

\begin{itemize}
\item (i) If we attempt to fit the spectrum of the GBM and/or the BAT data with a BBPL model, we shall get average $kT$
in the initial bins. As the lower BB is absent in the later bins of higher energy detectors, we expect a break 
during the transition of lower BB at $E_{\rm low}$ (see Figure~\ref{ch6_f8}). This break is expected to 
be smooth, as the higher part of the lower BB spectrum will have some residual effect during the transition. 
Note that, if the two BBs have their corresponding breaks, there can be a variety of averaging effects. In 
particular, the average temperature in the initial bins can show even an increasing temperature evolution. 

\item (ii) More interestingly, the lower BB can show up in the later XRT spectrum. Provided that the XRT 
data is available in the late prompt emission phase, this should be detectable. If this BB is indeed 
the lower BB of the initial time bins, the lower BB temperature, $kT_l$ as found by fitting 2BBPL model 
to the initial GBM and/or BAT spectrum should smoothly join with the $kT$ of the XRT data.
\end{itemize}

In the following, we shall look for these two observational effects. For the first effect to be seen, we 
require a GRB with single pulse and a break in $kT$ evolution (when fitted with a BBPL model). For the 
later effect, we require a GRB with long duration so that the XRT data is available in the final phase.
For this kind of GRB, we shall look at the falling part of the final pulse, firstly because we expect
a smooth cooling in the falling part, and second, only the last pulse is smoothly connected with the 
XRT observation, provided that the two observations overlap at all.

\begin{figure}\centering

{

\includegraphics[width=3.4in]{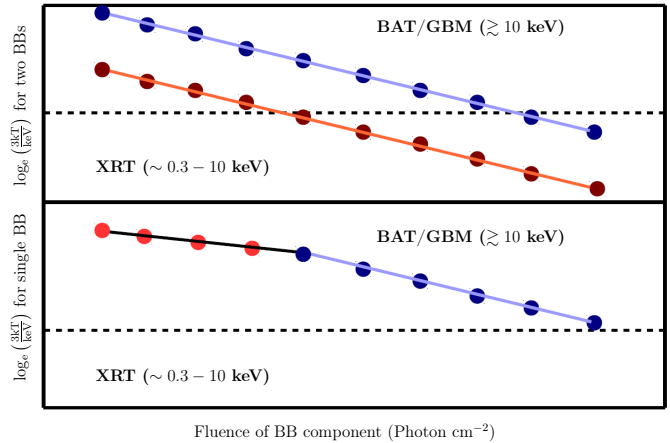} 

}
\caption[A hypothetical case: temperature ($kT$) evolution as a function of ``running fluence''
in a GRB with smooth single pulse]
{A hypothetical case: temperature ($kT$) evolution as a function of ``running fluence''
in a GRB with smooth single pulse. As the two $kT$ of two BBs evolve simultaneously, the lower 
BB can transit from the BAT/GBM lower range to the XRT band (upper panel). If one tries to fit BBPL 
the $kT$ will show a phenomenological break (lower panel). If the XRT data is available, the lower 
BB can be detected (see text).}
\label{ch6_f8}
\end{figure}

\begin{table*}\centering
 \caption{Parameters of linear fit to the fluence-log$_{\rm e}$($kT$) data}

\begin{tabular}{cccc}
\hline
Model & Intercept & Slope & $\chi^2$ (dof) \\
\hline
\hline
BBPL$^{(a)}$ & $3.02_{-0.29}^{+0.30}$ & $(-4.56\pm1.84) \times 10^{-2}$ & 0.46 (4) \\
2BBPL & $3.10_{-0.22}^{+0.23}$ & $(-5.10\pm1.48) \times 10^{-2}$ & 1.244 (7)\\
\hline
\end{tabular}
\vspace{0.1in}

\begin{footnotesize}
 \textbf{Notes:} $^{(a)}$ Fitted to the data in the falling part of the pulse
\end{footnotesize}
\label{ch6_t4}
\end{table*}

\section{Data In The Initial Time Bins}
In order to obtain an observational effect due to the averaging in the initial time bins, we use 
GRB 080904. We choose this GRB because, this is one of the brightest GRBs with a single pulse, and more importantly,
it shows a temperature evolution with a break. The time-resolved data of this GRB can be fitted with a BBPL
model with similar $\chi^2$ as a Band functional fit. In the falling part of the pulse, the BB temperature ($kT$)
falls off monotonically with ``running fluence'' (LK96-like behaviour). However, in the initial bins $kT$
has an increasing trend. The evolution of $kT$ as obtained by BBPL fitting is shown in Figure~\ref{ch6_f9}
(orange open circles). If we invoke the photospheric emission and an adiabatic cooling, a break in the 
evolution is indeed expected if the photosphere occurs at a higher radius than the saturation radius
($r_{\rm ph}>r_{\rm s}$). However, the evolution before the break is expected to be constant. If the 
majority of the energy is carried by magnetic field, then we expect a decreasing evolution of $kT$ before the 
break (\citealt{Drenkhahn_Spruit_2002}). But, the increase in $kT$ cannot be explained within these scenarios.
Let us investigate whether a 2BBPL model can give a phenomenological explanation. We note that the 
temperature at the fourth bin (where the turn over of $kT$ evolution occurs) 
is $12.73_{-0.98}^{+1.11}$ keV i.e., a peak at $\sim 36$ keV. If we assume 
a ratio of the two BB temperatures as $\sim 3$, the lower BB has $kT_l \sim 4.2$ keV, or a peak at $\sim12$ keV.
Hence, it is already outside the BAT lower energy range (15 keV), and almost outside the sensitive energy band
of the GBM ($>8$ keV). It is possible that the lower BB is not required for the GBM/BAT data from the fourth 
bin onwards. Now, we try to fit 2BBPL in the initial three bins. Though we find that a 2BBPL model is not 
statistically required over a BBPL model, but addition of a second BB with $kT_l \sim \frac{1}{3} kT_h$,
pushes the temperature higher (black filled circles in Figure~\ref{ch6_f9}). This higher temperature ($kT_h$)
is now fully consistent with $kT$ evolution of the later phase. Note that the data in the later phase 
is fitted only with a BBPL model. Hence, it is not necessary that the single temperature evolution of BBPL model
should follow the $kT_h$ evolution of a 2BBPL model fit. From a phenomenological point of view, the evolution 
of the later data is ``unaware'' of the initial evolution. The fact that $kT$ of later part is smoothly 
connected with $kT_h$ of the initial part requires that both of them are driven by a single emission. It 
is only because the lower BB temperature ($kT_l$) goes below the sensitive energy band of the GBM/BAT detector
that we do not require this BB in the later phase of the GBM/BAT data. In other words, both the BBs are 
present throughout the burst duration, but can appear in different energy bands. As the lower BB is outside
the GBM/BAT energy band, a BBPL model gives a consistent temperature evolution in the falling part, 
whereas we require to put this BB in the initial bins in order to get a consistent evolution throughout.

To quantity the evolution, we fit the fluence-log$_{\rm e} kT$ data with a linear function. 
The best-fit value of the slope and intercept are shown in Table~\ref{ch6_t4}. We first fit the data 
after the break to find the evolution of $kT$ as obtained by a BBPL fit. Next, we include $kT_h$ 
of the first three bins as obtained by 2BBPL model fitting, and perform a linear fit to all the data 
points. Note that for this linear fit, the initial data points are temperature of the 
higher BB ($kT_h$), while the later points are $kT$ of a single BB.
It is clearly seen from Figure~\ref{ch6_f9}, and Table~\ref{ch6_t4} that the two linear fits match
quite well with each other (black dotted line and red solid line, respectively). Hence, we conclude that 
the actual evolution is two BBs with correlated temperatures, and the break in $kT$ evolution could be simply 
because the lower BB affects the spectrum in the initial phase, while it goes below the sensitivity
at a later phase. However, we do not rule out that both the BBs can have breaks due to the transition
at the saturation radius.

\begin{figure}\centering
{

\includegraphics[width=3.4in]{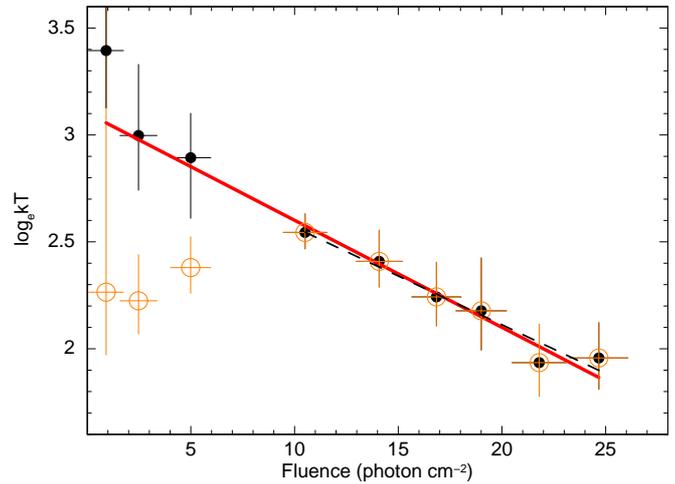} 

}
\caption[The evolution of temperature ($kT$) as a function of ``running fluence'' in single pulse 
GRB 080904]{The evolution of temperature ($kT$) as a function of ``running fluence'' in single pulse 
GRB 080904 is shown --- (i) $kT$ evolution of BBPL model fit (orange open circles), (ii) Higher 
BB temperature, $kT_h$ of 2BBPL fitted to the initial three bins, while BBPL fit is retained for 
the rest (filled circles). Both the evolution are fitted with a linear function (dotted line for 
the falling part of case (i), solid red line for case (ii)). The evolution match quite well at 
the overlapping region (see text for detail). Source: \cite{Basak_Rao_2012_germany}}
\label{ch6_f9}
\end{figure}

\section{The Lower BB In The XRT Window}

It is clear from the above discussion that the lower BB can affect the spectrum in the 
initial bins giving a phenomenological break. As the lower BB temperature goes below the 
GBM/BAT sensitivity it is expected to be seen in the low energy detector, namely 
the \textit{Swift}/XRT. This is a more direct proof of the evolution of the two BBs.
However, this opportunity is very rare as the XRT observation starts 
with some delay. GRB 090618 is one of such cases where the XRT observation started during 
the late phase of the prompt emission. In Figure~\ref{ch6_f10}, we have shown the count rate 
lightcurve of the BAT and the XRT data. Note that the two observations overlap at the late prompt 
phase. The XRT observation starts at 125 s, where the falling part of the last pulse is still 
visible in the BAT energy band.

\begin{figure}\centering
{

\includegraphics[width=3.4in]{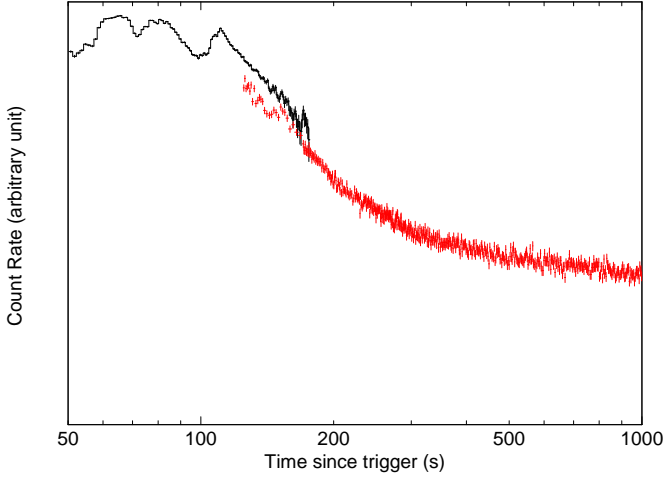} 

}
\caption[Lightcurve of GRB 090618 in the BAT and the XRT observation]
{Lightcurve of GRB 090618. Count rate (arbitrary unit) is plotted as a function of time ---
BAT (black points), XRT (red points). Note the overlap of observation time of the two instruments.
The XRT data is taken from $http://www.swift.ac.uk/xrt\_curves/$ (see \citealt{Evansetal_2007,Evansetal_2009}).
}
\label{ch6_f10}
\end{figure}

\begin{sidewaystable}\centering
 \caption{Parameters of BBPL and 2BBPL model fit to the time-resolved data of GRB 090618 in 116.95-130.45 s}

\begin{footnotesize}

\begin{tabular}{c|ccccccc|ccccc}
\hline
Interval & \multicolumn{7}{c}{2BBPL} & \multicolumn{5}{|c}{BBPL}\\
\cline{2-13}
 & $kT_h$ & $N_h$ & $kT_l$ & $N_l$ & $\Gamma$ & $N_{\Gamma}$ & $\chi^2$ & $kT$ & $N$ & $\Gamma$ & $N_{\Gamma}$ & $\chi^2$\\
\hline
\hline
&  &  &  &  &  &  &  &  &  &  &  & \\
$116.95-118.05$ & $23.8_{-5.7}^{+7.1}$ & $2.9_{-1.4}^{+1.3}$ & $9.0_{-2.6}^{+2.5}$ & $1.9_{-1.0}^{+1.0}$ & $-2.4_{-0.3}^{+0.2}$ & $541_{-215}^{+530}$ & 76.5 (102) & $14.8_{-1.8}^{+1.8}$ & $1.7_{-0.5}^{+0.5}$ & $-2.2_{-0.1}^{+0.1}$ & $368_{-67}^{+90}$ & 79.8 (104) \\

$118.05-119.35$ & $21.4_{-5.8}^{+5.6}$ & $2.1_{-0.9}^{+1.2}$ & $7.6_{-1.5}^{+1.3}$ & $2.1_{-0.8}^{+0.9}$ & $-2.3_{-0.2}^{+0.4}$ & $304_{-137}^{+417}$ & 95.5 (102) & $10.2_{-1.1}^{+1.4}$ & $1.4_{-0.3}^{+0.3}$ & $-2.1_{-0.1}^{+0.1}$ & $243_{-49}^{+58}$ & 100.6 (104) \\

$119.35-120.95$ & $20.8_{-4.4}^{+4.4}$ & $2.0_{-0.9}^{+0.9}$ & $8.3_{-1.6}^{+1.7}$ & $1.5_{-0.7}^{+0.8}$ & $-2.4_{-0.2}^{+0.4}$ & $412_{-163}^{+571}$ & 85.0 (102) & $12.6_{-1.4}^{+1.5}$ & $1.4_{-0.3}^{+0.3}$ & $-2.2_{-0.1}^{+0.1}$ & $280_{-56}^{+73}$ & 88.7 (104) \\

$120.95-122.65$ & $14.8_{-1.4}^{+1.9}$ & $2.8_{-0.5}^{+0.6}$ & $5.2_{-1.0}^{+1.8}$ & $1.1_{-0.6}^{+0.7}$ & $-2.4_{-0.2}^{+0.3}$ & $276_{-163}^{+196}$ & 123.8 (102) & $13.4_{-0.8}^{+0.8}$ & $2.1_{-0.3}^{+0.4}$ & $-2.4_{-0.1}^{+0.1}$ & $370_{-83}^{+120}$ & 127.5 (104) \\

$122.65-124.85$ & $20.4_{-1.2}^{+1.5}$ & $2.9_{-0.5}^{+0.3}$ & $6.4_{-0.5}^{+0.5}$ & $2.5_{-0.5}^{+0.6}$ & $-3.6_{-0.8}^{+1.3}$ & $4650$ & 105.1 (102) & $14.6_{-2.7}^{+2.4}$ & $0.8_{-0.3}^{+0.3}$ & $-2.3_{-0.1}^{+0.1}$ & $338_{-74}^{+94}$ & 123.0 (104) \\

$124.85-127.25$ & $24.1_{-10.1}^{+19.2}$ & $1.0_{-0.6}^{+0.7}$ & $8.7_{-3.8}^{+1.6}$ & $1.3_{-0.9}^{+0.5}$ & $-2.6_{-0.3}^{+0.4}$ & $487_{-279}^{+687}$ & 91.3 (102) & $10.6_{-1.2}^{+1.4}$ & $0.9_{-0.4}^{+0.4}$ & $-2.3_{-0.1}^{+0.1}$ & $276_{-88}^{+136}$ & 98.5 (104) \\

$127.25-130.45$ & $15.3_{-3.4}^{+4.0}$ & $1.0_{-0.4}^{+0.4}$ & $5.2_{-0.9}^{+1.2}$ & $1.2_{-0.4}^{+0.5}$ & $-2.3_{-0.3}^{+0.4}$ & $123_{-87}^{+226}$ & 91.5 (102) & $8.4_{-1.3}^{+1.7}$ & $0.6_{-0.2}^{+0.2}$ & $-2.3_{-0.1}^{+0.1}$ & $189_{-45}^{+56}$ & 93.3 (104) \\
&  &  &  &  &  &  &  &  &  &  &  & \\
\hline
\end{tabular}

\end{footnotesize}
\vspace{0.1in}

\begin{scriptsize}
 \textbf{Notes:} The bins are obtained by requiring $C_{\rm min}=1500$. Errors are $1\sigma$.  
\end{scriptsize}
\label{ch6_t5}
\end{sidewaystable}

\begin{figure}\centering
{

\includegraphics[width=3.4in]{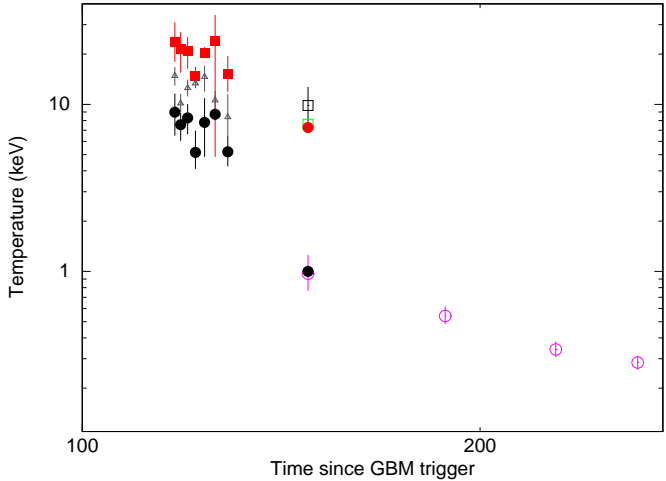} 

}
\caption[Evolution of BB temperature ($kT$) for different models]
{Evolution of BB temperature ($kT$) for different models are shown. The XRT data 
(open circles with magenta colour) is shown during 125-275 s post BAT trigger (last four points). 
The joint GBM-BAT data is fitted in the previous bins till 130.45 s post GBM trigger. Evolution of the 
following parameters are shown: (i) $kT$ of BBPL (grey open triangles), (ii) higher ($kT_h$) 
and lower ($kT_l$) BB temperature of 2BBPL model (red filled boxes, and black filled circles, 
respectively). Note that the evolution of $kT_l$ smoothly connects with the later evolution 
in the XRT. The data of the first XRT bin (125-165 s) is used for a joint BAT-XRT analysis. 
The corresponding $kT$ remarkably follows the two BB evolution of the previous bins 
(red filled circle for $kT_h$, and black filled circle for $kT_l$ in 125-165 s time bin; 
see text for detail).}
\label{ch6_f11}
\end{figure}

\cite{Pageetal_2011} have studied this GRB, and have found an evidence of BB component in the 
XRT spectrum. They have used four time bins from 125-275 s, and one large bin in 275-2453 s.
In all cases a BBPL model gives $>0.9999$ F-test significance as compared to a PL fitting.
The x-ray spectral fits include absorptions due to galactic and intrinsic neutral 
hydrogen ($N_{\rm H}=5.8 \times 10^{20}$ cm$^{-2}$, and $zN_{\rm H}=(1.82\pm0.08) \times 10^{21}$ cm$^{-2}$).
In Figure~\ref{ch6_f11}, we have shown the BB temperature in the XRT data by open circles 
with magenta colour (last four points).
The time axis is shifted by 3.192 s, as there is a delay of BAT trigger from the GBM 
trigger by this amount of time. In order to find the connection of this BB with that 
in the higher energies, we study the time-resolved GBM and/or BAT spectrum in 116.95-130.45 s.
The bins are obtained by requiring $C_{\rm min}=1500$ in the NaI detector 
having the highest count rate (n4). We have used both the GBM and the BAT for our 
analysis to constrain the parameters with good accuracy. In Table~\ref{ch6_t5},
we have shown the best-fit values of BBPL and 2BBPL model fit.

In Figure~\ref{ch6_f11}, we have marked the values of $kT$ as obtained by BBPL model
fitting by grey triangles. The values of $kT_h$ (red filled boxes) and $kT_l$ (black
filled circles) of 2BBPL model are also marked. Note that $kT$ of BBPL model in these 
initial bins will largely over-predict those in the later XRT data. Hence, these two BBs 
are not connected. In contrast, the lower BB temperature, $kT_l$ in the initial bins
clearly shows the trend of later $kT$ evolution in the XRT data (compare the black filled circles 
in the initial 7 bins to the magenta open circles in the last 4 bins). 

In order to find how the higher BB evolves, we fit the 125-165 s data of the BAT and the GBM 
data with a BBPL model. Note that if we get a BB temperature consistent with the 
evolution of $kT_h$ in this bin, then it shows that the lower BB must have gone below the 
GBM/BAT sensitivity level. We fit the data of this bin using BAT, joint BAT-GBM and joint
BAT-XRT, all of which are found to be consistent with each other. The corresponding 
points are shown in Figure~\ref{ch6_f11} by black open box, green open box, 
and red filled circle, respectively. Note that the BB temperature of this bin is clearly 
consistent with the trend of $kT_h$ evolution in the previous bins (compare with red filled 
boxes of the initial 7 bins). The corresponding lower BB temperature as found from the joint 
BAT-XRT analysis is shown by black filled circle in this time bin. The value of this lower 
BB temperature in this bin is consistent with that obtained by fitting a BBPL model in this 
bin (magenta open circle below the black symbol). To emphasize again, the GBM/BAT data and the
XRT data of 125-165 s are fitted independently by using single BB models for either of them. 
It is quite remarkable that these two temperatures are fully consistent with the trends of 
$kT_h$ and $kT_l$, respectively, and also with the joint BAT-XRT data. Hence, the findings 
are consistent with the hypothesis that there are indeed two evolving BBs during the prompt 
emission phase.

\section*{Part III: A Simplistic Model}
\vspace{0.3in}

\section{The Physical Model}

\subsection{List Of Observations:}

Let us first list down the important features we have observed.

\begin{itemize}
 \item (i) The prompt emission spectrum is not totally non-thermal. A Band only function either gives 
unphysical spectral index, or sometimes additional components are required. Most importantly, the additional 
parameters seem to show up for high flux case. Consider, e.g., GRB 090902B and GRB 100724B. The former 
can be fitted by a smoothly varying BB (multi-colour BB) along with a PL, while the later requires a separate BB component 
on top of a Band function. But, each of the spectra can be fitted with a 2BBPL model. We have also found that 
these GRBs are not special, all the bright GRBs with GeV photons, with single pulse, or multiple separable 
pulses are consistent with 2BBPL model. Though it is entirely possible that each GRB has a characteristic 
spectrum, but the fact that a single model can fit to all possible variety of GRBs strongly indicates a 
common radiation mechanism.

\item (ii) We have also found that the two BBs of 2BBPL model are highly correlated. Both the temperature 
and normalization of these BBs have a ratio in the range $\sim3-5$. 

\item (iii) The PL component of 2BBPL model is physically reasonable for synchrotron radiation. It is 
also found that the PL index becomes lower at the later phase. For example, GRB 081221 has the following 
evolution of indices in the rising and falling part of the constituent pulses --- 
pulse 1: $-1.74_{-3.04}^{+\infty}$ to $-2.04_{-0.14}^{+0.28}$, pulse 2: $-1.94_{-0.11}^{+0.16}$
to $-2.15_{-0.08}^{+0.09}$. Hence, the spectrum becomes softer with time.

\item (iv) GRBs with high GeV emission have a delayed onset of the PL component. This component becomes 
important at the later phase. On the other hand, GRBs with similar brightness in the GBM band, but an 
order of magnitude lower GeV emission have coupled PL and total flux evolution. The evolution of the PL
in each case remarkably mimics that of the corresponding GeV emission. 

\end{itemize}

\subsection{A Spine-sheath Jet Model:}

The model that we use to explain the observations listed above is a spine-sheath structure of the jet.
This is one of the most attractive proposed jet structure based on both theoretical and observational 
requirement. On the theoretical side, a slowly moving sheath surrounding a fast inner spine is expected 
as the jet punctures through the envelop of the dying star. The material of the star forms a hot 
cocoon layer on the spine jet (\citealt{Woosleyetal_1999, Meszaros_Rees_2001, Ramirez-Ruizetal_2002,
Zhangetal_2003, Zhangetal_2004, Mizutaetal_2006, Morsonyetal_2007}). Even if the cocoon is absent, 
a MHD jet can have a collimated proton spine with wider neutron sheath (\citealt{Vlahakisetal_2003, 
Pengetal_2005}). On the observational ground a shaeth is invoked in a GRB jet to explain 
several observations, e.g., shallow decay phase of x-ray afterglow (\citealt{Granotetal_2006, Panaitescuetal_2006,
Jinetal_2007, Panaitescu_2007}), observation of jet break in radio as well as in optical and x-ray afterglow 
(\citealt{Lipunovetal_2001, Bergeretal_2003_spinesheath, Shethetal_2003, Liang_Dai_2004, Huangetal_2004, 
Wuetal_2005, Hollandetal_2012}). For example, \cite{Bergeretal_2003_spinesheath} have studied the afterglow of 
GRB 030329 ($z=0.1685$) in the radio wavelength. They have found a jet break at 9.8 day, which together 
with the observed flux corresponds to a jet opening angle $\theta_{\rm sheath}\sim 17^{\circ}$. 
However, the optical and x-ray lightcurve also have an achromatic break at $\sim0.55$ day corresponding 
to $\theta_{\rm spine}\sim 5^{\circ}$. The optical data also show a re-brightening corresponding to 
the peak of the second jet component. The ratio of collimation corrected energy of the two components 
is found to be $E_{\rm sheath}/E_{\rm spine}\sim 5$ i.e., a larger fraction of energy is carried by a 
much wider jet component. Recently, \cite{Hollandetal_2012} have studied the optical afterglow data 
of GRB 081029 ($z=3.8479$). The opening angle of the two components as required by the data is 
much smaller than GRB 030329 --- $\theta_{\rm spine}\sim 0.86^{\circ}$, and $\theta_{\rm sheath}\sim 1.4^{\circ}$. 
In addition, unlike GRB 030329, the energetics of the two components are comparable to each other.
It is not clear whether there are two classes of GRBs with completely different properties of the 
spine-sheath components. 

Recently, \citet[][I13 hereafter]{Itoetal_2013} have used a \textit{Monte Carlo} simulation to study the emergent 
spectrum from a spine-sheath jet. They have considered the opening angle of spine and sheath as
$0.5^{\circ}$ and $1^{\circ}$, respectively. They have injected thermal photons at a region of high optical 
depth and followed each photon till it escapes at the photosphere. With varying ratio of $\eta$ 
between the two components (in the range 1-4), and for different viewing angles ($0.25^{\circ}-0.75^{\circ}$), 
they have obtained some spectra which have the signature of the two BB components (Figure 5 of I13). 
The two thermal components are most prominent for a viewing angle near the spine sheath 
boundary. In addition to the BB components, I13 have found a PL with a high energy cut-off. The extra PL 
component is argued to be the extra hard component as seen for GRBs like 090902B. 

\subsection{Origin Of The Spectral Components}
Let us consider the spine-sheath model of I13. We shall give some order of magnitude estimates as required by 
our observations. The assumptions are as follows. The coasting bulk \textit{Lorentz} factor of spine 
and sheath regions are $\eta_{\rm sp}$ and $\eta_{\rm sh}$, respectively, and $\eta_{\rm sp}>\eta_{\rm sh}$. 
As $\eta=L/\dot{M}c^2$, the difference in its value in the two regions can occur due the difference in 
jet luminosity $L$, or mass flow rate $\dot{M}$. It is reasonable to assume that as the sheath should be 
baryon dominated, it has higher mass flow rate. The jet luminosity can be assumed equal. 

\subsubsection{A. Origin Of Two Blackbodies}
The two BBs in this model appear from two photospheres of the spine and sheath.
As the photospheric radius, $r_{\rm ph} \propto L\eta^{-3}$, we find that the photosphere of the spine 
($r_{\rm ph,sp}$) occurs lower than that of the sheath ($r_{\rm ph,sh}$). Also, as the saturation 
radius, $r_{\rm s} \propto r_{\rm i}\eta$, where $r_{\rm i}$ is the initial injection radius, the 
saturation of spine occurs above the sheath ($r_{\rm s, sp}>r_{\rm s, sh}$). Now, we know that 
during the adiabatic expansion, the evolution of both bulk Lorentz factor ($\Gamma$) and co-moving 
temperature ($kT'$) have a break at $r_{\rm s}$. The evolution of $\Gamma_{\rm sp}$ and $\Gamma_{\rm sh}$ 
are similar to each other ($\propto r/r_{\rm i}$) till $r_{\rm s, sh}$. After this radius the sheath 
stops accelerating and coasts with its corresponding value, $\eta_{\rm sh}$. The spine, 
however, accelerates till $r_{\rm s, sp}$ (note that $r_{\rm s, sp}>r_{\rm s, sh}$).
The evolution of $kT'$ can be written as follows (see I13).

\begin{equation}
 kT'\propto \left(\frac{L}{r_{\rm i}^2}\right)^{1/4}\times \left\{ \begin{array}{ll}
 (r/r_{\rm i})^{-1} & \mbox{$r < r_{\rm s}$} \\
 (r_{\rm s}/r_{\rm i})^{-1} (r/r_{\rm s})^{-2/3} &\mbox{$r > r_{\rm s}$}
       \end{array} \right.
\label{i13_1}
\end{equation}
\vspace{0.1in}

Here $r$ is the radial distance from the centre of explosion in lab/observer frame.
As an observer sees a boosted temperature from the photosphere, the temperature before the break is a 
constant and does not degrade due to the adiabatic cooling. However, after the saturation the observer 
temperature drops as $r^{-2/3}$, As the temperature degrades above $r_{\rm s}$, it 
is evident that the spine has a brighter and efficient photospheric emission than the sheath.

Now, if we consider that the phtosphere in both components occur above the saturation, then 
from equation~\ref{i13_1}, we get the observed temperature at $r_{\rm ph}$ as 
$kT \propto r_{\rm i}^{1/6} \eta^{8/3} L^{-5/12}$. The peak luminosity is found to 
be $L_{\rm p} \propto r_{\rm i}^{2/3} \eta^{8/3} L^{1/3}$. Note that both these 
quantities strongly depend on the coasting bulk \textit{Lorentz} factor, $\eta$.
Considering the same radius of initial energy injection ($r_{\rm i}$), and similar jet 
kinetic luminosity ($L$) for spine and sheath, the ratio of temperature and peak 
luminosity are $\propto \eta^{8/3}$. Note that the peak luminosity is a representation 
of the normalization of the two BBs in our analysis. Hence, from the above discussion,
we need a ratio of $\eta \approx1.5-1.8$, to get a ratio of temperature and normalization 
$\sim3-5$. On the other hand, if we assume that the jet kinetic luminosities are not 
similar, then other ratios are possible. For example, I13 have assumed that $\dot{M}$
is the same for the two components, then the ratio of $L$ scales with $\eta$. Hence, 
we require a ratio $\eta\approx1.6-2.0$ for temperature ratio of $3-5$. However, the 
corresponding normalizations have a ratio $4-8$. Note that the observation of 
\cite{Hollandetal_2012} indicates that the kinetic luminosity of the two components should be
rather comparable. If the kinetic luminosity of the spine is lower than the sheath 
by a factor of $\sim5$ (as required by \citealt{Bergeretal_2003_spinesheath}), and 
the mass flow rate is lower by a factor of 10, then the ratio of $\eta$ is $\sim2$.
Hence, if the sheath is indeed the cocoon, the relative values of jet kinetic energy
and mass flow rate ensures a moderate ratio of $\eta$, and hence, a reasonable 
temperature ratio $\sim3-5$ is attained.

\subsubsection{B. Origin Of The Power-law}
As discussed, from the region above $r_{\rm s, sh}$, the sheath coasts with the value 
$\eta_{\rm sh}$, while the spine continues to accelerate. Hence a strong velocity 
shear occurs in the region $>r_{\rm s, sh}$. The photons crossing the spine-sheath 
boundary are Comptonized by electrons moving in the flow. Depending on the angle of 
incidence ($\theta_{\rm i}$) and scattering angle ($\theta_{\rm s}$), the photon 
either gains energy (up-scattered; $\theta_{\rm s}<\theta_{\rm i}$), or looses 
energy (down-scattered; $\theta_{\rm s}>\theta_{\rm i}$). I13 have shown that the 
average gain in up-scattering and in down-scattering are 
$\frac{1}{2}\left[1+\left(\frac{\Gamma_{\rm sp}}{\Gamma_{\rm sh}}\right)\right]$
and $\frac{1}{2}\left[1+\left(\frac{\Gamma_{\rm sh}}{\Gamma_{\rm sp}}\right)\right]$, 
respectively. Note that as $\Gamma_{\rm sp}>\Gamma_{\rm sh}$, the up-scattering 
denotes gain, while down-scattering denotes loss. The overall gain of this process 
is $>1$. As a result the synthetic photon spectrum is a power-law 
(see Figure 5 of I13). However, the acceleration is limited by the 
efficiency of Compton scattering. When the co-moving energy of the photon approaches 
the rest mass energy of electron, the Compton scattering approaches the \textit{Klein-Nishina}
regime, where the reaction cross section rapidly drops, leading to a very inefficient 
acceleration. I13 have obtained a sharp cut-off at $\sim 100 $ MeV, which corresponds to
a \textit{Lorentz} factor ($\Gamma$) of $\sim 200$. Note that for higher $\Gamma$,
the cut-off can occur at higher energy. In Figure~\ref{ch6_f12}, we have shown a 
schematic picture of a spine-sheath jet (the darker region shows the spine).
We have marked the photospheres of the two components. Note that the 
spine photosphere is lower than that of the sheath. The prompt emission 
can occur in the photosphere and/or in the IS region with a delay. 
Photons crossing the spine-sheath boundary layer are Comptonized. Hence, these 
photons will also form a non-thermal component in the higher energy band.

\begin{figure*}\centering
{

\includegraphics[width=5.9in]{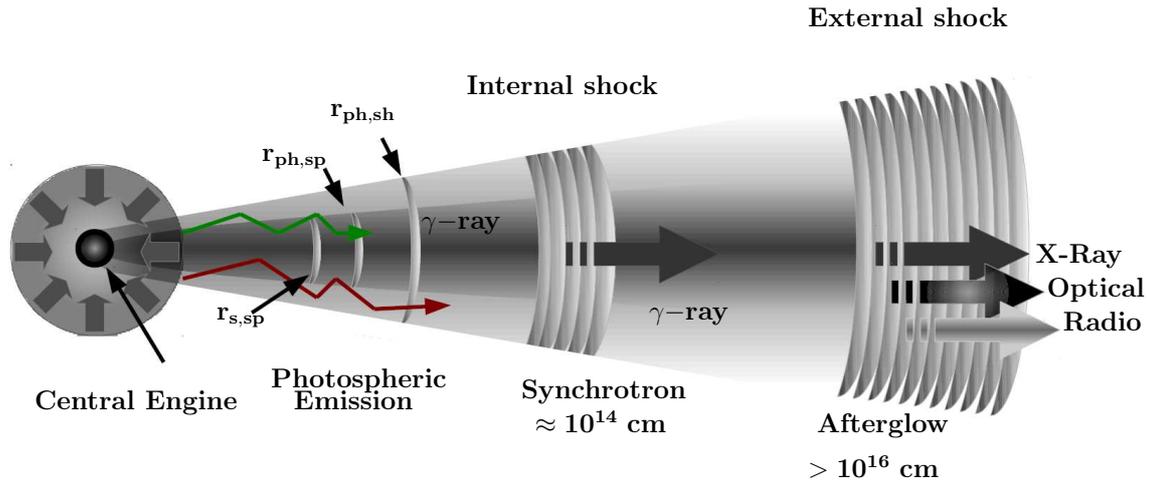} 

}
\caption[A schematic diagram of spine-sheath jet model]
{A schematic diagram of spine-sheath jet model. The spine efficiently 
emits thermal photon at the the photosphere ($r_{\rm ph, sp}$), while the sheath 
produces a second thermal component. The photons crossing the boundary layer (red and 
green zig-zag lines) are Inverse-Comptonized. The IS is delayed from these emissions.}
\label{ch6_f12}
\end{figure*}

\subsection{Explanation Of The Observations:}

Let us now use the spine-sheath structure of jet, and examine whether it is consistent 
with the observations we have shown.

\begin{figure}\centering
{

\includegraphics[width=3.4in]{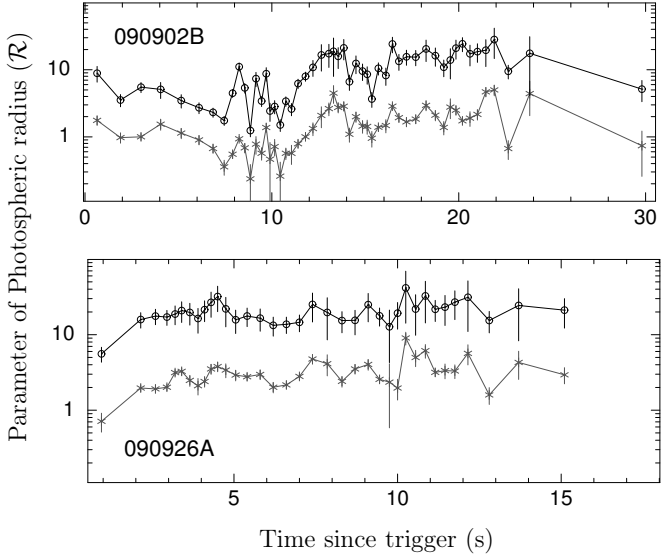} 

}
\caption[Evolution of the parameter $\cal{R}$ as a function of time for GRB 0909002B and GRB 090926A]
{Evolution of the parameter $\cal{R}$ as a function of time for GRB 0909002B (upper panel)
and GRB 090926A (lower panel). The actual photospheric temperature can be directly determined from 
$\cal{R}$. The symbols are: open circles (lower BB), and stars (higher BB). Note that higher BB has 
a lower $\cal{R}$, and the values are correlated throughout each burst. Source: \cite{Raoetal_2014}.}
\label{ch6_f13}
\end{figure}

\begin{itemize}
\item (i) First note that the model naturally produces two BB spectra from the two 
photospheres. If we assume that both spine and sheath components have reached the 
saturation, then the final coasting value of their $\Gamma$ remains constant for 
the later part, with a ratio $\eta_{\rm sp}/\eta_{\rm sh}$. Now, 
the co-moving temperature ($kT'$) of both spine and sheath undergoes adiabatic cooling.
Hence, the later observed temperature ($kT\propto \Gamma kT'$) should be correlated. 
Note that the sheath, having much lower $\eta$, has definitely reached the saturation 
before reaching the photosphere ($r_{\rm ph, sh}$). The spine, on the other hand, can 
reach $r_{\rm s, sp}$ after $r_{\rm ph, sp}$. In this case, the observed temperatures
($kT_h$ and $kT_l$) may not be correlated in the initial bins. Such cases can lead 
to diverse features in the temperature evolution before the break (as discussed in 
section 6.9). 
 
\item (ii) As discussed, the spine-sheath jet can effectively Inverse-Comptonize the 
photons crossing the boundary layer in the velocity shearing region. However, the emergent 
non-thermal component is found to be cut-off power-law (CPL) rather than a full power-law (PL). 
Hence, we shall probably need synchrotron (and /or IC) in the IS to account for a PL with 
no break in the observed GBM-LAT band. Note that the values of the PL index as found in 
our analysis remain within fast cooling regime of an optically thin synchrotron emission. 
Hence, the PL component can be a combination of both the processes.

\item (iii) It is interesting to compare the location of photosphere of the two 
thermal components and check with the data. For this purpose, let us define the 
dimensionless quantity ${\cal{R}}$ (see \citealt{Ryde_Pe'er_2009}) as follows.

\begin{equation}
 {\cal{R}} (t)=\left[\frac{F_{\rm Th}(t)}{\sigma T(t)^4} \right]^{1/2}
\end{equation}
\vspace{0.1in}

Here $F_{\rm Th} (t)$ and $T(t)$ are respectively the flux and temperature of one of the 
BBs at a given observer time, $t$. As the physical photosphere is directly proportional 
to ${\cal{R}}$, an evolution of ${\cal{R}}$ shows the evolution of the actual photosphere. 
In Figure~\ref{ch6_f13}, we have shown the evolution of ${\cal{R}}$ (in units of $10^{-19}$) 
as a function of time for two GRBs, namely GRB 090902B (upper panel), and GRB 090926A (lower panel). 
\cite{Ryde_Pe'er_2009} have calculated ${\cal{R}}$ for a sample of bursts, and have shown 
that the quantity either increases or remains constant throughout the prompt emission. 
The later behaviour is found in these two GRBs --- the photospheres of both the BBs 
remain steady throughout the bursts (see Figure~\ref{ch6_f13}). Note that the value of 
${\cal{R}}$ for higher BB (stars) is always lower than that of the lower BB (open circles), 
and the values are correlated. 

Let us calculate the ratio of the photospheric radius based on the ratio of temperature 
and normalization.

\begin{equation}
\frac{\cal{R}_{\rm spine}}{\cal{R}_{\rm sheath}}=\left(\frac{N_{\rm spine}}{N_{\rm sheath}}.\frac{T_{\rm sheath}^4}{T_{\rm spine}^4} \right)
\end{equation}
\vspace{0.1in}

With a ratio of 3 for both the temperature and normalization, we get $\cal{R}_{\rm spine}\approx$ 0.2$\cal{R}_{\rm sheath}$. 
This result is consistent with the spine-sheath jet model for a ratio of $\eta\sim1.5$, and assuming similar kinetic 
luminosity of both the components. 
 
\item (iv) Note that compared to the radius of IS region, the photospheric radius ($r_{\rm ph}$) of the 
two components are of the same order. Hence, the BB spectrum are nearly simultaneous. In our analysis
we have found that the normalization (and flux) of the two BBs are highly correlated. Also note 
that the IC of photons within $r_{\rm ph}$ will be simultaneously detected with these two 
BBs. Hence, all these events are nearly simultaneous. As discussed, the non-thermal component 
of this emergent spectrum is a PL with a cut-off near 100 MeV. Hence, the GRBs with these three 
components will have less GeV emission. In our analysis, we have indeed found such GRBs 
(see Figure~\ref{ch6_f6} --- case C and D). Note that these two GRBs, despite having high 
GBM count, have very little LAT emission. Also note from this figure that the non-thermal 
component of both these GRBs have coupled total and PL evolution. Hence, the spectra of 
these GRBs are probably composed of three components --- two BBs and a PL with a much steeper 
slope in the GeV range. All these components are nearly simultaneous as supported by the 
spine-sheath model as well as our observation.

\item (v) The radius where the ISs are formed is at least three order of magnitude higher than 
the photospheric radius. Hence, the IS will be delayed. As the PL of 2BBPL model is always consistent with 
the fast cooling synchrotron spectrum, this emission is possibly dominant in generating the PL 
of 2BBPL model. As this spectrum extends to GeV energies, the GRBs accompanying efficient 
IS will produce GeV emission. Note that the PL component, in this case, will be delayed from 
the thermal emission. Hence, the model is consistent with our finding that GRBs with delayed PL
emission are LAT bright. Note that we expect both synchrotron and IC in the IS regions, and 
the corresponding spectrum can be approximated as PL. However, for GRB 0909002B, we have shown 
that the required circumburst density is too high to accommodate SSC as a dominant component.
Note that for this GRB we have found that the spectral slope in the keV-MeV data is fully 
consistent with that in the GeV data. Hence, the synchrotron spectrum does not have a break 
in the observed band. However, for GRB 090926A, there is a break at a very high energy.

\end{itemize}

\section{Discussion}
\balance
In this final chapter, we have proposed a simplistic model to explain the observations of prompt emission 
spectrum in conjunction with the observation of GeV and x-ray data. We 
would like to point out that though the existence of the 2BBPL model has strong observational supports, 
the physical model which we have suggested may not be unique. Perhaps the model may be a simplification of 
a more complex and fundamental phenomenon. A modification of the simplistic model, or a completely different 
physical process is not ruled out. For example, consider the Cannonball model (\citealt{Dadoetal_2002,
Dar_Rujula_2004, Dadoetal_2007}). In this scenario, the central engine releases the ejecta 
in a sequence of ordinary matter (``cannon-balls''). The energy is released via particle-particle 
interaction rather than shock generation or photospheric emission. In this model, the cannon-balls 
can give rise to blue-shifted bremsstraulung and IC of the ambient photons. If the two BBs 
found in our prompt emission study are related to the coexisting glories of two photon fields,
then the higher BB temperature can be identified with a typical photon bath in the pre-supernova 
region (for $\Gamma\sim100$, a few eV photon is boosted by a $\Gamma^2$ factor, producing 
$\sim100$ keV), while the lower BB can be produced by some other mechanism like bremsstrahlung.
Note that as the same cannonball produces both these emissions, the temperatures should be 
correlated. Such a very different model, in principle, can conform with our observation. Even in the context 
of the standard fireball model, the IS can produce an effective pair photosphere. Then the thermal 
emission is likely to be Comptonized BB emission from the photosphere. A variety of such complications 
over the simple two-component model can be conceived.

The origin of the two components of GRB jet is also an open question. In the cocoon model of sheath, 
the opening angle is about three times larger than the spine, while the coasting \textit{Lorentz} 
factor ($\eta$) of the spine is about five times higher. In our analysis, we require a much lower 
ratio of $\eta$. It is important to note that as the observations indicate $L_{\rm sh}>L_{\rm sp}$ 
(\citealt{Bergeretal_2003_spinesheath}), and we expect $\dot{M_{\rm sh}}>\dot{M_{\rm sp}}$, it is rather possible that the 
ratio of $\eta_{\rm sp}/\eta_{\rm sh}$ ($\eta \sim L/\dot{M}$) is moderate, and not too high. 
This is consistent with our finding. The observation of the prompt emission does not put any constraint 
on the opening angle. Hence, the afterglow observation can give further clue on the origin of the structured jet.
Note that observation of \cite{Bergeretal_2003_spinesheath} and \cite{Hollandetal_2012} require very different 
jet structure both in terms of energetics and opening angle. A much wider sheath (\citealt{Bergeretal_2003_spinesheath})
is probably an indication of the cocoon structure. 

We would like to mention that the spine-sheath boundary is not necessarily as sharp as it is assumed in the 
simplistic model. There can be smooth roll-over between the two regions. A full roll-over (Gaussian
jet, \citealt{Zhangetal_2004_XRF}) will indeed produce a multicolour BB. Recently, \cite{Lundmanetal_2014} have 
considered a jet with a uniform spine, but a graded sheath i.e., the \textit{Lorentz} factor
of the sheath falls off gradually as a PL function of the angle from the jet axis (Spine: uniform $\Gamma_0$, sheath: 
$\Gamma(\theta)\propto \theta^{-p}$). They have studied the polarization properties in such a structured 
jet, and have found that for a narrow jet ($\theta_j\sim1/\Gamma_0$) and steep gradient ($p\gtrsim4$), the 
polarization can reach $\sim 40\%$. In recent years, high degree of polarization are indeed reported in a few 
cases (\citealt{Yonetokuetal_2012, Tomaetal_2012, Gotzetal_2013, Mundelletal_2013, Wiersemaetal_2014}). 
The polarization measurement in the optical afterglow (e.g., \citealt{Wiersemaetal_2014}) is argued to be a definite 
signature of synchrotron emission. However, as discussed, the finding of polarization in the 
prompt emission phase (e.g., \citealt{Gotzetal_2013}) is not necessarily associated with a synchrotron 
radiation. \cite{Lundmanetal_2014} have also shown that the spectrum below the thermal peak appears 
as a non-thermal spectrum due to aberration of light. The higher energy part produces a cut-off 
PL due to the similar mechanism as considered by I13.

\chapter{Summary and Future Directions} \label{ch7}

\section{The First Phase Of The Final Stage}
A gamma-ray burst (GRB) appears as an intense flash in the otherwise dark $\gamma$-ray sky.
While one class of GRBs mark the cataclysmic event at the final evolutionary stage of a massive 
star (long GRBs), the other class, namely the short GRBs are probably the outcome of NS-NS, or 
NS-BH mergers. Our knowledge about the progenitors of these two classes is based on the environment of 
their formation, and other observational evidences. Despite the differences in the progenitor and environment, 
the mechanism of the prompt emission and the subsequent afterglow is similar for the two classes of GRBs. It 
is quite fortuitous that the first phase of a GRB is so luminous in none other than the $\gamma$-ray 
band so that the open $\gamma$-ray detectors can detect the events occurring at random directions
from the edge of the universe. For example, compare a GRB with a supernova, which is bright in 
the optical wavelengths, and is detected by continuous scanning of the sky in this band. Due to the 
absorption of optical light a supernova cannot be observed farther than redshift, $z\sim1.7$, with
the majority found at much lower redshifts. In comparison GRBs can be observed at high redshift 
(the highest two are $z=9.4$, and $z=8.2$).

This thesis is primarily aimed at understanding the prompt emission phase of the long GRB class. 
In spite of the phenomenal discovery of the cosmological distance, and understanding a great deal 
of the environment, classes, and the afterglow phase, the emission mechanism of the very first phase 
of a GRB is still a matter of intense debate. It is 
important to understand the prompt emission phase in order to understand the progenitor, central 
engine, and the jet launching mechanism. In the absence of a detectable gravitational wave, 
the prompt emission characteristics seem to be the only direct signature of the central object. 
In addition, prompt emission provides the initial condition for the subsequent afterglow emissions, 
and is directly connected with the x-ray and GeV emission. Through extensive study of the prompt 
emission data provided by \textit{Swift} and \textit{Fermi} satellite, we have obtained several 
interesting results, and have found strong indication of such connections. 
In this final chapter, we shall summarize the results of our analysis, our understanding of the 
GRB phenomenology in general, and the possible future extension of our study.

\section{Summary and Conclusions}

Let us first summarize the main conclusions discussed in the thesis.

\subsection{Simultaneous Timing And Spectral Description}
We have attempted to combine the timing description, $F(t)$, and the spectral description, $F(E)$ to 
describe GRBs simultaneously in time and energy domain, $F(t, E)$. Such description is solely motivated 
by the need for using the full information of the data to capture rapid spectral evolution during the 
prompt emission phase. The constituent broad pulses of a GRB are chosen for this purpose as the spectral 
evolution is a pulse property, rather than a burst characteristic. We have used GRB 090618, one of the 
brightest GRBs in the \textit{Fermi} era, and having four broad pulses with a precursor. We assume that 
the lightcurve of each pulse can be described by an empirical model with exponential rise and decay part
(\citealt{Norrisetal_2005}) --- $F(t)=A_{\rm n} f_{\rm n}(t, \tau_1, \tau_2, t_{\rm s})$. The instantaneous 
spectrum is assumed as a Band function (\citealt{Bandetal_1993}) --- $F(E)=A_{\rm b} f_{\rm b}(E, \alpha, \beta, E_{\rm peak})$.
In addition, the evolution of $E_{\rm peak}$ is assumed to have a hard-to-soft evolution
(\citealt{LK_1996}) --- $E_{\rm peak}=f_{\rm LK}(\phi(t), E_{\rm peak, 0}, \phi_0)$,
where, $\phi(t)$ is the time integrated flux from start of the pulse to time $t$, called 
``running fluence''. This quantity relates the spectral and timing properties and leads to 
the simultaneous description. We have developed a {\tt XSPEC} table model for each pulse to 
determine the model parameters --- initial peak energy ($E_{\rm peak, 0}$), and characteristic 
evolution parameter ($\phi_0$). 

The best-fit model parameters are used to reconstruct each pulse. These are shifted and added to 
generate a 3-dimensional model (flux as a function of time and energy) of the entire burst. Such a 3D 
description has immense flexibility as the derived timing and spectral parameters can be derived and 
checked against the data. In particular, 
\begin{itemize}
\item (i) we construct the lightcurve of the GRB in various 
energy bands, and find remarkable similarity of the data with the synthetic lightcurves.

\item (ii) We derive the width variation of each pulse with energy, and find agreement with the data. 
It is expected that the width should increase with decreasing energy. The first two pulses 
indeed show such width variation. However, the data of the last two pulses indicate an ``anomalous'' 
width broadening in the lower end of the GBM energy band. This phenomenon is also shown by our 
pulse model. It is quite remarkable that such a minute detail of pulse characteristic is adequately 
captured, and explained as due to some particular combination of a few model parameters.

\item (iii) We also deduce the delay of the high energy bands with respect to the lowest energy 
band for each pulse. We always find a soft delay, which is a direct consequence of the assumption 
of hard-to-soft spectral evolution. The spectral lag as calculated by our model also conforms with 
that calculated using the data directly. 

\end{itemize}

The fact that all the derived parameters agree with those obtained by using the data directly
confirms that the pulse description is correct. In addition, we get a better handle on the 
data as the timing and spectral description can be simultaneously obtained with any desired 
resolution. The limitation of our model is that the functions used to achieve it are all 
empirical. In spite of this limitation, the fact that the model correctly predicts the pulse 
characteristics, possibly indicates a fundamental process responsible for a broad pulse generation.
We would like to emphasize again that the pulse description is quite generic. Any spectral and timing 
model, whether empirical or motivated by theory, can be connected by means of a description of the 
spectral evolution. 
   
\subsection{Improved GRB Correlation}
The crucial step for analyzing the prompt emission data is the realization of the broad 
pulse structure in the lightcurve. The rapid variability on the broad pulses are considered 
to be independent. As we have shown that a full burst can be re-generated by adding the 
individual broad pulses, it is interesting to study each pulse separately. In particular,
we have studied GRB correlations in the individual pulses. For this 
purpose, we have collected a sample of GRBs with secure redshift measurement, and have used 
the pulses for Amati correlation (\citealt{Amatietal_2002}, $E_{\rm peak}-E_{\gamma, \rm iso}$).
We have found that the pulse-wise Amati correlation is significantly better (Pearson correlation,
$r=0.89$, $P_{\rm r=}2.95\times10^{-8}$, Spearman rank correlation, $\rho=0.88$, $P_{\rho}=4.57\times10^{-8}$)
than both the time-integrated ($r=0.80$, $P_{\rm r=}9.6\times10^{-3}$, $\rho=0.75$, $P_{\rho}=2.0\times10^{-1}$),
and time-resolved correlation which is not accounted for the broad pulses 
($r=0.37$, $P_{\rm r=}9.5\times10^{-3}$, $\rho=0.45$, $P_{\rho}=3.0\times10^{-4}$).

As we have developed a simultaneous pulse model, it is interesting to use this model to see the 
improvement in the GRB correlation. The $E_{\rm peak}$ used in the pulse-wise correlation is a 
pulse-averaged quantity, and consequently the information of spectral evolution is lost. Hence, 
it is important to replace $E_{\rm peak}$ by $E_{\rm peak,0}$ which is a constant characterizing 
the peak energy at the beginning of a pulse. We have found that a pulse-wise 
$E_{\rm peak,0}-E_{\gamma, \rm iso}$ correlation is better considering the Pearson correlation
($r=0.96$, $P_{\rm r=}1.6\times10^{-12}$). However, the Spearman rank correlation, which is a robust 
estimator of the correlation (\citealt{Macklin_1982}), is similar ($\rho=0.87$, $P_{\rho}=1.43\times10^{-7}$).
The intrinsic data scatter ($\sigma_{\rm int}$) per data points are also similar.
In addition, we have found in later analysis that the HTS spectral evolution is not universal.
Hence, in a later study, we have used the pulse-wise Amati correlation using $E_{\rm peak}$ instead 
of $E_{\rm peak,0}$ for a larger set of GRBs (19 GRBs with 41 pulses). We have found a reasonable 
correlation ($r = 0.86, P_{\rm r} = 1.50 \times 10^{-13}; \rho= 0.86, P_{\rho} = 7.47 \times 10^{-14}$).
We have studied the possible redshift evolution of the correlation parameters, and have found it 
insignificant (within $\triangle \chi^2=1.0$). To find possible bias due to the hardness at the 
beginning of a GRB, we have studied the correlation in the first/single pulses and the rest of 
the pulses. We again find no statistically significant difference. Improvement of GRB correlation 
within the constituent pulses is a significant step in understanding the pulse emission mechanism. 
It possibly indicates that the pulses are independent episodes of the prompt emission.
In addition, the pulse-wise correlation signifies that the Amati-type correlations are real and devoid 
of selection effects. However, due to large intrinsic scatter even such a tight correlation is still 
far from being usable as cosmological luminosity indicators (Arabsalmani et al., in preparation). 
In future, if a pule-wise correlation with an order of magnitude lower intrinsic scatter is indeed 
available, it can give a good constraint on the inferred redshift of a GRB, and the cosmological 
parameters due to the application of different pulses of the same GRB.

\subsection{Alternative Models Applied To GRBs With Single Pulses}
The 3D pulse model is developed with the assumption that the instantaneous spectrum is a 
Band function. Though the Band function is statistically the most appropriate standalone model of the prompt 
emission spectrum, it is a completely phenomenological model. In the internal shock model of 
a GRB, the shocks are produced at a larger radius from the photosphere. The electrons accelerated 
by Fermi process in the shock gyrate around the aligned magnetic field (also developed by the 
shock) to produce synchrotron emission. Hence, it is natural to expect that the emergent spectrum 
should follow the synchrotron predictions. It is shown that the value of the low energy 
index ($\alpha$) of the Band function often crosses the limit due to synchrotron radiation.
As we expect the electrons to cool fast during the prompt emission phase, the predicted 
index should be softer than -3/2, i.e., $\alpha>-3/2$ is forbidden (the so called 
``fast cooling line of death''). Even for slow cooling, the index should be less than 
-2/3 (``slow cooling line of death''). However, the spectral slope are often higher than 
both these values (\citealt{Preeceetal_1998, Crideretal_1999, Kanekoetal_2006}). 

Our inability to associate Band function with a physical model instigates the search 
for alternative models. We have used three alternative models for our purpose : (i) blackbody 
with a power-law (BBPL), (ii) multi-colour blackbody with a power-law (mBBPL), and (iii) 
two BBs with a PL (2BBPL). Note that all the alternative models segregate the spectrum into a thermal 
and a non-thermal part. The thermal part is either as narrow as a Planck spectrum (BB), or
have a surface profile (mBB), or have two BBs. We have applied these models on 
high flux GRBs with single pulses. Such a set is chosen in the hope of understanding a single 
pulse, and then use the knowledge for a complex GRB with multiple pulses.  

\begin{itemize}

\item (i) We have found, based on $\chi^2$ values, that all the models are better than the BBPL model, 
though we could not find which among the other three is the best. Based on the value of $\alpha$ 
we have found that the Band function does not conform with a synchrotron interpretation. 
On the other hand, the PL index of the BBPL model ($\Gamma$) is lower than -3/2. As $\Gamma$ 
of mBBPL and 2BBPL are always found lower than that of BBPL, the corresponding values of 
$\Gamma$ are well within the fast cooling regime of synchrotron radiation. Based on our analysis,
we suggest that either mBBPL or 2BBPL is the preferred model for the prompt emission spectral data.
It is interesting to find that the two BBs of 2BBPL model are highly correlated in terms of their 
temperature and normalization. Hence, if the 2BBPL model is the correct spectral model, the origin 
of the two BBs should be connected.

\item (ii) We have also studied the evolution of $E_{\rm peak}$ and $kT$ with time, and have identified 
two classes of GRBs --- hard-to-soft (HTS) and intensity tracking (IT). As some of the GRBs with 
single pulse are indeed IT, such a spectral evolution must be physical (at least in some cases) 
rather than a superposition effect due to the preceding pulse. 

\item (iii) Important differences are found between the two classes e.g., HTS have generally 
higher values of $\alpha$ than the IT GRBs. We have  also found that the PL component of three 
HTS bursts have \textit{delayed} onset than the thermal component, and this component \textit{lingers} 
at the final phase of the prompt emission. Interestingly, all these GRBs have reported GeV emission 
in the \textit{Fermi}/LAT observation.

\end{itemize}

\subsection{Parametrized Joint Fit: The 2BBPL Model}
Based on the $\chi^2$ values and physical arguments, we have seen that the most preferred models
of the prompt spectral data are mBBPL and 2BBPL, rather than the Band or BBPL model. Application of 
various models on GRBs with single pulse is the first step towards understanding the emission mechanism.
Though single pulses are ideal for analysis purpose, majority of GRBs have either clean multiple pulses 
or largely overlapping pulse structure. It is not clear whether the difference between these classes are 
due to the difference in fundamental radiation process or the difference of the number of episodes.

In order to find the correct model, we have developed a new technique of spectral analysis, namely 
``Parametrized Joint Fit''. The main motivation of this technique is to reduce the number of parameters 
in the time-resolved spectroscopy by parameterizing and tying model parameters over certain time interval.
We note e.g., the evolution of $kT$ of BBPL model can be parameterized as $kT\propto t^{\mu}$, and the 
PL index ($\Gamma$) can be tied separately in the rising and the falling part of a pulse. Such parametrization 
and tying scheme is applied for all models to achieve similar number of free parameters for all of them.
For our analysis, we have chosen 2 bright GRBs with clean multiple pulses and 3 bright GRBs with highly variable 
lightcurves. For the latter class, we have dropped the parametrization scheme.

The results of our analysis is summarized as follows.

\begin{itemize}
\item (i) The data agrees with mBBPL and/or BBPL model with similar or better $\chi^2$ than the Band function 
at the rising part. Also, the value of $\alpha$ of Band function is higher than -2/3, which disfavours a 
synchrotron origin. The PL index of the alternative models are within fast cooling regime of synchrotron. 
Hence, the rising part of each pulse has both a thermal and a non-thermal component. In the falling 
part, however, the Band function is preferred in terms of $\chi^2$, and the value of $\alpha$ is less than
-2/3. The PL index of the other models also lower than that in the rising part. Hence, the spectrum
has a definite transition from a thermal to synchrotron domination. 

\item (ii) As there is a definite transition of the spectrum, and the second pulse repeats the similar 
behaviour, it can be inferred that the pulses are independent, and possibly represent two episodes of 
the central engine activity.

\item (iii) A comparison with 2BBPL model with all other model shows that this model is either better or 
comparable to mBBPL and Band function. 2BBPL model, being an extension of the BBPL model shows high 
significance of adding the extra BB. More importantly, while the $F$-test significance level of the other 
models (mBBPL, Band) in comparison with BBPL model decreases with finer bin size (e.g., significance 
of mBBPL/BBPL changes from $2.55\sigma$ to $1.56\sigma$ in the rising part of pulse 2), the significance 
of 2BBPL over BBPL model remains similar ($9.29\sigma$ to $9.66\sigma$). This signifies that the second 
BB is required to capture the spectral evolution. The PL index of 2BBPL model has similar characteristic
as the BBPL and mBBPL models.

\item (iv) For bright GRBs with highly variable lightcurve, we obtain similar conclusions. The time-resolved 
spectra of GRB 090902B shows a definite improvement in the residual while fitted with a 2BBPL model as compared 
to the Band and BBPL models. The mBBPL model gives a similar fit statistics as the 2BBPL model.

\end{itemize}

\subsection{Consequences and Predictions of the 2BBPL Model}
In the final chapter, we have discussed about various consequences and predictions of the 2BBPL model.
The fact that the spectral analysis of the prompt keV-MeV data has predictive powers for both lower and 
higher energy data, gives us the confidence that the spectral model is correct. In the following we 
list the consequences and predictions.

\begin{itemize}
\item (i) The evolution of the two BBs (with or without a break) can lead to an increasing temperature
during the early part of a pulse. In the standard fireball model the temperature can have a break due 
to the transition from an accelerated to a coasting phase, and provided that the photosphere occurs 
below this transition radius. The evolution below the break is expected to be constant (for radiation 
dominated fireball), or decreasing (for magnetic field dominated fireball). An increasing temperature
is unexplained. It is possible that an averaging of the temperature of two BBs, both having characteristic 
evolution, give rise to such a phenomenological evolution. In the later part, the lower BB may go below the 
bandwidth of the GBM/BAT, showing no effect on the evolution. We have illustrated such a hypothetical situation 
with a single pulse GRB.

\item (ii) If the lower BB indeed goes below the bandwidth of the higher energy detectors, it should become 
visible in the low energy detector like the \textit{Swift}/XRT, provided the data is available in the late 
prompt emission phase. At a much later time, the emission may be afterglow dominated. Hence, it is important
to obtain the XRT data as early as possible. For GRB 090618, we have found that the XRT data in WT mode is 
available from 125 s after the trigger. During this time the falling part of the last pulse is still 
visible in the GBM/BAT band. The GBM/BAT data in the falling part of the pulse fitted with a BBPL model 
gives a much higher trend of temperature evolution compared to that found in the XRT data. However, 
the lower BB of the 2BBPL model fitted to the same data shows a impressive similarity. In addition, 
a BBPL model fitted to the GBM/BAT data of the first XRT bin falls perfectly on the predicted evolution 
of the higher BB of 2BBPL model. Such a finding confirms the presence of a separate BB component in the 
prompt emission spectrum.

\item (iii) In addition to the data in the lower energy band, we find a remarkable prediction
for the very high energy band covered by the \textit{Fermi}/LAT. First we investigate the basic difference 
of the prompt keV-MeV spectrum of GRBs with similar GBM brightness, but an order of magnitude difference in
LAT brightness. We fit the spectral data obtained only from the GBM with a 2BBPL model, and try
to predict which among these classes should lead to high GeV emission. We find that the LAT-bright GRBs 
have a delayed onset of the PL, and this component becomes progressively important at a later phase. GRBs 
with similar GBM count, but having lower LAT count do not show such a behaviour. In other words, GRBs which show 
a delayed and lingering non-thermal component in the prompt emission should accompany high GeV emission.
The current strategy of re-pointing \textit{Fermi} to observe GeV emission is to target the GBM-bright GRBs.
We propose that the re-pointing criteria should include moderately bright bursts with delayed PL 
component. Such an inclusive strategy may increase the LAT sample size.

\item (iv) One of the puzzles in the comparative study of GeV emission and keV-MeV emission is
the poor correlation between these components. It is not clear why GRBs with similar GBM brightness 
produce seemingly different GeV photons. As the PL component of the 2BBPL model shows a similar evolution 
as the GeV emission, we study a correlation between the fluence of the PL component with that of the GeV component. 
We find that the correlation is strong (Spearman rank correlation coefficient, $\rho=0.81$, 
$P_{\rho}=9.23\times 10^{-5}$), and it is unaffected by the presence of another variable, 
namely total GBM fluence ($D=2.3$). 
                                                                                                     
\end{itemize}

\subsection{Physical Origin: A Spine-sheath Jet}
Based on the observations, 2BBPL model seems to be the most preferred model for the prompt emission spectrum. 
We propose a spine-sheath jet structure as a simplistic physical model to explain the observations. Let us 
summarize the major points.

\begin{itemize}
\item (i) The two BBs are natural consequence of the spine-sheath jet. If the coasting bulk \textit{Lorentz}
factor of the spine and sheath are $\eta_{\rm sp}$ and $\eta_{\rm sh}$ respectively, then the observed temperature 
and normalization above the saturation radius will always show a ratio $(\eta_{\rm sp}/\eta_{\rm sh})^{8/3}$. As
$\eta\equiv L/\dot{M}c^2$, and we generally expect the spine to have higher $\dot{M}$, the sheath will show higher 
temperature and normalization. Also, note that the jet luminosity ($L$) as found by observation of the two 
components are either comparable, or sheath has a higher value. Hence, we do not expect a large ratio of $\eta$.
For a nominal ratio of temperature and normalization  as found in our observation, we require the ratio 
of $\eta=1.5-2.0$. 

\item (ii) In the velocity shear region, the jet structure Comptonize the photons crossing the boundary layer.
As the photons on an average gain in energy, the emergent spectrum is a PL. This spectrum also has a cut-off 
due to decreasing cross-section at very high energy $\sim 100$ MeV. Hence, it is likely that some GRBs should have 
a cut-off PL (CPL) as the non-thermal component. However, GRBs with PL at very high energies (GeV) require 
other processes. As these processes are delayed, the GeV emission should also be delayed. This supports the 
observation that GRBs with delayed PL component have high GeV emission. 

\item (iii) Using the data of two GRBs, and with an order of magnitude calculation, we have also shown that 
the photosphere of the spine should occur at a lower radius than the sheath, which is also required by the model. 

\end{itemize}

\section{Future Directions}

\subsection{Analyzing The GBM Data To Predict The LAT Data}
One of the remarkable results of our analysis is finding a connection between the GeV and keV-MeV emission.
The fact that the delayed onset of the PL component indicates high GeV emission can be used for targeting 
this category of GRBs for the LAT observation.
For off-axis GRBs from the LAT field of view, this strategy can increase the sample size of GRBs with LAT detection, 
and can shed light on the ongoing research in identifying potential GRBs with high energy emission. Such data 
needs to be quickly handled, and the analysis results should be immediately available. As our analysis is dependent 
only on the lower energy spectrum, only the GBM data is targeted. From 2012 November 26, the GBM is supplying 
continuous time-tagged event data (TTE) with $2\mu$s resolution. The data are sent through TDRSS message
within a few seconds to the GBM's Instrument Operation Center (GIOC). 
The GBM software is implemented to analyze the characteristics of the event, classify the object, and provide a crude
event location. If a GRB is detected, the sky location is sent to the LAT. If the LAT monitoring reveals an increasing 
$\gamma$-ray flux, it sends an autonomous re-point recommendation (ARR) to the spacecraft. Generally the spacecraft 
accepts this request and performs a slew to the sky location of the target. An observation of a GRB by the LAT is 
performed for the next 2.5 hour including earth occultation. 

In order to implement our finding for the LAT observation, we require science data of the GBM rather than the preliminary 
flux and localization data. For the data analysis, an automatized software should be developed. 
This software should analyze the GBM data in nearly real time, and it should report any signature of delayed 
PL component in the prompt emission data. For moderately bright GRBs, such an analysis can indicate a high 
GeV emission, which otherwise might have been missed based only on the flux level. Note that such a procedure 
crucially depends on the quickness of receiving the GBM science data which in practice can take several hours.
Even if the procedure may not be used in the current strategy used by the \emph{Fermi} satellite, it may 
prove useful to get an insight on the radiation process nonetheless, and may be used for later missions.

\subsection{Physical Model Of The Prompt Emission}
Through extensive data analysis, we have shown that the preferred model of the prompt emission is 
a 2BBPL model. Though this model is applicable for all variety of GRBs, it is important to establish 
a physical model for such a functional form. We have suggested a tentative physical picture within the 
frame work of the standard fireball model, namely a spine-sheath jet structure. We have shown that 
a fast moving spine with a slow sheath layer can indeed give rise to two BB components. It also 
produces a PL component by inverse Comptonizing the photons crossing the spine-sheath boundary.
However, it is clear that shocks outside the photosphere are also required to explain 
a PL with no break in the LAT energy band. Both the mechanisms, in principle, can operate 
for the production of the non-thermal spectral component.

It is important to find out the relative contribution 
of the two components in order to constraint the physical parameters e.g., \emph{Lorentz} factor,
energy in the magnetic field etc. In this context, it is useful to employ physical function of 
synchrotron model (see e.g., \citealt{Burgessetal_2011}). Such functions require electron 
distribution, \emph{Lorentz} factor, magnetic field etc. Hence, the fitting directly gives these 
quantities. Various components in addition to the simple synchrotron model can be easily 
conceived, e.g., one can implement inverse Compoton (IC) in the internal shock to account for the 
high energy component. Using such models can give further insight in the production of the non-thermal 
component of the spectrum. However, these models are computationally expensive. As we are more interested in 
studying spectral evolution, the models should be made fast to be usable for time-resolved spectroscopy 
with parameterized-joint fitting. 

\subsection{Afterglow Observations}
In this thesis, we have primarily discussed about the prompt emission phase. It is important 
to connect the inferences of the prompt emission to those of the afterglow. Though the afterglow 
emission is mostly related to the circumburst environment, the prompt emission sets the 
initial conditions in terms of total energy, energy in the thermal and non-thermal component 
and energy in the two components of the jet. Hence, a full prompt-afterglow analysis can constraint 
these quantities and establish the connection. For example, an early indication of delayed PL 
emission indicates GeV emission. Hence, it is interesting to perform afterglow observation in other 
wavelengths for this class of GRBs. As a two-component jet model naturally explains the observations
during the prompt emission, it is important to establish this model on the basis of afterglow observations.
 
A two component jet has interesting observational signature e.g., re-brightening of the optical flux and 
double jet break in the x-ray/optical and radio (e.g., \citealt{Bergeretal_2003, Hollandetal_2012}).
Such signatures may give important constraints on the jet opening angle and the correct energetics of 
each component. While we obtain the $\gamma$-ray energy from the observation of prompt emission, the 
correct jet kinetic energy can be obtained by afterglow data and the observed jet break. It is worthwhile 
to point out again that the opening angle and jet kinetic energy obtained for two GRBs can be very different
For example, \cite{Bergeretal_2003} have found opening angles: $\theta_{\rm spine}\sim5^{\circ}$, 
$\theta_{\rm sheath}\sim17^{\circ}$, and $E_{\rm sheath}/E_{\rm spine}\sim5$. \cite{Hollandetal_2012},
on the other hand, have found $\theta_{\rm spine}\sim0.86^{\circ}$, $\theta_{\rm sheath}\sim1.4^{\circ}$, 
and $E_{\rm sheath}/E_{\rm spine}\sim1$. Observations of a variety of such GRBs might provide important 
clues on the energy budget of each component, and the jet launching mechanism in general.

It is interesting to note that jet-breaks are relatively rare in \emph{Swift} era ($\sim12\%$;
\citealt{Liangetal_2008, Racusinetal_2009}). Such observations even challenge the consensus of 
a GRB jet. Alternatively, the unobserved jet-break may be attributed to the uncertainties and 
bias in the observations (e.g., \citealt{Curranetal_2008, Kocevski_Butler2008, Racusinetal_2009}),
and/or smoother break due to off-axis viewing angle (e.g., \citealt{vanEertenetal_2010}). Recently,
\cite{Zhangetal_2014_openingangle} have investigated whether a jet-break is indeed present at a
much later time and at a much deeper level of sensitivity than the \emph{Swift}/XRT ($10^{-14}$ erg 
cm$^{-2}$ s$^{-1}$). They have used \emph{Chandra}/Advanced CCD Imaging Spectrometer (ACIS) for a 
set of 27 GRBs, and have found 56\% of jet-break. They have also performed a Monte Carlo simulation 
to show that the off-axis effects are indeed important to interpret a jet-break. It is interesting 
to investigate how the interpretations of an observed jet-break change with two components in the jet. 
For example, the proposed break will be affected by the presence of a much wider sheath component.

\subsection{Clues From Other Objects}

Additional information of the spectral model and jet structure can be obtained by comparing the observations 
of GRBs with other astrophysical objects. Though each of the objects would show a characteristic property, the 
underlying jet launching mechanism and radiation process should provide important clue on whether the processes 
are similar, diverse or follows a continuity. In the following, we have listed some of the objects which might 
prove very useful in studying these properties in a general sense.

\subsubsection{I. X-ray Flashes (XRFs)} 
\balance
XRFs are lower energy cousins of GRBs and are believed to be a subset of 
the LGRB class, due to the similarity of LC, spectrum and the SN association in a few cases (\citealt{Soderbergetal_2005}). 
XRFs were extensively studied with \emph{High Energy Transient Explorer 2 (HETE-2)} (\citealt{Heiseetal_2001, Kippenetal_2003}). 
The peak energy ($E_{\rm peak}$) and hardness of XRFs are found to be lower than the typical GRBs. \cite{Liang_Dai_2004} 
using a combined sample of GRB and XRF have found a tentative bi-modality in the $E_{\rm peak}$ distribution, which 
has peaks at $\lesssim30$ keV and $\sim160-250$ keV. Various intrinsic and extrinsic models are suggested to explain the 
mechanism of XRFs (see \citealt{Zhang_2007_review} for a detailed discussion). Various speculations in the intrinsic 
differences are (i) XRFs are fainter and wider jets than GRBs (\citealt{Lambetal_2005}), (ii) `dirty fireballs'' 
(\citealt{Dermer_Mitman_1999}) are possibly XRFs, (iii) XRFs are clean but inefficient fireballs (\citealt{Barraudetal_2005}),
(iv) both are photosphere dominated and follows a continuous distribution of $E_{\rm peak}$ (e.g., \citealt{Lazzatietal_2011}), 
(v) they have completely different progenitors e.g., the progenitor of XRF 060218 (\citealt{Campanaetal_2006})
is possibly a neutron star (\citealt{Soderbergetal_2006_XRF_060218, Mazzalietal_2006}).
In the extrinsic models the two classes are similar objects with differences in the observer's point 
of view. One of most important suggestions is that XRFs are off-axis 
GRBs. However, we require a structured jet to account for such a model (\citealt{Zhang_2007_review}).
There are two types of structured jet discussed in the literature. The first one is the spine-sheath 
structure, which we have discussed in the previous chapter. Another possible structure is a smooth 
symmetric variation of $\Gamma$ with angle from the jet axis ($\theta$). Such a jet is modelled by 
power-law (\citealt{Rossietal_2002}), or a Gaussian function (\citealt{Zhangetal_2004_XRF}). It is 
important to note that the structured jets are completely theoretical. Only the observation of 
jet structure can find out which among these proposals are indeed the correct one. It is also 
possible that the difference between the two classes are both intrinsic and extrinsic. For example, 
consider the spine-sheath jet. It is quite possible that the spine, being relatively baryon 
free has mainly photospheric emission. As the photospheric emission is efficient, the spine may 
get exhausted. The sheath, having wider angle, and higher baryon load mainly produces the 
afterglow via shocks, with a little photospheric emission during the prompt phase. While an 
on-axis observer sees a prompt emission with high $E_{\rm peak}$, an off-axis observer sees 
a XRF. Investigation of such combinations may give important clues for the GRB/XRF connection 
as well as jet structure of the two classes.

\subsubsection{II. Blackhole Binaries And BL Lacertae}

It is worthwhile to mention that a spine-sheath structure is also claimed in other astrophysical 
jets e.g., Cyg X-1 (\citealt{Szostek_Zdziarski_2007}), and BL Lacertae objects (\citealt{Ghisellinietal_2005}). 
In this context, GRBs provide the best opportunity to study the launching mechanism of jet. A 
universal structure in a wide range of jets may provide crucial information on the mechanism of 
jet propagation, and can give clues to the common radiation process.

\subsection{GRB Correlation Using Physical Models}
We have seen that simple assumption of empirical spectral evolution can lead to a better handle 
on the data, and improvement of a GRB correlation. It is interesting to see how we can be benefited from 
a physical understanding. However, a simple replacement of Band function with 2BBPL model for a simultaneous
timing and spectral description may not be useful. Firstly because the temperature evolution may not be always
hot-to-cold. Either or both of the BB components can have break in the temperature evolution depending on the 
relative baryon load. Such evolutions cannot be distinguished by statistical means. Secondly, the 2BBPL model 
can be an approximation of a more complex and physical picture e.g., a non-trivial structured jet. Such 
complications may not be understood by analyzing only the prompt emission spectrum. A detailed analysis of 
a set of GRBs with valuable inputs from afterglow data, circumburst environment, and theoretical understanding
may enrich our understanding of the physical mechanism and thereby lead to using GRBs as cosmological candles.

\chapter{Annexure} 

\begin{figure}
\includegraphics[scale=0.35]{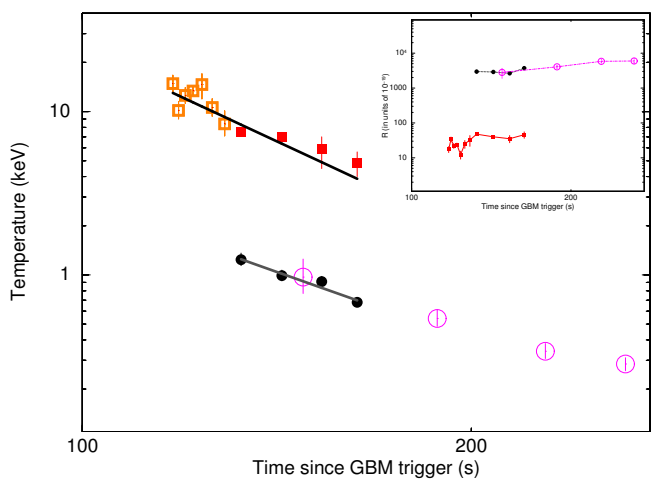}
\caption[Evolution of the thermal components in the falling part of the last pulse of GRB 090618]
{Evolution of the thermal components in the falling part of the last pulse of GRB 090618 is shown.
The symbols are: (i) open squares denote the temperature of the BBPL model fitted to the BAT-GBM joint data,
(ii) open circles denote the temperature of the BBPL model fitted to the XRT data (from \citealt{Pageetal_2011}),
(iii) filled squares and filled circles denote the temperatures of the higher and lower temperature blackbodies 
obtained by 2BBPL model fitted to the joint BAT-XRT data. It is evident that the evolution of the higher temperature 
blackbody is similar to case (i), while the evolution of the lower temperature blackbody is similar to case (ii).
Hence, the two blackbodies are always present and shows up in two different detectors (BAT and XRT) in the overlapping 
observation. The inset shows the evolution of $\cal{R}$, which is a proxy of the photospheric 
radius, for the two components. A smooth evolution is seen (see text for explanation).}
\label{ch8_fig1}
\end{figure}

This annexure gives updated information we have obtained by re-analyzing the data presented in Section 6.10.
A comparison of the fit statistics of the 2BBPL and BBPL model fitted to the time-resolved BAT-GBM joint 
data (Table 6.5) shows that the 2BBPL model is not required based on the $\chi^2_{\rm red}$ values. 
However, when we fit the joint BAT-XRT data in the overlapping region (125-165\,s) a 2BBPL model is 
required at a high significance ($\chi^2$ for BBPL and 2BBPL: 787.0 (283) and 307.1 (281), respectively).
As the flux evolution in this overlapping observation has a sharp variation, we have re-analyzed the data 
by using finer time bins (four uniform bin in 125-165\,s). The data is fitted with 2BBPL model. The temperature 
of the two blackbodies are shown graphically in Figure~\ref{ch8_fig1}. The filled squares represent the higher 
temperature blackbody, while the filled circles represent the lower temperature blackbody. The time-resolved data 
before the XRT observation is fitted with a BBPL model, and the corresponding temperature is shown by open squares. 
The temperature of the blackbody found in the XRT data (from \citealt{Pageetal_2011}) are also shown (open circles).
Note that the temperature of both the blackbodies of the 2BBPL model are consistent with those of the single blackbodies 
found in the higher (BAT/GBM) and lower energy (XRT) detectors. Hence, we conclude that the 
two blackbodies are always present in the data. The low blackbody is not visible in the early data because it 
is outside the lowest energy coverage of the higher energy detectors, and the XRT observation is absent during this time.
We have also calculated the $\cal{R}$, which is a proxy of the photospheric radius, for both the components. The evolutions 
are shown in the inset of Figure~\ref{ch8_fig1}. The photosphere of the lower temperature blackbody is about 65 times higher 
than the higher temperature blackbody. From this ratio, one finds that the required ratio of the coasting bulk 
\emph{Lorentz} factor ($\eta$) of the spine and sheath is $\sim8$, provided that the jet luminosity scales with $\eta$.

\appendix
\chapter{A Few Acronyms}

\section{Astrophysical Objects}
\begin{tabular}{ll}
BH & Blackhole \\
GRB & Gamma-Ray Burst \\
LGRB & Long Gamma-Ray Burst \\
SGRB & Short Gamma-Ray Burst \\
NS & Neutron Star \\
SN & Supernova \\
\end{tabular}

\section{Timing And Spectral Features}
\begin{tabular}{ll}
LC & Lightcurve \\
FRED & Fast Rise Exponential Decay \\
BB &  Blackbody \\
PL & Power-law \\
2BBPL & Two blackbodies and a power-law \\
mBBPL & Multicolour blackbody and a power-law \\
HTS & Hard-to-soft \\
IT & Intensity Tracking \\
IC & Inverse Compton \\
IS & Internal Shock \\
ES & External Shock \\
\end{tabular}

\section{General Astronomy}
\begin{tabular}{ll}
CCD & Charge Coupled Device \\
FITS & Flexible Image Transport System \\
FOV & Field Of View \\
FWHM & Full Width at Half Maximum \\
HEASARC & High Energy Astrophysics Science Archive Research Center \\
PSF & Point Spread Function \\
PHA & Pulse Height Amplitude \\
TTE & Time-tagged Event \\
\end{tabular}

\section{Instruments And Missions}
\begin{tabular}{ll}
CGRO & Compton Gamma Ray Observatory \\
BATSE & Burst and Transient Source Experiment \\
EGRET & Energetic Gamma Ray Experiment Telescope \\
FGST & Fermi Gamma-ray Space Telescope \\
GBM & Gamma-ray Burst Monitor \\
LAT & Large Area Telescope \\
MIDEX & Medium Explorer Program \\
BAT & Burst Alert Telescope \\
XRT & X-ray Telescope \\
UVOT & Ultraviolet/Optical Telescope \\
MET & Mission Elapsed Time \\
\end{tabular}

\bibliographystyle{apj}
\bibliography{thesis}

\end{document}